\def\Mode{2} %2-column [preprint2] %single-spaced lines, embedded figs & tables
\newcommand{\Lir}{${\rm L_{ir\ }}$} % Lir
\newcommand{\Lsun}{$L_{\odot}$} % L sun
\newcommand{\Msun}{$M_{\odot}$} % M sun
\newcommand{\ts}{\thinspace}

\ifnum\Mode=0
\documentclass[manuscript]{aastex}
\fi
\ifnum\Mode=1
\documentclass[preprint]{aastex}
\singlespace
\fi
\ifnum\Mode=2
\documentclass[iop]{emulateapj}% Works best; use for astroph
\fi
\received{}
\accepted{29 July 2012}
\slugcomment{To appear in the Astronomical Journal; draft version 15 Aug 2012}
\shorttitle{Dual AGNs, Outflows, Shocks, and Young Star Clusters in Mrk 266}
\shortauthors{Mazzarella et al.}

\usepackage{graphicx}
\usepackage[ovp]{psfragx} %For \parbox
\usepackage{multirow}

\begin{document}
\pdfoutput=1

\renewcommand{\thefigure}{\arabic{figure}} %Avoid x.y style figure numbers

\ifnum\Mode=0
\title{
Investigation of Dual Active Nuclei, Outflows, Shock-Heated Gas, and 
Young Star Clusters in Markarian 266
}
\else
\title{
Investigation of Dual Active Nuclei, Outflows, Shock-Heated Gas, and 
Young Star Clusters in Markarian 266\\
~~\\ %Force start of ABSTRACT to next page
~~\\ %Force start of ABSTRACT to next page
}
\fi

\author{
J. M. Mazzarella,\altaffilmark{1}
K. Iwasawa,\altaffilmark{2}
T. Vavilkin,\altaffilmark{3}
L. Armus,\altaffilmark{4}
D.-C. Kim,\altaffilmark{5,6}
G. Bothun,\altaffilmark{7}
A. S. Evans,\altaffilmark{5,6}
H. W. W. Spoon,\altaffilmark{8}
S. Haan,\altaffilmark{4,9}
J. H. Howell,\altaffilmark{4}
S. Lord,\altaffilmark{10}
J. A. Marshall,\altaffilmark{4,11}
C. M. Ishida,\altaffilmark{12}
C. K. Xu,\altaffilmark{10}
A. Petric,\altaffilmark{4,11}
D. B. Sanders,\altaffilmark{13}
J. A. Surace,\altaffilmark{4} 
P. Appleton,\altaffilmark{10}
B.H.P. Chan,\altaffilmark{1}
D. T. Frayer,\altaffilmark{6}
H. Inami,\altaffilmark{4}
E. Ye. Khachikian,\altaffilmark{14}
B. F. Madore,\altaffilmark{15,1}
G. C. Privon,\altaffilmark{5}
E. Sturm,\altaffilmark{16}
Vivian U,\altaffilmark{13}
S. Veilleux\altaffilmark{17}
}
\altaffiltext{1}{Infrared Processing \& Analysis Center, MS 100-22, California Institute of Technology,
Pasadena, CA 91125; mazz,bchan@ipac.caltech.edu}
\altaffiltext{2}{ICREA and Institut del Ci\`encies del Cosmos (ICC), Universitat de Barcelona (IEEC-UB), Mart\'i i Franqu\`es 1, 08028 Barcelona, Spain; kazushi.iwasawa@icc.ub.edu}
\altaffiltext{3}{Department of Physics and Astronomy, 
State University of New York at Stony Brook, Stony Brook, NY 11794-3800;
vavilkin@grad.physics.sunysb.edu}
\altaffiltext{4}{Spitzer Science Center, MS 314-6, California Institute of Technology, Pasadena,
CA 91125; lee,jhhowell,jason,inami,haan@ipac.caltech.edu}
\altaffiltext{5}{Department of Astronomy, 
University of Virginia, Charlottesville, VA 22904-4325;
aevans,gcp8y@virginia.edu}
\altaffiltext{6}{National Radio Astronomy Observatory, 
520 Edgemont Road, Charlottesville, VA 22903-2475;
dkim,dfrayer@nrao.edu}
\altaffiltext{7}{University of Oregon, Physics Department, Eugene OR, 97402;
nuts@bigmoo.uoregon.edu}
\altaffiltext{8}{Department of Astronomy, Cornell University, Ithaca, NY 14853; 
spoon@isc.astro.cornell.edu}
\altaffiltext{9}{CSIRO Astronomy \& Space Science, ATNF, Marsfield NSW 2122, Australia;
 sebastian.haan@csiro.au}
\altaffiltext{10}{NASA Herschel Science Center, MS 100-22, California Institute of Technology,
Pasadena, CA 91125; lord,apple,cxu@ipac.caltech.edu}
\altaffiltext{11}{Astronomy Department, California Institute of Technology, MS 249-17, 
Pasadena, CA 91125; jason.mashall@caltech.edu,ap@astro.caltech.edu}
\altaffiltext{12}{Subaru Telescope, National Astronomical Observatory of Japan, 
Hilo, HI 96720; cmishida@mac.com}
\altaffiltext{13}{Institute for Astronomy, University of Hawaii, 2680
Woodlawn Drive, Honolulu, HI 96822; sanders,vivian@ifa.hawaii.edu}
\altaffiltext{14}{Byurakan Astrophysical Observatory,
Academician of National Academy of Sciences of Armenia,
Byurakan, 378433 Aragatsodn Province, Republic of Armenia;
khache@bao.sci.am}
\altaffiltext{15}{The Observatories, Carnegie Institution of Washington, 
813 Santa Barbara Street, Pasadena, CA 91101; barry@obs.carnegiescience.edu}
\altaffiltext{16}{Max-Planck-Institut fŸr extraterrestrische Physik, 
Postfach 1312, 85741 Garching, Germany; sturm@mpe.mpg.de}
\altaffiltext{17}{Department of Astronomy, University of Maryland, College Park, 
MD 20742; veilleux@astro.umd.edu}

\begin{abstract}

Results of observations with the {\it Spitzer, Hubble, GALEX, Chandra}, 
and {\it XMM-Newton} space telescopes are presented for the  
Luminous Infrared Galaxy (LIRG) merger Markarian 266.
The SW (Seyfert 2) and NE (LINER) nuclei reside in galaxies with 
Hubble types SBb (pec) and S0/a (pec), respectively.
Both companions are more luminous than $L^{*}$ galaxies and they are
inferred to each contain a $\approx 2.5\times10^{8}$ \Msun\ black hole.
Although the nuclei have an observed hard X-ray flux ratio of $f_X(NE)/f_X(SW)=6.4$,
Mrk 266 SW is likely the primary source of a bright Fe K$\alpha$ line detected from
the system, consistent with the reflection-dominated
X-ray spectrum of a heavily obscured AGN. Optical knots embedded in an arc with
aligned radio continuum radiation, combined with luminous $\rm H_2$ line emission, provide 
evidence for a radiative bow shock in an AGN-driven outflow surrounding the NE nucleus.
A soft X-ray emission feature modeled as shock-heated plasma with $T \sim10^7$ K
is co-spatial with radio continuum emission between the galaxies. 
Mid-infrared diagnostics provide mixed results, but overall suggest
a composite system with roughly equal contributions of AGN and
starburst radiation powering the bolometric luminosity.
Approximately 120 star clusters have been detected, with most having 
estimated ages less than 50 Myr.
Detection of 24 \micron\ emission aligned with soft X-rays, radio continuum and
ionized gas emission extending $\sim$34\arcsec\ (20 kpc) north of the galaxies 
is interpreted as $\sim2\times10^7$ \Msun\ of dust entrained in an outflowing superwind.
At optical wavelengths this Northern Loop region is resolved into a fragmented morphology 
indicative of Rayleigh-Taylor instabilities in an expanding shell of ionized gas.
Mrk 266 demonstrates that the dust ``blow-out'' phase can begin in a LIRG well 
before the galaxies fully coalesce during a subsequent ULIRG phase,
and rapid gas consumption in luminous dual AGNs with kpc-scale separations
early in the merger process may explain the paucity of detected binary QSOs 
(with pc-scale orbital separations) in spectroscopic surveys.
An evolutionary sequence is proposed representing a progression from dual
to binary AGNs, accompanied by an increase in observed 
$\rm L_{x}/L_{ir}$ ratios by over four orders of magnitude.\\
%Force blank line before Keywords; this only works inside the Abstract
\end{abstract}


%%%%%%%%%%%%%%%%%%%%%%%%%%%%%%%
\section{Introduction}\label{sec:intro}

\subsection{LIRGs, ULIRGs and GOALS}\label{subsec:LIRGs}

Luminous Infrared Galaxies (LIRGs; $10^{11}{\ts}L_\odot \le L_{ir} < 10^{12}{\ts}L_\odot$)
and Ultraluminous Infrared Galaxies (ULIRGs; $L_{ir} \ge 10^{12}{\ts}L_\odot$) 
are intriguing objects with widespread implications for galaxy evolution. 
They contain the highest known rates of star formation, they exhibit a high frequency of 
active galactic nuclei (AGNs) and large-scale outflows (superwinds), and mounting 
evidence indicates that the majority of local (U)LIRGs represent 
an evolutionary process involving the transformation of major mergers between dusty, 
gas-rich disk galaxies into elliptical galaxies hosting
classical UV-excess QSOs or powerful radio galaxies 
\citep[e.g.,][]{1996ARA&A..34..749S, 2006NewAR..50..701V}.
At redshifts of $\rm z\sim 1$, LIRGs have a higher space density than ULIRGs and
dominate the total star-formation density at that epoch 
\citep{2005ApJ...632..169L}. At $\rm z\sim2$ the contributions of LIRGs and ULIRGs 
to the total IR luminosity density are about equal \citep[e.g.,][]{2007ApJ...660...97C}. 
Since LIRGs and ULIRGs were much more common in the early 
universe,  these populations are fundamental in understanding both 
star formation and galaxy evolution.

The Great Observatories All-Sky LIRG Survey (GOALS)\footnote{http://goals.ipac.caltech.edu/} 
is utilizing imaging and spectroscopic data from NASA's 
{\it Spitzer, Hubble, Chandra}  and {\it GALEX} space observatories, in
combination with ground-based observations, in a comprehensive study of
more than 200 of the most luminous infrared galaxies in the local universe 
\citep{2009PASP..121..559A}. 
The sample consists of 181 LIRGs and 23 ULIRGs that form a statistically complete
subset of the flux-limited {\it IRAS} Revised Bright Galaxy Sample  (RBGS), which itself
is comprised of 629 extragalactic objects with 60~\micron\ flux densities above 5.24 Jy 
and Galactic latitudes $\vert b \vert > 5^\circ$ \citep{2003AJ....126.1607S}.

\subsection{Dual AGNs in Major Mergers}\label{subsec:dualAGNs}

A key scientific driver for GOALS is the investigation of
the relative time-scales and energetics of active star formation and AGN phenomena
during different phases of the merger sequence. It is now widely accepted that all 
massive galaxies likely have a supermassive black hole (SMBH) in their centers 
\citep[e.g.,][]{2005SSRv..116..523F}, with a SMBH mass proportional to the mass of 
the stellar bulge \citep[$\rm M_{SMBH}/M_{bulge}\approx0.2$\%;][]{2003ApJ...589L..21M}.
However, many uncertainties remain regarding the fueling
of paired SMBHs during major mergers. 
While the nuclei are still separated by kpc scales, how common is it for both SMBHs to have 
accretion rates high enough to produce energetically significant {\it dual AGNs},
as opposed to one or both nuclei being powered predominantly by a burst of star formation? 
How do AGN characteristics depend on properties of the 
host galaxies and dynamics of the encounter?
Are the fuel supplies and accretion rates sufficient to sustain two luminous
AGNs well into a true {\it binary} stage (e.g., binary QSOs), when the 
SMBHs are closely bound ($\rm r < 100~pc$) in Keplerian orbits inside a
dynamically relaxed (elliptical) merger remnant \citep[e.g.,][]{2009arXiv0906.4339C}?
Is there something special about the physical conditions in systems that host 
dual AGNs, or is their lack of detection in some (U)LIRGs merely a matter of 
observing them during the right stage prior to merging (i.e., an evolutionary timing coincidence),
or accounting for differences in nuclear dust obscuration?
What is the relative importance of AGNs and star formation in the energetics
of the ``superwind'' phenomena that appears to be ubiquitous in (U)LIRGs
\citep[e.g., see the review by][]{2006NewAR..50..701V}?

Although (U)LIRGs are predominantly systems involving major mergers,
extensive observations utilizing optical and infrared spectroscopy 
have so far turned up surprisingly few systems containing confirmed dual AGNs.
This has a number of possible explanations. 
First, (U)LIRGs contain large quantities of centrally concentrated dust that can 
effectively obscure one or both AGNs at optical, near-infrared, and even 
mid-infrared wavelengths \citep[e.g.][]{2000A&A...359..887L,2004ApJS..154..184S}. 
Therefore, circumnuclear star formation can dominate the observed spectra
while in reality a powerful AGN may be buried inside.
Second,  about 30\% of observed LIRGs \citep[e.g.,][]{1995ApJS...98..171V},
and $\sim$40\% of ULIRGs \citep[e.g.,][]{1999ApJ...522..113V} are 
classified as LINERs based on optical spectroscopy. 
It has proven difficult to distinguish between 
low-luminosity AGNs (photoionization from radiation due to 
accretion onto a SMBH), photoionization from very hot stars, 
and shock heating (induced by SNe and starburst-driven superwinds) 
as the primary source of ionization in LINERs
\citep[e.g.,][]{1999ApJ...522..113V}. 

Recent results indicate that most nearby LINERs are dominated by
photo-ionization rather than shock heating, and that they are an important class 
of AGNs distinguished primarily by a lower accretion rate than in Seyfert 
nuclei \citep{2006MNRAS.372..961K,2008ARA&A..46..475H}.
This implies that the frequency of dual AGNs may be much higher than inferred to date.
Only recently, with the capability of the {\it Chandra X-ray Observatory} 
to penetrate their extensive dust cocoons with high-resolution hard X-ray
observations, have suspected dual AGNs been confirmed in a small number 
of (U)LIRGs. Perhaps the best known example is NGC 6240, 
in which hard X-rays and strong Fe K$\alpha$ emission lines indicate
the presence of two AGNs  with a projected separation of 1\farcs4 (1 kpc)
\citep{2003ApJ...582L..15K}. A second case is the ULIRG Mrk 463,
a system first pointed out as an active double-nucleus galaxy by
\citet{1978Afz....14...69P}. The eastern component undoubtedly hosts a
dust-enshrouded Type 1 AGN that is powering apparently 
young radio jets \citep{1991AJ....102.1241M}, but conflicting 
optical spectral classifications from various studies left the nature of 
the western nucleus in doubt (LINER, Seyfert 2, or starburst/AGN composite). 
Dual AGNs in Mrk 463 have been confirmed recently via 
detection of two compact, luminous sources of hard X-rays
using {\it Chandra} \citep{2008MNRAS.386..105B}. 
To date, very few systems in the GOALS sample have been confirmed 
to contain dual AGNs via X-ray observations. The first system is NGC 6240, 
and the second is Arp 299 (NGC 3690 + IC 694) \citep{2004ApJ...600..634B}.
This article presents an in-depth investigation of a third system in this rare 
class of objects that has considerable importance for understanding 
the evolution of galaxy mergers and their remnants.

\subsection{Complex Phenomena in Mrk 266}\label{subsec:background}

One of the most remarkable LIRGs in the local Universe is Mrk 266 (NGC 5256).
The system was first called out as an extraordinary object in the 
Markarian Survey (First Byurakan Sky Survey) of ultraviolet-excess 
galaxies due to the presence of two nuclei within what appeared to be
a single galaxy \citep{1978Afz....14...69P,1979Afz....15..209P}.
In a detailed spectroscopic investigation at optical wavelengths,
\citet{1980Afz....16..621P} showed that the nuclei have 
a radial velocity difference of 280 $\rm km~s^{-1}$, and based 
on the assumption that  the nuclei revolve about a common center of gravity,
masses of $\rm 7\times10^9~and~10\times10^9$\Msun~were estimated for the SW
and NE nuclei, respectively. As for all objects in the Markarian Survey, 
Mrk 266 stood out because its UV-to-optical flux ratio is higher 
than in normal galaxies. However, Mrk 266 is also a LIRG, with an
infrared luminosity of $\rm L_{ir} = 3.6\times10^{11} L_\odot$ 
\citep[8 - 1000 \micron, as estimated from flux densities in all four 
{\it IRAS} bands;][]{2009PASP..121..559A}.
As shown in the current study (\S \ref{subsec:SEDanalysis}), the global
spectral energy distribution indicates the bulk of its energy is emitted
in the mid- to far-infrared; much more energy is being 
emitted at the peak of the SED than is escaping in UV radiation 
($\rm \nu L_{\nu}(70\micron)/\nu L_{\nu}(0.2\micron) \approx 20$).

Mrk 266 contains a filamentary nebula of ionized hydrogen 
$\sim$30 kpc in diameter \citep{1990ApJ...364..471A} surrounding the two nuclei. 
A luminous X-ray nebula $\sim$100 kpc in extent detected by {\it ROSAT}
(also surrounding the two nuclei) and complex kinematics derived from 
optical spectroscopy revealed one of the most
spectacular examples of an outflowing, starburst-driven 
superwind \citep{1997ApJ...474..659W}.
In addition to radio continuum emission from the two nuclei,
a bright radio source located between the
nuclei was detected and interpreted as enhanced synchrotron emission induced by 
extensive shocks at the interface of the merging galaxies \citep{1988ApJ...333..168M}.
Imaging in $\rm H\alpha + [N II]$ \citep{1988ApJ...333..168M,1990ApJ...364..471A}, 
[O III] $\lambda 5007$ \citep{1988AJ.....96.1227H} and soft
X-rays \citep{1998AA...336L..21K} also revealed a bright loop of 
emission extending $\sim$24\arcsec (17 kpc) to the north-east 
and connected to the SW (Seyfert 2) nucleus \citep{1997ApJ...474..659W}. 
Recent {\it Chandra} observations have resolved X-ray emission
from both nuclei, and detected diffuse emission associated with the 
northern feature and between the nuclei \citep{2007MNRAS.377.1439B}.
The nature of the northern emission region (hereafter called the Northern Loop) 
has remained controversial, with suggestions including a component of the 
superwind \citep{1997ApJ...474..659W}, a ``jet'' from an AGN \citep{1998AA...336L..21K},
and a tidal tail \citep{2007MNRAS.377.1439B}.

Mrk 266 is worthy of further detailed study because it manifests complex processes that are
rarely detected simultaneously, presumably due to their relatively short time scales: 
vigorous star formation, a candidate dual AGN (depending on the ionizing source 
of the LINER), a powerful large-scale superwind, and substantial radio continuum 
and X-ray emission between the nuclei.
The relative proximity of the system permits close-up investigation of an important 
stage in the evolution of major mergers that may involve the transformation of LIRGs into 
ULIRGs, AGN feedback with the ISM, and implications for fueling (or quenching) 
of accreting SMBHs which galaxy evolution models 
predict to have occurred in large numbers during earlier epochs.

In this study, new observations and re-processed archival data from the
{\it Spitzer, Hubble, Chandra, XMM-Newton} and {\it GALEX} space telescopes
are analyzed in combination with multi-wavelength ground-based data.
Imaging, photometric, and spectroscopic data are presented in \S\ref{sec:Observations}.
In \S\ref{sec:Discussion} the new data are interpreted to obtain new insights 
into the nature of the colliding/merging system (\S\ref{subsec:Galaxies}), 
the nuclei and circumnuclear regions (\S\ref{subsec:Nuclei}),
the region between the nuclei (\S\ref{subsec:Center}),
the extensive superwind (\S\ref{subsec:Superwind}),
the SEDs of the major components (\S\ref{subsec:SEDanalysis}),
newly detected star clusters (\S\ref{subsec:SCs}), and the
molecular gas properties (\S\ref{subsec:MolGas}).
In \S\ref{subsec:Implications}, Mrk 266 is examined in comparison with other 
interacting systems with strong radiation sources between the galaxies, and a 
sequence is proposed that may represent an evolutionary progression from 
dual AGNs (with kpc-scale separations) to binary AGNs (with parsec-scale orbits).
Due to the large number of new measurements
spanning many spectral regions, the basic observations and data reductions are 
described in \S\ref{sec:Observations}, but most of the corresponding figures and tables are 
displayed alongside their interpretation and analyses in \S\ref{sec:Discussion}.
Conclusions are summarized in \S\ref{sec:Summary}.
A systemic heliocentric recession velocity of
$\rm 8353~\pm~13~km~ s^{-1}$ \citep{1991deVaucouleurs} is adopted, corrected 
to $\rm 8825 \pm~22~km~s^{-1}$ via the flow model of  \citet{Mould2000a,Mould2000b}
that accounts for three major attractors in the local Universe. 
We adapt cosmological parameters $\rm H_{\rm o} = 70~km~s^{-1}~Mpc^{-1}$,
$\Omega_{\rm M} = 0.28$, and $\Omega_{\rm V} = 0.72$ \citep{2009ApJS..180..225H}. 
The corresponding luminosity distance to Mrk 266 is 129 Mpc (distance modulus 35.55 mag), 
and the spatial scale is 0.59 kpc/arcsec\footnote{\it Provided by NED at http://ned.ipac.caltech.edu/}.

%%%%%%%% Begin Figure %%%%%%%%%%%
%Mrk 266 with HST ACS
\def\figcapHSTcolor{
\footnotesize 
{\it HST} ACS I band and B band images combined to form a color composite image 
of the Mrk 266 system and its tidal features. North is up and east is left. 
The image field of view is 1\farcm4~x~1\farcm6, with a scale bar in the lower left corner.
The color table has been chosen to maximize contrast for faint extended structures.
The limiting surface brightness is 25.0 and 24.5 $\rm mag~arcsec^{-2}$ in the 
B and I band, respectively. Labels identify major structures that are studied throughout this article.
%\vskip 0.02truein %Because a page beak does not work to omit a single line of text below this figure
}
\ifnum\Mode=0 %Insert Figure/Table here only in [preprint] or [preprint2] modes
\placefigure{fig:HST_ACS4x}
\begin{verbatim}fig01\end{verbatim}
\else
%For preprint
\ifnum\Mode=2 
\begin{figure*}[p]
\else
\begin{figure}[p]
\fi
\includegraphics[width=1.0\textwidth,angle=0]{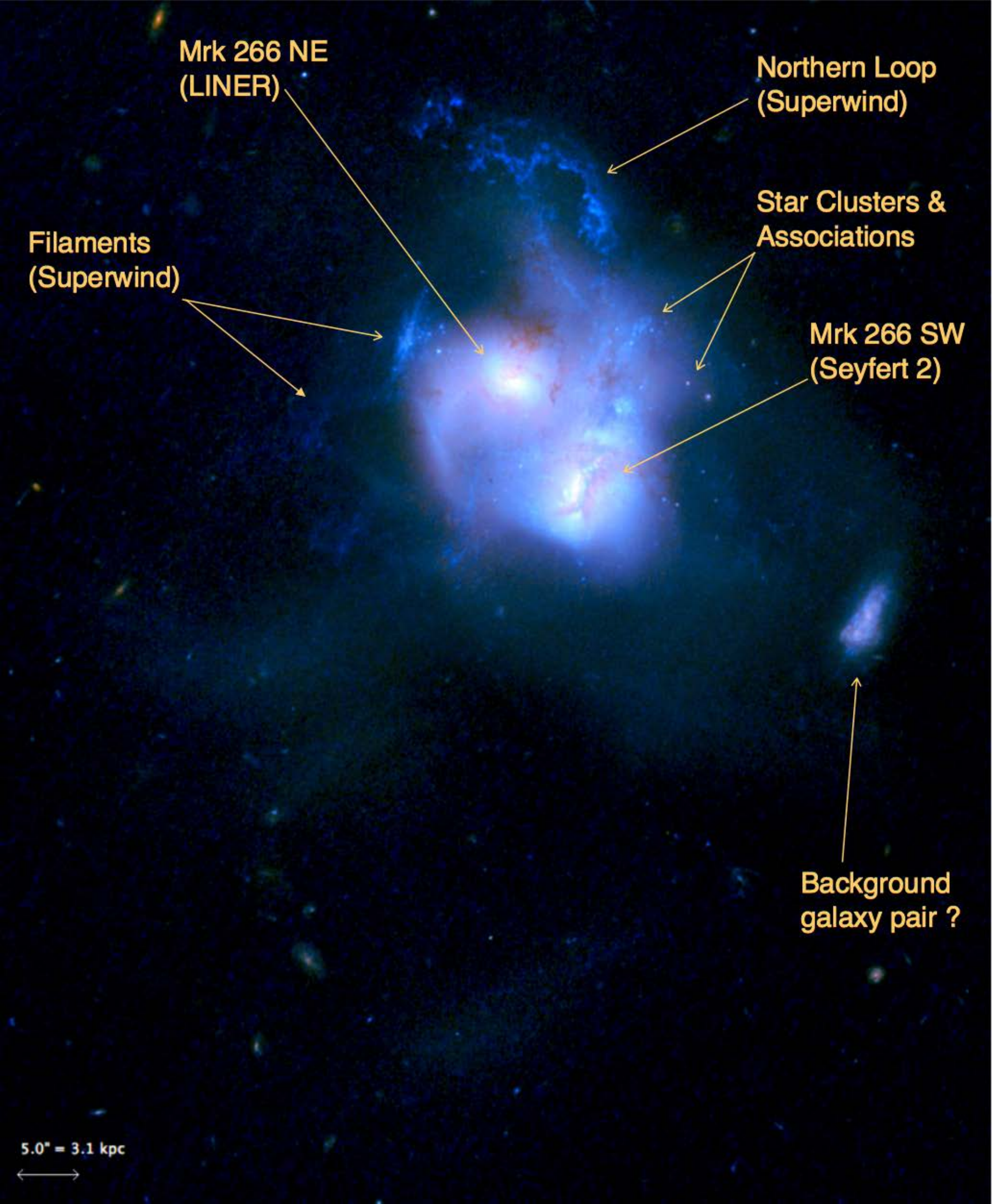}
\caption{\figcapHSTcolor \label{fig:HST_ACS4x}}
\ifnum\Mode=2 
\end{figure*}
\else
\end{figure}
\fi %close \ifnum\Mode=2
\fi %close \ifnum\Mode=0 
%%%%%%%% End Figure %%%%%%%%%%%

\section{Observations}\label{sec:Observations}

\subsection{Optical and Near-Infrared Imaging with HST}\label{subsec:HSTimaging}

Mrk 266 was observed 2005 November 17  with the 
Advanced Camera for Surveys (ACS) Wide Field Channel (WFC) as part 
of a {\it HST} Cycle 14 GOALS program to investigate 
88 LIRGs with $\rm L_{ir} > 10^{11.4} L_{\odot}$ (PID 10592, PI A. Evans).
The system was imaged with an integration time of 22.5 minutes in the F435W filter
(B band, 0.435 \micron, image \#J9CV48010) and 13 minutes in the 
F814W filter (I band, 0.814 \micron, image \#J9CV48020).
The WFC has a field of view of 202\arcsec~x~202\arcsec\ with pixels subtending 0\farcs05.
The images were processed through the standard STScI pipeline to remove instrumental 
signatures, subtract a dark frame, apply flat fielding, remove geometric distortion, 
and apply flux calibration. Residual cosmic ray removal, background subtraction,
and accurate astrometric calibration using sources from {\it 2MASS} 
\citep{2006AJ....131.1163S} were also performed using procedures 
detailed by Evans et al. (2012, in preparation).
Figure \ref{fig:HST_ACS4x} presents a composite color image constructed 
using the B and I band data, and labels identify major features that are 
analyzed throughout this article.

Near-infrared observations at 1.6 \micron\ (F160W, H band) with the 
NIC2 aperture of NICMOS were acquired on 1997 September 13 
in Cycle 7 (PID 7328, PI M. Malkan). 
The image (\#N44B40010) was retrieved from the HST archive,
corrected for bad pixels, cosmic rays were removed, 
and astrometric calibration was applied by aligning the galaxies 
with the ACS images. The central regions of the SW and NE 
galaxies in these three {\it HST} band are displayed
in context with their analysis in \S\ref{subsec:Galaxies}.

\subsection{Deep B+V+I Band Ground-Based Imaging}\label{subsec:UH88imaging}

To complement the {\it HST} high-resolution imagery and reveal the faintest 
emission possible, Figure \ref{fig:UH88image} shows the result of 
spatial registration and stacking of B, V, and I band images obtained on 
1998 December 22 and 2000 May 04 with the University of Hawaii 2.2 m 
telescope (effective seeing $\sim$1\farcs5 FWHM).
Details of the data acquisition and reduction are given by \citet{2004PhDT........18I}. 
The data are presented using Gaussian equalization
to contrast the low and high surface-brightness features in the same view.
Highly asymmetric, low surface-brightness emission can be seen spanning 
$\approx$103 kpc (2\farcm9) from the SE to NW extremities, which is approximately 
twice the extent of the faintest emission visible in the {\it HST} ACS data, 
SDSS, and previously published imagery.

%%%%%%%% Begin Figure %%%%%%%%%%%
\def\figcapUH{
\footnotesize 
The result of stacking B, V, and I band images obtained with the University of Hawaii 
2.2 m telescope (effective seeing $\sim$1\farcs5 FWHM).
The field of view is 150 kpc x 150 kpc (4\farcm2 x 4\farcm2), and a scale bar
is provided in the lower left. Faint, low surface-brightness emission spanning
2\farcm9 ($\approx 103$ kpc) is detected.
}
\ifnum\Mode=0 %Insert Figure/Table only in [preprint] or [preprint2] modes
\placefigure{fig:UH88image}
\begin{verbatim}fig02\end{verbatim}
\else
%For preprint
\ifnum\Mode=2 
\begin{figure}[]
\center
\includegraphics[width=1.0\columnwidth,angle=0]{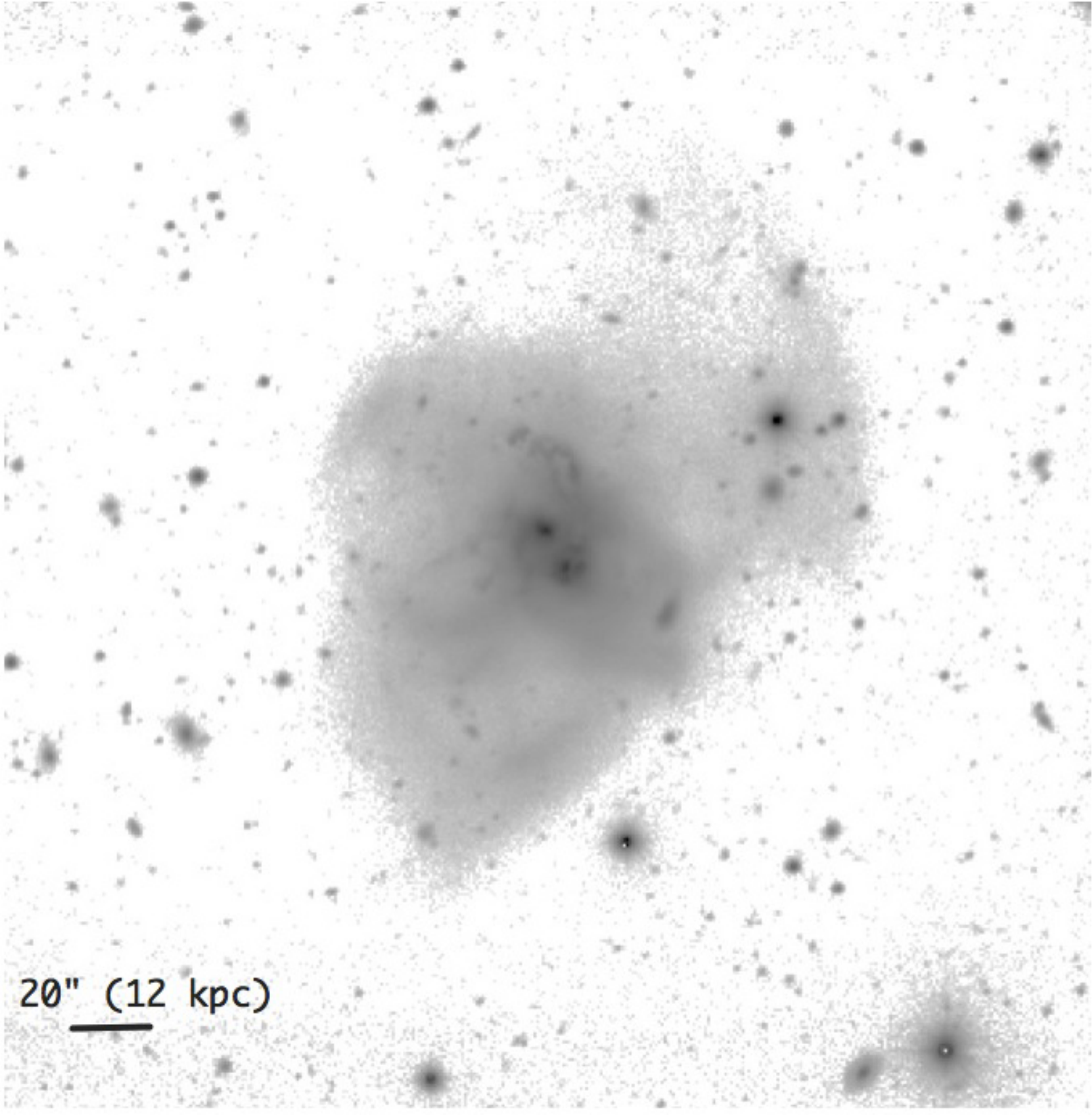} 
\else
\begin{figure}[t]
\center
\includegraphics[scale=0.6,angle=0]{fig02} 
\fi
\caption{\figcapUH \label{fig:UH88image}}
\end{figure}
\fi %close \ifnum\Mode=0 
%%%%%%%% End Figure %%%%%%%%%%%

\subsection{X-Ray Observations with Chandra and XMM-Newton}\label{subsec:ChandraXMM}

\subsubsection{Chandra Observations}\label{subsubsec:ChandraObs}

{\it Chandra} observations were obtained with the 
Advanced CCD Imaging Spectrometer (ACIS-S) 
on 2001 November 02 (PI A. Read) with a total integration time of 19.95 ks.
The spatial resolution is $\sim1$\arcsec\ FWHM.
These observations were first presented by \citet{2007MNRAS.377.1439B}.
In order to make refined measurements, highlight additional features, and 
analyze the X-ray properties alongside observations at other wavelengths, the data were 
taken from the {\it Chandra} Data Archive\footnote{\it http://cxc.harvard.edu/cda/} 
(OBSID \#2044) and re-processed using standard procedures in CIAO 4.0.
The data were corrected for detector response and converted to flux density ($\rm f_{\nu}$) units.
Details of the data reduction procedures can be found in \citet{2011A&A...529A.106I}.
Figure \ref{fig:ChandraImages} shows the resulting full band (0.4 - 7 keV) 
image, the hard band (2-7 keV) image, a smoothed version of the full band 
data, and the 20 cm radio continuum image from \citet{1988ApJ...333..168M}.
Figure \ref{fig:XrayRegions} illustrates apertures with parameters 
listed in Table \ref{tbl:XrayRegions} that were used to 
define regions of interest, and Figure \ref{fig:ChandraSpectra} shows the 
X-ray spectra constructed in these regions.

%%%%%%%% Begin Figure %%%%%%%%%%%
%Unsmoothed Full band (0.4 - 7 keV), unsmoothed Hard band (2 - 7 keV), 
% Smoothed full band, 20cm
%Chandra_Full_Hard_FullSmooth_20cm
\def\figcapChandraImages{
\footnotesize 
The X-ray and radio continuum emission in Mrk 266:
(a) {\it Chandra} 0.4 - 7 keV (full band) X-ray image;
(b)  2 - 7 keV (hard band)  X-ray image;
(c) 0.4 - 7 keV X-ray data (same as panel a) 
smoothed with a 2-pixel Gaussian kernel; and
(d) 20 cm radio continuum image (VLA) from \citet{1988ApJ...333..168M}.
The scale bar indicates 10\arcsec.
}
\ifnum\Mode=0 %Insert Figure/Table only in [preprint] or [preprint2] modes
\placefigure{fig:ChandraImages}
\begin{verbatim}fig03a_03d\end{verbatim}
\else
%For preprint
%\ifnum\Mode=2 \onecolumn \fi %4-panel fig is too small in 2-column mode
\begin{figure}[h]
\center
\ifnum\Mode=2 
\includegraphics[width=1.0\columnwidth,angle=0]{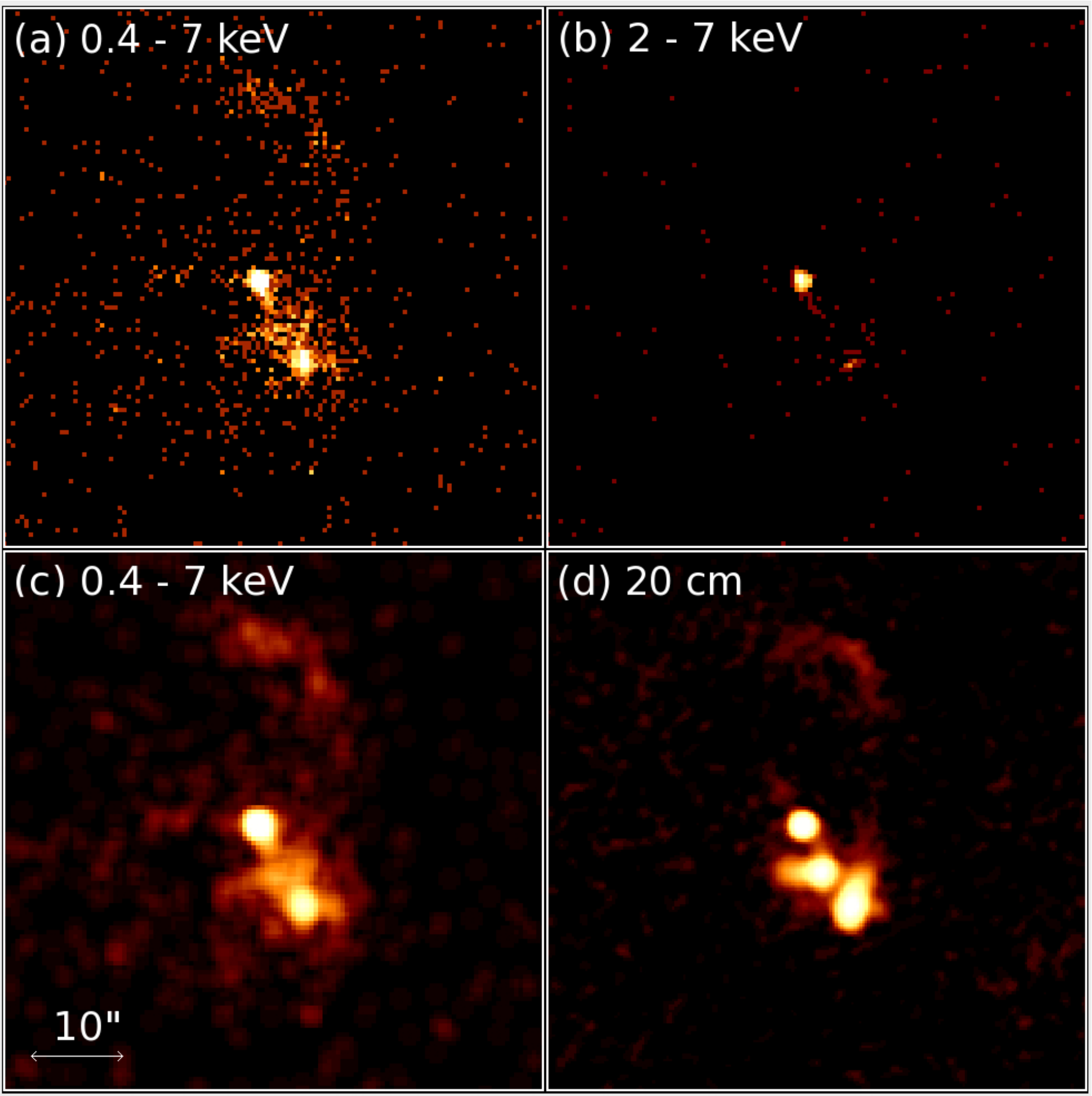}
\else
\includegraphics[scale=0.6,angle=0]{fig03a_03d}
\fi
%Because the background is black, we must label the panels in the image.
\caption{\figcapChandraImages \label{fig:ChandraImages}}
\end{figure}
\fi %close \ifnum\Mode=0 
%%%%%%%% End Figure %%%%%%%%%%%

%%%%%%%% Begin Figure %%%%%%%%%%%
%XrayRegions 
\def\figcapXrayRegions{
\footnotesize 
Elliptical apertures used to define regions of interest for extraction of X-ray spectra from the
{\it Chandra} data: Mrk 266 NE (``NE''), Mrk 266 SW (``SW''), the central region between 
the nuclei (``Cen''), the Northern Loop  (``Loop''), and surrounding diffuse emission (``Dif'').
The two ellipses marked with red lines indicate regions subtracted from the ellipse labeled ``Dif''
to estimate the diffuse component.  Parameters for the regions are listed in Table \ref{tbl:XrayRegions}.
A circular aperture of radius 33\arcsec\ used to measure the integrated system flux is not illustrated.
}
\ifnum\Mode=0 %Insert Figure/Table only in [preprint] or [preprint2] modes
\placefigure{fig:XrayRegions}
\begin{verbatim}fig04\end{verbatim}
\else
\begin{figure}[h]
\center
\ifnum\Mode=2 
\includegraphics[width=1.0\columnwidth,angle=0]{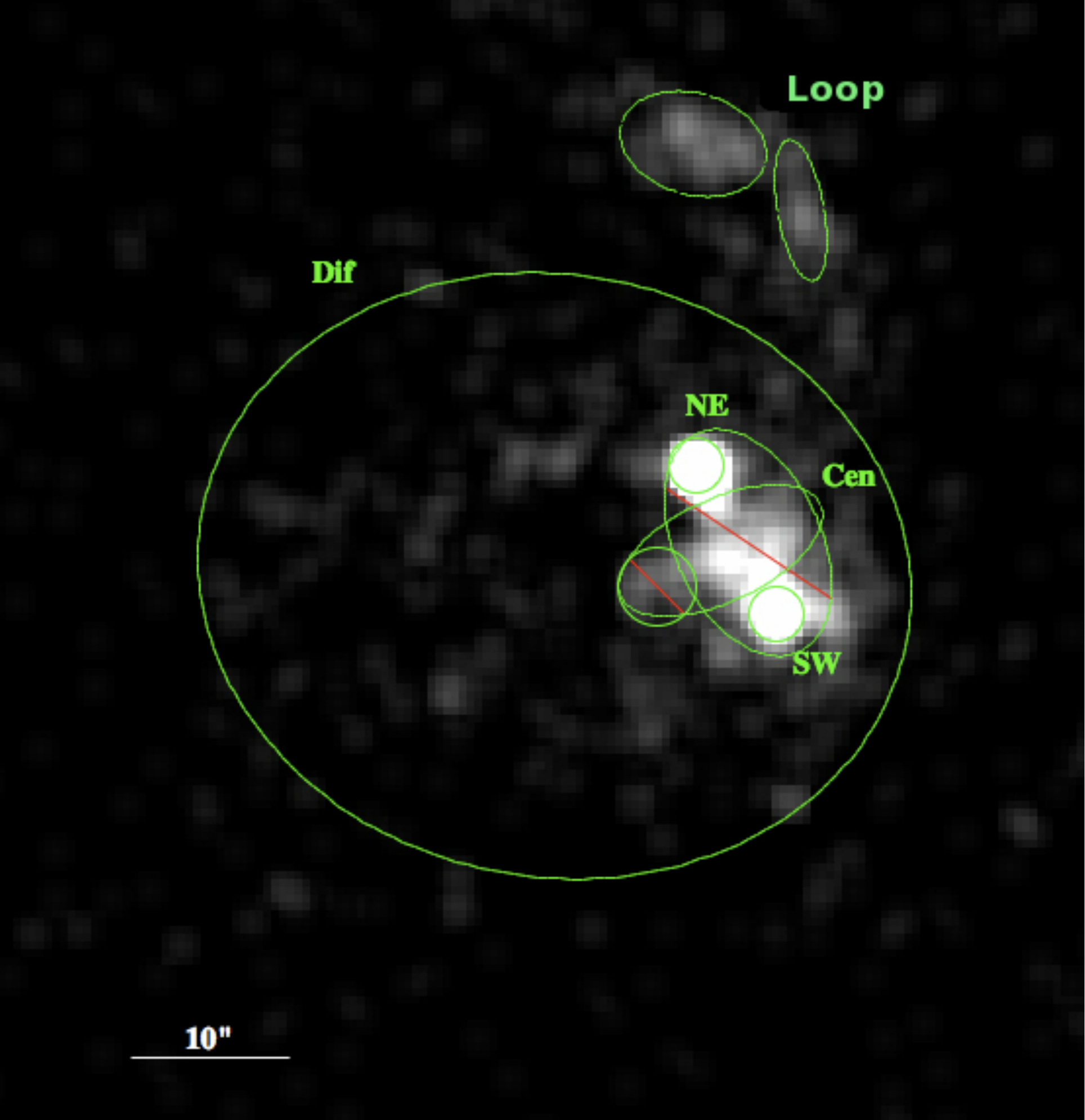} 
\else
\includegraphics[scale=0.5,angle=0]{fig04} 
\fi
\caption{\figcapXrayRegions \label{fig:XrayRegions}}
\end{figure}
\fi %close \ifnum\Mode=0 
%%%%%%%% End Figure %%%%%%%%%%%

%%%%%%%% Begin Figure %%%%%%%%%%%
%X-ray Spectra of Nuclei, Center and Arm 
%fden_five.ps (was n5256cxosp)
\def\figChandraSpectra{
\footnotesize
{\it Chandra} X-ray spectra covering 0.35 - 7 keV for the regions 
illustrated in Figure \ref{fig:XrayRegions}: (a) Mrk 266 NE (``NE''); (b) Mrk 266 SW (``SW'');
(c) the central region between the nuclei (``Cen''); (d) the Northern Loop (``Loop''); 
(e) the diffuse emission (``Dif''); (f) and the total system emission (``Total'').
The horizontal lines indicate the variable-width energy bands, 
and the vertical lines are $1\sigma$ error bars. 
}
\ifnum\Mode=0 %Insert Figure/Table only in [preprint] or [preprint2] modes
\placefigure{fig:ChandraSpectra}
\begin{verbatim}fig05a_05f\end{verbatim}
\else
\ifnum\Mode=2 
\begin{figure*}[ht]
\else
\begin{figure}[ht]
\fi
\center
\includegraphics[width=6.5truein]{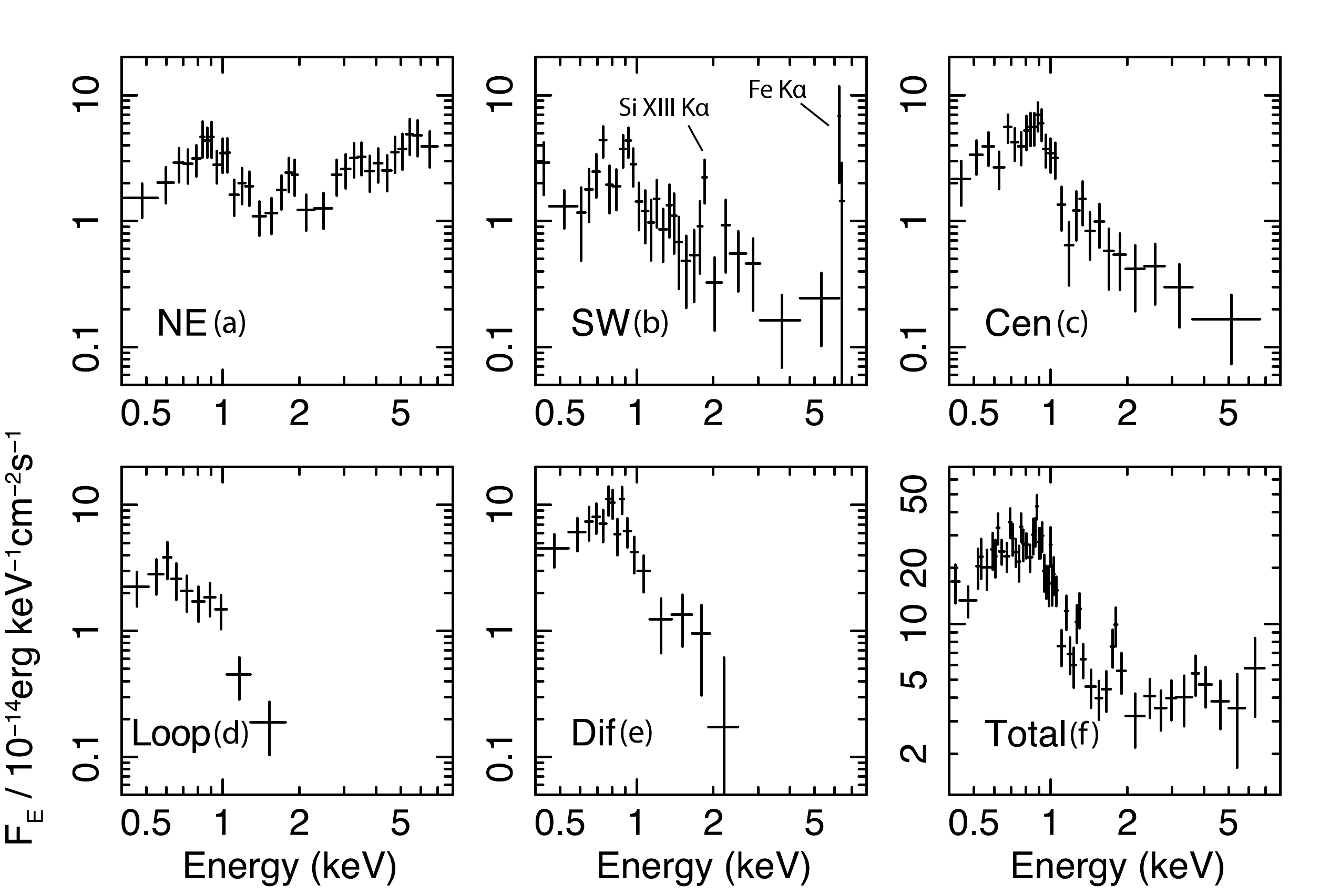}
\caption{\figChandraSpectra \label{fig:ChandraSpectra}}
\ifnum\Mode=2
\end{figure*}
\else
\end{figure}
\fi
%\ifnum\Mode=2 \twocolumn \fi %[preprint2] mode only
\fi %close \ifnum\Mode=0 
%%%%%%%% End Figure %%%%%%%%%%%

\subsubsection{XMM-Newton Observations}\label{subsubsec:XMMObs}

X-ray observations with {\it XMM-Newton}, reaching significantly higher
energies (10 keV) than {\it Chandra} (7 keV), were acquired with the
European Imaging Photon Camera (EPIC) on 2002 May 15 (PI A. Read).
The data obtained with the EPIC pn-CCD camera were utilized, providing an 
angular resolution of $\sim$6\arcsec\ FWHM and a spectral resolution of ($E/\delta E) \sim 40$.
The observation was performed with the Thin Filter in the extended full-window imaging mode. 
The data set (\#0055990501) was taken from the {\it XMM-Newton}
Science Archive\footnote{\it http://xmm.esac.esa.int/xsa/} and processed with 
standard procedures in the software packages SAS 8.1 and HEASoft 6.6 
utilizing the latest calibration files (as of 2009 January). 
Time intervals with high particle background levels were discarded, 
leaving useful data with an effective exposure time of 13.2 ks. 
The count rate in the 0.4-10 keV band is 0.19 count s$^{-1}$.
The X-ray emission is only marginally resolved by {\it XMM}, therefore the 
spectrum for the total system emission within a large aperture of 45\arcsec\ radius 
is produced here. Figure \ref{fig:XMMSpectrum} shows the resulting {\it XMM} spectrum
of the integrated emission from the system. 
Key features include Si XIII  K$\alpha$ (1.853 keV) emission,
as also observed in the {\it Chandra} spectra of the two nuclei 
(Fig. \ref{fig:ChandraSpectra}), and strong Fe K$\alpha$ line emission at 6.4 keV.

%%%%%%%% Begin Figure %%%%%%%%%%%
%XMM-Newton Spectrum
%fden_pn.ps (was n5256xmmsp)
\def\figcapXMMSpectrum{
\footnotesize 
The 0.5 - 10 keV X-ray spectrum of Mrk 266 (integrated system) measured
with the EPIC instrument onboard {\it XMM-Newton}. Primary features include the 
Fe K$\alpha$ emission line at 6.4 keV and an Fe K absorption edge at 7 keV.
}
\ifnum\Mode=0 %Insert Figure/Table only in [preprint] or [preprint2] modes
\placefigure{fig:XMMSpectrum}
\begin{verbatim}fig06\end{verbatim}
\else
%For preprint
%\ifnum\Mode=2 \onecolumn \fi %4-panel fig is too small in 2-column mode
\begin{figure}[!t]
\center
\ifnum\Mode=2 
\includegraphics[width=1.0\columnwidth,angle=0]{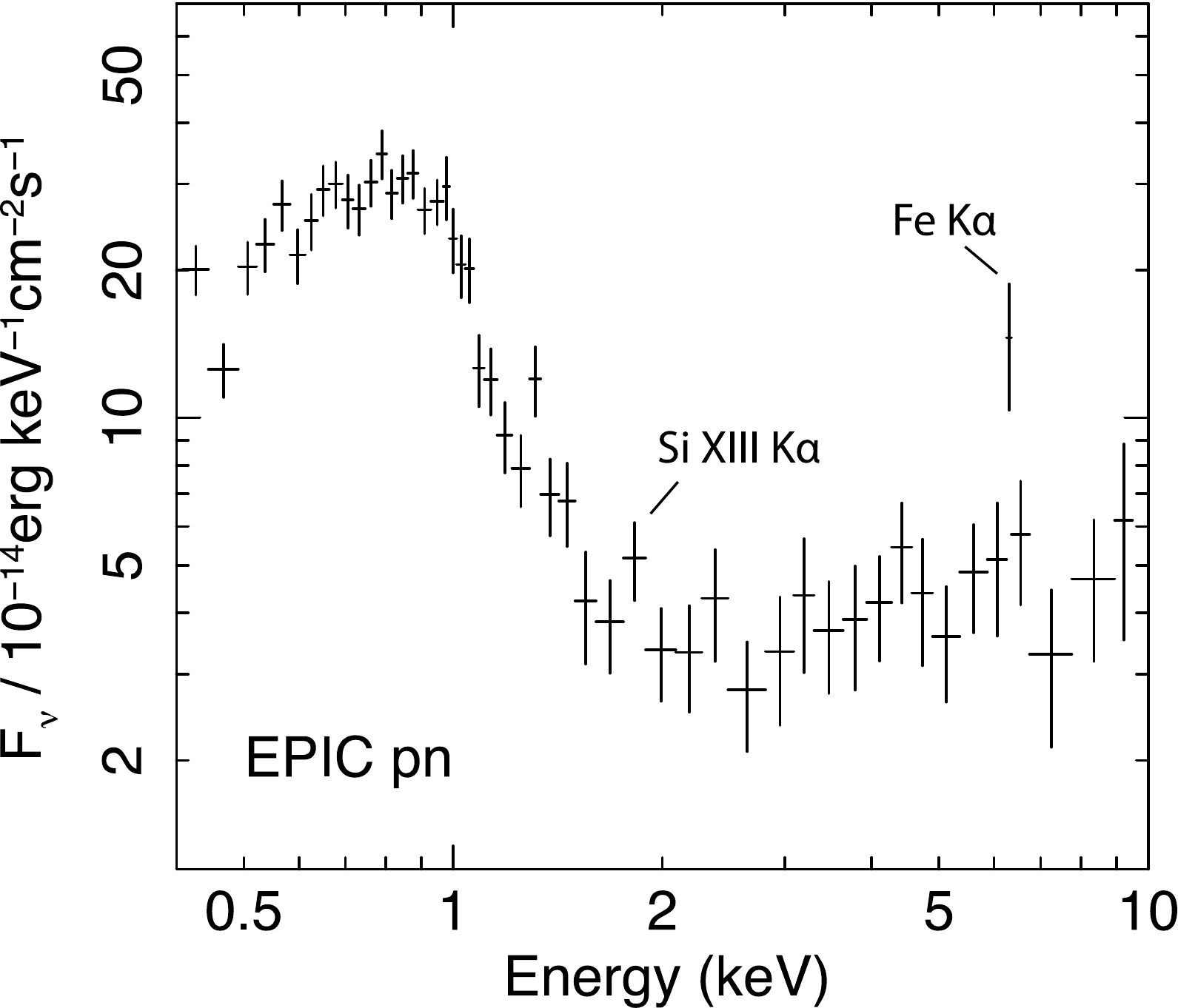}
\else
\includegraphics[scale=1.0,angle=0]{fig06}
\fi
\caption{\figcapXMMSpectrum \label{fig:XMMSpectrum}}
\end{figure}
\fi %close \ifnum\Mode=0 
%%%%%%%% End Figure %%%%%%%%%%%

\subsubsection{X-ray Measurements}\label{subsubsec:XrayMeasurements}

%%%%%%%% Begin Table %%%%%%%%%%%
\def\tableXrayRegions{
\begin{center}
\begin{deluxetable}{lllll}
\ifnum\Mode=2
\renewcommand\arraystretch{0.5}% (MyValue=1.0 is for standard spacing)
\tabletypesize{\scriptsize} 
\setlength{\tabcolsep}{0.001in} %Tighten up the columns. See AASTeX FAQ
\tablewidth{\columnwidth}
\fi
\renewcommand\arraystretch{1.0}% (MyValue=1.0 is for standard spacing)
\setlength{\tabcolsep}{0.25in} %Tighten up the columns. See AASTeX FAQ
%\tablewidth{\textwidth}
\tablecaption{Elliptical Apertures for X-ray Measurements \label{tbl:XrayRegions}}
\tablehead{
\colhead{Region} & \colhead{RA, DEC (J2000)} & \colhead{a\arcsec} & \colhead{b/a} & \colhead{PA\arcdeg}\\
\colhead{(1)} & \colhead{(2)} & \colhead{(3)} & \colhead{(4)} & \colhead{(5)} 
}
\tablecolumns{5}
\startdata
SW      & {\tt \phm{-}(13:38:17.31,+48:16:31.9)} & {\tt \phn1.7} & {\tt 1.00} & \nodata\\
NE      & {\tt \phm{-}(13:38:17.82,+48:16:41.2)} & {\tt \phn1.7} & {\tt 1.00} & \nodata \\
Center & {\tt \phm{-}(13:38:17.66,+48:16:35.8)} & {\tt \phn6.9} & {\tt 0.47} & {\tt 155}\\
Loop A  & {\tt \phm{-}(13:38:17.15,+48:16:57.2)} & {\tt \phn4.5} & {\tt 0.32} & {\tt \phn10}\\
Loop  B  & {\tt \phm{-}(13:38:17.83,+48:17:01.4)} & {\tt \phn4.7} & {\tt 0.68} & {\tt \phn75}\\
Diffuse\tablenotemark{a} & {\tt  \phm{-}(13:38:18.71,+48:16:34.2)} & {\tt 22.5} & {\tt 0.84} & {\tt \phn80}\\
Diffuse -1    & {\tt -(13:38:17.49,+48:16:36.3)} & {\tt \phn4.6} & {\tt 0.61} & {\tt \phn25}\\
Diffuse -2    & {\tt -(13:38:18.07,+48:16:33.6)} & {\tt \phn2.5} & {\tt 1.00} &\nodata \\
Total                             & {\tt \phm{-}(13:38:17.66,+48:16:35.8)} & {\tt 33.0} & {\tt 1.00} & \nodata
\enddata
%\caption{Caption X\label{tbl:Xrays}}
\tablenotetext{a}{The diffuse emission is approximated by the flux within this 
large ellipse after subtraction of the flux in the two small ellipses in the 
following two rows, which are proceeded by a minus sign in this table and 
marked with red lines in Figure \ref{fig:XrayRegions}.}
%\tablecomments{Column descriptions are given in the text (\S\ref{sec:rbgs}).}
\end{deluxetable}
\end{center}
}
\ifnum\Mode=0
\placetable{tbl:XrayRegions}
\else
\tableXrayRegions
\fi
%%%%%%%% End Table %%%%%%%%%%%

%%%%%%%% Begin Table %%%%%%%%%%%
\def\tableXrayData{
\begin{deluxetable}{lllll}
\tabletypesize{\footnotesize} 
%\tiny(5pt);\scriptsize(7pt);\footnotesize(8pt);\small(9pt);\normalsize(10pt)
%\rotate
%\tablenum{1}
\ifnum\Mode=2
\renewcommand\arraystretch{0.5}% (MyValue=1.0 is for standard spacing)
\tabletypesize{\scriptsize} 
\setlength{\tabcolsep}{0.01in} %Tighten up the columns. See AASTeX FAQ
\tablewidth{\columnwidth}
\fi
\renewcommand\arraystretch{1.0}% (MyValue=1.0 is for standard spacing)
\tabletypesize{\normalsize} 
\setlength{\tabcolsep}{0.25in} %Tighten up the columns. See AASTeX FAQ
%\tablewidth{\textwidth}
\tablecaption{X-Ray Measurements \label{tbl:XrayData}}
\tablehead{
\colhead{Region} & \colhead{F(Fe K$\alpha)$} & \colhead{kT (soft)} & \colhead{$\rm F_x (soft)$} & \colhead{$\rm F_x (hard)$}\\
\colhead{(1)} & \colhead{(2)} & \colhead{(3)} & \colhead{(4)} & \colhead{(5)}
}
\tablecolumns{5}
\startdata
Mrk 266 SW & \phm{$<$}$1.2^{+0.7}_{-0.5}$ & $0.77^{+0.07}_{-0.08}$    & $\phn2.2$  & $\phn2.2$\\
Mrk 266 NE  & $< 1.6$                      & $0.72^{+0.11}_{-0.10}$    & $\phn3.0$  & $14$\phd\phn\\
Between Nuclei  & \nodata                        & $0.66^{+0.06}_{-0.07}$  & $\phn2.6$  & $\phn1.0$\\
Northern Loop       & \nodata                         & $0.35^{+0.05}_{-0.05}$  & $\phn1.4$  & \nodata \\
Diffuse Region   & \nodata                         & $0.54^{+0.04}_{-0.04}$  & $\phn4.1$  & \nodata \\
Total ({\it Chandra})  & \nodata     & \nodata                                           & $18\phd\phn$   & $20\phd\phn$ \\
Total ({\it XMM}) & \phm{$<$}$1.5^{+0.8}_{-0.6}$ & \nodata & $18\phd\phn$   & $23\phd\phn$
\enddata
\tablecomments{
\footnotesize
Measurements of the {\it Chandra} spectra in regions of interest; 
they are not corrected for absorption.
Column (1): Region of interest, as illustrated in Fig. \ref{fig:XrayRegions}. 
Column (2): Flux of the Fe K$\alpha$ line in units of $\rm 10^{-6}~ph~cm^{-2}~s^{-1}$.
The rest-frame energy measured from the XMM spectrum 
for this feature is $6.50^{+0.10}_{-0.05}$ keV. 
The value for the NE nucleus is a $2\sigma$ upper limit.
Column (3): Temperature (kT in units of keV) of the soft (0.5-2 keV) 
X-ray emission with $1\sigma$ uncertainties.
Column (4): Flux in the soft  (0.5-2 keV) X-ray band in units of 
$\rm 10^{-14}~erg~cm^{-2}~s^{-1}$; the estimated relative uncertainty is 20\%.
Column (5): Flux in the hard  (2-7 keV) X-ray band in units of 
$\rm 10^{-14}~erg~cm^{-2}~s^{-1}$; the estimated relative uncertainty is 30\%.
No flux is detected above 2 keV in the Northern Loop or in the Diffuse Region. 
Mrk 266 SW shows some emission above 2 keV, but most of the detected flux 
appears to originate in the Fe K$\alpha$ line. 
The NE and SW galaxies also both show strong Si XIII emission at 1.8-1.9 keV.}
\end{deluxetable}
}
\ifnum\Mode=0
\placetable{tbl:XrayData}
\else
\tableXrayData
\fi
%%%%%%%% End Table %%%%%%%%%%%

Spectral measurements and results from continuum model fitting of the {\it Chandra} 
and {\it XMM-Newton} observations are presented in Table \ref{tbl:XrayData}.
The sum of the emission in the elliptical regions misses some low surface-brightness 
emission that is recovered in measurements of the total system in a circular aperture 
with a radius of 33\arcsec. The total system fluxes
measured by {\it Chandra} are, within the uncertainties 
(i.e., 20\% in $F_x (soft)$ and 30\% in $F_x (hard)$), in agreement with
the XMM total flux measurements listed in Table \ref{tbl:XrayData}.
There may be some contribution of photoionized gas (by the AGNs) in the 
soft X-ray emission, but the S/N is insufficient to fit this component; therefore, 
the temperatures given in Table \ref{tbl:XrayData} are obtained from a model fit 
consisting of thermal emission due to star formation.
Mrk 266 NE has an AGN component, which is modeled by an absorbed power-law 
with photon index 1.8 (fixed) and $\rm N_H = 8.2^{+1.8}_{-1.3}\times10^{22}~cm^{-2}$;
correcting for this absorption suggests an intrinsic hard-band flux of 
$\rm 2.7\times10^{-13}~erg~cm^{-2}~s^{-1}$. There is a tail of hard X-ray emission
in the source between the nuclei, but its spectral shape is not well constrained; 
the 2-7 keV flux given here is mostly due to this hard tail.

\subsection{Ultraviolet Imaging with GALEX and XMM-Newton OM}\label{subsec:UVimaging}

Ultraviolet images  of Mrk 266 were obtained on 
2004 June 04 with an integration time of 111 sec in the
NUV (2271 \AA) and FUV (1528 \AA) bands
during the All Sky Imaging Survey (AIS) conducted by the 
Galaxy Evolution Explorer (GALEX). The data were
extracted from MAST (tile AISCHV3\_104\_01135) as provided
by the standard GALEX data processing pipeline, and photometry was
performed directly from these images using procedures outlined by
\citet{2010ApJ...715..572H}.

Contemporaneously with the X-ray observations by {\it XMM-Newton}
summarized above,  the Optical/UV Monitor Telescope (XMM-OM) 
obtained observations of Mrk 266 in the U (3000 - 3900 \AA), 
UVW1 (2450-3200 \AA), and UVW2 (1800-2250 \AA) filters with
exposure times of 800, 1000 and 1400 sec, respectively. 
The XMM-OM instrument and data characteristics are described by 
\citet{2001AA...365L..36M}.
These images have significantly better spatial resolution
($\sim2$\arcsec\ FWHM) and higher sensitivity than the AIS observations 
from GALEX, and they also considerably extend the UV coverage.
The reduced and calibrated XMM-OM images, as produced using
the methods documented by \citet{2008PASP..120..740K},
were obtained from MAST(dataset \#0055990501).
Photometric measurements (presented in \S\ref{subsubsec:SEDdata}) were 
calibrated using zero-points in Table 19 of the {\it XMM-Newton} Users 
Handbook,\footnote{http://heasarc.gsfc.nasa.gov/docs/xmm/uhb/XMM\_UHB.html}
which are 19.1890, 18.5662 and 16.5719 AB mag in the 
U, UVW1 and UVW2 filters, respectively. 
Figure \ref{fig:UV_GALEX_XMM} displays the UV imagery of 
Mrk 266 in the FUV band of GALEX and in three filters of XMM-OM.

%%%%%%%% Begin Figure %%%%%%%%%%%
%UV images: GALEX FUV, XMM-OM U, UVW1, UVW2
\def\figcapUVGALEXXMM{
\footnotesize 
UV imagery of Mrk 266.
(a) FUV (1528 \AA) band of GALEX; 
(b) UVW2 (1800-2250 \AA) band of XMM-OM;
(c) UVW1 (2450-3200 \AA) band of XMM-OM;
(d) U (3000 - 3900 \AA) band of XMM-OM.
}
\ifnum\Mode=0 %Insert Figure/Table only in [preprint] or [preprint2] modes
\placefigure{fig:UV_GALEX_XMM}
\begin{verbatim} fig07a_07d\end{verbatim}
\else
\begin{figure}[!htb]
\center
\ifnum\Mode=2 
\includegraphics[width=1.0\columnwidth,angle=0]{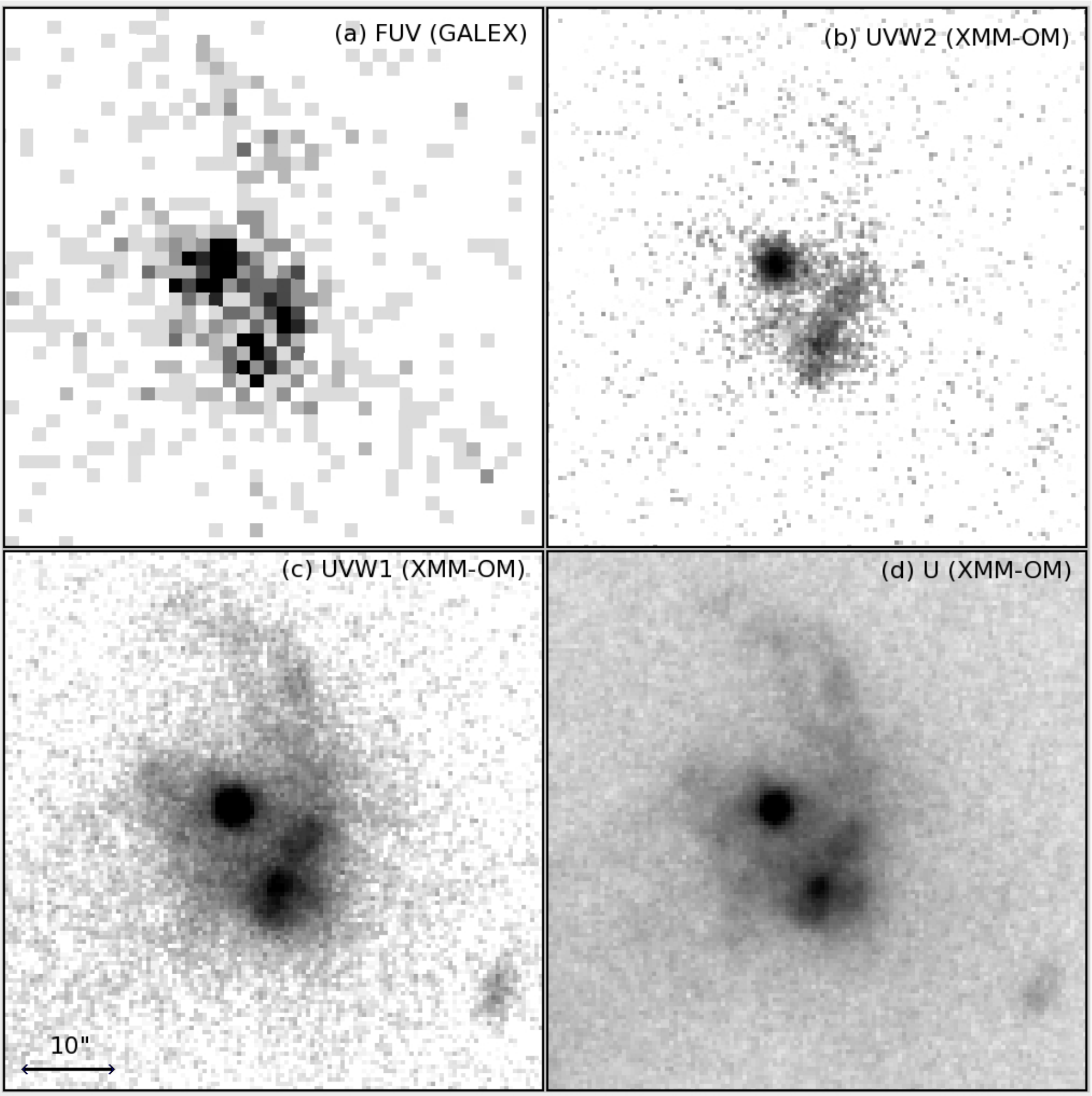} 
\else
\includegraphics[scale=0.5,angle=0]{fig07a_07d} 
\fi
\caption{\figcapUVGALEXXMM \label{fig:UV_GALEX_XMM}}
\end{figure}
\fi %close \ifnum\Mode=0 
%%%%%%%% End Figure %%%%%%%%%%%

\subsection{Mid- and Far-Infrared Imaging with Spitzer}
\label{subsec:SpitzerImaging}

Imaging observations of Mrk 266 were obtained with the {\it Spitzer Space Telescope}
as part of the Cycle 1 General Observing program for GOALS (PID 3672; PI J. Mazzarella).
The Infrared Array Camera \citep[IRAC,][]{2004ApJS..154...10F} onboard {\it Spitzer} 
was used to image the source at 3.6, 4.5, 5.8 and 8.0 \micron\ (AOR \#12305408). 
Due to the presence of bright galactic nuclei, the 
high dynamic range (HDR) mode was utilized in all four bands.
The HDR frames were comprised of 1-2 second exposures to 
calibrate pixels that may have saturated in the 30-second exposures.
The IRAC observations involved dithering to achieve
redundancy to correct for cosmic ray hits and bad pixels.
The Multiband Imaging Photometer for Spitzer
\citep[MIPS,][]{2004ApJS..154...25R} was utilized to image Mrk 266
at 24, 70 and 160 \micron\ using the Photometry
and Super Resolution mode (AOR \#12349184).
Each observation consisted of multiple 3-second
integrations to insure adequate redundancy.  Mapping
cycles were used at each wavelength to correct the data for transients
in the detectors. Details of the post-pipeline data processing 
and calibration of these data are given in Mazzarella et al. 2012 (in preparation).
Various {\it Spitzer}  images and photometric measurements
for Mrk 266 are incorporated into the following sections in
conjunction with the analyses of data in other spectral regions.

\subsection{Mrk 266 from Radio Through X-Rays}
\label{subsec:MultiWavelength}

\subsubsection{Multi-Waveband Imagery}
\label{subsubsec:MultiImages}

%%%%%%%% Begin Figure %%%%%%%%%%%
%Mrk 266 in 9 Bands
\def\figcapNineBands{
\footnotesize 
Imagery of Mrk 266 spanning radio through X-rays.  The images are 
displayed at their native resolution (no PSF matching) and registered 
co-spatially within the uncertainty of the astrometry in each bandpass. 
Each panel shows the same 32\farcs5 X 43\farcs7 field of view. 
Labeled in each panel are the bandpass and telescope (upper left),
and a scale bar indicating  5\arcsec (3.1 kpc).
From the upper left to the lower right, each panel shows contours of the
6 cm radio continuum emission \citep{1988ApJ...333..168M} superposed 
on a grayscale representation of the following: 
(a) 20 cm radio continuum from the VLA; 
(b) 70 \micron\ {\it Spitzer} MIPS;
(c) 24 \micron\ {\it Spitzer} MIPS;
(d) 8 \micron\ {\it Spitzer} IRAC;
(e) 1.6 \micron\ {\it HST} NICMOS;
(f) I band (0.81 \micron, F814W) {\it HST} ACS;
(g) B band (0.44 \micron, F435W) {\it HST} ACS;
(h) U band (0.35 \micron) {\it XMM-OM};
(i) X-rays (0.4 - 2 keV) {\it Chandra}.
}
\ifnum\Mode=0 %Insert Figure/Table only in [preprint] or [preprint2] modes
\placefigure{fig:NineBands}
\begin{verbatim}fig08a_08i\end{verbatim}
\else
\ifnum\Mode=2 
\begin{figure*}[t]
\else
\begin{figure}[t]
\fi
\center
\includegraphics[width=6.4truein,angle=0]{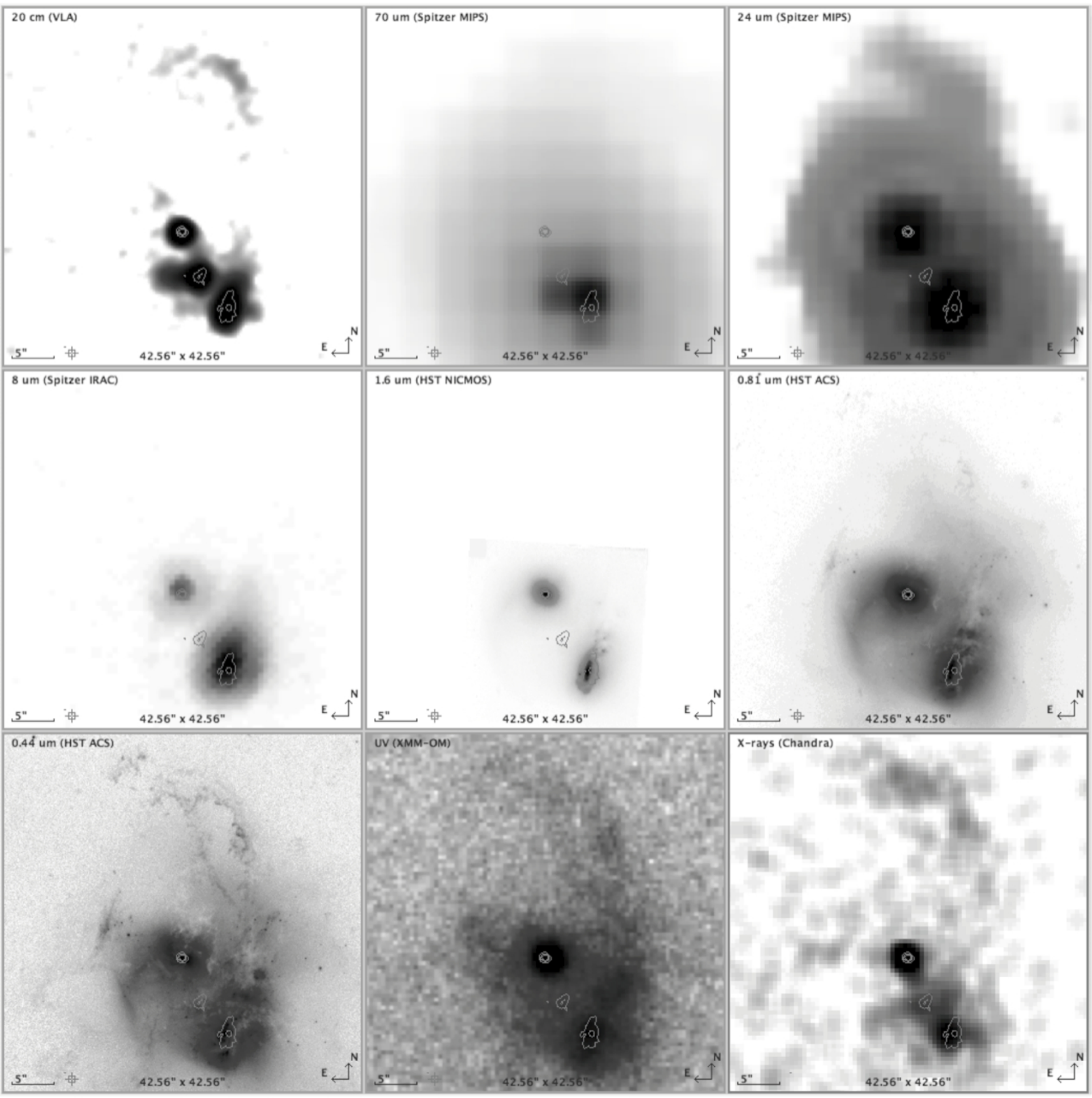}
%These \parbox coordinates go with \includegraphics[width=6.4truein]
\put(-322,450) {\parbox{10cm}{(a)}} \put(-170,450) {\parbox{10cm}{(b)}} \put(-15,450) {\parbox{10cm}{(c)}}
\put(-322,295) {\parbox{10cm}{(d)}} \put(-170,295) {\parbox{10cm}{(e)}} \put(-15,295) {\parbox{10cm}{(f)}}
\put(-322,142) {\parbox{10cm}{(g)}} \put(-170,142) {\parbox{10cm}{(h)}} \put(-15,142) {\parbox{10cm}{(i)}}
\caption{\figcapNineBands \label{fig:NineBands}}
\ifnum\Mode=2
\end{figure*}
\else
\end{figure}
\fi
%\ifnum\Mode=2 \twocolumn \fi %[preprint2] mode only
\fi %close \ifnum\Mode=0 
%%%%%%%% End Figure %%%%%%%%%%%

Figure \ref{fig:NineBands} shows the structure of Mrk 266 in nine spectral 
regions spanning radio through X-rays using the observations discussed above.
All images are displayed with the same orientation and scale to 
facilitate comparison. In each panel, contours of the 6 cm 
(4.885 GHz, 0\farcs3 x 0\farcs4 beam) image from \citet{1988ApJ...333..168M} 
are superposed, primarily for astrometric reference. It is noteworthy that
the elongated radio continuum structure with double peaks  observed between
the galaxy nuclei, as well as the elongated morphology of the
SW galaxy, was confirmed in more sensitive 8.4 GHz observations 
with 0\farcs27 x 0\farcs22 resolution \citep[][]{2006A&A...455..161L}.
%Fig. A.3

\subsubsection{Photometry and Spectral Energy Distributions} \label{subsubsec:SEDdata}

Using the data displayed in Figure \ref{fig:NineBands} and other 
passbands discussed above, aperture photometry was performed
in elliptical regions parameterized in the Appendix 
(Table \ref{tbl:Regions} and Figure \ref{fig:Regions}). However, an alternative 
measurement procedure was necessary for the lower resolution far-infrared data.
The spatial resolution of the MIPS images, determined by the size of the
{\it Spitzer} PSF ($\lambda$/D) are 6\arcsec, 18\arcsec, and 40\arcsec\ FWHM 
at 24, 70 and 160~\micron\ respectively.
The 24~\micron\ IRAC image (Fig. \ref{fig:NineBands}c) clearly resolves the 
two galaxies, and it also reveals dust emission from the Northern Loop
which is analyzed in \S\ref{subsec:SuperwindDust}.
The 10\arcsec\ angular separation of the two galaxies
is too small to be resolved by MIPS at 70 and 160~\micron.
However,  as shown in Figure \ref{fig:NineBands}b, 
the centroid of the 70 \micron\ emission clearly indicates that the SW
galaxy dominates the far-infrared emission from the system.
The centroid of the 160~\micron\ emission is also consistent
with the SW galaxy providing the majority of the flux.

The 70 and 160 \micron\ flux densities for the two galaxies 
were estimated as follows: The relation derived by \citet{2010ApJ...713L.330X},
{\footnotesize
\begin{equation}\label{eqn:Lir}
\rm log(L_{ir}) = log(L_{24}) + 0.87(\pm0.03) + 0.56(\pm0.09)\times log(L_{8}/L_{24}), 
\end{equation}
}
\noindent was used with measured monochromatic luminosities at 8 and 24 \micron\ to 
derive \Lir\ (1 - 1000 \micron) estimates of $2.3\times10^{11}$, $6.6\times10^{10}$,
and $3.4\times10^{11}$ \Lsun\ for Mrk 266 SW, NE, and the total system respectively. 
Then, the ratios of $\rm L_{70}/L_{ir} = 0.58~and~L_{160}/L_{ir} = 0.19$ 
measured for the total system were applied to estimate the 70 and 160 \micron\ flux 
densities for the two galaxies (Table \ref{tbl:SED_Regions}).
The estimated uncertainty in the fluxes derived using this procedure is $\approx$20\%.
Comparison of the summation of the estimated component flux densities with the 
measured total 70 and 160 \micron\ flux densities for the total system 
(Table \ref{tbl:SED_Tot}) indicates consistency within the uncertainties.

Appendix Tables \ref{tbl:SED_Tot} and \ref{tbl:SED_Regions} present
photometry for the total system and  for the components (SW, NE, 
the central region between the nuclei, and the Northern Loop).
Figure \ref{fig:SEDs} displays the resulting Spectral Energy 
Distributions (SEDs).

%%%%%%%% Begin Figure %%%%%%%%%%%
%SEDs
\def\figcapSEDs{
\footnotesize 
SEDs for Mrk 266 SW (red), Mrk 266 NE (blue), the Central Region (green) between the nuclei, 
the Northern Loop (purple), and the integrated emission from the Total System (brown). 
The legend shows the color coding used to identify the various components. 
Most of the photometric measurements are presented here for the first time
and some data were selected from NED, as listed in
Tables \ref{tbl:SED_Tot} and \ref{tbl:SED_Regions}.
The monochromatic fluxes plotted are computed via $\rm log(\nu f_{\nu} [Jy Hz])$.
Conversions to other units are as follows:
$\rm log(\nu f_{\nu} [W m^{-2}]) = log(\nu f_{\nu} [Jy Hz]) - 26.00$;
$\rm log(\nu L_{\nu} [W]) = log(\nu f_{\nu} [W m^{-2}]) + 50.30$;
$\rm log(\nu L_{\nu} [L_{\odot}]) = log(\nu f_{\nu} [W]) - 26.58$.
}
\ifnum\Mode=0 %Insert Figure/Table only in [preprint] or [preprint2] modes
\placefigure{fig:SEDs}
\begin{verbatim}fig09\end{verbatim}
\else
\ifnum\Mode=2
\begin{figure*}
\center
\includegraphics[width=1.0\textwidth,angle=0]{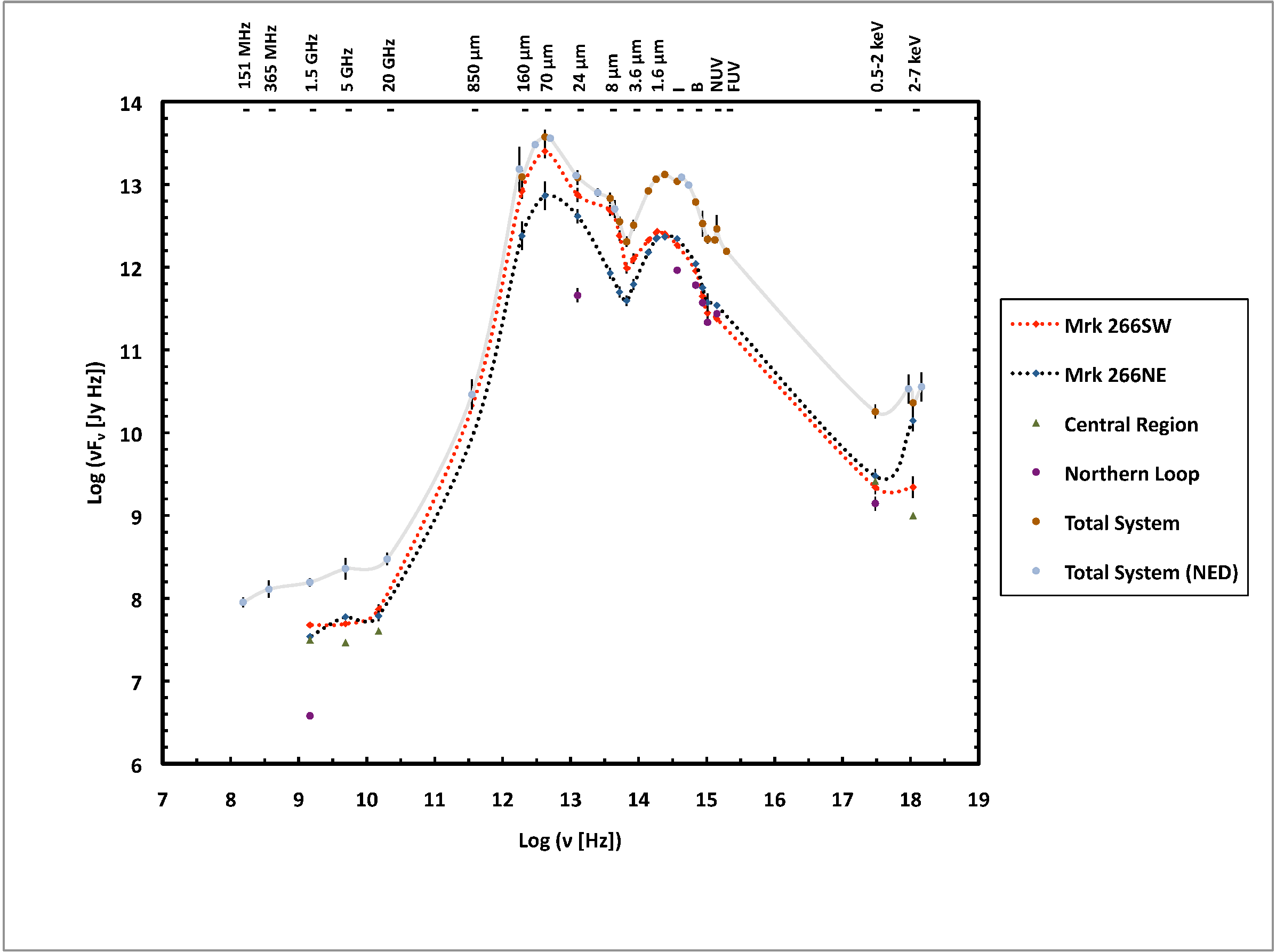} 
%Use with {figure*}
\else
\begin{figure}[!t]
\center
\includegraphics[width=0.95\textwidth,angle=0]{fig09} 
%Use with {figure}
\fi
\caption{\figcapSEDs \label{fig:SEDs}}
\ifnum\Mode=2
\end{figure*}
\else
\end{figure}
\fi
\fi %close \ifnum\Mode=0 
%%%%%%%% End Figure %%%%%%%%%%%

\subsection{Mid-Infrared Spectroscopy with Spitzer}
\label{sec:IRSdata}

Mid-infrared spectroscopy of Mrk 266 obtained in staring mode with the
Infrared Spectrograph \citep[IRS,][]{2004ApJS..154...18H} onboard {\it Spitzer} 
have been published by \citet{2006ApJ...653.1129B}, \citet{2007ApJ...664...71D}, 
and \citet{2009ApJS..184..230B}. Unfortunately, these observations suffer from 
aperture affects that resulted in missing emission 
from some components in different spectral regions.
%In the low spectral resolution (R $\sim$100) data, the Short-Low (5.2-14.5 \micron) 
%slit contained the NE nucleus (off center) but missed the SW nucleus and the central 
%radio/X-ray source, and the Long-Low (14--38 \micron) slit contained the SW nucleus 
%and the radio/X-ray source between the nuclei but missed the NE nucleus.
%Likewise, in the high spectral resolution (R$\sim$600) data, the Short-High 
%(9.9 - 19.6 \micron) and Long-High (18.7-37.3~\micron) slits contained the 
%SW nucleus and the central radio source, but missed the NE nucleus.
Appendix \S\ref{subsec:PublishedIRSdata} provides clarification of which 
emission features have been measured for each component of Mrk 266.

In order to overcome the limited spatial extent of the staring-mode observations,
the IRS spectral map obtained on 2005 January 10 available in the {\it Spitzer} 
archive (program P03269; PI J. Gallimore; AOR \#12459264) was used 
to construct spectra for each nucleus and for the total system. 
Spectral mapping with the SL module consisted of 13 integrations,
each 6 seconds in duration and stepped perpendicular to the slit with 
an interval of 1\farcs8 (half the slit width). Mapping with the LL module
was performed with 5 integrations, each 6 seconds in duration and 
stepped perpendicular to the slit with an interval of 4\farcs85 (half the slit width).
The individual calibrated spectra produced by version S15.3 of the 
{\it Spitzer} Science Center pipeline were used to construct a spectral cube with the 
CUBISM package \citep{2007PASP..119.1133S}. In all cases, background observations 
were assembled from the non-primary slit and subtracted from the data.
CUBISM was used to construct images over specific wavelength ranges and spectral 
features, including 5.5, 10, and 14 \micron\ continuum and continuum-subtracted 
6.2, 7.7, and 11.3 \micron\ PAH, 9.7 \micron\ $\rm H_2$ S(3), and 
[S IV] 10.5 \micron\ emission lines. These images are presented
in Figure \ref{fig:IRScube}, where the first panel illustrates rectangular apertures
used in CUBISM to extract 1-d spectra.

%%%%%%%% Begin Figure %%%%%%%%%%%
%Image slices from the IRS low-res cube
%IRScube.ps
\def\figcapIRScube{
\footnotesize 
Image slices through the 2-d spectral IRS map. Panels (a) - (h) show images
in the indicated spectral features (pure line or continuum emission).
All images have north up and east left, and have
the same scale indicated by the bar in panel (a). The colored rectangles overlaid on the 
5.5 \micron\ image in panel (a) indicate the regions of the 2-d spectral cube
used to extract 1-d spectra.
Brown box: ``total'' emission in the SL$+$LL modules plotted in Figure 
\ref{fig:IRScube_SLLL_Total} (Mrk 266NE was not fully covered by the LL aperture); 
orange box: ``total'' emission in the SL module plotted in 
Figure \ref{fig:IRScube_SL}a;
red box: emission from the NE nucleus in the SL module plotted in 
Figure \ref{fig:IRScube_SL}b;
blue box: emission from the SW nucleus in the SL module plotted in 
Figure \ref{fig:IRScube_SL}c.
}
\ifnum\Mode=0 %Insert Figure/Table only in [preprint] or [preprint2] modes
\placefigure{fig:IRScube}
\begin{verbatim} fig10a_10h\end{verbatim}
\else
\begin{figure*}[!tb]
\center
\ifnum\Mode=2 
\includegraphics[scale=0.46,angle=0]{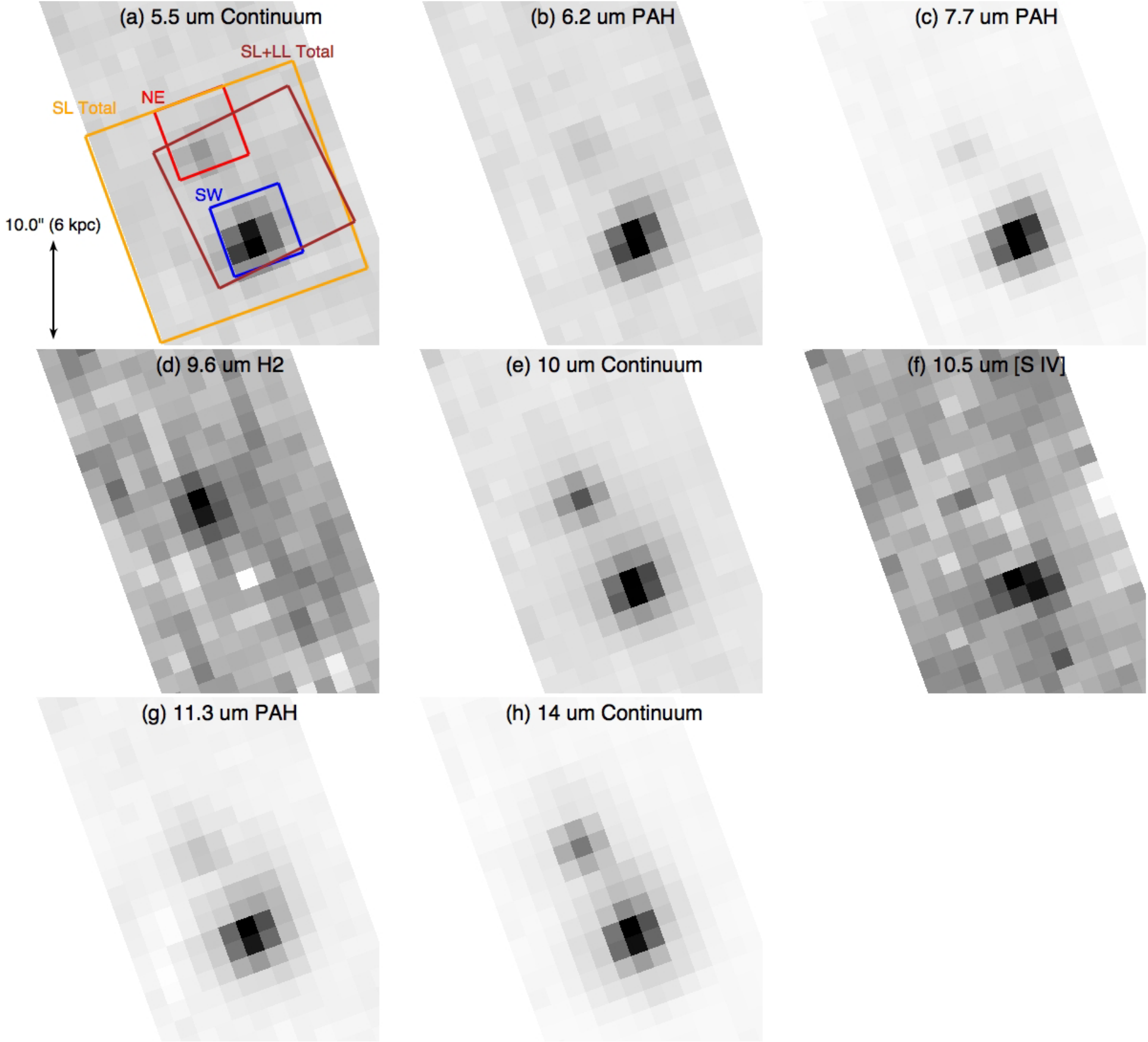} %with \begin{figure*}
\else
\includegraphics[scale=0.80,angle=0]{fig10a_10h} 
\fi
\caption{\figcapIRScube \label{fig:IRScube}}
\end{figure*}
\fi %close \ifnum\Mode=0 
%%%%%%%% End Figure %%%%%%%%%%%

%%%%%%%% Begin Figure %%%%%%%%%%%
%IRS Low-Res 1-d extractions 
\def\figcapIRScubeSLLLTotal{
\footnotesize  
1-d spectrum of Mrk 266 extracted from a low-resolution spectral map 
constructed from the SL and LL modules of the IRS. The region
used to produce this spectrum is shown as a brown box in
Figure \ref{fig:IRScube}. This is our best attempt to represent
the integrated emission for both galaxies across the full 5-38 \micron\
range covered by the low-resolution modules of the IRS, 
with the caveats noted in the text. The colors in this spectrum
represent the various orders in the IRS instrument and are not
related to the colors in Fig.  \ref{fig:IRScube}.
}
\ifnum\Mode=0 %Insert Figure/Table only in [preprint] or [preprint2] modes
\placefigure{fig:IRScube_SLLL_Total}
\begin{verbatim}fig11\end{verbatim}
\else
\begin{figure}[!htb]
\center
\ifnum\Mode=2 
\includegraphics[scale=0.45,angle=0]{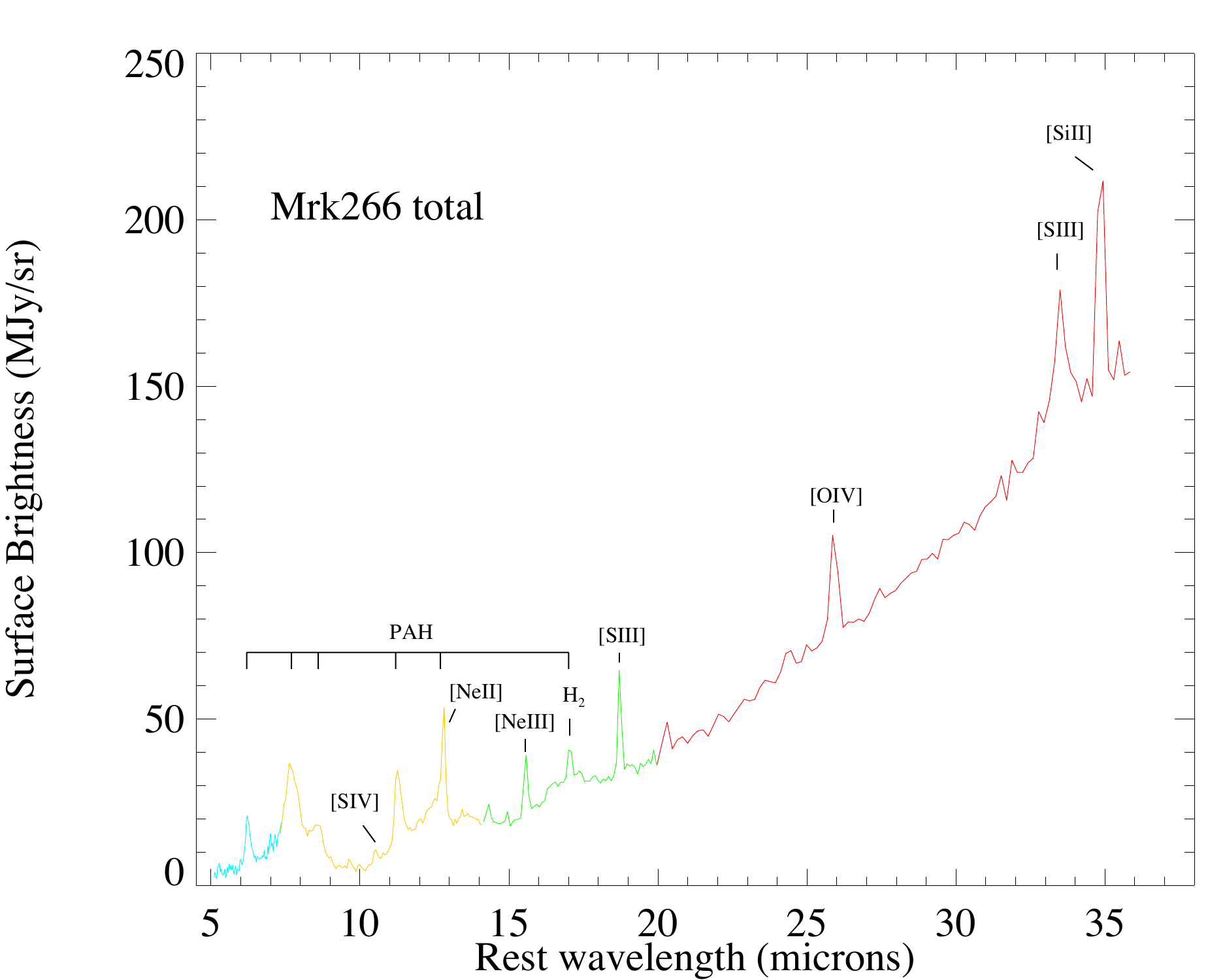} 
\else
\includegraphics[scale=0.8,angle=0]{fig11} 
\fi
\caption{\figcapIRScubeSLLLTotal \label{fig:IRScube_SLLL_Total}} 
\end{figure}
%\ifnum\Mode=2 \twocolumn \fi %[preprint2] mode only
\fi %close \ifnum\Mode=0 
%%%%%%%% End Figure %%%%%%%%%%%

%%%%%%%% Begin Figure %%%%%%%%%%%
%IRS Low-Res 1-d extractions 
%mrk266_spectra2B
\def\figcapIRScubeSL{
\footnotesize 
The spatial resolution of the SL observations permit a clean separation of 
the two galaxies that is not possible with the LL observations. Each panel is a 
1-d extraction from the SL spectral map presented in Figure \ref{fig:IRScube}:
(a) orange region covering the entire region; 
(b) red region containing Mrk 266 NE; 
(c) blue region containing Mrk 266 SW.
}
\ifnum\Mode=0 %Insert Figure/Table only in [preprint] or [preprint2] modes
\placefigure{fig:IRScube_SL}
\begin{verbatim}fig12a_12c\end{verbatim}
\else
%For preprint
%\ifnum\Mode=2 \onecolumn \fi %4-panel fig is too small in 2-column mode
\begin{figure}[!hbt]
\center
\ifnum\Mode=2 
\includegraphics[width=1.0\columnwidth,angle=0]{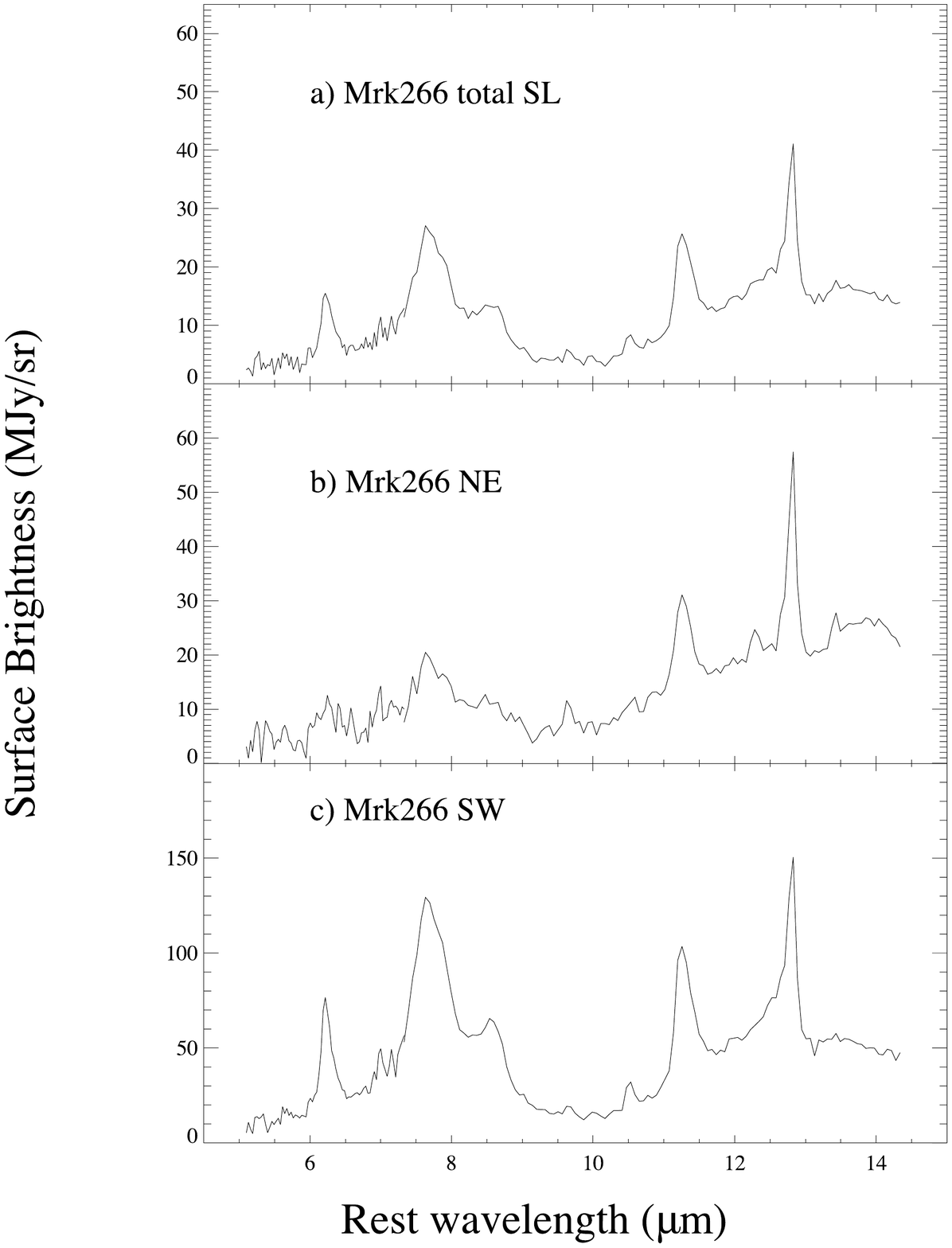}
%In 2-column mode, full-scale works fine
\else
\includegraphics[scale=0.8,angle=0]{fig12a_12c}%In 1-column mode, it must be scaled down
\fi
\caption{\figcapIRScubeSL \label{fig:IRScube_SL}}
\end{figure}
%\ifnum\Mode=2 \twocolumn \fi %[preprint2] mode only
\fi %close \ifnum\Mode=0 
%%%%%%%% End Figure %%%%%%%%%%%

%%%%%%%% Begin Table %%%%%%%%%%%
\def\tableIRScube{
\begin{deluxetable}{lll}
\ifnum\Mode=2
\renewcommand\arraystretch{0.5}% (MyValue=1.0 is for standard spacing)
\tabletypesize{\scriptsize}
\setlength{\tabcolsep}{0.0in} %Tighten up the columns. See AASTeX FAQ
\tablewidth{\columnwidth}
\else
\renewcommand\arraystretch{0.75}% (MyValue=1.0 is for standard spacing)
\tabletypesize{\footnotesize} 
\setlength{\tabcolsep}{0.25in} %Tighten up the columns. See AASTeX FAQ
%\tablewidth{\textwidth}
\fi
\tablecaption{Measurements from Low-Res Spectral Mapping \label{tbl:IRScube}}
\tablehead{
\colhead{Feature} & 
\colhead{Flux ($\rm 1\times 10^{-20}~W~cm^{-2}$)} &
\colhead{EQW (\micron)} \\
\colhead{(1)} &
\colhead{(2)} &
\colhead{(3)} \\
}
\tablecolumns{3}
\startdata
%%%%%%%%%
\cutinhead{Mrk 266 Total\tablenotemark{a}}
$\rm 6.2 \mu m~PAH$&	$\rm \phn 40.2\phn\phn\phn (\pm1.8)$\tablenotemark{p}	&\rm $0.47 (\pm0.03)$\tablenotemark{s}	\\
$\rm H_2~S(5)~6.91~\mu m$&	$\rm \phn\phn 0.66\phn\phn (\pm0.50)$\tablenotemark{p}	&\nodata	\\
$\rm [Ar~II]~6.99 \mu m$&	$\rm \phn\phn 2.34\phn\phn (\pm0.72)$\tablenotemark{p}	&\nodata	\\
$\rm 7.7 \mu m~PAH~Complex$&	$\rm  163.0\phn\phn\phn (\pm6.1)$\tablenotemark{p}	&\rm $0.46 (\pm0.02)$\tablenotemark{s}	\\
$\rm 8.3 \mu m ~PAH$&	$\rm \phn 13.7\phn\phn\phn (\pm1.8)$\tablenotemark{p}	&\nodata	\\
$\rm 8.6 \mu m~PAH$&	$\rm \phn 26.8\phn\phn\phn (\pm1.5)$\tablenotemark{p}	&\rm $0.15 (\pm0.01)$\tablenotemark{s}	\\
$\rm [Ar~III]~9.02 \mu m$&	$\rm \phn\phn 0.80\phn\phn (\pm0.49)$\tablenotemark{p}	&\nodata	\\
$\rm H_2~S(3)~9.65 \mu m$&	$\rm \phn\phn 1.42\phn\phn (\pm0.38)$\tablenotemark{p}	&\nodata	\\
$\rm [S~IV]~10.51 \mu m$&	$\rm \phn\phn 1.59\phn\phn (\pm0.29)$\tablenotemark{p}	&\nodata	\\
$\rm 11.3 \mu m~PAH~Complex$&	$\rm \phn 31.9\phn\phn\phn (\pm1.2)$\tablenotemark{p}	&\rm $0.51 (\pm0.01)$\tablenotemark{s}	\\
$\rm 12.0 \mu m~PAH$&	$\rm \phn 11.85\phn\phn (\pm0.83)$\tablenotemark{p}	&\nodata	\\
$H2~S(2)~12.27 \mu m$&	$\rm \phn\phn 0.89\phn\phn (\pm0.15)$\tablenotemark{p}	&\nodata	\\
$\rm 12.6 \mu m~PAH~Complex$&	$\rm \phn 23.78\phn\phn (\pm0.73)$\tablenotemark{p}	&\nodata	\\
$\rm [Ne~II]~12.81 \mu m$&	$\rm \phn\phn 6.78\phn\phn (\pm0.18)$\tablenotemark{p}	&\nodata	\\
$\rm 13.6 \mu m~PAH$&	$\rm \phn\phn 8.86\phn\phn (\pm0.70)$\tablenotemark{p}	&\nodata	\\
$\rm 14.2 \mu m~PAH$&	$\rm \phn\phn 2.80\phn\phn (\pm0.52)$\tablenotemark{p}	&\nodata	\\
$\rm [Ne~III]~15.56 \mu m$&	$\rm \phn\phn 3.57\phn\phn (\pm0.24)$\tablenotemark{p}	&\nodata	\\
$\rm 16.4 \mu m~PAH$&	$\rm \phn\phn 2.08\phn\phn (\pm0.38)$\tablenotemark{p}	&\nodata	\\
$\rm H_2~S(1)~17.04 \mu m$&	$\rm \phn\phn 1.61\phn\phn (\pm0.19)$\tablenotemark{p}	&\nodata	\\
$\rm 17 \mu m~PAH~Complex$&	$\rm \phn 19.4\phn\phn\phn (\pm1.7)$\tablenotemark{p}	&\nodata	\\
$\rm 17.4 \mu m~PAH$&	$\rm \phn\phn 1.01\phn\phn (\pm0.36)$\tablenotemark{p}	&\nodata	\\
$\rm [S~III]~18.71 \mu m$&	$\rm \phn\phn 4.16\phn\phn (\pm0.28)$\tablenotemark{p}	&\nodata	\\
$\rm [O~IV]~25.9 \mu m$&	$\rm \phn\phn 4.52\phn\phn (\pm0.14)$\tablenotemark{p}	&\nodata	\\
$\rm [S~III]~33.51 \mu m$&	$\rm \phn\phn 3.71\phn\phn (\pm0.28)$\tablenotemark{p}	&\nodata	\\
$\rm [Si~II]~34.86 \mu m$&	$\rm \phn\phn 5.69\phn\phn (\pm0.39)$\tablenotemark{p}	&\nodata	\\
%%%%%%%%%
\cutinhead{Mrk 266 SW\tablenotemark{b}}
$\rm 6.2 \mu m~PAH$&	$\rm \phn 17.78\phn\phn (\pm0.35)$\tablenotemark{s}	&\rm $0.67 (\pm0.03)$\tablenotemark{s}	\\
$\rm 7.7 \mu m~PAH~Complex$&	$\rm \phn 30.46\phn\phn (\pm0.35)$\tablenotemark{s}	&\rm $0.54 (\pm0.01)$\tablenotemark{s}	\\
$\rm 8.6 \mu m~PAH$&	$\rm \phn\phn 5.45\phn\phn (\pm0.14)$\tablenotemark{s}	&\rm $0.17 (\pm0.01)$\tablenotemark{s}	\\
$\rm [S~IV]~10.51 \mu m$&	$\rm \phn\phn 0.95\phn\phn (\pm0.13)$\tablenotemark{s}	&\nodata	\\
$\rm 11.3 \mu m~PAH~Complex$&	$\rm \phn 10.376\phn (\pm0.088)$\tablenotemark{s}	&\rm $0.73 (\pm0.01)$\tablenotemark{s}	\\
$\rm [Ne~II]~12.81 \mu m$&	$\rm \phn\phn 6.15\phn\phn (\pm0.99)$\tablenotemark{s}	&\nodata	\\
%%%%%%%%%
\cutinhead{Mrk 266 NE\tablenotemark{b}}
$\rm 6.2 \mu m~PAH$&	$\rm \phn\phn 2.418\phn (\pm0.019)$\tablenotemark{s}	&\rm $0.4\phn (\pm0.1)$\tablenotemark{s}	\\
$\rm 7.7 \mu m~PAH~Complex$&	$\rm \phn\phn 2.86\phn\phn (\pm0.28)$\tablenotemark{s}	&\rm $0.25 (\pm0.03)$\tablenotemark{s}	\\
$\rm 8.6 \mu m~PAH$&	$\rm \phn\phn 0.69\phn\phn (\pm0.11)$\tablenotemark{s}	&\rm $0.10 (\pm0.02)$\tablenotemark{s}	\\
$\rm H_2~S(3)~9.65 \mu m$&	$\rm \phn\phn 0.52\phn\phn (\pm0.14)$\tablenotemark{s}	&\nodata	\\
$\rm 11.3 \mu m~PAH~Complex$&	$\rm \phn\phn 2.170\phn (\pm0.078)$\tablenotemark{s}	&\rm $0.34 (\pm0.02)$\tablenotemark{s}	\\
$H2~S(2)~12.27 \mu m$&	$\rm \phn\phn 0.308\phn (\pm0.063)$\tablenotemark{s}	&\nodata	\\
$\rm [Ne~II]~12.81 \mu m$&	$\rm \phn\phn 1.84\phn\phn (\pm0.29)$\tablenotemark{s}	&\nodata	
\enddata
\tablenotetext{a}{
Spectral measurements for the integrated emission from both galaxies in Mrk 266.
Features in the 5.2-14.5 \micron\  (SL) and 14.5-38 \micron\ (LL) regions were 
measured from the 1-d spectrum plotted in Figure \ref{fig:IRScube_SLLL_Total}; 
this region is 4x4 LL pixels in size, where each pixel is 5\farcs1 x 5\farcs1,
as outlined in brown in Figure \ref{fig:IRScube}(a).}
\tablenotetext{b}{
Spectral features measured from the SL spectra of Mrk 266 SW and NE
as plotted in Figure \ref{fig:IRScube_SL}c and \ref{fig:IRScube_SL}b.
These 1-d spectra were extracted from the blue and red regions of the 
spectral map shown in Figure \ref{fig:IRScube}, which have dimensions 
of 5x5 SL pixels, where each pixel is 1\farcs8 x 1\farcs8. 
}
\tablenotetext{p}{
These line fluxes were measured using PAHFIT and include an extinction
correction based on a fit with a $\tau_{9.7\mu m} = 0.30$ dust screen, 
which corresponds to $\rm A_V \approx 5.7$ mag.
}
\tablenotetext{s}{
EQW measurements for the PAH features were measured using a spline-fitting
technique for estimating the local continuum. Spline fitting was also used to estimate
fluxes of features in the SL spectra of the individual galaxies which could not be measured
with PAHFIT. See text for details.
}
\end{deluxetable}
}
\ifnum\Mode=0
\placetable{tbl:IRScube}
\else
\tableIRScube
\fi
%%%%%%%% End Table %%%%%%%%%%%

With a pixel size of 5\farcs1x5\farcs1 and a spatial resolution of 10\arcsec\ 
on the red end, we cannot use the LL observations (14-38 \micron)
to separate the two galaxies as we can with the SL data (5.2-14.5 \micron). 
Further, the LL map did not completely cover the NE nucleus.  
A ``total'' SL$+$LL 1-d spectrum was extracted that contains most of the 
integrated emission of the system over 5-38 \micron\ in a region 4x4 LL 
pixels in size (outlined in brown in Figure \ref{fig:IRScube}a).
Another 1-d spectrum was extracted that fully covers both galaxies 
in the 5-14 \micron\ SL spectrum in a region 13x13 SL pixels in size, 
where each pixel is 1\farcs8x1\farcs8.
The resulting 1-d spectra, representing the sum of the emission 
from both galaxies, are presented in
Figure \ref{fig:IRScube_SLLL_Total} (SL$+$LL) and in 
Figure \ref{fig:IRScube_SL}a (SL only), respectively.
In addition, 1-d extractions from the SL map were constructed that provide 
a clean separation between the NE and SW nuclei; these regions are 
shown as red and blue rectangles (each covering 5x5 SL pixels) in 
Figure \ref{fig:IRScube}a, and the resulting 1-d spectra are presented 
in Figures \ref{fig:IRScube_SL}b and \ref{fig:IRScube_SL}c.
Due to the very high mid-infrared surface brightness of the galaxies, with a 
spatial resolution of 3\farcs6 (two 1\farcs8 pixels), even the SL 
spectral map does not enable a reliable separation of the nuclear emission 
from the diffuse emission surrounding the nuclei.

This is the first time mid-infrared spectra are available for each nucleus in 
Mrk 266 separately, as well as the integrated emission from both galaxies.
Table \ref{tbl:IRScube} lists measured line fluxes and equivalent widths for
the total system and, where possible, the individual galaxies.
The fluxes were converted from surface brightness units ($\rm W~m^{-2}~sr^{-1}$) 
to $\rm W~cm^{-2}$ by multiplying by the extraction aperture.
Most of the spectral features in the SL$+$LL spectrum of the integrated emission 
from the galaxy pair (Fig. \ref{fig:IRScube_SLLL_Total}) were measured 
using the PAHFIT package \citep{2007ApJ...656..770S}, which includes 
decomposition of various spectral components and an extinction correction 
based on a fit with a $\tau_{9.7\mu m} = 0.30$ dust screen, corresponding 
to $\rm A_V \approx 5.7$ mag.
However, since the nuclei were resolved only in the SL data (Figure 
\ref{fig:IRScube_SL}b and \ref{fig:IRScube_SL}c), and PAHFIT produces most 
reliable continuum and feature fits when SL and LL are combined, we have used 
a spline-fitting technique  \citep{2001A&A...370.1030H,2002A&A...390.1089P} 
to measure the SL features listed in Table \ref{tbl:IRScube} for the SW and NE galaxies.
This further allows for direct comparison to other sources in the literature where this 
technique has been employed.
Spline-fitting was also utilized for the PAH EQW measurements to estimate 
the local continuum and the silicate absorption strength. 
As noted by \citet[][]{2007ApJ...656..770S}, spline fitting systematically 
underestimates PAH fluxes compared to full spectral decomposition via PAHFIT. 
The various measurements are identified in Table \ref{tbl:IRScube}
with superscripts $p$ (PAHFIT) and $s$ (spline fitting), respectively.

\section{Analysis \& Discussion}
\label{sec:Discussion}

Mrk 266 is a complex collection of many different physical subsystems, each of which are identified in 
Fig. \ref{fig:HST_ACS4x} and discussed in the following subsections.  At the macro level, Mrk 266 is clearly a merging system with two distinct AGNs coupled with strong spatial differences in current star formation, outflow, merger dynamics and AGN heating.  But within that framework there are several specific questions that we can directly address using our multiwavelength data set:  What are the structural parameters and underlying stellar populations of the two host galaxies?  Is there evidence for dynamic outflow from each nucleus? Is the NE LINER a genuine AGN or is it powered by star formation?  Is the origin of the observed diffuse emission tidal or due to galactic superwinds? 
What are the properties of the suspected star clusters that are forming in the merging system? 
Is Mrk 266 a unique dual AGN system, or are we observing a short-lived phase of an evolutionary
process that commonly occurs when two massive, dusty disks merge?

\ifnum\Mode=2 
\pagebreak
\fi

\subsection{Interacting/Merging Galaxies Revealed}
\label{subsec:Galaxies}

\subsubsection{Morphology and Bulge/Disk/Bar Decomposition}
\label{subsubsec:galfit}

%%%%%%%% Begin Table %%%%%%%%%%%
\def\tableGALFIT{
% GALFIT parameters
\begin{deluxetable}{rrlrrrrr}
\ifnum\Mode=2
\renewcommand\arraystretch{0.5}% (MyValue=1.0 is for standard spacing)
\tabletypesize{\tiny}
\setlength{\tabcolsep}{0.01in} %Tighten up the columns. See AASTeX FAQ
\tablewidth{\columnwidth}
\else
\renewcommand\arraystretch{0.8}% (MyValue=1.0 is for standard spacing)
\tabletypesize{\footnotesize} 
\setlength{\tabcolsep}{0.20in} %Tighten up the columns. See AASTeX FAQ
%\tablewidth{\textwidth}
\fi
\tablecaption{GALFIT Results \label{tbl:galfit}}
\tablehead{
\colhead{Band} &
\colhead{$\rm \frac{f}{f_{tot}}$(\%)} & 
\colhead{C(n)} & 
\colhead{$\frac{B}{D}$} &
\colhead{$\rm r_e$} & 
\colhead{$\rm \frac{b}{a}$} & 
\colhead{$\rm PA\arcdeg$} & 
\colhead{$\rm mag_{AB}$} \\
\colhead{(1)} & 
\colhead{(2)} & 
\colhead{(3)} & 
\colhead{(4)} & 
\colhead{(5)} & 
\colhead{(6)} &
\colhead{(7)} &
\colhead{(8)} 
}
\tablecolumns{8}
\startdata
\cutinhead{Mrk 266 SW\tablenotemark{a}}
0.44\micron & $68.2$ & B(4) &  \nodata &  $9.99$ & $0.79$ & $-38$ & \nodata \\
0.44\micron & $68.2$ & T &  \nodata &  \nodata & \nodata & \nodata & $14.40\pm0.02$ \\
0.81\micron  & $63.0$ & B(4)&  \nodata &  $8.28$ & $0.79$ & $-19$ & \nodata \\
0.81\micron  & $63.0$ & T &  \nodata &  \nodata & \nodata & \nodata & $13.13\pm0.01$ \\
1.6\micron & $41.8$ & B(4) &  \nodata &  $4.92$ & $0.49$ & $  -13$ & \nodata \\
1.6\micron & $9.6$  & D(1)  &  $0.23$  &  $2.45$ & $0.74$ & $-88$ & \nodata \\
1.6\micron & $2.4$  & b  &  \nodata &  $0.57$ & $0.27$ & $  -9$ & \nodata \\
1.6\micron & $0.0$  & P &  \nodata &  \nodata & \nodata &  \nodata  & \nodata \\
1.6\micron &  $53.8$ & T &  \nodata &  \nodata & \nodata & \nodata & $12.55\pm0.03$ \\
\cutinhead{Mrk 266 NE\tablenotemark{b}}
0.44\micron  & $31.8$ & D(1) &  \nodata &  $3.11$ & $0.70$ & $-91$ & \nodata \\
0.44\micron  & $31.8$ & T &  \nodata &  \nodata & \nodata & \nodata & $15.23\pm0.02$ \\
0.81\micron   & $13.9$ & B(4) &  \nodata &  $1.43$ & $0.60$ & $-113$ & \nodata \\
0.81\micron   & $23.1$ & D(1)&  $0.58$  &  $3.98$ & $0.69$ & $-63$ & \nodata \\
0.81\micron   & $37.0$ & T &  \nodata &  \nodata & \nodata & \nodata & $13.71\pm0.02$ \\
1.6\micron & $46.1$ & B(4) & \nodata &  $3.46$ & $0.81$ & $-113$ & \nodata \\
1.6\micron & $0.2$   & P & \nodata &  $0.00$ & $1.00$ & $   0$ & \nodata \\
1.6\micron &  $46.3$ & T &  \nodata &  \nodata & \nodata & \nodata & $12.71\pm0.03$ \\
\cutinhead{Mrk 266 Total (SW+NE)\tablenotemark{c}}
0.44\micron &  $1.0$ & T &  \nodata &  \nodata & \nodata & \nodata & $13.97\pm0.02$ \\
0.81\micron   &  $1.0$ & T &  \nodata &  \nodata & \nodata & \nodata & $12.60\pm0.02$ \\
1.6\micron  &  $1.0$ & T &  \nodata &  \nodata & \nodata & \nodata & $11.88\pm0.04$ \\
\enddata
\tablenotetext{a}{The 1.6 \micron\ image of Mrk 266 SW is modeled as the sum of 
$\rm Point source + Bar + Disk~(n=1) + Bulge~(n=4)$ components. 
The 0.44 and 0.81 \micron\ images are too confused by extinction and 
H II regions to derive reliable models of the underlying stars; 
simple $\rm Bulge~(n=4)$ models were fit at these wavelengths 
only to examine the residual images for fine structure in the 
star-forming regions (Fig. \ref{fig:galfitBIHres}).}
\tablenotetext{b}{The 1.6 \micron\ image of Mrk 266 NE is modeled as the sum of 
$\rm Point source + Bulge~(n=4)$ components; there are no significant Bar or Disk 
components detected for this galaxy.}
\tablenotetext{c}{The total flux (apparent magnitude) accounted for in the GALFIT model 
for the galaxy pair in each {\it HST} passband.}
\tablecomments{
\footnotesize
Fitting parameters and results of modeling the HST data using GALFIT.
Column (1): HST Filter: B = F435W (0.44\micron); 
I = F814W (0.81\micron); H = F160W (1.6\micron).
Column (2): Percentage of the total flux of the GALFIT 
model of the {\it galaxy pair} accounted for by 
the indicated component of the model image at 0.44, 0.81 or 1.6 \micron.
Column (3): Model components coded as follows: 
B=Bulge (Sersic index n=4); D=Disk (Sersic index n=1); b=bar;
P=Point source based on Tiny Tim \citet{1993ASPC...52..536K}; 
T = Total flux from all GALFIT components for the galaxy.
Column (4): Bulge-to-Disk ratio.
Column (5): Effective radius of the Sersic fit (kpc units).
Column (6): Axis ratio of the component fit.
Column (7): Position angle of the component fit (degrees east of north).
Column (8): Apparent AB magnitude of the sum of the GALFIT model 
components for each galaxy, where $f_{\nu} (Jy) = 10^{0.4(8.9-m_{AB})}$. 
Conversions between the AB  and Vega systems
(http://www.stsci.edu/hst/acs/analysis/zeropoints/) are as follows: 
$\rm B_{AB} - B_{Vega} = -0.11; I_{AB} - I_{Vega} = 0.42; H_{AB} - H_{Vega} = 1.32$.
}
\end{deluxetable}
}
\ifnum\Mode=0
\placetable{tbl:galfit}
\else
\tableGALFIT
\fi
%%%%%%%% End Table %%%%%%%%%%%

High resolution  {\it HST} imagery (Fig. \ref{fig:HST_ACS4x}) has revealed a wealth of morphological
structures within Mrk 266. At optical wavelengths (B and I) there are two distinct galaxies within a 
diffuse and highly asymmetric envelope. Patchy dust obscuration and clumpy star formation 
regions are evident on a variety of scales. The underlying structure of the two galaxies is best 
revealed in the NICMOS H band image. The SW galaxy (Fig. \ref{fig:Mrk266SW}) is a barred spiral 
with two arms and predominant dust features, and the estimated Hubble type is SBb (pec). 
The NE galaxy (Fig. \ref{fig:Mrk266NE}) has no discernible spiral arms and appears to be 
Hubble type S0 or S0/a (pec).

%%%%%%%% Begin Figure %%%%%%%%%%%
%Mrk266SW_BI_H_B_I   & Mrk266NE_BI_H_B_I
\def\figcapSW{
\footnotesize 
A close-up of HST imagery of Mrk 266 SW showing a color composite of 
ACS B+I bands (a); NICMOS H band (b);  ACS B band (c); and ACS I band (d). 
The images are displayed with an emphasis on the high surface brightness 
emission near the nucleus. The spatial resolution is very similar in all 4 bands,
but there has been no PSF matching. The scale bar indicates 2\arcsec.
}
\def\figcapNE{
\footnotesize 
HST imagery of Mrk 266 NE displaying the same bands as in Figure \ref{fig:Mrk266SW}.
}
\ifnum\Mode=0 %Insert Figure/Table only in [preprint] or [preprint2] modes
\placefigure{fig:Mrk266SW}
\begin{verbatim}fig13a_13d\end{verbatim}
\else
%For preprint
\ifnum\Mode=2
  \begin{figure}[!htb]
\else
  \begin{figure}[!htb]
\fi
\center
\includegraphics[width=1.0\columnwidth,angle=0]{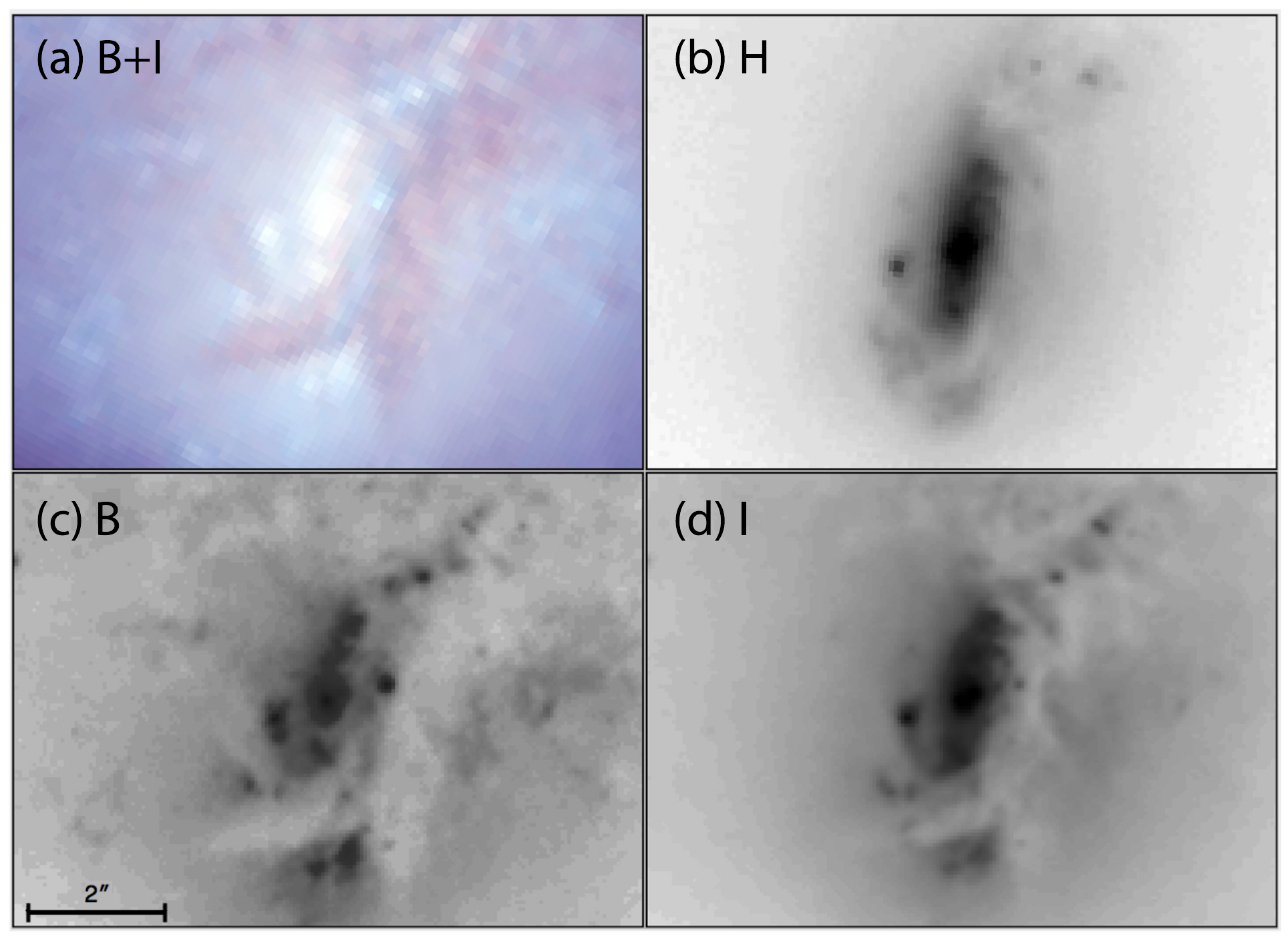} 
\caption{\figcapSW  \label{fig:Mrk266SW}}
\ifnum\Mode=2
  %In 2-column mode only, insert here to force on to the same page 
  \includegraphics[width=1.0\columnwidth,angle=0]{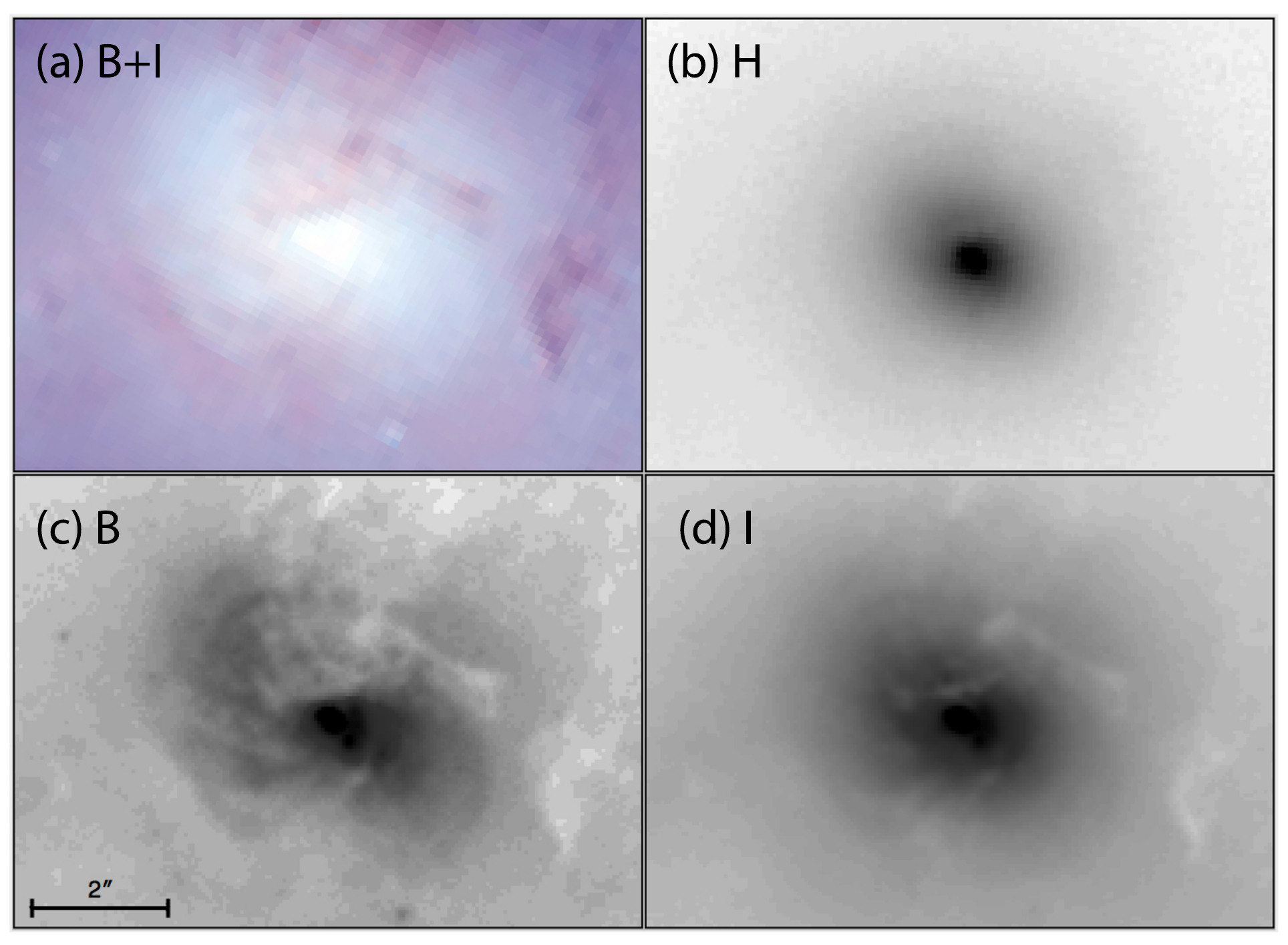} 
  \caption{\figcapNE  \label{fig:Mrk266NE}}
  \end{figure}
\else
  \end{figure}
\fi
%\ifnum\Mode=2 \twocolumn \fi %[preprint2] mode only
\fi %close \ifnum\Mode=0 
%%%%%%%% End Figure %%%%%%%%%%%

%%%%%%%% Begin Figure %%%%%%%%%%%
\ifnum\Mode=0 %Insert Figure/Table only in [preprint] or [preprint2] modes
\placefigure{fig:Mrk266NE}
\begin{verbatim}fig14a_14d\end{verbatim}
\else
%For preprint
\ifnum\Mode=2
  %Empty figure here, because in 2-column mode it was inserted above with Mrk 266 SW 
  \begin{figure*}[!htb]
  \end{figure*}
\else
  \begin{figure}[!htb]
  \center
  \includegraphics[width=6.0truein,angle=0]{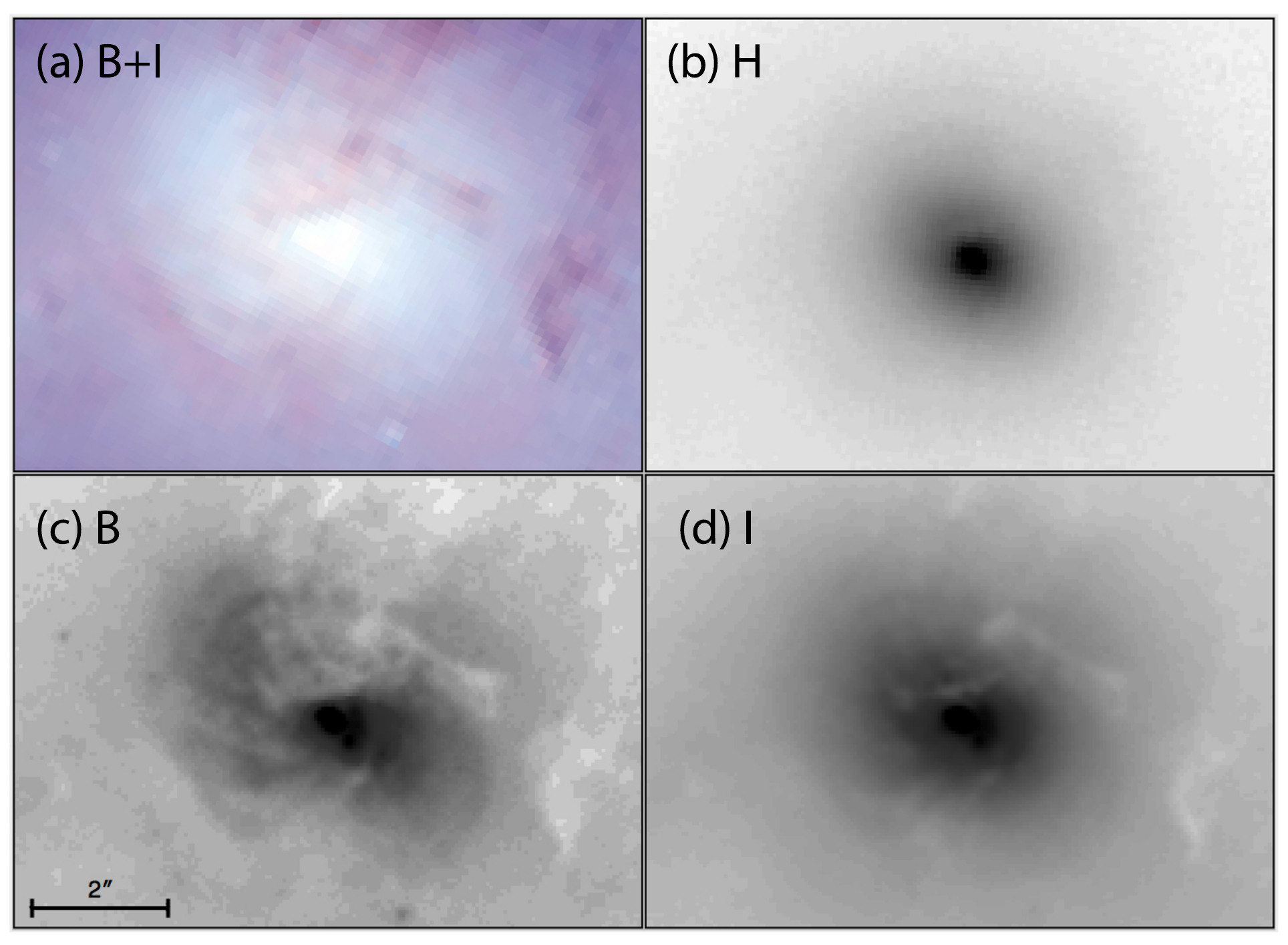} 
   \caption{\figcapNE  \label{fig:Mrk266NE}}
  \end{figure}
\fi%close \ifnum\Mode=2 \twocolumn \fi %[preprint2] mode only
\fi %close \ifnum\Mode=0 
%%%%%%%% End Figure %%%%%%%%%%%

%%%%%%%% Begin Figure %%%%%%%%%%%
%GALFIT H-band results
%mrk266_galfit_Hb
\def\figcapGalfitH{
\footnotesize
Results of GALFIT modeling of the 1.6 \micron\ NICMOS image of Mrk 266.
The top row shows (a) the direct image, (b) the GALFIT model fit, and (c) the 
residual image (direct image minus model fit). 
The center row shows the components of the model for the SW galaxy:
(d) the disk, bulge (e), and stellar bar (f).
The bottom row shows the components of the model for the NE galaxy:
(g) the bulge and (h) the point source modeled with Tiny Tim. The scale
bar indicates 10\arcsec\ (6 kpc).
}
\ifnum\Mode=0 %Insert Figure/Table only in [preprint] or [preprint2] modes
\placefigure{fig:galfitH}
\begin{verbatim}fig15a_15h\end{verbatim}
\else
\ifnum\Mode=2
\begin{figure}[!th]%\begin{figure*}[!ht]
\else
\begin{figure}[!th]
\fi
\center
\includegraphics[scale=0.5,angle=0]{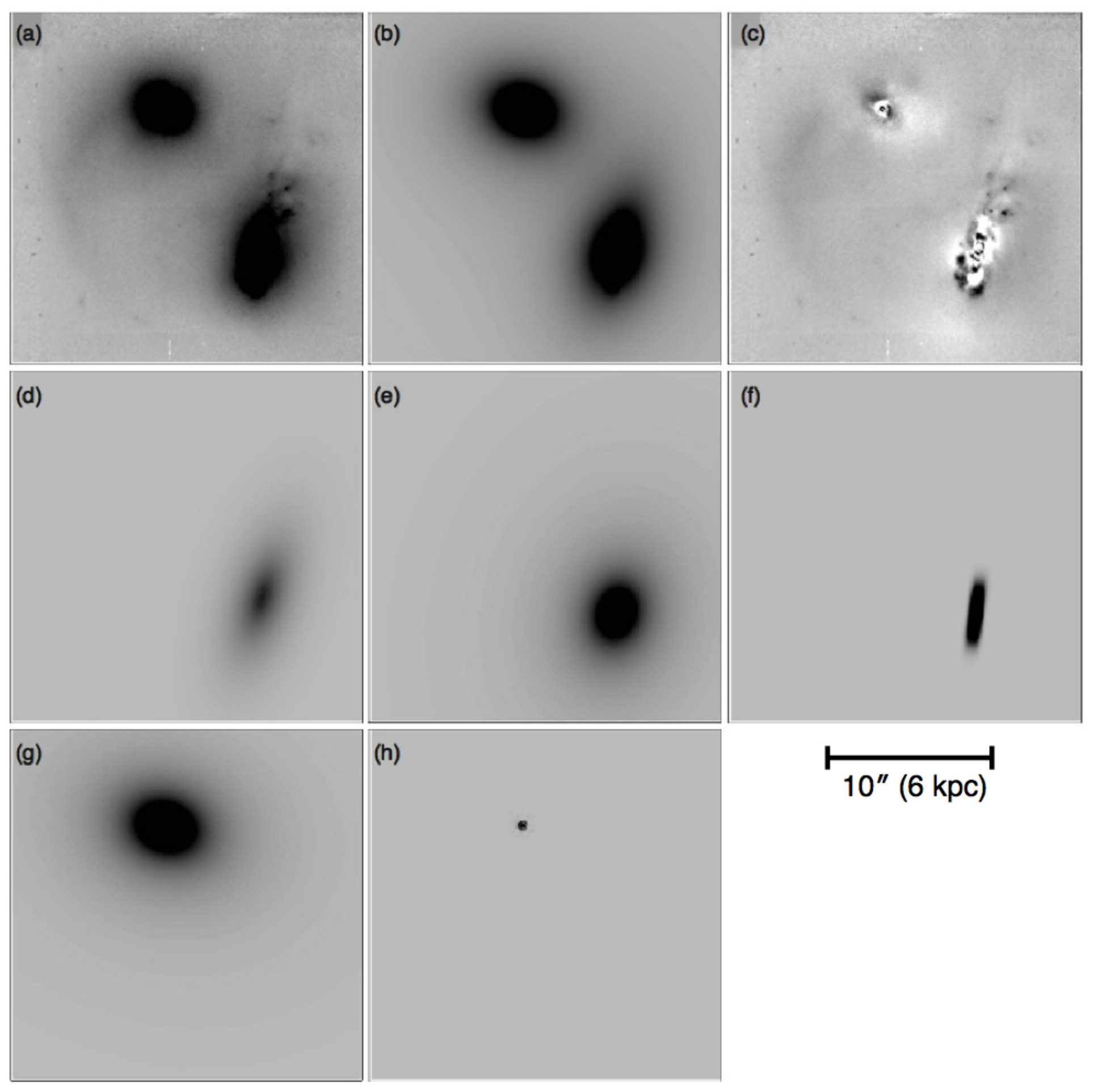} 
\caption{\figcapGalfitH \label{fig:galfitH}}
\ifnum\Mode=2
\end{figure}%\end{figure*}
\else
\end{figure}
\fi
%\ifnum\Mode=2 \twocolumn \fi %[preprint2] mode only
\fi %close \ifnum\Mode=0 
%%%%%%%% End Figure %%%%%%%%%%%

In the extremities of the system are three notable features at optical wavelengths 
(Fig. \ref{fig:HST_ACS4x}):
(1) Extending $\sim$25\arcsec\ (15 kpc) to the north is a fragmented, filamentary feature known as 
the Northern Loop; this structure is examined in detail in \S\ref{subsec:Superwind}.
(2) Faint emission extending $\sim$60\arcsec\ (36 kpc) to the south and $\sim$40\arcsec\ (24 kpc) to the
south-east from the center of the system appears to be tidal debris. The deep B+V+I image 
(Fig. \ref{fig:UH88image}) reveals much fainter, asymmetric emission spanning $\approx$103 kpc
(2\farcm9). The peculiar morphology and vast extent of this emission are consistent with numerical
simulations of tidal debris created during a major merger
\citep[e.g.,][]{1999Ap&SS.266..195M,2006MNRAS.373.1013C}.
(3) Approximately 25\arcsec\ to the south-west are two objects that could be either knots near the end of a
tidal tail or background galaxies. The latter interpretation is favored due to what appear to be compact 
nuclei and spiral arms. If these are background galaxies with similar physical size and separation 
as Mrk 266,  their projected diameters ($\approx$3\arcsec) and separation ($\approx$2\arcsec) imply
they are $\sim$5 times further away than Mrk 266 (distance $\sim$650 Mpc, $z\sim 0.15$); 
redshift measurements are needed for confirmation.

%%%%%%%% Begin Figure %%%%%%%%%%%
%GALFIT residuals at B, I, H
%mrk266_galfit_BIH2
\def\figcapGalfitBIHres{
\footnotesize
Residuals resulting from subtraction of GALFIT 
models parameterized in Table \ref{tbl:galfit} from the
direct HST images. The left column shows results
for the NE galaxy at 0.44 \micron\ (a), 0.81 \micron\ (b) and 1.6 \micron\ (c).
The right column shows results for the SW galaxy at 0.44 \micron\ (d), 
0.81 \micron\ (e) and 1.6 \micron\ (f). Black (white) regions represent positive 
(negative) flux with respect to the image model fits. The tick marks are
separated by 0\farcs5.
}
\ifnum\Mode=0 %Insert Figure/Table only in [preprint] or [preprint2] modes
\placefigure{fig:galfitBIHres}
\begin{verbatim}fig16a_16f\end{verbatim}
\else
%For preprint
\ifnum\Mode=2 %force the figure to span both columns
\begin{figure}[!th]%\begin{figure*}[!th]
\center
\includegraphics[scale=0.45,angle=0]{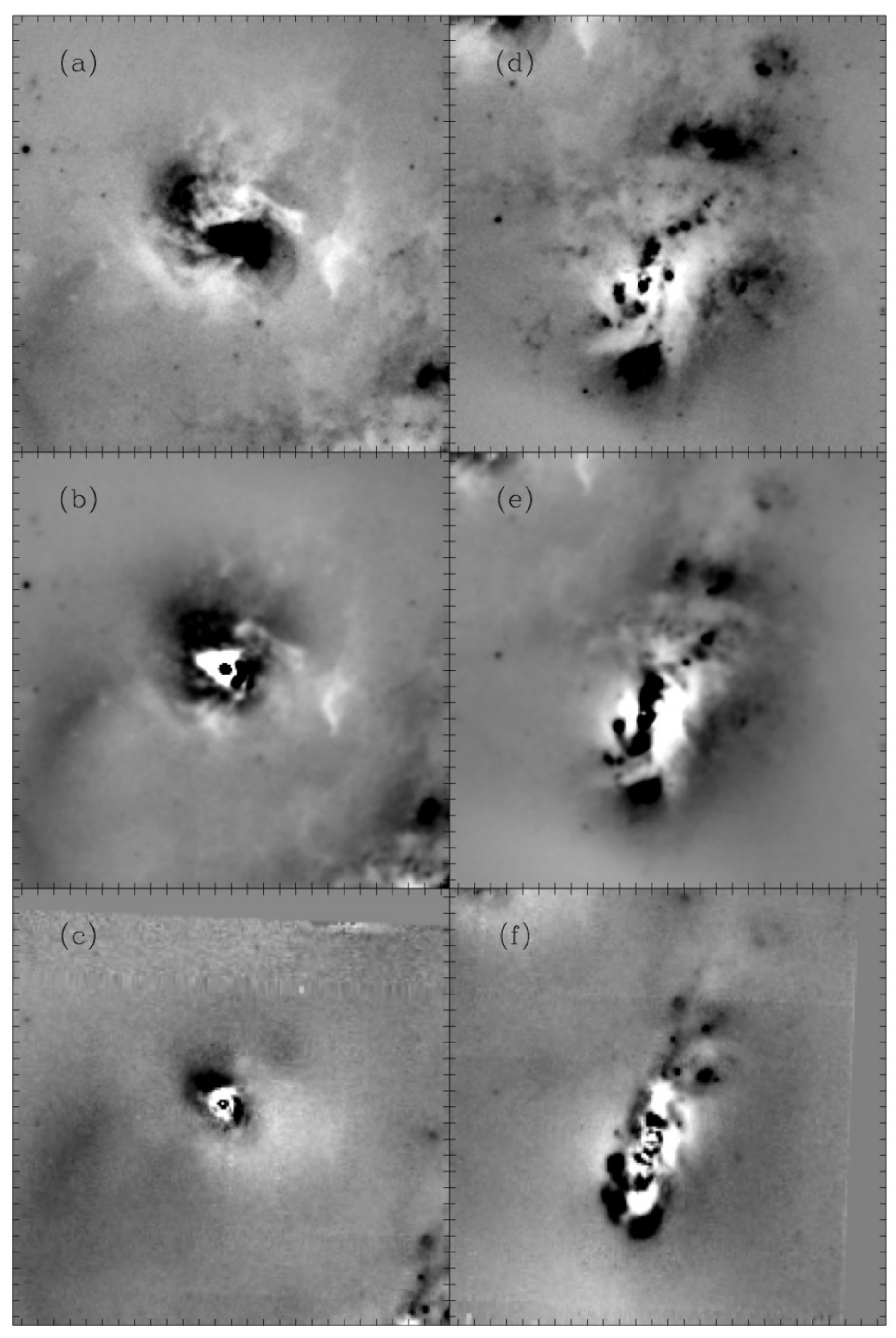} 
\else
\begin{figure}[!th]
\center
%In single-column mode, scale=0.6 is the maximum without floating it to the end
\includegraphics[scale=0.5,angle=0]{fig16a_16f} 
\fi
\caption{\figcapGalfitBIHres \label{fig:galfitBIHres}}
\ifnum\Mode=2 
\end{figure}%\end{figure*}
\else
\end{figure}
\fi
%\ifnum\Mode=2 \twocolumn \fi %[preprint2] mode only
\fi %close \ifnum\Mode=0 
%%%%%%%% End Figure %%%%%%%%%%%

Using GALFIT \citep{2002AJ....124..266P}, quantitative isophotal analysis was performed on both
component galaxies in the three {\it HST} bands. At 1.6 \micron\ (H band) where dust obscuration 
is minimal and thus permits the most reliable decomposition of the underlying galaxies, 
various trial fits involving possible point source, bulge, disk and/or bar components were made; 
potential nuclei were modeled using a point spread function generated from the Tiny Tim package 
\citep{1993ASPC...52..536K}\footnote{http://www.stsci.edu/software/tinytim/tinytim.html}.
Figure \ref{fig:galfitH} shows the components of the best-fitting models in the H band.
Due to extensive, patchy dust obscuration and complex star-forming regions, only models with 
possible bulge and disk components were considered in the B and I bands.
Subtraction of the best-fitting models from the direct images are useful to elucidate further fine 
structures, as shown visually for all three bands in Figure \ref{fig:galfitBIHres}. 
The results of this structural analysis at H band are the following: 
a) The SW galaxy has no detected point source emission; its nucleus is therefore obscured by dust
even at 1.6 \micron.
b) The NE galaxy exhibits no significant disk component and therefore can be modeled using 
only a point source plus bulge profile.
c) Approximately 10\% of the total H-band light is in the form of diffuse emission which is 
likely tidal in origin (Fig. \ref{fig:galfitH}a). 
In the optical bands the residual images reveal the following: 
a) The SW galaxy has many ÒknotsÓ of emission which are likely star clusters or 
associations; these features are analyzed in detail in \S\ref{subsec:SCs}. 
b) The largest features in the B-band residual image appear to be giant star-forming regions
(superassociations) situated on both ends of the stellar bar. This phenomenon, in which 
superassociations occur on both sides of stellar bars much more frequently than they 
occur on a single side, was pointed out by \citet{1991SvAL...17..447M}. 
c) The NE galaxy has numerous radial, asymmetric filaments (most pronounced in the B band) indicative 
of bi-conical outflow from the AGN; this region is analyzed more fully in \S\ref{subsubsec:Outflows}. 
Parameters from the GALFIT modeling are listed in Table \ref{tbl:galfit}. 
Adopting a distance modulus of 35.55 mag (see \S\ref{subsec:background}),
the apparent H magnitudes (Vega system) of Mrk 266 SW and NE (Table \ref{tbl:galfit}) correspond to
remarkably similar absolute magnitudes of  $\rm M_H (SW) = -24.32$ and $\rm M_H (NE) = -24.16$.

\subsubsection{Galaxy Luminosities and Derived Stellar Masses}
\label{subsubsec:Masses}

The H-band (1.6 \micron) luminosity function of galaxies in the Coma Cluster has been approximated 
as a Schechter function with $\rm L^*_H = -23.9$ mag \citep{1998ApJ...503L..45D}. Therefore, 
Mrk 266 SW and NE are $\approx$50\% and 30\% more luminous than the typical $\rm L^{*}_{H}$ 
galaxy in the local universe. Applying the mean mass-to-light ratio at H-band of 
4.6 (\Msun/\Lsun) for spiral galaxies (including early types S0 and S0/a; 
\cite{1996A&A...312..397G}) results in total galaxy mass 
estimates of $6.3\times10^{10}$ \Msun\ (SW) and $5.3\times10^{10}$ \Msun\ (NE).
Estimates of the stellar mass in old stars derived from the mean ratio of
$\rm M_{*}/\nu L_{\nu}(2.2\micron) = 5.6 (\pm1.5)$ based on model predictions by 
\citet{2008MNRAS.385.1155L} produces similar results of $\rm M_{*} = 6.1\times10^{10}$ \Msun\ (SW) 
and $\rm M_{*} = 4.4\times10^{10}$ \Msun\ (NE).\footnote{The model curve for $\rm z = 0$ galaxies in
Figure 13c of \citet[][]{2008MNRAS.385.1155L}, adjusted to h $=$ 0.70, 
corresponds to a range of 4.5 - 7.0 in $\rm M_{*}/\nu L_{\nu}(2.2\micron)$.
A significantly lower ratio of $\rm M_{*}/L_{2.2\micron} \approx 1$ 
was found by \citet{2007A&A...476..137A} and \citet{2001MNRAS.326..255C}. 
The nature of this discrepancy is beyond the scope of this article.}
SED model fitting can improve on stellar masses estimated from
monochromatic luminosities \citet{2011U}. 
Next we turn attention to the nuclei and their circumnuclear regions.

\ifnum\Mode=2 
\pagebreak
\fi

\subsection{The Nuclei and Circumnuclear Regions}
\label{subsec:Nuclei}

\subsubsection{Optical and Infrared Spectral Diagnostics}
\label{subsubsec:SpectralDiagnostics}

Using optical emission-line diagnostics, the NE nucleus has been classified as a LINER and the SW 
nucleus as a Seyfert 2 \citep[e.g.,][]{1988AJ.....96.1227H,1993ApJS...85...27M,2000PASJ...52..185I}.
At optical wavelengths, the NE nucleus has brighter continuum emission and less extinction 
than the SW nucleus. \citet{2010ApJ...709..884Y} classified Mrk 266 NE as a 
composite AGN/starburst nucleus, and they concluded that the NE and SW nuclei have 
similar relative AGN and starburst contributions within the LINER and Seyfert 2 
branches of the [O III]5007/$H\beta$ vs. [O I]6300/H$\alpha$ diagnostic 
diagram.\footnote{The identifications of the NE and SW nuclei of Mrk 266 were mistakenly 
swapped in the tables and figures of \citet{1995ApJS...98..129K} 
and \citet{1995ApJS...98..171V}, and thus also in Table 2 of \citet{2010ApJ...709..884Y}. 
The identifications are corrected here.}
Since the IRS apertures encompass an area of 5.4 x 5.4 kpc$^{2}$ (Fig. \ref{fig:IRScube}), 
whereas published optical diagnostics were obtained through smaller apertures of 
$\sim$1 x 1 kpc$^{2}$, nuclear emission is more diluted by extranuclear emission 
in the mid-IR spectra than in the optical spectra. However, since the mid-IR spectra penetrate
much more dust than the optical spectra, if there is sufficient centrally concentrated dust 
heated by an embedded AGN, this emission can potentially overpower the 
extended star formation (as in many ULIRGs).
The IRS spectra of the two nuclei (Fig. \ref{fig:IRScube_SL}) show them
to be similarly dominated by PAH emission, and the large PAH 
equivalent widths  (Table \ref{tbl:IRScube}) indicate the prominence 
of star-formation \citep[e.g.,][]{2007ApJ...656..148A}. 
The NE source has lower PAH fluxes and equivalent widths, indicating
excess warm dust emission compared to the SW source.

%%%%%%%% Begin Table %%%%%%%%%%%
\def\tableAGNfractions{
\ifnum\Mode=2
\begin{deluxetable*}{lrrrrrrrrr}[!th]
\tabletypesize{\normalsize}
\setlength{\tabcolsep}{0.01in} %Tighten up the columns. See AASTeX FAQ
\renewcommand\arraystretch{0.5}% (MyValue=1.0 is for standard spacing)
\tablewidth{\textwidth}
\else
\begin{deluxetable}{lrrrrrrrrr}
\setlength{\tabcolsep}{0.04in} %Tighten up the columns. See AASTeX FAQ
\renewcommand\arraystretch{1.0}% (MyValue=1.0 is for standard spacing)
\tabletypesize{\footnotesize} 
%\tablewidth{\textwidth}
\fi
\tablecaption{Estimated AGN Contributions to the Mid-Infrared and Bolometric Luminosity \label{tbl:Q4}}
\tablehead{
\colhead{Diagnostic} & 
\multicolumn{3}{c}{SW Galaxy} & \multicolumn{3}{c}{NE Galaxy} & \multicolumn{3}{c}{Total System} \\
\cline{2-4} \cline{5-7} \cline{8-10} 
\colhead{} & 
\colhead{Value} & 
\colhead{$\rm \frac{L_{AGN}}{L_{Diag}}$} &
\colhead{$\rm \frac{L_{AGN}}{L_{Bol}}$} &
\colhead{Value} & 
\colhead{$\rm \frac{L_{AGN}}{L_{Diag}}$} &
\colhead{$\rm \frac{L_{AGN}}{L_{Bol}}$} &
\colhead{Value} & 
\colhead{$\rm \frac{L_{AGN}}{L_{Diag}}$} &
\colhead{$\rm \frac{L_{AGN}}{L_{Bol}}$} \\
\colhead{(1)} & 
\colhead{(2)} & 
\colhead{(3)} &
\colhead{(4)} &
\colhead{(5)} & 
\colhead{(6)} &
\colhead{(7)} &
\colhead{(8)} & 
\colhead{(9)} &
\colhead{(10)}
}
\tablecolumns{10}
\startdata
%%%%%%%%%
%1 & 2 & 3 & 4 & 5 & 6 & 7 & 8 & 9 & 10 
(1) $[O~IV]~25.89\micron/[Ne~II]~12.81\micron$\tablenotemark{a}  & 0.93 & 23\% & 73\%     & \nodata  & \nodata & \nodata   & 0.67      & 17\%    & 64\% \\
(2) $[Ne~V]~14.32\micron/[Ne~II]~12.81\micron$\tablenotemark{a} & 0.14 & 11\% & 53\%     & \nodata   & \nodata & \nodata  & \nodata & \nodata & \nodata \\
(3) PAH $6.2\micron$ EQW [\micron]\tablenotemark{b}                                   & 0.67 &   0\% & \nodata & 0.4\phn   & 30\%     & \nodata  &  0.47     & 10\%    & \nodata \\
(4) PAH $7.7\micron$ EQW [\micron]\tablenotemark{b}                                   & 0.54 & 69\% & 43\%     & 0.25        & 79\%    & 56\%       & 0.46      & 71\%    & 46\% \\
(5) $f(PAH~6.2\micron)/f_{5.5\micron}~vs.~f(15\micron)/f_{5.5\micron}$\tablenotemark{b}  & 1.6,8.6 & 47\% & 9\% & 0.56,6.4 & 81\% & 33\% & 2.0,2.9 & 52\% &  11\% \\
(6) $f_{\nu}(30)/f_{\nu}(15)$\tablenotemark{c}                                                  & 3.9   & 89\% & 82\%     & 3.3          & 92\%    & 87\%       & 4.8         & 85\%  & 76\% \\
~~~                                                                                                     & ~~~ & ~~~    & ~~~       & ~~~        & ~~~      & ~~~         & ~~~        & ~~~    & ~~~ \\
Mean\tablenotemark{d}                                                                   &         & 40\%  & 52\%     &                & 71\%   & 59\%       &                & 47\%  & 49\% \\
Std. Dev.                                                                                         &         & 35\%  & 29\%      &               & 28\%   & 27\%        &                & 33\%  & 28\% 
\enddata
\tablenotetext{a}{
\footnotesize
Based on published high-resolution, staring-mode data  (Appendix Table \ref{tbl:IRSpublished}).
The NE nucleus was not fully covered in these observations.
}
\tablenotetext{b}{
\footnotesize
Based on new low-resolution measurements from the IRS spectral map (Table \ref{tbl:IRScube}).
}
\tablenotetext{c}{
\footnotesize
The continuum flux density ratio estimated via interpolation in the SED of the global system,
$f_{\nu}(30\mu m)/f_{\nu}(15 \mu m)=4.8$, is within 8\% of the value 
$f_{\nu}(30\mu m)/f_{\nu}(15 \mu m)=5.4$ computed from the {\it IRS} (SL+LL) spectrum.
Given this close agreement, since the 15 and 30 \micron\ fluxes could not be 
measured for the individual galaxies with the {\it IRS}  (because they were 
unresolved by the LL module), the flux ratios estimated by interpolation from 
the broad-band photometry in the SEDs for the individual galaxies 
(Table \ref{tbl:SED_Regions}) were substituted.
}
\tablenotetext{d}{
\footnotesize
The mean $\rm L_{AGN}/L_{Bol}$ values omit the 6.2 \micron\ PAH EQW
because there is no bolometric correction available for this diagnostic.
}
\tablecomments{
\footnotesize
Column (1): Mid-infrared spectral diagnostic used to estimate the 
fractional contribution of radiation from an AGN.
Columns (2)-(4): For the SW galaxy, the value of the diagnostic followed by 
the percentage of AGN contribution to the diagnostic luminosity and 
the estimated percentage of AGN contribution to $\rm L_{bol}$. 
The latter uses the zero-points and bolometric corrections from 
\citet[][]{2009ApJS..182..628V}; see \S\ref{subsubsec:SpectralDiagnostics}.
Columns (5)-(7): The same as columns (2)-(4), but for the NE galaxy.
Columns (8)-(10): The same as columns (2)-(4), but for the total system.
}
%The last two rows are the mean and standard deviation of the 5 methods from
%\citet[][]{2009ApJS..182..628V} calibrated as fractions of $\rm L_{bol}$:
\ifnum\Mode=2
\end{deluxetable*}
\else
\end{deluxetable}
\fi
}
\ifnum\Mode=0
\placetable{tbl:Q4}
\else
\tableAGNfractions
\fi
%%%%%%%% End Table %%%%%%%%%%%

The 6.2 \micron\ PAH EQWs and 9.7 \micron\ silicate absorption 
strengths\footnote{The empirical method of \citet{2007ApJ...654L..49S} 
was applied to the SL spectrum (5-14 \micron\ region; Fig. \ref{fig:IRScube_SL}) to measure 
silicate strengths of $\rm -0.62~(\pm 0.06), -0.69~(\pm 0.12), and~-0.66~(\pm 0.07)$ 
for the SW nucleus, the NE nucleus, and the total system, respectively. A similar absorption 
strength, $-0.60~(\pm 0.06)$, was measured using the 5-40 \micron\  SL$+$LL data for the 
total system (Fig. \ref{fig:IRScube_SLLL_Total}).}
place the SW galaxy on the bottom right vertex of the 
silicate depth versus PAH EQW plot of \citet{2007ApJ...654L..49S}, a region
dominated by galaxies with high star formation rates; the 
NE galaxy (with similar Si strength but a lower 6.2 \micron\ EQW of 0.4 \micron\ 
compared to 0.67 \micron\ for the SW galaxy) 
is in a region populated with AGN and ULIRGs having substantial 
dust heating by non-stellar radiation. The observed silicate strength in the SW 
galaxy ($-0.60~\pm 0.06$) is intermediate between the values observed for the 
face-on galaxy NGC 7714 and the edge-on galaxy M82, which is consistent with the 
derived disk axial ratio of $\approx 42$\arcdeg for Mrk 266 SW (Table \ref{tbl:galfit}).
The observed silicate absorption strengths correspond to 
$\rm N_H = 2.2\times 10^{22}~and~2.4\times 10^{22} ~cm^{-2}$ for the 
SW and NE nuclei, respectively; these values result from applying
the relationships $\rm N_H = \tau_{9.7\mu m}\times 3.5\times10^{22}~cm^{-2}$ and 
$\rm A_{9.7\mu m}/A_v = 0.06$ \citep{1984MNRAS.208..481R}.
However, for clumpy dust configurations expected in AGN tori, emission and absorption 
cannot be disentangled along the line of sight and therefore these are lower limits 
to the true silicate column to the central sources. 

Various empirical diagnostics have been published to quantify the relative contributions 
to dust heating from AGN and starburst sources \citep[e.g.,][]{2007ApJ...656..148A}. 
We focus primarily on the suite of methods applied to local ULIRGs and QSOs by 
\citet[][]{2009ApJS..182..628V}, including application of their bolometric corrections 
to express the estimated AGN dust heating contributions as a fraction of $\rm L_{bol}$. 
All methods assume that the line emitting regions 
are subject to the same amount of extinction within the aperture, which may be 
questionable, especially for Mrk 266 SW.  
Since the NE galaxy was not observed in the high-resolution IRS mode, 
methods that involve the fine structure lines can only be applied to the SW galaxy.  
The results are summarized in Table \ref{tbl:Q4}. For the SW galaxy, the 
[O IV]/[Ne II] ratio indicates an overall AGN contribution to $\rm L_{bol}$ of 73\%, 
whereas the [Ne V]/[Ne II] ratio suggests 53\%. 
The 6.2 \micron\ EQW indicates that the mid-IR luminosity of the SW component is 
due entirely to a starburst, and a 30\% AGN contribution is inferred for the NE galaxy. 
This diagnostic suggests a 10\% AGN contribution to the mid-IR luminosity of the
total system.\footnote{The CAFE SED fitting package
provides an independent, model-dependent estimate of the AGN contribution 
to the total infrared luminosity. See \S\ref{subsec:SEDanalysis}.}
The AGN fractions estimated from the 6.2 \micron\ EQW are substantially lower than those 
obtained using the EQW of the 7.7 \micron\ PAH feature (see Table \ref{tbl:Q4}). 
This discrepancy may be due to greater difficulty measuring the 7.7 \micron\ 
PAH feature because of confusion from surrounding features (8.6 \micron\ PAH 
and silicate absorption) and complications in defining the proper continuum level.

The continuum shape over 3-16 \micron\ provides another diagnostic. 
The physical basis is that dust grains located near an 
AGN can reach temperatures as high as 500-1000 K 
(before they sublimate), and such high dust temperatures would completely 
dominate the continuum emission compared to dust heated by a starburst. 
This method was developed by \citet{2000A&A...359..887L}, adapted by 
\citet{2004ApJ...613..986P} for interpretation of mid-IR spectra from the 
{\it Infrared Space Observatory (ISO)}, and modified further for {\it Spitzer} IRS data by
\citet{2007ApJ...656..148A}. Continuum fluxes in the relevant passbands were 
combined with the 6.2 \micron\ PAH fluxes, and the formulation of 
\citet[][]{2009ApJS..182..628V}
was used to compute AGN fractions before and after bolometric corrections. 
The results suggest $\rm L_{AGN}/L_{bol}$ is 33\% for the NE galaxy,
but only about 10\% for the SW galaxy and the total system. 
The continuum flux density ratio $\rm f_{\nu}(30\mu m)/f_{\nu}(15\mu m)$ 
provides our final empirical diagnostic; \citet{2009ApJS..182..628V} found this to be
an effective substitute for the PAH-free and silicate-free MIR/FIR ratio. 

On average, with a standard deviation of nearly 30\%, these diagnostics indicate 
approximately equal contributions of AGN and starburst heating 
of the dust in both galaxies (Table \ref{tbl:Q4}).
This is similar to the average AGN contribution to $\rm L_{bol}$ found for ULIRGs in the 
QUEST sample with Seyfert 2 spectral classifications \citep{2009ApJS..182..628V}.
The variation in these diagnostics may be due, in part, to the different optical depths
probed by different wavelengths. 

Finally, there is a significant difference in the PAH properties of the galaxies. 
The modeling of \citet{2001ApJ...551..807D} combined with 
the ratios of the PAH features ($\rm f_{11.3\mu m}/f_{7.7\mu m}$ and  
$\rm f_{6.2\mu m}/f_{7.7\mu m}$) indicate a dominance of relatively warm, 
ionized PAHs in the SW component and cooler, neutral PAHs in the NE component. 
The PAH line flux ratios for the total system reflect the fact that the SW 
galaxy dominates the total mid- and far-infrared emission from Mrk 266.
The $\rm f_{11.3\mu m}/f_{7.7\mu m}$ PAH ratio in Mrk 266 NE (0.76) is substantially 
larger than the ratio in the SW galaxy (0.34). The former is similar to the ratios observed in 
other galaxies containing AGNs, and the latter is near the mean value (0.25) observed in 
galaxies with pure starburst (H II) spectra \citep{2007ApJ...656..770S}.
Following the inferences proposed by Smith et al., this suggests that the hard radiation 
field surrounding the partially obscured AGN in the NE galaxy may be preferentially destroying
PAH molecules small enough to emit at 7.7 \micron, thus increasing the 11.3\micron/7.7\micron\  
PAH ratio compared to the SW nucleus which has a more heavily obscured AGN and 
a higher star formation rate. 

\subsubsection{Warm Molecular Gas}
\label{subsubsec:WarmH2}

In the NE galaxy, the IRS spectral maps 
(Figs. \ref{fig:IRScube} and \ref{fig:IRScube_SL})
reveal bright rotational lines of molecular hydrogen, specifically
$\rm H_2~S(3)~9.7$\micron\ and $H_2~S(2)~12.3$\micron.
In the SW galaxy, the $H_2~S(3)~9.7$\micron\  emission is much weaker, 
and any $H_2~S(2)~12.3$ \micron\ emission that may be present intrinsically 
is overshadowed by the blue wing of the strong 12.6 \micron\ PAH complex. 
In contrast, the SW galaxy has bright $\rm [S~IV]~10.51$~\micron\ emission, 
whereas the NE nucleus is undetected in this feature. Give that the NE galaxy
has 8 and 24 \micron\ flux densities that are 1.8 and 5.7 times fainter 
than the SW galaxy (see Table \ref{tbl:SED_Regions} and Fig. \ref{fig:SEDs}), 
it is surprising that the $\rm H_2$ line emission in Mrk 266 NE is 
substantially more luminous than in Mrk 266 SW. 
This suggests that the majority of the detected $\rm H_2$ line emission in the 
NE galaxy has a different origin than in the SW galaxy. 
A prime candidate for the enhanced $\rm H_2$ emission
in Mrk 266 NE is shock excitation.

Analysis of near-infrared
spectra of Mrk 266 led \citet{2000ApJ...535..735D} to conclude that fast C shocks 
($\rm v_s \sim 40~km~s^{-1}$) are responsible for $\approx$70\% of the 
$\rm H_2$ 1-0 S(1) emission in the two nuclei, with fast J shocks 
($\rm v_s \sim 300~km~s^{-1}$) and UV fluorescence responsible 
for the remaining $\approx$30\%. Davies et al. demonstrated that the powerful
radio continuum and 1-0 S(1) emissions observed in Mrk 266 NE cannot be explained 
with pure starburst models, and they are likely generated predominantly by shocks.
Furthermore, the presence of 1-0 S(1) emission extended out to $\sim$1 kpc
from the NE nucleus, with a large equivalent width and low S(1)/B$\gamma$ ratio
that rule out star formation, provides additional evidence that the $\rm H_2$ 
emission is excited by spatially extended shocks. Davies et al. speculated that
Mrk 266 NE may itself be a very close pair of galactic nuclei with shocked 
molecular gas settled between them, as observed in NGC 6240.
Our interpretation of the latest data from {\it Spitzer, HST} and {\it Chandra} favors
an AGN-driven outflow as the origin of strong shocks in 
Mrk 266 NE (see \S\ref{subsubsec:Outflows}). 

\subsubsection{AGN Black Hole Masses}
\label{subsubsec:SMBHs}

\citet{2003ApJ...589L..21M} established an empirical relation between H-band bulge luminosity and
nuclear black hole mass.  Combining measurements of the total H-band absolute magnitudes with the
GALFIT image decomposition results showing that 78\% and 99\% of the total H-band light is in the bulge
components of the SW and NE galaxies (\S\ref{subsubsec:galfit}), the calibration of Marconi \& Hunt
(Group 1 Galaxies) yields black hole masses of $2.3\times10^{8}$ (SW) and 
$2.6\times10^{8}$ \Msun\ (NE). An independent diagnostic is available from the high-ionization, mid-IR
emission lines detected in Mrk 266 SW. Using the approach of \citet{2008ApJ...674L...9D}, the 
$\rm [Ne~V]~14.32$\micron\ ($\rm [O~IV]~25.89$\micron) emission line luminosity of
$\rm 1.58~(10.5) \times10^{41}~erg~s^{-1}$ computed from published
high-resolution {\it IRS} measurements (Appendix Table \ref{tbl:IRSpublished}) corresponds to 
$\rm M_{BH} = 1.7~(3.2)\times~10^{8}$\Msun~for the SW nucleus, which is consistent with the IR
photometry. The inferred black hole masses are therefore in the typical range known to power 
other luminous AGNs.

\subsubsection{Evidence for Nuclear Outflows and Shocks}
\label{subsubsec:Outflows}

%%%%%%%% Begin Figure %%%%%%%%%%%
%Mrk266_18cmHiResOnB; Mrk266_18cmHiResOnBe
\def\figcapNEoutflow{
\footnotesize
Comparison of new 0.44 \micron\ ({\it HST} ACS) imagery with 
published 18 cm radio continuum (MERLIN) emission \citep[][Fig. 1]{2001MNRAS.327..369T} 
in the circumnuclear region of Mrk 266 NE. 
Top (a): {\it HST} ACS B-band image in grayscale;
the same B-band data are contoured (blue) and overlaid here and in the following two panels.
Center (b): 18 cm radio continuum interferometry with natural weighting 
and spatial resolution 0\farcs5 x 0\farcs3 HPBW (thick black contours) overlaid with 
ACS B-band (thin blue contours).
Bottom  (c): 18 cm radio continuum interferometry with uniform weighting 
and spatial resolution 0\farcs2 x 0\farcs1 HPBW  (thick black contours) overlaid with
ACS B-band (thin blue contours).
The plus signs in panels (b) and (c) are from \citep[][see their Fig. 1]{2001MNRAS.327..369T} 
and they mark the optical nuclear positions, where the marker extent indicates 
the $2\sigma$ positional uncertainty. These positions are in good agreement with the 
astrometry we derived for the ACS images.
The scale bar indicates 1\arcsec, and the field of view is 6\farcs0 x 4\farcs3 for all three panels.
}
\ifnum\Mode=0 %Insert Figure/Table only in [preprint] or [preprint2] modes
\placefigure{fig:NEinBand18cm}
\begin{verbatim}fig17a_17c\end{verbatim}
\else
%For preprint
\begin{figure}[!htb]
\center
\ifnum\Mode=2
\includegraphics[width=0.9\columnwidth,angle=0]{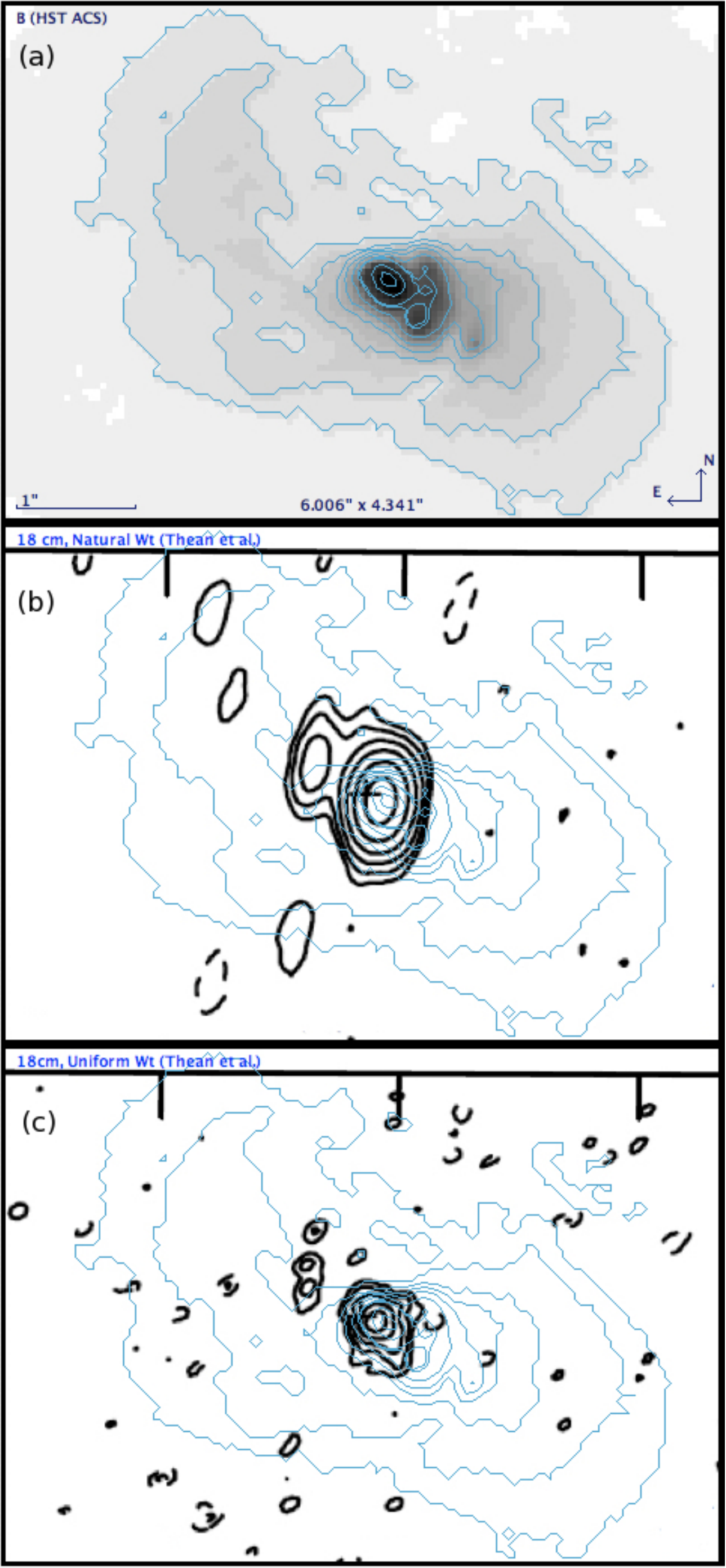} 
\else
\includegraphics[scale=0.55,angle=0]{fig17a_17c}
\fi
\caption{\figcapNEoutflow \label{fig:NEinBand18cm}}
\end{figure}
\fi %close \ifnum\Mode=0 
%%%%%%%% End Figure %%%%%%%%%%%

In the previous section the radiation field heating the dust in the two nuclear environments was 
established.  In this section evidence is presented for dynamical outflows from the nuclei. Within a radius
of 2\arcsec\ (1.2 kpc) around the NE nucleus (Figs. \ref{fig:Mrk266NE}c,d and \ref{fig:galfitBIHres}a), 
the {\it HST} optical imaging reveals structure that is strongly suggestive of a bi-conical outflow with patchy
dust extinction. Evidence for this interpretation rather than spiral arms is as follows: (a) Multiple filamentary
structures extend radially and asymmetrically from the nucleus rather than in a symmetric spiral pattern.
(b) There is the lack of improved definition in any spiral arm pattern in the I-band image relative to B-band.
(c) Photometric deconvolution (\S\ref{subsubsec:galfit}) indicates the NE galaxy is bulge-dominated. 
There are also conspicuous ÒknotsÓ of concentrated emission along a concave arc located  0Ó.4 W of the
nucleus.  Table \ref{tbl:BowKnots} lists photometric measurements of the 3 most prominent optical knots.
The individual knot luminosities of $L_B\approx 3\times10^8$\Lsun are substantially higher than other
luminous star clusters in Mrk 266 (see \S\ref{subsec:SCs}, Fig. \ref{fig:ClustersColorMag}), and their
orientation in an arc embedded within a triangular bow-like structure suggests they may be energized by
an AGN-driven outflow in a situation similar to M51 \citep{1988ApJ...329...38C}
rather than by unusually massive star clusters. 

A comparison of the B-band emission with published high-resolution 18 cm radio continuum observations
\citep{2001MNRAS.327..369T} is shown in Figure \ref{fig:NEinBand18cm}. The 18 cm MERLIN map
exhibits limb-brightening at the location of the optical arc and knots 0\farcs4 (240 pc) west of the nucleus,
and a secondary radio continuum peak is located 0\farcs6 (360 pc) NE of the nucleus. An axis at position
angle $56\arcdeg$ intersects the radio/optical nucleus, the radio emission 0\farcs6 NE of the nucleus, and
B-band knots situated 0\farcs4 and 0\farcs9 SW of the nucleus. Bi-conical AGN outflows with aligned radio
continuum and optical emission-line radiation with dimensions of $\la$1 kpc are commonly observed in
Seyfert galaxies and in some LINERS, especially those that are nearly face-on such as NGC 1068 and
M51. (See, for example, the review of \citet[][]{2005ARA&A..43..769V}.)
The optical morphology of the NE nuclear environment is remarkably similar to the radiative bow shock
observed 5\arcsec (230 pc) south of the nucleus of M51, which is also a LINER 
\citep{1988ApJ...329...38C,1985ApJ...293..132F}. 

The combined data suggest that the AGN in Mrk 266 NE is ejecting radio-emitting plasma along an axis
with PA $\approx 56\arcdeg$, which is also aligned with a bi-conical ionization cone that extends to a 
radius of 1.2 kpc from the nucleus.  The 6 cm radio power ($\rm 4.2\times10^{22}~W~Hz^{-1}$) and
spectral index ($\alpha_{20,6,2cm} = 0.61\pm0.04$) of Mrk 266 NE are consistent with optically 
thin synchrotron emission \citep{1988ApJ...333..168M}, and these parameters are similar to other
luminous AGNs in the local universe \citep[e.g.,][]{1996ApJ...473..130R}. The radio jet appears to be
running into dense material concentrated 0\farcs4 (240 pc) SW from the nucleus, but it is able to escape to
a greater distance in the opposite direction (Fig. \ref{fig:NEinBand18cm}). This hypothesis is consistent
with the velocity field measured for the molecular gas (\S\ref{subsec:MolGas}) which shows a rotating disk
aligned roughly orthogonal to this putative ionization cone. The B-I color map from the HST data (Figure
\ref{fig:BminusI}) reveals that dust is distributed primarily along the NS direction, which again is roughly
orthogonal to the AGN jet/outflow axis.  Regions inside a radius of $\sim$2\arcsec\ from the nucleus of 
Mrk 266 NE (Fig. \ref{fig:BminusI}) are as blue as the Northern Loop, which is known to be dominated 
by bright [O III] $\lambda$5007 and H$\alpha$ line emission 
\citep{1988AJ.....96.1227H,2000PASJ...52..185I}\footnote{The [0 III] $\lambda$5007 line emission 
around Mrk 266 NE \citep[][]{1988AJ.....96.1227H} is elongated along the same position angle as the 
{\it HST} B-band image contoured here in Figure \ref{fig:NEinBand18cm}. This provides evidence that the
{\it HST} B-band image is dominated by emission-line gas in this region. The passband of the F435W filter
omits the [O III] lines, but it includes the [O II] $\lambda$3727 line observed in the spectrum of Mrk 266 NE
\citep[][]{1995ApJS...98..129K} and also at large radii within the ionization cones of many AGNs 
\citep[e.g.][]{2008NewAR..52..227T}.}. \citet{2000ApJ...535..735D} provided independent evidence for
shock-heated gas in this region in the form of the radial profile of the 1-0 S(1) $\rm H_2$ 2.12 \micron\ 
emission line. Furthermore, the $\rm Br\gamma$ emission-line image presented by 
\citet{2000ApJ...535..735D} shows a bimodal structure oriented at PA $\approx 50^\circ$ 
centered on the NE nucleus, providing additional support for our hypothesis.
While most of this evidence is indirect, its combined weight favors our interpretation that the
extranuclear emission around Mrk 266 NE is dominated by an AGN-powered outflow 
that is generating spatially extend shocks.

Similar events may be occurring in the SW galaxy.  The {\it HST} images show numerous compact
structures over a 1 kpc size region that share the same NS orientation as the prominent stellar bar seen 
in the H-band image.  It is highly likely that these compact structures are young stellar clusters 
(see \S\ref{subsec:SCs}).   Observations at 1.6 GHz \citep{2001MNRAS.327..369T} show sub-arsecond
structure aligned NW-SE with respect to the SW nucleus, similar to previous 5 Ghz measurements 
\citep{1988ApJ...333..168M}.  Comparing the new HST images with the extended radio emission maps
suggests the circumnuclear emission in this case is dominated by star formation.  An AGN outflow 
in the SW galaxy may be present but is difficult to detect due to its relatively high inclination 
($\approx 42\arcdeg$). It is very difficult to disentangle starburst-driven winds from AGN-driven winds 
in galaxies containing both a Seyfert 2 and a powerful starburst \citet[e.g.,][]{2005ARA&A..43..769V}.
Although evidence for an AGN-driven outflow in the circumnuclear region of Mrk 266 SW is weak, 
there is ample evidence for a starburst-driven superwind on a larger scale based on the filamentary 
$\rm H\alpha$ and [O III] line emission that comprise the Northern Loop, and from the fainter 
emission-line filaments labeled ``eastern arm-like region'' and ``west knot'' in Figure 2 of 
\citet{2000PASJ...52..185I}, all of which appear to connect or align with the SW nucleus. Properties of 
the kpc-scale outflow revealed by the new data are discussed in \S\ref{subsec:Superwind}.

%%%%%%%% Begin Table %%%%%%%%%%%
\def\tableBowKnots{
\begin{deluxetable}{llcc}
\ifnum\Mode=2
\renewcommand\arraystretch{0.5}% (MyValue=1.0 is for standard spacing)
\tabletypesize{\scriptsize}
\setlength{\tabcolsep}{0.00in} %Tighten up the columns. See AASTeX FAQ
\tablewidth{\columnwidth}
\else
\renewcommand\arraystretch{0.6}% (MyValue=1.0 is for standard spacing)
\tabletypesize{\normalsize} 
\setlength{\tabcolsep}{0.30in} %Tighten up the columns. See AASTeX FAQ
%\tablewidth{\textwidth}
\fi
\tablecaption{Photometry of Knots Embedded in the Arc 0\farcs4 west of the NE Nucleus \label{tbl:BowKnots}}
\tablehead{
\colhead{R.A} & 
\colhead{Dec} & 
\colhead{$\rm m_{F435W}$} & 
\colhead{$\rm m_{F814W}$} \\
\colhead{hh:mm:ss.sss} & 
\colhead{dd:mm:ss.ss} & 
\colhead{mag} &
\colhead{mag} \\
\colhead{(1)} & 
\colhead{(2)} & 
\colhead{(3)} &
\colhead{(4)}
}
\tablecolumns{4}
\startdata
%%%%%%%%%
13:38:17.75 & 48:16:40.9 & $19.94 \pm 0.03$ & $19.02 \pm 0.05$ \\ 
13:38:17.75 & 48:16:41.1 & $20.09 \pm 0.03$ & $18.89 \pm 0.04$ \\
13:38:17.75 & 48:16:41.3 & $20.27 \pm 0.03$ & $19.06 \pm 0.04$
\enddata
\tablecomments{
\footnotesize
Columns (1) and (2): J2000 R.A. and Dec, with $1\sigma$ uncertainty in the absolute 
astrometry of 0\farcs5. Columns (3) and (4): Apparent magnitudes and uncertainties measured 
in the HST/ACS F435W (B) and F814 (I) bands, respectively.
}
\end{deluxetable}
}
\ifnum\Mode=0
\placetable{tbl:BowKnots}
\else
\tableBowKnots
\fi
%%%%%%%% End Table %%%%%%%%%%%

\subsubsection{Hard X-Ray Properties of the Nuclei}
\label{subsubsec:XrayAGNs}

%%%%%%%% Begin Figure %%%%%%%%%%%
%Hard X-ray Hardness Ratio Map: S=1.5-4 keV; H=4-7 keV 
%n5256cxohxhrmap
%Soft X-ray Hardness Ratio Map: S=0.4-0.8 keV; H=0.8-1.5 keV 
%n5256cxosxhrmap
\def\figcapXrayHardness{
\footnotesize 
Maps of the hardness of the {\it Chandra} X-ray emission. 
(a): The hardness ratio between 1.5 and 7 keV, defined as 
$\rm HR_{hard} = (F_{4-7 keV} - F_{1.5-4 keV})/(F_{4-7 keV} + F_{1.5-4 keV})$.
The hard X-ray emission from the SW nucleus is too faint to measure this flux ratio,
with the exception of the peak pixel which has $\rm HR_{hard} = -0.4$ (blue).
(b): The hardness ratio between 0.4 and 1.5 keV, defined as 
$\rm HR_{soft} = (F_{0.8-1.5 keV} - F_{0.4-0.8 keV})/(F_{0.8-1.5 keV} + F_{0.4-0.8 keV})$;
this illustrates the temperature variation over the diffuse emission.
$\rm HR_{hard}$ and $\rm HR_{soft}$ are plotted using colors as indicated 
in their respective legends. 
The soft band X-ray data are overlaid as contours
drawn with linear intervals between 1.4\% and 12\% of the peak brightness 
(in the NE nucleus) to emphasize the extended, diffuse emission.
}
\ifnum\Mode=0 %Insert Figure/Table only in [preprint] or [preprint2] modes
\placefigure{fig:ChandraHardness}
\begin{verbatim}fig18a, fig18b\end{verbatim}
\else
\begin{figure}[!ht]
\center
\ifnum\Mode=2 %stacked vertically
\includegraphics[width=0.80\columnwidth,angle=0]{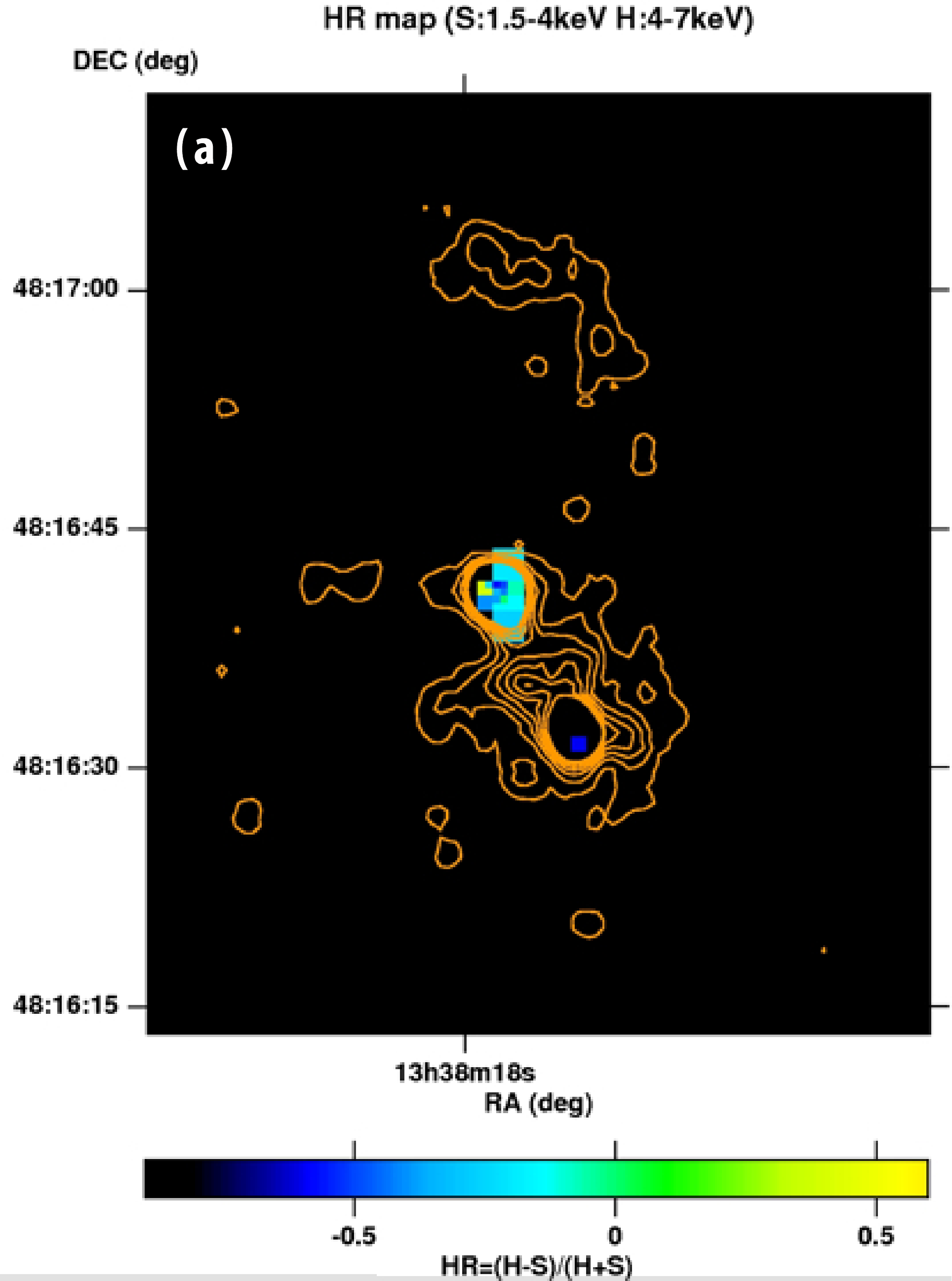}
\vskip 0.3in
\includegraphics[width=0.80\columnwidth,angle=0]{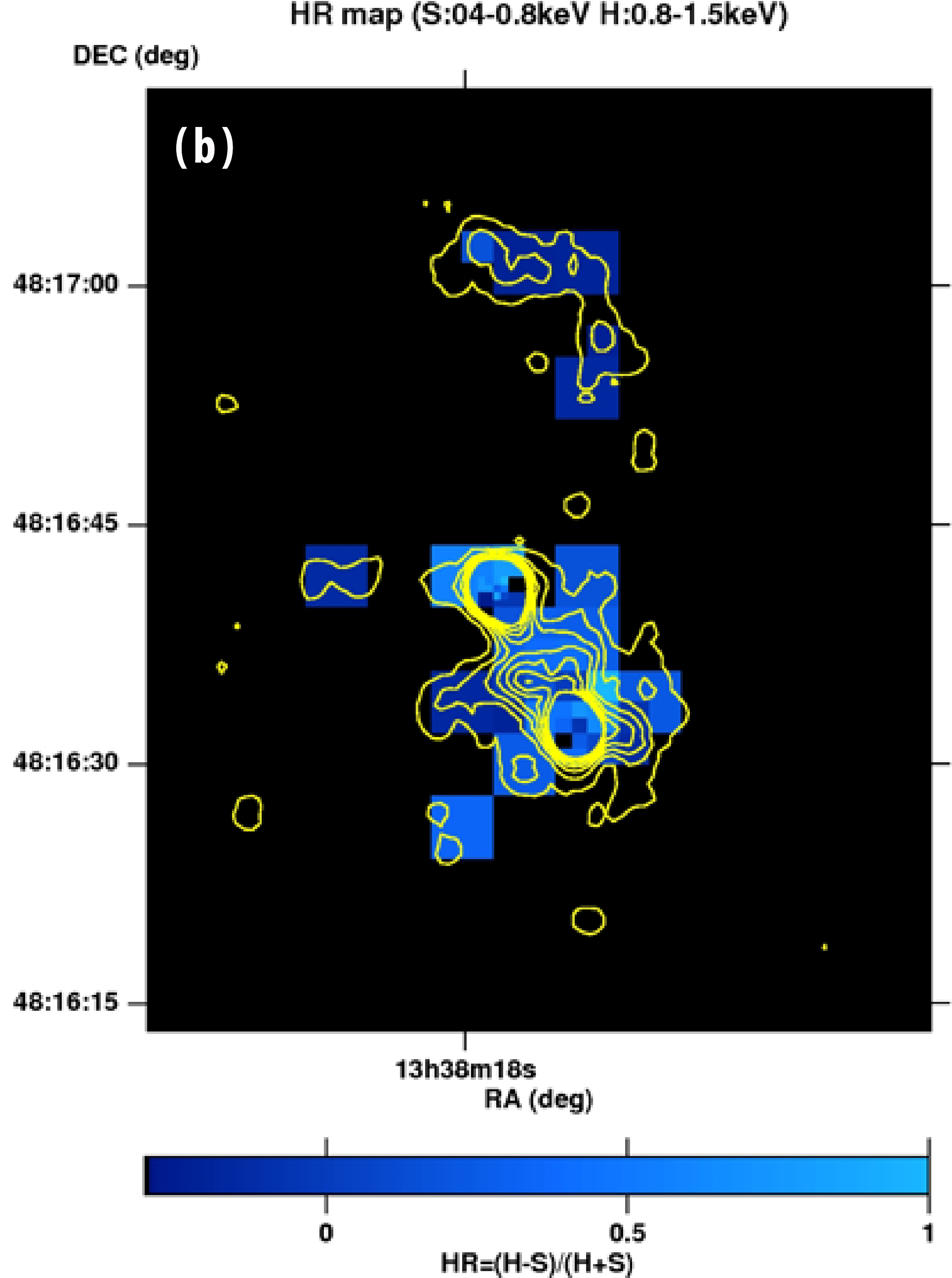} 
\else
\includegraphics[width=0.4\columnwidth,angle=0]{fig18a}
%\vskip 0.2in
\includegraphics[width=0.4\columnwidth,angle=0]{fig18b} 
\fi
\caption{\figcapXrayHardness \label{fig:ChandraHardness}}
\end{figure}
%\ifnum\Mode=2 \twocolumn \fi %[preprint2] mode only
\fi %close \ifnum\Mode=0 
%%%%%%%% End Figure %%%%%%%%%%%

Basic properties of the {\it Chandra} data for Mrk 266 were presented by \citet{2007MNRAS.377.1439B}.
We have performed additional processing to reveal previously unknown structural details and physical
properties. As shown in Figure \ref{fig:ChandraHardness}a, the SW (Seyfert 2) nucleus has a much softer
X-ray spectrum and lower luminosity than the NE (LINER or composite AGN/Starburst) nucleus. The
luminosity and hard X-ray spectrum of the NE nucleus provide a clear signature of an obscured AGN.
The X-ray spectrum obtained of the integrated system with XMM (Fig. \ref{fig:XMMSpectrum})
shows strong Fe K$\alpha$  emission at 6.4 keV. This high-ionizaton feature has been observed in 
the ULIRGs Arp 220 \citep{2005MNRAS.357..565I}, in the LIRG NGC 3690 \citep{2004ApJ...600..634B}, 
and more recently in the integrated (stacked) spectrum of 30 LIRGs in the GOALS
sample that are relatively faint in the hard X-ray band \citep{2009ApJ...695L.103I}. Modeling the global
XMM spectrum of Mrk 266 with a thermal (soft, extended) component due to star formation and an
absorbed power-law (hard, compact) component from the AGNs results in an absorption column of 
$\rm N_H = 6.5\times 10^{22}~ cm^{-2}$, which is close to the lower limits derived from the 
9.6 \micron\ silicate absorption feature (\S\ref{subsubsec:SpectralDiagnostics}). 

Unlike {\it XMM}, {\it Chandra} was able to resolve the nuclei and has detected two separate sources of
(2-7 keV) X-ray emission, with the bulk of the combined emission emanating from the NE component 
(Fig. \ref{fig:ChandraImages}). The physical origin of the bright Fe K$\alpha$ line observed in the {\it XMM}
spectrum of the integrated system is less clear. The {\it Chandra} spectrum of the total emission from 
Mrk 266 (Fig. \ref{fig:ChandraSpectra}f) does not show the Fe K$\alpha$ line. However, the large aperture
(33\arcsec\ radius) used to produce this spectrum introduces substantial noise in the background in the
higher energy bands where there is very little extended emission. Detailed inspection of the {\it Chandra}
data within small apertures centered on the nuclei provides a strong clue that the Fe K$\alpha$ line
originates primarily from the SW nucleus\footnote{The 6-7 keV (Fe K) image contains 10 counts from
the NE nucleus and 4 counts from the SW nucleus, while the neighboring 3-6 keV band shows a much
larger contrast, with 79 counts from the NE nucleus and only 1 count from the SW nucleus. In other words,
the spectrum of the SW nucleus has almost no detected flux at energies $> 4$ keV, but it shows a
sudden rise at the Fe K band (Fig. \ref{fig:ChandraSpectra}). This suggests that the hard band spectrum of
the SW nucleus is dominated by flux from the Fe K$\alpha$ line.}, indicative of a reflection-dominated
spectrum of a heavily obscured AGN. Since optical spectra reveal the presence of Seyfert 2 activity in the
SW nucleus, the weakness of the hard X-ray emission can be understood as a consequence of
suppression of X-rays by Compton thick absorption. Heavy obscuration towards the SW nucleus 
as inferred from the {\it Spitzer} IRS spectrum and the large column density of obscuring material
estimated from the CO (1-0) observations (\S\ref{subsec:MolGas})  
support this interpretation of the X-ray data. 

If the 6-7 keV photons in the {\it Chandra} spectrum of the SW nucleus are due to Fe K$\alpha$ line
emission, as our analysis suggests, the flux of $1.2^{+0.7}_{-0.5}\times 10^{-6}$ ph cm$^{-2}$ s$^{-1}$
accounts for the majority of the Fe K$\alpha$ line flux detected in the XMM-Newton spectrum. 
This means that the NE nucleus accounts for most of the escaping hard X-ray continuum, but it emits 
little Fe K emission. The equivalent width of Fe K expected from a source absorbed by 
$\rm N_{H}\approx (6-8)\times 10^{22}$~cm$^{-2}$, as inferred for the NE nucleus 
(Table \ref{tbl:XrayData}), is at most 0.1 keV 
\citep[e.g.][]{1991PASJ...43..195A,1994ApJ...420L..57K,1994MNRAS.267..743G}, 
while the equivalent width measured in the {\it XMM-Newton} spectrum is $\sim 0.5$ keV. 
This mismatch can be readily explained if the bulk of the Fe K$\alpha$ line emission is from the 
SW nucleus. The Fe K$\alpha$ line luminosity is $\rm 3\times10^{40}~erg~s^{-1}$.
Using the relation derived in \citet{2005MNRAS.362L..20I} for the Type 2 AGN 
in the HyLIRG IRAS F15307+3252,  the intrinsic 2-10 keV luminosity of the obscured AGN in 
Mrk 266 SW is $\rm\sim1\times10^{42}~f^{-1}~erg~s^{-1}$, where f is the (unknown) visible fraction 
of the reflecting matter in the line of sight. The observed $\rm F_x/F_{[O~III]}$ ratio for the 
SW nucleus also indicates a highly absorbed X-ray source with $\rm N_H \ga 10^{24}~ cm^{-2}$.
(See Appendix \S\ref{subsec:FeKdetails}.)

Finally, it should be emphasized that the Fe K$\alpha$ line detections are marginal. 
However, we gain confidence through independent $2\sigma$ detections for 
the global system with {\it XMM} and for the SW galaxy with {\it Chandra}, 
combined with line flux values that agree within the uncertainties in the 
calibrations and S/N (Table \ref{tbl:XrayData}). A detection of Fe 
has also been reported at the 90\% confidence level with
ASCA GIS observations \citep{2000A&A...357...13R}.
Confirmation requires more sensitive observations.

\subsubsection{The LINER in Mrk 266 NE}
\label{subsubsec:LINER}

The relative strength of the PAH features, silicate absorption, and continuum shape indicate 
Mrk 266 NE is typical of an ``IR LINER" or ``Transition" between IR-luminous LINERs 
and ``Type 1'' LINERs. Although the mid-IR data for Mrk 266 NE are only sufficient to compute 
3 of the 6 parameters in diagnostic diagrams (Fig. 2 of \citet{2006ApJ...653L..13S}),
the relatively small 6.2\micron/11.2\micron\ PAH flux ratio (1.1) is on the low end of the
distribution observed in ``IR LINERs'' (where the transition to Type 1 occurs), whereas
the continuum flux ratios $\rm f_{\nu}(15\micron)/f_{\nu}(6\micron) = 16$ and
$\rm f_{\nu}(30\micron)/f_{\nu}(6\micron) = 54$ place Mrk 266 NE on the extreme 
end of the distribution of ``IR LINERs'' with very red, warm dust continua. 
Mid-IR spectral data alone do not permit a clear distinction between the various kinds 
of energy sources that can power a LINER 
\citep[e.g.,][]{2006MNRAS.372..961K,2008ARA&A..46..475H}. 
However, there are at least four indicators that Mrk 266 NE is energized primarily by an AGN, 
in addition to shock excitation: 
(1) small PAH equivalent widths and a deficiency of 6.2 and 7.7 \micron\  PAH 
emission relative to 11.3 \micron\  PAH emission (\S\ref{subsubsec:SpectralDiagnostics});
(2) morphological evidence for a radiative bow shock within an ionization cone aligned with 
radio plasma in an outflow (\S\ref{subsubsec:Outflows});
(3) spectral properties of the luminous, hard X-ray point source (\S\ref{subsubsec:XrayAGNs}); and 
(4) a high $\rm H_2(1-0)~S(1)~to~Br\gamma$ flux ratio \citep{2009AJ....137.3581I}. 
Indeed, the strength and spatial extent of $\rm H_2$ emission indicate the presence
of extensive shocks in the region (\S\ref{subsubsec:WarmH2}) that 
may be triggered by outflow from the AGN (\S\ref{subsubsec:Outflows}).

\subsection{Between the Colliding Galaxies}
\label{subsec:Center}

\subsubsection{A Search for Counterparts to the Radio Emission}
\label{subsubsec:CenterSearch}

The bright radio continuum source located between the nuclei of Mrk 266 
(Figs. \ref{fig:NineBands}a and \ref{fig:ChandraVsRadio})
was interpreted as a source of enhanced synchrotron emission induced by shocking of the ISM at 
the interface of the merging galaxies \citep{1988ApJ...333..168M}.
Direct imaging of this area with {\it HST} ACS and NICMOS, {\it Spitzer} IRAC 
and MIPS (\S\ref{subsubsec:MultiImages}), and spectral mapping with {\it Spitzer} IRS 
(\S\ref{sec:IRSdata}) reveales no obvious source at optical continuum,
near-IR, or mid-IR wavelengths. This strongly rules out any ÒthirdÓ (dust-obscured) galaxy 
nucleus or a luminous off-nuclear star-forming region as observed in other interacting systems
such as the Antennae \citep{1998A&A...333L...1M} and II Zw 96 \citep{2010AJ....140...63I}.
Additonally, there is no detectable emission from this region in the near-UV or far-UV 
(Fig.  \ref{fig:UV_GALEX_XMM} and \ref{fig:NineBands}h), and there is no detection 
of warm molecular gas in the continuum-free
$\rm H_2~S(3)~9.67$ \micron\ emission-line image (Fig. \ref{fig:IRScube}).
Although a bridge of CO (1-0) emission has been detected between the 
galaxies \citep{2009AJ....137.3581I}, it is shown below (\S\ref{subsec:MolGas})
that this material is located to the north of the central radio continuum emission. 
Next we turn attention to the only other waveband where a source of radiation has 
been detected that is spatially correlated with the radio continuum emission between the galaxies.

\subsubsection{X-Rays Between the Nuclei}
\label{subsubsec:CenterXrays}

Substantial soft X-ray emission has been detected between the nuclei of 
Mrk 266 by {\it Chandra}. Analysis by \citet{2007MNRAS.377.1439B}
indicates that the X-ray emitting gas in this region has an intrinsic luminosity of 
$\rm 9.8 (\pm 0.5)x10^{40}~erg~s^{-1}$ and has a temperature  of 1.07 keV, approximately 
twice the temperature of much more diffuse emission surrounding 
the two galaxies (omitting the nuclei and northern source). Brassington et al. suggested that 
this X-ray emission is likely associated with the collision between the two galaxies,
as proposed previously for the radio continuum emission \citep{1988ApJ...333..168M}.
The present work extends these results using reprocessing of the {\it Chandra} data to 
enhance the definition of structure in the diffuse X-ray emission and to further 
characterize its physical properties. 

%%%%%%%% Begin Figure %%%%%%%%%%%
%Chandra full-band (0.4 - 7 keV) smoothed X-ray image in pseudocolor, 
% VLA 20 cm in white contours, and VLA 6 cm in black contours
%Xrays_20cm_6cm
\def\figcapChandraVLA{
\footnotesize 
Mrk 266 {\it Chandra} 0.4 - 7 keV (full band) X-ray image in pseudocolor, 
VLA 20 cm  radio continuum in white contours, 
and VLA 6 cm radio continuum in black contours. 
The X-ray data are based on image processing described in 
\S\ref{subsubsec:ChandraObs} of this study, 
and the radio data are from \citet{1988ApJ...333..168M}.
The brightest 20 cm contours have been omitted so as not to
obscure the 6 cm and X-ray emission. (See also Fig. \ref{fig:NineBands}.)
The field of view is 45" x 45" and the coordinate grid is in J2000 equatorial coordinates. 
}
\ifnum\Mode=0 %Insert Figure/Table only in [preprint] or [preprint2] modes
\placefigure{fig:ChandraVsRadio}
\begin{verbatim}fig19\end{verbatim}
\else
%For preprint
\begin{figure}[!ht]
\center
\ifnum\Mode=2
   \includegraphics[width=0.9\columnwidth,angle=0]{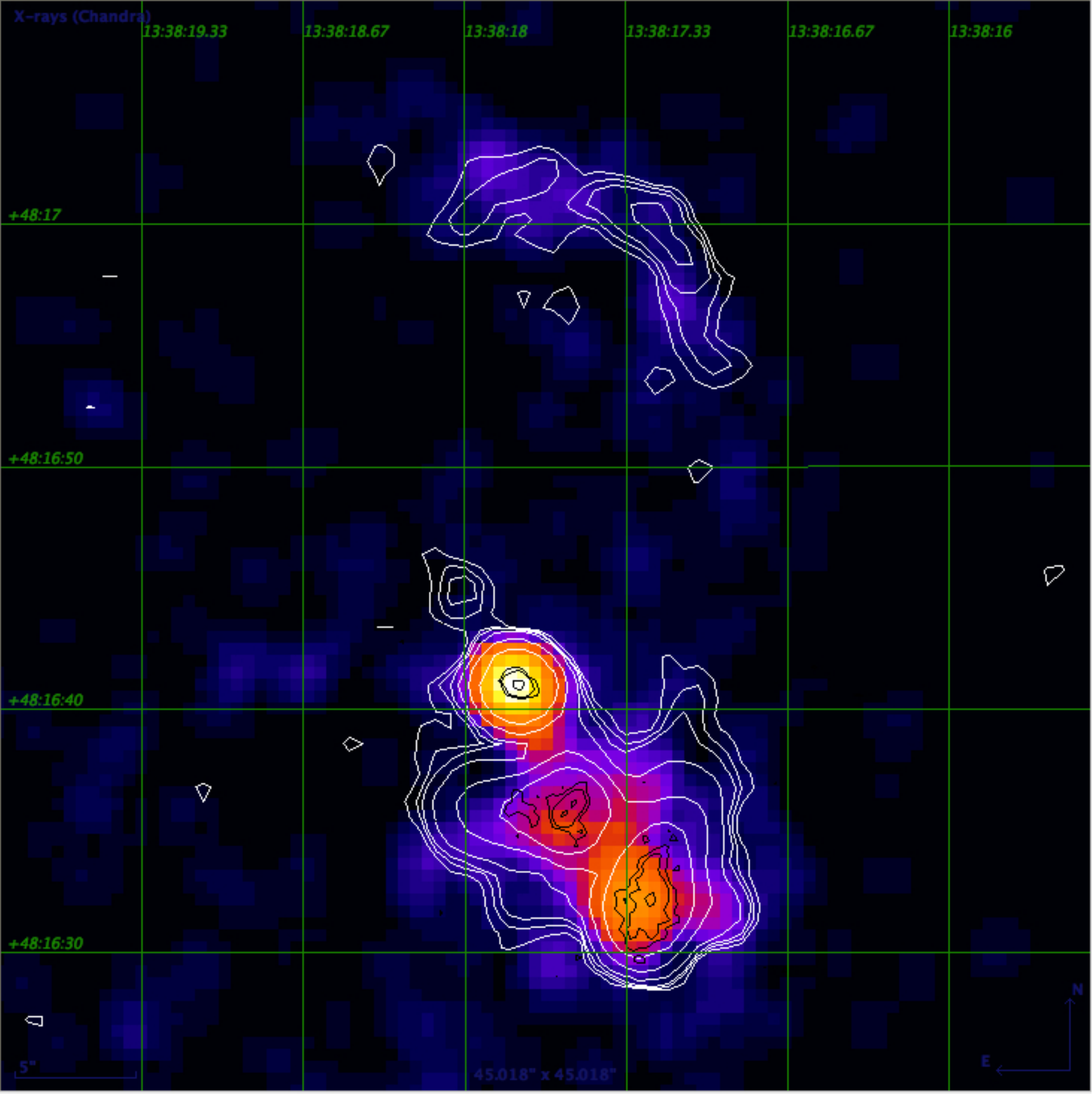} 
\else
  \includegraphics[scale=0.5,angle=0]{fig19} 
\fi
\caption{\figcapChandraVLA \label{fig:ChandraVsRadio}}
\end{figure}
%\ifnum\Mode=2 \twocolumn \fi %[preprint2] mode only
\fi %close \ifnum\Mode=0 
%%%%%%%% End Figure %%%%%%%%%%%

To begin with, Figures \ref{fig:ChandraImages} and \ref{fig:ChandraVsRadio} illustrate the 
remarkable correspondence between the 0.4-7.0 keV X-ray emission and the
20 cm radio continuum emission. As shown in Figures \ref{fig:NineBands} and 
\ref{fig:ChandraVsRadio}, the higher resolution 6 cm radio continuum emission 
(0\farcs3x0\farcs4 FWHM), resolves the 20 cm source into a flattened structure with two 
sub-components.  These components are coincident with the brightest knot of soft (0.4-4.0 keV) 
X-ray emission located between the galaxies. The contours of the full-band (0.4-7.0 keV)
X-ray image in Figure \ref{fig:ChandraHardness} further delineates the structure of the 
diffuse emission between the nuclei and to the south-west. Figure \ref{fig:ChandraHardness}b 
presents the hardness of the soft X-ray spectrum ($\rm HR_{soft}$), which 
indicates temperature variations within the extended emission. 

Figure \ref{fig:CenterXrays} provides a detailed view of the enhanced soft X-ray emission
between the two galaxies.  The X-ray spectra in the three circled 
areas show that  the SE region has a softer spectrum indicating a lower temperature 
(kT $\approx0.35$ keV) than the other two regions (kT $\approx0.8$ keV).
These differences are also reflected in the $\rm HR_{soft}$ map, where the SE region has 
$\rm HR_{soft} = -0.16$ and the region at the radio continuum peak and 4\arcsec\ NE of that 
location have $\rm HR_{soft}$ ranging from 0.2 to 0.4. Therefore,  the X-ray brightness certainly 
increases at the radio peak, but the temperature of the X-ray emitting gas at that location is similar 
to the region 4\arcsec\ (2.4 kpc) to the NW. 
A caveat in this analysis is that when dust obscuration is also varying spatially, 
$\rm HR_{soft}$ may not translate directly to real variations in temperature.

%%%%%%%% Begin Figure %%%%%%%%%%%
%Close up of the X-ray emission between the nuclei
%Mrk266_CenterXrays
\def\figcapCenterXrays{
\footnotesize
A close-up view of the {\it Chandra} soft-band (0.4 - 7 keV) X-ray
emission in the region between the nuclei and to the SW.
The three circled regions centered on the radio continuum peak (70 counts),
4\arcsec\ to the NW (70 counts), and 4\arcsec\ SE (39 counts), are used to
analyze the hardness of the X-ray emission to deduce differences in the gas temperature.
}
\ifnum\Mode=0 %Insert Figure/Table only in [preprint] or [preprint2] modes
\placefigure{fig:CenterXrays}
\begin{verbatim}fig20\end{verbatim}
\else
%For preprint
%\ifnum\Mode=2 \onecolumn \fi %4-panel fig is too small in 2-column mode
\begin{figure}[!htb]
\center
\ifnum\Mode=2 
\includegraphics[width=0.8\columnwidth,angle=0]{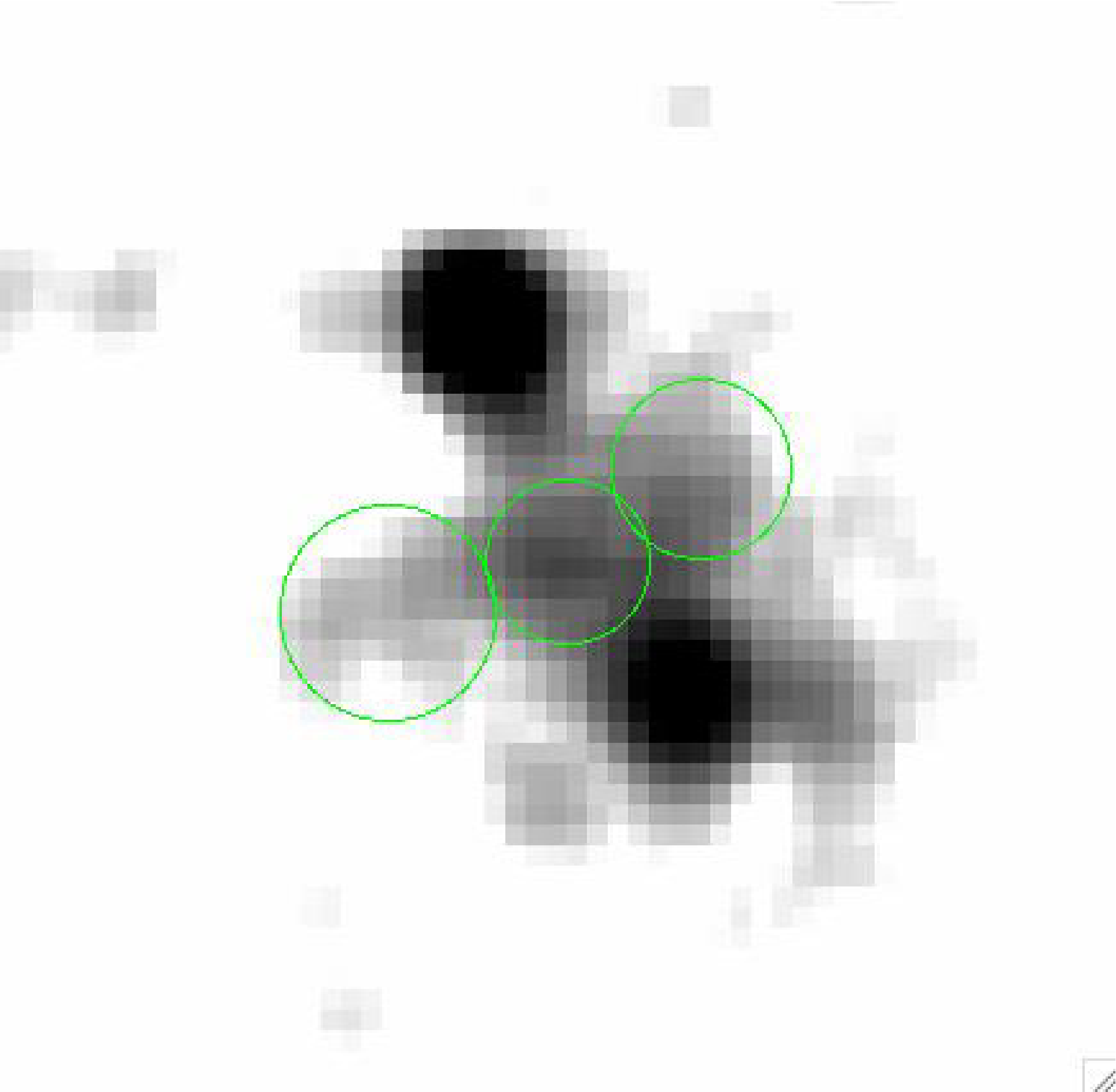} 
\else
\includegraphics[scale=0.5,angle=0]{fig20} 
\fi
\caption{\figcapCenterXrays \label{fig:CenterXrays}}
\end{figure}
\fi %close \ifnum\Mode=0 
%%%%%%%% End Figure %%%%%%%%%%%

Table \ref{tbl:CenterXrays} presents physical parameters estimated from spectral analysis of the soft X-ray
emission illustrated in Figure \ref{fig:CenterXrays}.  The most straightforward interpretation is that the
X-ray emission is the result of gas being shock-heated to $\rm T \sim 10^7$.  This physical
process would also give rise to the observed enhanced synchrotron radiation \citep{1988ApJ...333..168M}.
\citet{2000ApJ...535..735D} interpreted the lack of $\rm H_2$ 1-0 S(1)
emission from the central region of Mrk 266 as evidence for very fast shocks with
relative velocities of $\rm\approx1000~km~s^{-1}$ produced at the interface
of superwinds emerging from the two galaxies. Such fast shocks would generate the
observed non-thermal radio continuum emission and also heat the gas to $\ga10^6$ K, far above 
the dissociation temperature of $\rm H_2$. This would explain the lack of detected $\rm H_2$ 
emission lines between the galaxies in the near-infrared observations of Davies et al. and 
in the mid-infrared IRS spectroscopy presented in the current study (\S\ref{sec:IRSdata}). 
This physical scenario also predictes higher temperatures in the shocked region compared with 
the diffuse superwind. This is exactly what our refined analysis of the {\it Chandra} data reveals.

The estimated cooling time in this region is only 4 Myr.  This is very short compared 
to the dynamical time scale of the merger (a few hundred million years), and therefore it is 
likely that we are observing this emission during a short-lived phase of the interaction. 
This would explain why so few interacting systems have been found with 
radio continuum and X-ray emission enhanced between the galaxies to the same
extent as in Mrk 266. However, as shown below, there is evidence that gas in this region may 
be replenished from the powerful starburst-driven superwinds (\S\ref{subsec:Superwind}), 
or from gravitational infall of material toward the center of mass of the system
(\S\ref{subsec:MolGas}).

%%%%%%%% Begin Table %%%%%%%%%%%
\def\tableCenterXrays{
\begin{deluxetable}{lrl}
\ifnum\Mode=2
\renewcommand\arraystretch{0.5}% (MyValue=1.0 is for standard spacing)
\tabletypesize{\scriptsize}
\setlength{\tabcolsep}{0.00in} %Tighten up the columns. See AASTeX FAQ
\tablewidth{\columnwidth}
\else
\renewcommand\arraystretch{1.0}% (MyValue=1.0 is for standard spacing)
\tabletypesize{\normalsize} 
\setlength{\tabcolsep}{0.50in} %Tighten up the columns. See AASTeX FAQ
%\tablewidth{\textwidth}
\fi
\tablecaption{Spectral Analysis of X-rays Between the Nuclei \label{tbl:CenterXrays}}
\tablehead{
\colhead{Parameter} & 
\colhead{Value} & 
\colhead{Units} \\ 
\colhead{(1)} & 
\colhead{(2)} & 
\colhead{(3)}
}
\tablecolumns{3}
\startdata
%%%%%%%%%
0.5-2 keV Flux  & $\rm 2.7x10^{-14}$ & $\rm erg~s^{-1}~cm^{-2}$ \\
Temperature (kT)  & $\rm 0.72^{+0.11}_{-0.09}$ & keV \\
Luminosity\tablenotemark{a} & $\rm 1.4x10^{41}$ &  $\rm erg~s^{-1}$ \\
Volume & $\rm 8.1x10^{65}$ &  $\rm cm^{3}$ \\
Electron Density & $\rm 1.1x10^{-2}$ &  $\rm cm^{-3}$ \\
Thermal Energy & $\rm 1.8x10^{55}$ &  $\rm erg$ \\
Gas Mass & $\rm 1.2x10^{7}$ &  \Msun \\
Cooling Time& $\rm 4x10^{6}$ &  yr \\
 \enddata
\tablenotetext{a}{
The bolometric luminosity of this thermal gas, as deduced via extrapolation
from fitting the data at energies of 0.01-10 keV.\\
}
\tablecomments{
\footnotesize
Column (1): Physical parameter computed from analysis of the soft band (0.5-2.0 keV) {\it Chandra} emission
in the region between the colliding disks of Mrk 266. Column (2): Value. Column (3): Units.
The computations assume an angular-scale distance of 129 Mpc and the scale of 
0.59 kpc/\arcsec\ (see \S\ref{subsec:background}), and the emitting region is approximated by a tube 
4\arcsec\ in diameter and 13\arcsec\ in length at a position angle of $-45\arcdeg$.
}
\end{deluxetable}
}
\ifnum\Mode=0
\placetable{tbl:CenterXrays}
\else
\tableCenterXrays
\fi
%%%%%%%% End Table %%%%%%%%%%%
%\ifnum\Mode=2 
%\pagebreak
%\fi

\subsection{The Superwind and Northern Loop}
\label{subsec:Superwind}

An outstanding characteristic of Mrk 266 is the extensive filamentary 
nebula of ionized hydrogen $\sim$30 kpc in diameter \citep{1990ApJ...364..471A}
contained within an X-ray nebula $\sim$100 kpc in extent. Optical emission-line
diagnostics and kinematics indicate this is one of the most energetic examples 
of an outflowing, starburst-driven superwind \citep{1997ApJ...474..659W}. 
Mrk 266 has the third highest $\rm H\alpha$ luminosity in a sample of 40 local
infrared-bright galaxies studied by \citet{1990ApJ...364..471A}.
\citet{1998AA...336L..21K} presented deep observations with 
{\it ROSAT}~HRI that revealed luminous soft X-ray emission 
spatially correlated with a loop of H$\alpha$ and [O III] $\lambda$5007 line
emission $\sim$14\arcsec\ long by 7\arcsec\ wide extending $\sim$10-34\arcsec\ (6-20 kpc) 
to the north of the system (herein called the Northern Loop),
leading these authors to interpret the X-ray emission in this region
as hot post-shock gas in a high-velocity ``jet'' (most likely emanating from the SW nucleus).
\citet{1998AA...336L..21K} argued also that the X-ray emission is too luminous and the opening 
angle of the X-ray/[O III] emitting region is too small to be explained by a superwind.
A third possibility, suggested by \citet{2007MNRAS.377.1439B}, is that the northern 
structure is a tidal tail containing in situ star formation that is responsible for the soft X-ray emission 
and H$\alpha$ emission. In the subsections below we exploit the new multiwavelength data to 
make the case that the Northern Loop is the most luminous component of a superwind energized 
by star formation and one (or both) AGNs.

%%%%%%%% Begin Figure %%%%%%%%%%%
%Mrk266_DiffuseXrays 
\def\figcapDiffuseXrays{
\footnotesize 
{\it Chandra} 0.4 - 7 keV (full band) X-ray data contrasting the high
surface-brightness emission (highest resolution) with the large-scale diffuse emission.
Left (a): Processed for maximum spatial resolution and smoothed with a 2-pixel
Gaussian kernel (same as Fig. \ref{fig:ChandraImages}, bottom left).
Right (b): The same data smoothed with a 6-pixel Gaussian kernel
to highlight diffuse emission.
}
\ifnum\Mode=0 %Insert Figure/Table only in [preprint] or [preprint2] modes
\placefigure{fig:DiffuseXrays}
\begin{verbatim}fig21a_21b\end{verbatim}
\else
\begin{figure}[h]
\center
\ifnum\Mode=2 
\includegraphics[width=1.0\columnwidth,angle=0]{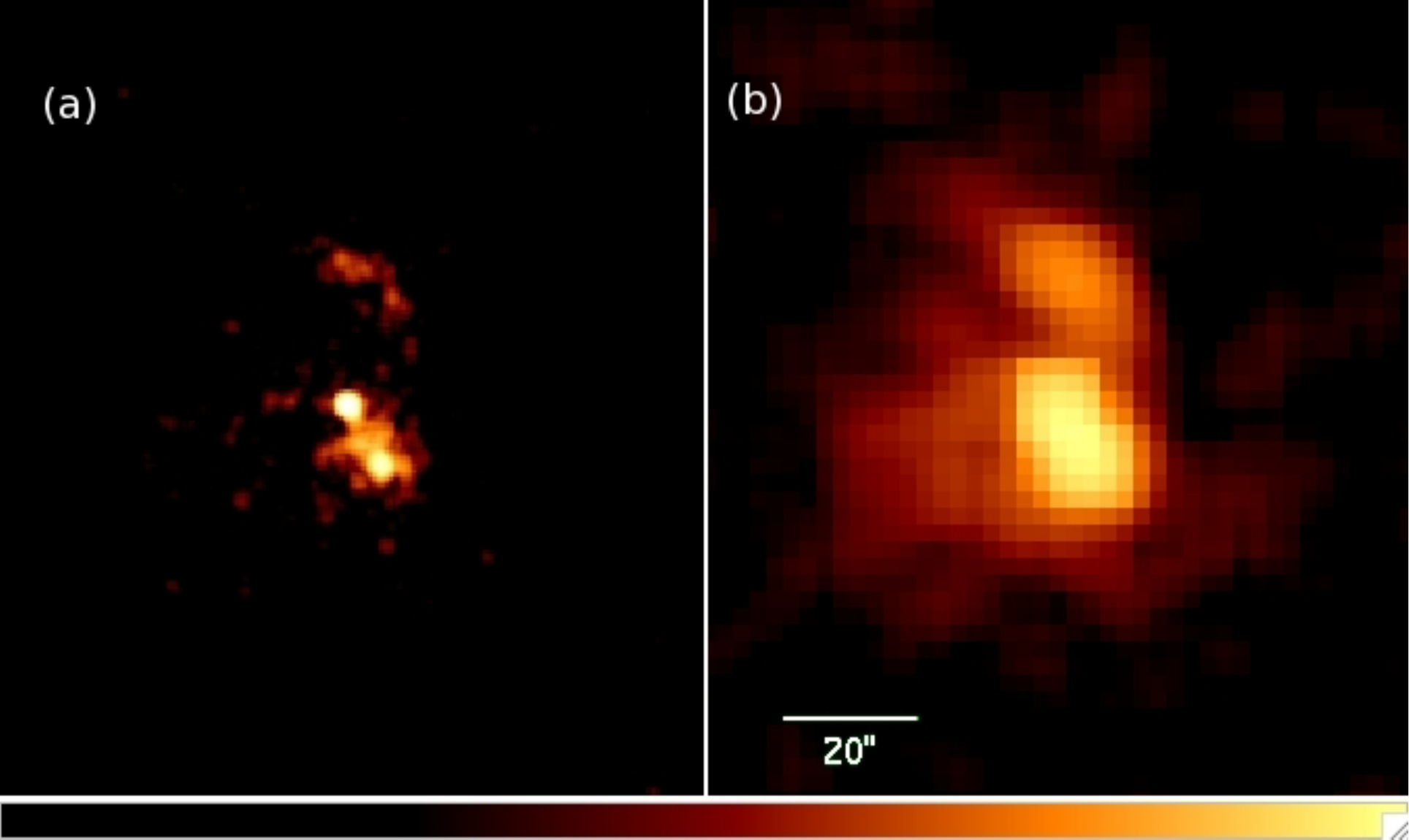}
\else
\includegraphics[scale=0.5,angle=0]{fig21a_21b}
\fi
\caption{\figcapDiffuseXrays \label{fig:DiffuseXrays}}
\end{figure}
%\ifnum\Mode=2 \twocolumn \fi %[preprint2] mode only
\fi %close \ifnum\Mode=0 
%%%%%%%% End Figure %%%%%%%%%%%

%%%%%%%% Begin Figure %%%%%%%%%%%
\def\figcapSoftXrays{
\footnotesize
Soft X-rays compared to other wavelengths:
(a): {\it Chandra} soft (0.4-1 keV) X-ray image re-binned with 2\arcsec x 2\arcsec\ pixels 
(grayscale). The contours are the same data smoothed with a 3-pixel Gaussian kernel to 
emphasize the faint, diffuse emission; the same contours are also overlaid in the other three panels. (b):  {\it HST}/ACS B-band (0.44 \micron, F435W) image (grayscale) with the 
0.4 - 1 keV X-ray data overlaid as magenta contours and the
VLA 20 cm radio continuum image \citep{1988ApJ...333..168M} overlaid as orange contours.
(c): H$\alpha$ image \citep{2000PASJ...52..185I} with the same contours as panel (b).
(d): 24 \micron\ {\it Spitzer} MIPS image with the same contours as panels (b) and (c).
Each field is 1\farcm2 x 1\farcm2 and the scale bar indicates 15\arcsec.
}
\ifnum\Mode=0 %Insert Figure/Table only in [preprint] or [preprint2] modes
\placefigure{fig:SmoothXrays}
\begin{verbatim}fig22a_22d\end{verbatim}
\else
%For preprint
%\ifnum\Mode=2 \onecolumn \fi %4-panel fig is too small in 2-column mode
\begin{figure}[!htb]
\center
\ifnum\Mode=2 
\includegraphics[width=1.0\columnwidth,angle=0]{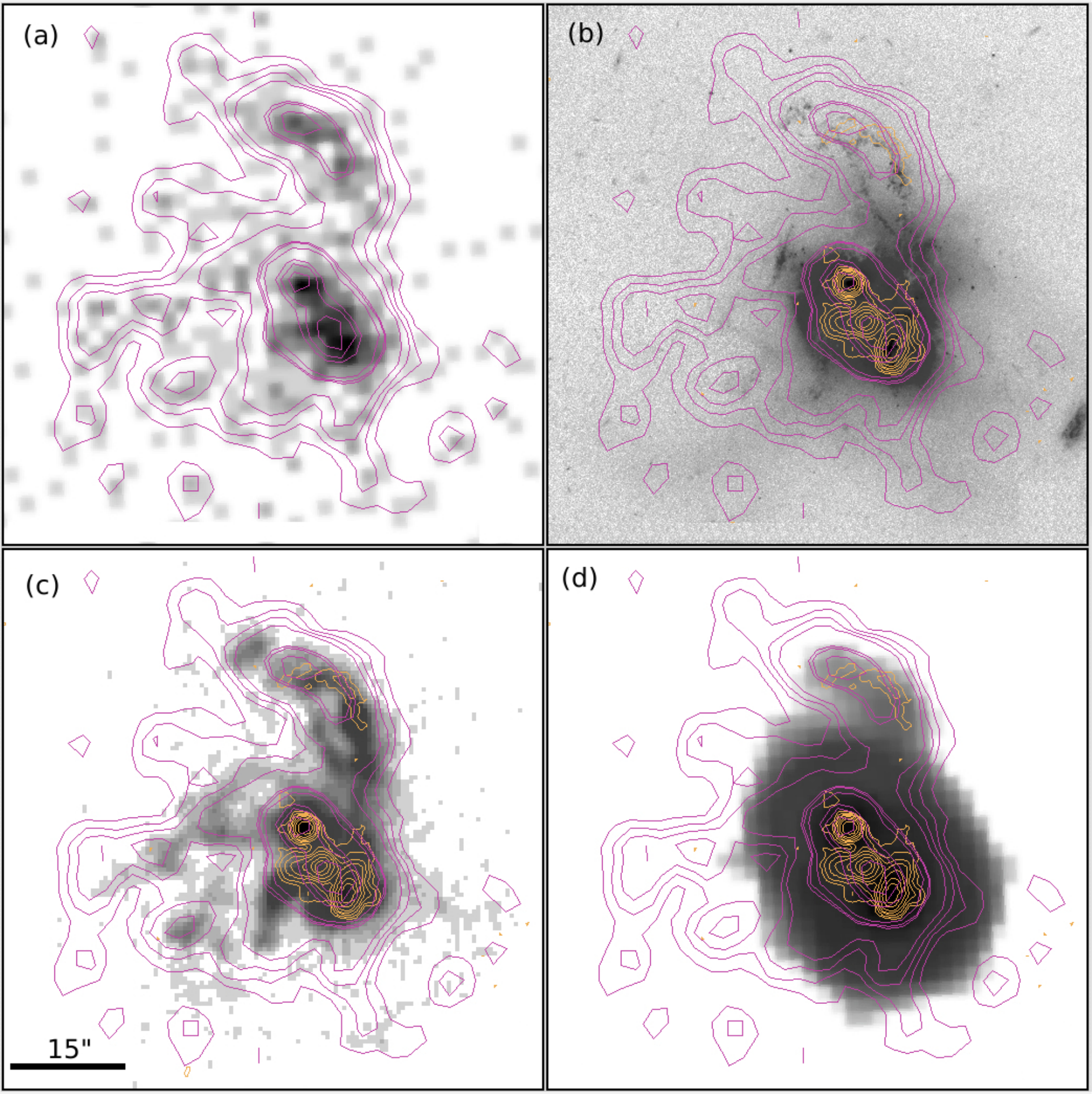} 
\else
\includegraphics[scale=0.7,angle=0]{fig22a_22d} 
\fi
\caption{\figcapSoftXrays \label{fig:SmoothXrays}}
\end{figure}
%\ifnum\Mode=2 \twocolumn \fi %[preprint2] mode only
\fi %close \ifnum\Mode=0 
%%%%%%%% End Figure %%%%%%%%%%%

\subsubsection{Spatially Correlated Soft X-rays and H$\alpha$ Filaments}
\label{subsec:SuperwindXrays}

Our reprocessing of the {\it Chandra} data (Fig. \ref{fig:DiffuseXrays}) reveals that 
the X-ray emission detected in the direct high-resolution data is a high-surface-brightness 
ridge embedded in a longer spur that extends an additional $\sim$20\arcsec\ (12 kpc) 
toward the NE. A map of the hardness of the soft X-ray spectrum 
(Fig. \ref{fig:ChandraHardness}) shows the northern region has 
$\rm HR_{soft} \approx -0.2$, indicating a relatively cool temperature of kT$\approx0.3$ keV.
If this is a region of in situ star formation, one would expect to find a 
high density of star clusters; however, the nearest detected optical star
clusters are more than 5\arcsec\ (3 kpc) SW from the brightest X-ray and radio continuum
emission in the Northern Loop (\S\ref{subsec:SCs}).  Therefore, the properties of this feature
must be ascribed to something other than starlight.
Figure \ref{fig:SmoothXrays} shows that the diffuse soft X-ray emission (0.4 - 1.0 keV)
corresponds remarkably well with observed H$\alpha$ filaments and knots that extend out to
$\sim$18 kpc (30\arcsec) to the north and to the east of the galaxies. In the northern region,
the soft X-rays have highest surface brightness near the outer edge of the Northern Loop 
detected in H$\alpha$ and in the B band (including [O II]$\lambda$ 3727 
line emission), which is also co-spatial with 20 cm radio continuum emission. 
Close spatial correlation between hot X-ray emitting gas and warm
H$\alpha$-emitting gas is a signature of winds in objects ranging from 
ULIRGs to dwarf starburst galaxies \citep{2005ApJ...628..187G}.

%%%%%%%% Begin Figure %%%%%%%%%%%
%Zoom in on the northern loop, HST B band
\def\figcapLoopZoom{
\footnotesize
{\it HST}/ACS B-band (0.44 \micron, F435W) image in grayscale with
a field of view and intensity mapping chosen to emphasize the filamentary
structure in the Northern Loop of Mrk 266. The arrows point out elongated structures 
0.5-1\arcsec\ (300-600 pc) in length on the western side of the loop that appear 
to be aligned radially with the NE nucleus. 
}
\ifnum\Mode=0 %Insert Figure/Table only in [preprint] or [preprint2] modes
\placefigure{fig:NorthernLoopZoom}
\begin{verbatim}fig23\end{verbatim}
\else
%For preprint
%\ifnum\Mode=2 \onecolumn \fi %4-panel fig is too small in 2-column mode
\ifnum\Mode=2 
\begin{figure}[!htb]
\center
\includegraphics[width=1.0\columnwidth,angle=0]{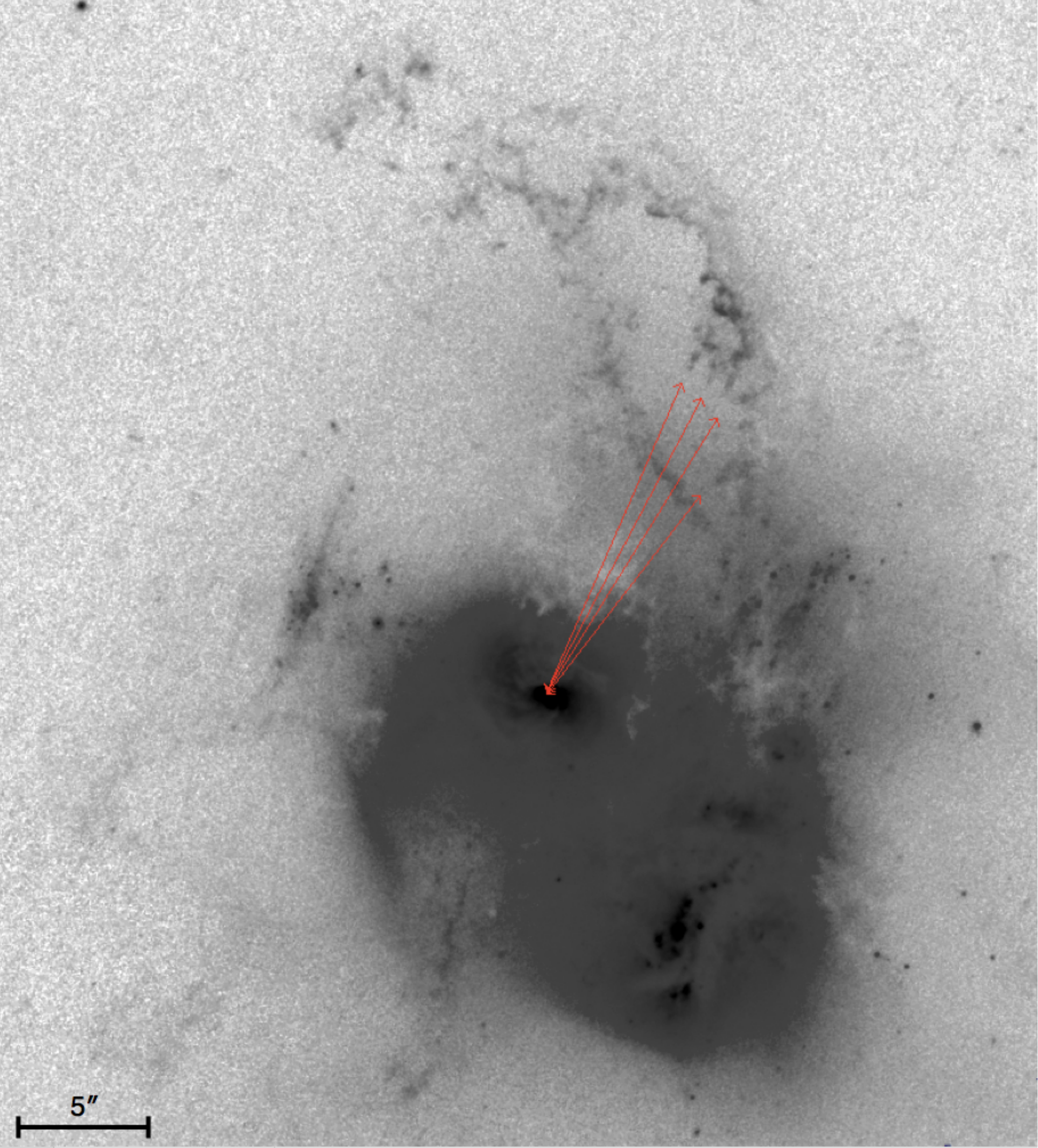} 
\else
\begin{figure}[!tb]
\center
\includegraphics[scale=0.5,angle=0]{fig23} 
\fi
\caption{\figcapLoopZoom \label{fig:NorthernLoopZoom}}
\end{figure}
%\ifnum\Mode=2 \twocolumn \fi %[preprint2] mode only
\fi %close \ifnum\Mode=0 
%%%%%%%% End Figure %%%%%%%%%%%

\subsubsection{Structure in the Northern Loop at $\sim$100 pc Scales}
\label{subsec:Superwind100pc}

The filamentary morphology of the Northern Loop revealed by the {\it HST} 
imagery is evidence for an expending shell or ``bubble'' that is fragmenting 
due to Rayleigh-Taylor instabilities in a starburst- or AGN-driven superwind.
Three dimensional hydrodynamic simulations of galactic winds
predict this phenomena \citep{2008ApJ...674..157C}. 
Elongated structures $\approx0.5-1$\arcsec\ (300-600 pc) in length protruding 
inward from the western, outer edge of the Northern Loop are 
highly suggestive of Rayleigh-Taylor fingers or bow shocks predicted by 
simulations \citep[e.g.,][]{1990ApJS...74..833H,2008ApJ...674..157C}. 
These protrusions appear to have a radial distribution centered on the NE nucleus 
(Fig. \ref{fig:NorthernLoopZoom}), indicating this AGN may be a 
primary energizing source for this component of the superwind.
\citet{1997ApJ...474..659W} and \citet{2000PASJ...52..185I} have shown 
that filaments throughout Mrk 266, including the Northern Loop,
have line ratios (e.g., $\rm [O~III]\lambda 5007/H\beta$) indicating photoionization 
from an AGN dominates over ionization from massive stars, and shocks
that could explain the LINER-like nebular spectrum have insufficient energy 
to account for the luminosity of the extended nebula.
However, the luminous starburst in the SW galaxy is also likely to be a major
contributor to the over all energetics of the superwind.

The morphological, kinematic and ionization properties of the prominent Northern Loop
can be interpreted with the superwind model of \citet{1990ApJS...74..833H}.
In this scenario, the star-burst driven gaseous outflow has inflated a ``bubble'' that is highly 
elongated along the direction of least resistance (maximum pressure gradient), 
and the closed loop of optical line emission indicates the superwind in Mrk 266 is 
in a relatively young radiative cooling phase that occurs before the wind expands
further during a ``blow-out'' phase.  This again is consistent with the idea that we are 
observing a relatively short lived and dynamic phenomena.

\subsubsection{Dust in the Wind}  \label{subsec:SuperwindDust}

The detection of 24\micron\ emission co-spatial with the Northern Loop 
(Fig. \ref{fig:SmoothXrays}d) is strong evidence for thermal continuum emission 
from dust grains entrained in the superwind.  By estimating the dust temperature 
and grain emissivity profile, the mass of dust can be derived following the single 
temperature thermal model of \citet{1983QJRAS..24..267H}, which is given by 
\begin{equation}\label{Mdust24a}
M_{dust} = [(4/3)~a~\rho~ / Q_{\nu}] [S_{\nu}~D^2 / B(T,\nu)],
\end{equation}
where $a$ is the average grain size (0.1 \micron), 
$\rho$ is the density of grain material,
$Q_\nu$ is the grain emissivity characterized as
$Q_{250\mu m}(250\mu m/\lambda)^{\beta}$,
$S_\nu$ is the flux density of the thermal emission at frequency $\nu$,
$D$ is the distance, and $B(T,\nu)$ is the Planck function.
Assuming $\beta = 1$ and other grain properties from
\citet{1983QJRAS..24..267H} results in
$[{(4/3)~a~\rho/Q_{24\mu m}}] = {\rm 9.6x10^{-3}~g~cm^{-2}}$ at 24 \micron.
In more convenient units, the above equation becomes
\begin{equation}\label{Mdust24}
M_{dust} = 0.016~S_{\nu}(24\mu m)~D^2~ [e^{(599.49/T_{dust})} -1]~M_{\odot},
\end{equation}
where $S_{\nu}(24\mu m)$ is in Jy, $D$ is in Mpc, $T$ is in K, 
and $M_{dust}$ is in \Msun.
The 24 \micron\ flux density measured for the Northern Loop
(17 mJy, or 2\% of the total flux) corresponds to a dust mass 
$\rm M_d (24\mu m) \approx 5\times10^7$\Msun.  
Using the calibration of \citet[][]{1999A&A...343...51A} based on 
the known case of M82, the total far-infrared luminosity ratio of
$\rm L_{ir}(Mrk~266)/L_{ir}(M82) = 18$ predicts a dust mass of 
$\rm\sim 2\times10^{7-8}~M_{\odot}$ in the entire wind of Mrk 266.
This would imply that the 24 \micron\ emission
from the Northern Loop contains at least 25\% (and possibly the majority)
of the total dust mass within the superwind of Mrk 266.

However, this result depends on the assumption that most of the observed flux is 
due to dust emission alone and that we have made a reasonable determination of 
$\rm T_{dust}$.  One possible source of alternate emission is the [O IV] 25.9 \micron\ 
emission line.  We estimate this contribution to be less than 10\%\footnote
{
This estimate is based on the assumption that the mixture of ionized gas and 
thermal dust in the superwind of Mrk 266 may resemble a scaled up version of a 
planetary nebula such as NGC 2346 \citep{2004ApJS..154..302S}.
The published IRS spectrum of Mrk 266 shows a bright [O IV] 25.9 \micron\
emission line, however it does not permit separation of emission 
from the SW nucleus and the diffuse emission in the Northern Loop.
}.
Since there is no detection in any other mid-IR or
sub-mm passband available to help constrain $\rm T_{dust}$,
a characteristic temperature of 37 K is assumed based on
modeling of sub-mm observations of the dusty outflow in M82 
\citep{1999A&A...343...51A}. 

The overall physical interpretation presented above adds strong support for the superwind
interpretation of \citet{1997ApJ...474..659W} in which stellar winds and supernovae are driving
a powerful outflow of gas and dust. The Northern Loop is likely the cross section of the
most luminous ``bubble'' of expanding ionized gas in the wind. 
The fragmented morphology and radial orientation of protrusions in this region 
suggest it is strongly influenced by radiation or a gaseous outflow energized by
the NE galaxy, and the detection of co-spatial 24 \micron\ emission 
provides strong evidence for dust mixed with the ionized gas in the superwind.
The new observations are difficult to reconcile with a jet (ejection from an AGN) as 
suggested by \citet{1998AA...336L..21K} or with a (stellar) tidal tail as hypothesized 
by \citet{2007MNRAS.377.1439B}.

\subsection{SED Analysis}
\label{subsec:SEDanalysis}

\subsubsection{Decomposition of Dust Emission Components}\label{subsubsec:SED_Dust}

Figure \ref{fig:SEDcafe} presents results of a multicomponent model decomposition
of the IRS spectrum (Fig. \ref{fig:IRScube_SLLL_Total}) combined with the broad-band 
SED (Table \ref{tbl:SED_Tot}, Fig. \ref{fig:SEDs}) of global measurements for 
the Mrk 266 system using the CAFE software package 
\citep{2007ApJ...670..129M}\footnote{
CAFE could not be used to constrain the fits for the individual galaxies because the 
IRS spectral map could not resolve the galaxies over the 14-38\micron\ range of
the LL module (\S\ref{sec:IRSdata}).}.
A very good fit to the SED is provided by a model consisting of
PAH emission, atomic and molecular line emission, plus
cold (26 K), cool (72 K), warm (235 K) and hot (1500 K) thermal dust 
components heated by the combination of an unobscured 
photospheric interstellar radiation field (ISRF), an obscured starburst (SB) 
component with a strength constrained by the observed PAH emission, 
and an AGN (accretion disk) component with a strength 
constrained by the fitted hot dust emission.
Uncertainties in the derived dust temperatures are $\pm 1$ K.

Table \ref{tbl:SEDfitting} lists the results of the fit.
The derived extinction-corrected PAH flux ratios appear 
to be fairly ``normal,'' located between the curves for ionized and 
neutral PAHs in Figure 16 of \citet{2001ApJ...551..807D}.
The parameter $\rm L_{source}$ is the sum of the derived 
intrinsic source luminosities $\rm L_{ISRF}+L_{SB}+L_{AGN}$, and
$\rm L_{dust}$ is the total dust luminosity.
The primary results are the relative fractions of $\rm L_{source}$,
the fractions of $\rm L_{dust}$, and the dust masses characterized as arising from cold, cool, 
warm and hot thermal dust components. As noted in \citet{2007ApJ...670..129M},
the four dust components used to model galaxies are typically heated to 
characteristic temperatures of $\rm T_{cold}\approx 35 K, T_{cool}\approx 80 K, 
T_{warm}\approx 200 K, and~T_{hot}\approx 1400 K$, where the actual temperatures 
(noted above and in Table \ref{tbl:SEDfitting} for Mrk 266) are determined from fitting 
of the radiation field energy density of the corresponding radiation sources.
Although the model suggests that a hot dust component accounts for 
$\approx$10\% of the total dust luminosity, this is uncertain because the fitting 
of this component is weighted heavily by the mid-IR wavelength region 
where emission from hot dust competes with stellar photospheric emission.
Reconciliation of this relatively low AGN contribution with spectral diagnostics that 
suggest a $\sim$50\% AGN contribution in the individual galaxies 
(\S\ref{subsubsec:SpectralDiagnostics}) may be due primarily to an aperture effect. 
The latter measurements were made in apertures that isolate the nuclear regions, 
and they are thus more sensitive to AGN signatures than the larger apertures 
used to construct the SED for the global system which encompass more spatially 
extended starburst emission.

%%%%%%%% Begin Figure %%%%%%%%%%%
%SED fitting with CAFE
%Mrk266.cafe.sed.mix
\def\figcapSEDcafe{
\footnotesize
Results of multicomponent fitting using the CAFE software package 
\citep{2007ApJ...670..129M} to model the SED of the global emission from Mrk 266.
Solid black circles are broad-band continuum measurements, and
open black squares are IRS spectral measurements. The solid orange curve 
represents the sum of the various fitted components identified in the legend.
The corresponding fit parameters and uncertainties are given in Table \ref{tbl:SEDfitting}.
}
\ifnum\Mode=0 %Insert Figure/Table only in [preprint] or [preprint2] modes
%\vskip 0.3in
%\fbox{\bf Figure 25 goes here.}
%\vskip 0.3in
\placefigure{fig:SEDcafe}
\begin{verbatim}fig24\end{verbatim}
\else
%For preprint
%\ifnum\Mode=2 \onecolumn \fi %4-panel fig is too small in 2-column mode
\begin{figure}[!tb]
\center
\ifnum\Mode=2
\includegraphics[width=1.0\columnwidth,angle=0]{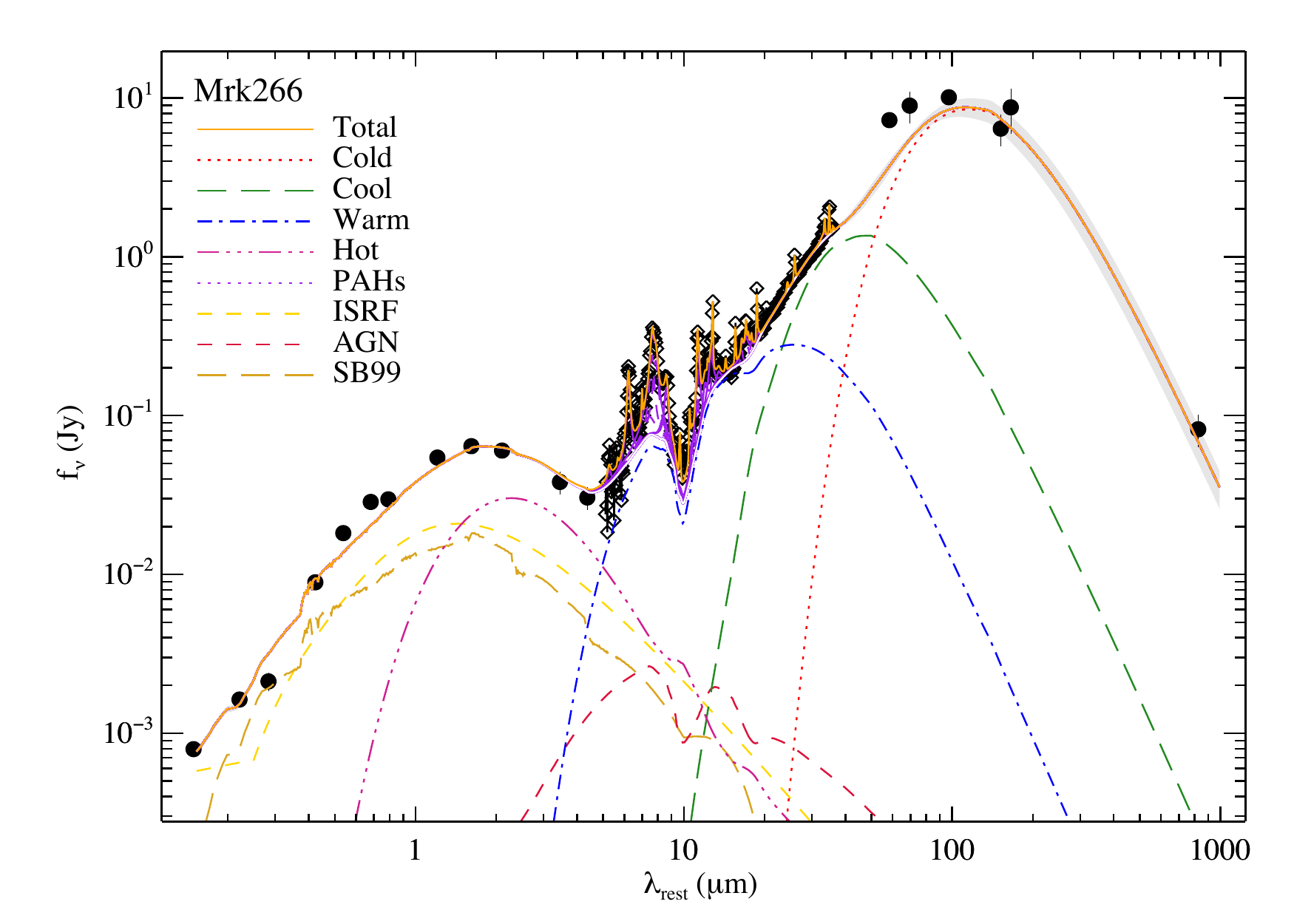}
\else
\includegraphics[width=4.8truein,angle=0]{fig24}
\fi
\caption{\figcapSEDcafe \label{fig:SEDcafe}}
\end{figure}
\fi %close \ifnum\Mode=0 
%%%%%%%% End Figure %%%%%%%%%%%

%%%%%%%% Begin Table %%%%%%%%%%%
\def\tableSEDfitting{
\begin{deluxetable}{ll}
\ifnum\Mode=2
\renewcommand\arraystretch{0.5}% (MyValue=1.0 is for standard spacing)
\tabletypesize{\scriptsize}
\setlength{\tabcolsep}{0.00in} %Tighten up the columns. See AASTeX FAQ
\tablewidth{\columnwidth}
\else
\renewcommand\arraystretch{0.6}% (MyValue=1.0 is for standard spacing)
\tabletypesize{\small} 
\setlength{\tabcolsep}{0.40in} %Tighten up the columns. See AASTeX FAQ
%\tablewidth{\textwidth}
\fi
\tablecaption{Results of SED Decomposition \label{tbl:SEDfitting}}
\tablehead{
\colhead{Parameter} & 
\colhead{ISRF+SB+AGN} \\ 
\colhead{(1)} & 
\colhead{(2)}
}
\tablecolumns{2}
\startdata
%%%%%%%%%
$\rm \tau_{warm} (9.7\mu m)$\tablenotemark{a}  & $\rm 2.7\phn \pm 0.1$ \\
$\rm PAH~6.2\mu m / PAH~7.7\mu m$   &  $\rm 0.28 \pm 0.03$ \\
$\rm PAH~11.3\mu m / PAH~7.7\mu m$  &  $\rm 0.28 \pm 0.02$ \\
$\rm L_{ISRF}/L_{source} $      & $\rm 0.19$ \\
$\rm L_{SB}/L_{source} $         & $\rm 0.71$ \\
$\rm L_{AGN}/L_{source} $      & $\rm 0.10$ \\
$\rm L_{source}~(L_{\odot})$   & $2.4 (\pm 0.2) \times 10^{11}$\\
$\rm L_{cold}~(26~K)/L_{dust} $         & $0.57 \pm 0.09$ \\
$\rm L_{cool}~(72~K)/L_{dust} $         & $0.20 \pm 0.03$ \\
$\rm L_{warm}~(235~K)/L_{dust} $\tablenotemark{b}  & $0.12 \pm 0.01$ \\
$\rm L_{hot}~(1500~K)/L_{dust} $           & $0.11 \pm 0.01$ \\
$\rm L_{dust} ~(L_{\odot})$    & $2.2 (\pm 0.2) \times 10^{11}$\\
$\rm M_{cold}~(26~K)~(M_{\odot})$   & $1.5 (\pm 0.4) \times 10^8$ \\
$\rm M_{cool}~(72~K)~(M_{\odot})$   & $1.5 (\pm 0.6) \times 10^5$ \\
$\rm M_{warm}~(235~K)~(M_{\odot})$ & $7.0 (\pm 0.2) \times 10^2$ \\
$\rm M_{hot}~(1500~K)~(M_{\odot})$     & $6.0 \phm{(\pm 0.2)} \times 10^{-2}$
\enddata
\tablenotetext{a}{These values are likely closer to the intrinsic physical attenuation 
than the much lower value of $\rm \tau_{warm} (9.7\mu m) = 0.7 \pm 0.1$ 
derived from spline fitting that estimates only the apparent strength of the 
silicate absorption feature \S\ref{subsubsec:SpectralDiagnostics}.
}
\tablenotetext{b}{
The $\rm L_{warm}/L_{dust} $ values are corrected for the attenuation by
$\rm 1 - e^{-\tau_{warm} (9.7\mu m)}/\tau_{warm} (9.7\mu m)$.
}
\tablecomments{
\footnotesize
Parameters resulting from SED decomposition using the sum of 
ISRF + starburst + AGN radiation sources, as
described in the text and plotted in Figure \ref{fig:SEDcafe}.
}
\end{deluxetable}
}
\ifnum\Mode=0
\placetable{tbl:SEDfitting}
\else
\tableSEDfitting
\fi
%%%%%%%% End Table %%%%%%%%%%%

It is illustrative to compare Mrk 266 with the galaxies modeled by 
\citet{2007ApJ...670..129M}.
Overall, the SED of Mrk 266 can be modeled as a starburst 
dominated LIRG, with weak AGN indicators from the thermal dust or PAH 
emission, as is also the case in NGC 6240 and NGC 2623.
While the cold dust component in Mrk 266 has essentially the same 
temperature as derived for NGC 6240 (29 K), 
the cool (72 K) and warm (235 K) dust components of Mrk 266 are 
significantly warmer than those in NGC 6240 (61 K and 193 K). 
Together the warm and hot components comprise 23\% of the total dust
luminosity of Mrk 266. For context, the sum of the (apparent) warm and hot dust 
fractions are 22\% in NGC 6240, 77\% in Mrk 463, 
and 82\% in the QSO PG $0804+761$.
Within the uncertainties, Mrk 266 has the same cold dust mass as NGC 6240.
Compared to the other two dual-AGN systems which have had
SED decomposition using this technique, the $\rm T_{dust}$ 
distribution in Mrk 266 is quite similar to NGC 6240, 
and both of these systems have much smaller fractions of
warm+hot dust than Mrk 463. The Mrk 266 SED fitting results also 
illustrate that only 700 \Msun\ of dust at $\rm T\ga 200~K$ 
can generate the same luminosity ($4.4\times10^{10}$ \Lsun) as
$1.5\times10^5$ \Msun\ of dust at $\rm T\approx 72~K$.

\subsubsection{Flux Ratios}\label{subsec:FluxRatios}

The physical nature of AGNs is often revealed via flux ratios.
We begin by determining the ``q'' parameter as defined by \citet{2001ApJ...554..803Y}, 
\begin{equation}\label{q24a}
{\rm\scriptstyle
q_{FIR} = log\left(\frac{FIR}{3.75\times10^{12}~[W~m^{-2}]}\right) - log\left(\frac{S_{1.4GHz}}{[W~m^{-2}~Hz^{-1}]}\right), 
}
\end{equation}
\noindent where $\rm FIR~[W~m^{-2}] = 1.26\times10^{-14}(2.58\cdot S_{60\mu m} [Jy] + S_{100\mu m}[Jy])$,
and the flux densities are those measured by {\it IRAS} for the global system.
In addition, the $\rm q_{FIR}$ parameter can be defined as 
\begin{equation}\label{Mdust24}
{\rm\scriptstyle
q_{24} = log[(S_{24\mu m})/(S_{1.4GHz})],
}
\end{equation}\label{Mdust24a}
\noindent where $\rm q_{24}$ is the 24 \micron\ flux density measured 
by {\it Spitzer} MIPS. The data in APPENDIX Tables \ref{tbl:SED_Tot}
and \ref{tbl:SED_Regions} give 
$\rm q_{FIR} = 2.0$ and $\rm q_{24} = 0.95$ for the global system,
and $\rm q_{24} = 1.77~(1.14)$ for Mrk 266 SW (NE).
Mrk 266 is typical of galaxies in the {\it IRAS} 2 Jansky Survey,
98\% of which fall in the range $\rm 1.64 \leq \langle q_{FIR}\rangle \leq 3.02$
(population mean $\rm \langle q_{FIR}\rangle = 2.34$). A useful comparison for the 
$\rm q_{24}$ values is the sample mean $\rm \langle q_{FIR}\rangle = 0.83 \pm 0.31$
($1\sigma$ dispersion) for over 770 galaxies in the {\it Spitzer} First Look Survey 
with redshifts less than 0.4 \citep{2007ApJ...663..218M}. The global system and
Mrk 266 NE are within $1\sigma$ of this sample mean, and although 
$\rm q_{24}$ for Mrk 266 SW is large compared to most LIRGs, it is within the 
$3\sigma$ dispersion of the FLS population. Therefore, Mrk 266 shows no significant
deviations from the infrared-radio correlation followed by star-forming galaxies
and radio-quiet AGNs.

The ratio of infrared to ultraviolet luminosities provides a diagnostic of the amount
of reprocessed versus escaping radiation from young, massive stars.
Mrk 266 has an observed ratio of $\rm L_{ir}/L_{UV} = 20$, 
where $\rm L_{UV}$ is the sum of the NUV and FUV luminosities measured by 
{\it GALEX}, and it has a ratio $\rm L_{UV}/(L_{ir}+L_{UV}) = 0.047$ (5\%). 
With a correction for the finding that $L_{ir}$ is powered roughly equally by the 
AGNs and star formation (\S\ref{subsubsec:SpectralDiagnostics}), 
about 10\% of the total star formation in Mrk 266 is observable directly as escaping 
UV photons, and 90\% is UV radiation reprocessed into thermal infrared emission.
Since the star formation rate (SFR) scales directly with the starburst component of
$\rm L_{UV} + L_{ir}$, the system has $\rm SFR_{UV}/SFR_{Total} = 10$\%, which  
is on the high end of the distribution for the GOALS sample ranging
0.2\% - 18\% with mean 3\% \citep{2010ApJ...715..572H}.

%%%%%%%% Begin Figure %%%%%%%%%%%
%HST ACS B image direct and after processing with SourceExtractor
%Mrk266_SSC_Bdirect
%Mrk266_ClustersBsubWithRegionsD
\def\figcapClustersBband{
\footnotesize
Star clusters (SCs) extracted by Source Extractor identified with small
circles overlaid on the 0.44 \micron\  (B, F435W) {\it HST} image 
before background subtraction (a) and  after background subtraction 
using a median-masking technique (b).
Panel (b) also identifies notable regions discussed in the text: 
three areas where most of the SCs are concentrated (large green ellipses),
each of which contains a smaller concentration of very blue SCs (blue ellipses; 
see also Fig. \ref{fig:BminusI}); regions with radius 5\arcsec\ centered on the 
nuclei of the two galaxies (black circles);
and a dearth of SCs bisecting the system at PA$\approx-47$\arcdeg\  (red band).
}
\ifnum\Mode=0 %Insert Figure/Table only in [preprint] or [preprint2] modes
\placefigure{fig:ClustersBband}
\begin{verbatim}fig25a, fig25b\end{verbatim}
\else
%For preprint
%\ifnum\Mode=2 \onecolumn \fi %4-panel fig is too small in 2-column mode
\ifnum\Mode=2 
\begin{figure*}[!tbh]
\else
\begin{figure}[!tbh]
\fi
\center
\includegraphics[width=3.0truein,angle=0]{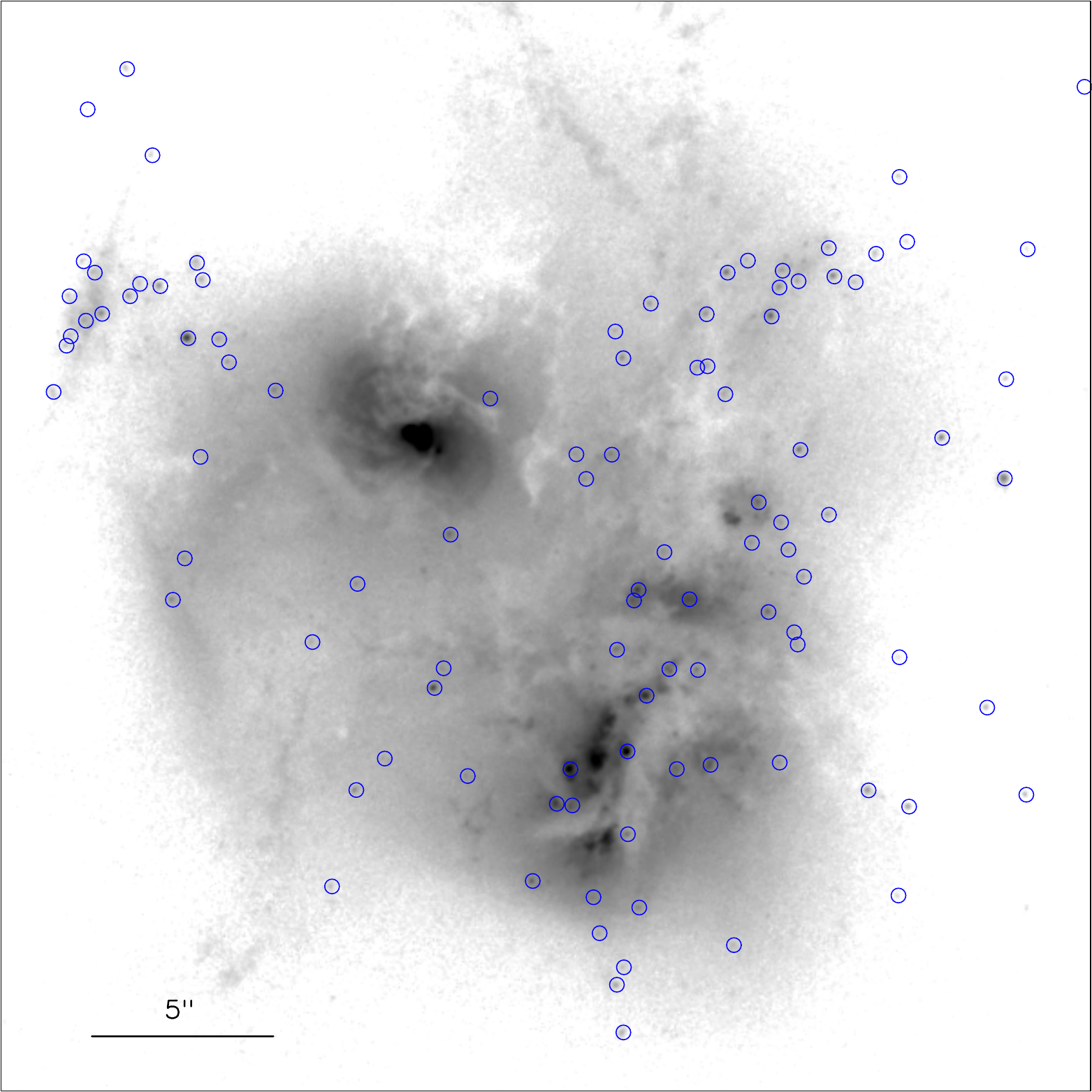}
\put(-20,10) {\parbox{10cm}{(a)}}
\hskip 0.05in
\includegraphics[width=3.0truein,angle=0]{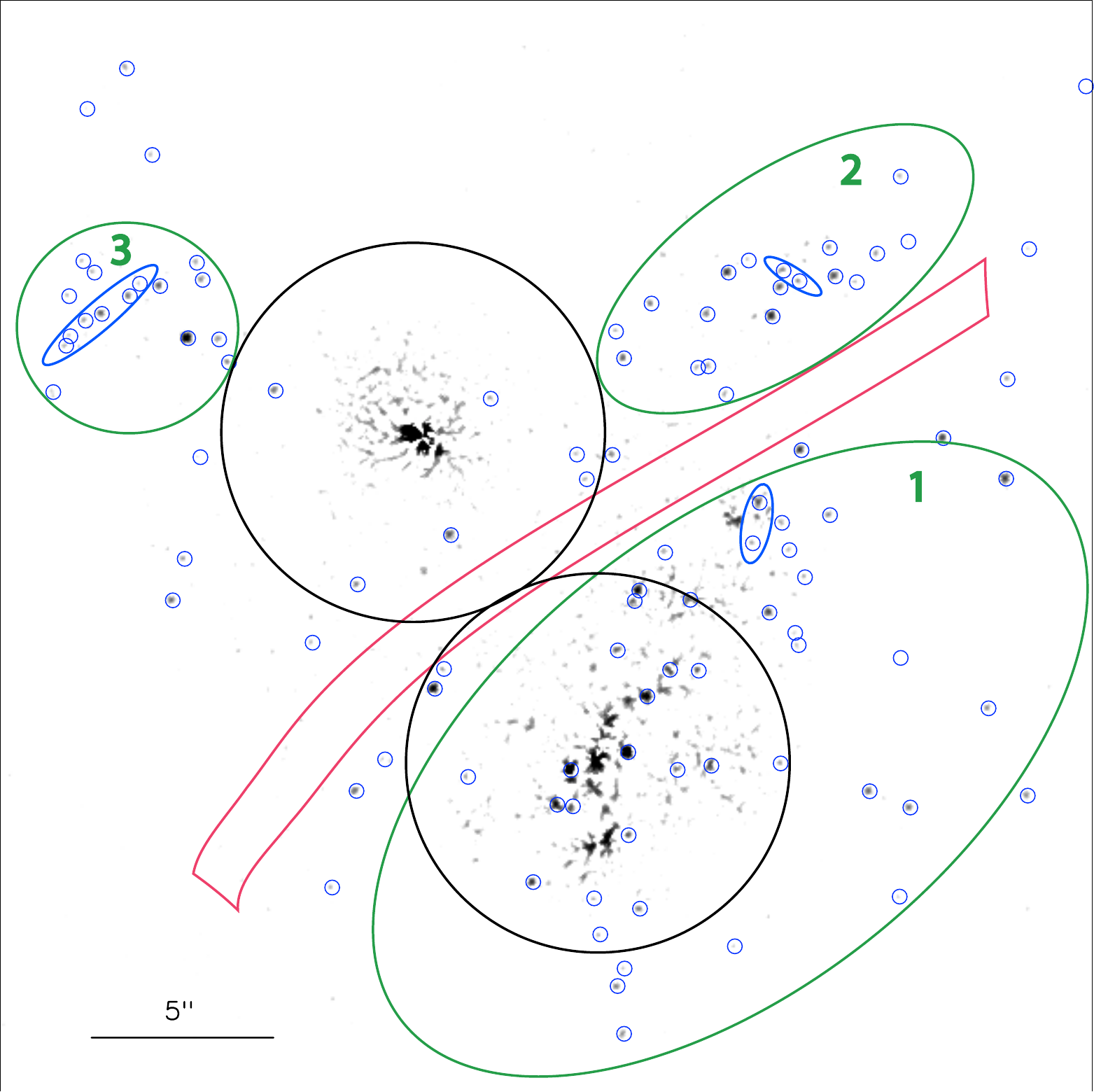}
\put(-20,10) {\parbox{10cm}{(b)}}\\
\caption{\figcapClustersBband \label{fig:ClustersBband}}
\ifnum\Mode=2 
\end{figure*}
\else
\end{figure}
\fi
\fi %close \ifnum\Mode=0 
%%%%%%%% End Figure %%%%%%%%%%%

Lastly we examine the ratio of hard X-rays to infrared flux.
Among 13 local ULIRGs studied by \citet{2007ApJ...656..148A}, those with 
AGN-dominated mid-IR spectra have a ratio of hard X-rays to infrared luminosity 
$\rm L_{HX}/L_{ir} \ga 10^{-3}$, whereas starburst-dominated ULIRGs 
have $\rm L_{HX}/L_{ir} < 10^{-4}$. 
The {\it XMM-Newton} measurement for the global Mrk 266 system gives 
$\rm L_{HX}/L_{ir} = 3.5\times10^{-4}$, and the higher resolution {\it Chandra} measurements 
give $\rm L_{HX}/L_{ir} = 5.0\times10^{-5}$ and $\rm L_{HX}/L_{ir} = 1.0\times10^{-3}$  
for the SW and NE nuclei, respectively. 
The component galaxies of Mrk 266 are therefore at extreme ends of the 
range observed to date in the local (U)LIRG population:
The value of $\rm L_{HX}/L_{ir}$ for Mrk 266 NE is 40\% larger 
than NGC 6240, which itself is similar to ratios found in optical/UV-selected
Seyfert galaxies and quasars \citep{2003ApJ...592..782P}.
Mrk 266 SW, on the other hand, has one of the smallest $\rm L_{HX}/L_{ir}$ ratios 
among local (U)LIRGs, similar to Arp 220 and IRAS 22491-1808 \citep{2007ApJ...656..148A}.
This is consistent with new evidence presented in \S\ref{subsubsec:XrayAGNs} 
that the X-ray source in Mrk 266 SW is Compton-thick.
The fact that the luminosity ratio of the SW and NE galaxies in hard X-rays
($\rm L_{HX}(SW)/L_{HX}(NE) = 0.16$) is close to the reciprocal of the ratio in 
the infrared ($\rm L_{ir}(SW)/L_{ir}(NE) = 3.3$) and in cold molecular gas 
($\rm L_{CO}(SW)/L_{CO}(NE) = 4.9$; \S\ref{subsec:MolGas}) is a strong
clue that the factor of $\approx 20$ disparity in the $\rm L_{HX}/L_{ir}$ ratios for these 
galaxies is mainly due to differences in the column density of material 
absorbing UV/X-ray photons and re-emitting the energy in the far-infrared,
rather than an order of magnitude difference in the intrinsic hard X-ray luminosities of the AGNs.

This same conclusion about high column density material as the responsible agent for the
variance of flux ratios can be arrived at by simply comparing the two nuclei.
The data presented in \S\ref{subsubsec:SEDdata} show that the luminosity ratio
of the two galaxies, $\rm L(Mrk~266~SW)/L(Mrk~266~NE)$, 
is $\approx 0.7$ in the far-UV and soft X-ray bands, 
transitions through 1.0 in the near-infrared (1.2 \micron),  increases to a 
maximum of 5.7 at 8 \micron, and drops to 3.5 at 
70 \micron\ (the SED peak) and 3.3 in the total infrared emission ($\rm L_{ir}$).
The $\rm L(Mrk~266~SW)/L(Mrk~266~NE)$ ratio drops further to 2.3 at 850 \micron,
and to 1.4 at 20 cm.
Thus, although Mrk 266 SW clearly dominates this merger in the mid-IR and far-IR,
the two galaxies have similar luminosities in the near-IR, UV, soft X-rays, and radio continuum.
It is only at the highest energies with available data for the separate galaxies where 
the situation is reversed: Mrk 266 NE is 6.4 times more luminous in the hard (2-7 keV) X-ray 
band than its companion Mrk 266 SW. As shown in \S\ref{subsubsec:XrayAGNs}, this is more 
likely due to a higher column density of absorbing material along our line of sight to the SW 
nucleus (compared to the NE nucleus) than to an intrinsic  difference in their power sources.

\subsection{Star Clusters}
\label{subsec:SCs}

\subsubsection{Cluster Detection and (B - I) Color Map}
\label{subsubsec:SCsBasic}

The {\it HST}/ACS images of Mrk 266 reveal numerous very
compact, marginally resolved sources scattered throughout the system.
Their over all properties indicate they are star clusters (SCs) and associations
similar to those observed in other galaxies undergoing extensive bursts of
star formation. The detection method followed the procedure described in detail by
\citet{2012AJ....143...16M} for SCs in the late stage LIRG merger IC 883.
First, masks outlining the galaxy were applied to the images. 
Then {\it Source Extractor} \citep[][]{1996A&AS..117..393B} was used to fit and subtract
the underlying galaxy and perform point source detection in the B and I band images.
The extracted objects were processed with IDL routines to perform photometry and apply
selection criteria that accepted only reliable sources with a FWHM of 1.7-4 pixels, 
have S/N $>$ 5, and are detected in both bands; details are given in \citet{2012AJ....143...16M}. 
This procedure resulted in the detection of 120 point sources with 
$\rm 21~mag \la B \la ~27~mag$ and $\rm 19.5~mag \la I \la ~25.7~mag$ 
(uncorrected for extinction); their coordinates and photometric properties are listed 
in the Appendix (Table \ref{tbl:SCdata}).

Figure \ref{fig:ClustersBband} shows the direct HST B-band image (a)
and the image after processing with {\it Source Extractor} to fit and subtract 
the local background and extract SCs (b). 
The density of SCs is highest in three extended regions 
identified with green ellipses in Figure \ref{fig:ClustersBband}b:
(1) within and surrounding the disk of the SW galaxy;
(2) along a band extending $\sim$5-15\arcsec\ (3-9 kpc) north-west from the 
NE galaxy, passing near the base of the Northern Loop; and
(3) in a group centered $\sim$8\arcsec\ (5 kpc) north-east of the NE nucleus.
There are 22 SCs within a projected radius of 5\arcsec (3 kpc) around the nucleus of
Mrk 266 SW, and only 6 SCs within the same radius around Mrk 266 NE 
(black circles in Fig. \ref{fig:ClustersBband}b).
The ratio of surface densities of detected clusters in the SW and NE galaxies (3.7) is nearly
the same as the ratio of their infrared luminosities (3.5, see \S\ref{subsec:SEDanalysis}), and 
it is also similar to the ratio of their cold molecular gas masses (4.9, see \S\ref{subsec:MolGas}).
This situation is consistent with the apparent Hubble types and 1.6 \micron\ 
bulge-to-disk ratios (\S\ref{subsubsec:galfit}) of the two galaxies (i.e., SBb disk galaxies 
contain more gas to fuel star formation than S0/a galaxies). 
Despite a nearly 4-to-1 ratio in the number of clusters detected in the central 
3 kpc of the SW and NE galaxies, across the entire system there is a nearly equal 
number of clusters on both sides of a notable gap about 2\arcsec\ (1.2 kpc) 
wide running between the galaxies at $\rm PA\approx-47$\arcdeg\ 
(red band in Fig. \ref{fig:ClustersBband}b).

A color map revealing the variation of (B - I) magnitudes throughout Mrk 266 is shown 
in Figure \ref{fig:BminusI}, with an overlay of symbols coded to indicate the (B - I) colors
of the SCs (uncorrected for extinction). Dust lanes present in the direct images 
(Figs. \ref{fig:HST_ACS4x}, \ref{fig:Mrk266SW} and \ref{fig:Mrk266NE}) are more 
clearly delineated in this color map as filamentary regions with (B - I) $\ga$ 2 mag. 
In the SW galaxy, the reddest dust lanes are on the east and west sides
of the nucleus, and aligned with the major axis of the underlying starlight 
detected in the 1.6 \micron\ image (Fig. 3); they are also aligned with 
the apparent rotation axis of the molecular gas (see \S\ref{subsec:MolGas}).
In the NE galaxy, the dust lane has an ``S'' shape that crosses the nucleus at 
PA$\approx-10^{\circ}$. Higher than average extinction explains the lack of 
detected clusters in these circumnuclear dust lanes. However, in the region 
between the galaxies the color map shows bluer (B - I) values, 
suggesting the apparent dearth of clusters along this gap 
(red band in Fig. \ref{fig:ClustersBband}b) may be intrinsic.

Some of the bluest SCs (blue circles in Fig. \ref{fig:BminusI}) are located in 
regions identified with blue ellipses in Figure \ref{fig:ClustersBband}b, including 
groupings centered $\sim$7\arcsec\ and $\sim$13\arcsec\ north-west of the SW 
nucleus (inside regions 1 and 2 noted above), and a notable chain $\approx$5\arcsec\
in length located north-east of the NE nucleus (inside region 3 noted above).
The detected clusters are a direct probe of star formation occurring in 
regions with relatively low extinction. However, the fact that the total luminosity of the 
unobscured clusters ($\rm M_B = -16.25$ mag, or $\rm L_B = 5\times10^8$ \Lsun) 
is only 0.1\% (1.6\%) of the total infrared (B-band) luminosity indicates that these
clusters provide a minor contribution to the total energy output of the system. 

%%%%%%%% Begin Figure %%%%%%%%%%%
%(B - I) with color-coded SC markers
%Mrk266_SSC_BminusI2
\def\figcapBminusI{
\footnotesize
A grayscale F435W - F814W (B - I) color map of Mrk 266, with star cluster positions 
color-coded by their individual (B - I) colors in ranges as follows:
[F435W] - [F814W] (B - I) = 0 - 0.5 mag: blue circles; 0.5 - 1 mag: green squares;
1-1.5 mag: yellow triangles; 1.5-2.0 mag: orange diamonds; (B-I) $>$ 2.0 mag: red stars.
The scale bar on the right indicates the (B-I) colors of the pixels as observed, 
uncorrected for extinction (reddening).
}
\ifnum\Mode=0 %Insert Figure/Table only in [preprint] or [preprint2] modes
\placefigure{fig:BminusI}
\begin{verbatim}fig26\end{verbatim}
\else
\ifnum\Mode=2 
\begin{figure}[!bt]
\center
\includegraphics[width=1.0\columnwidth,angle=0]{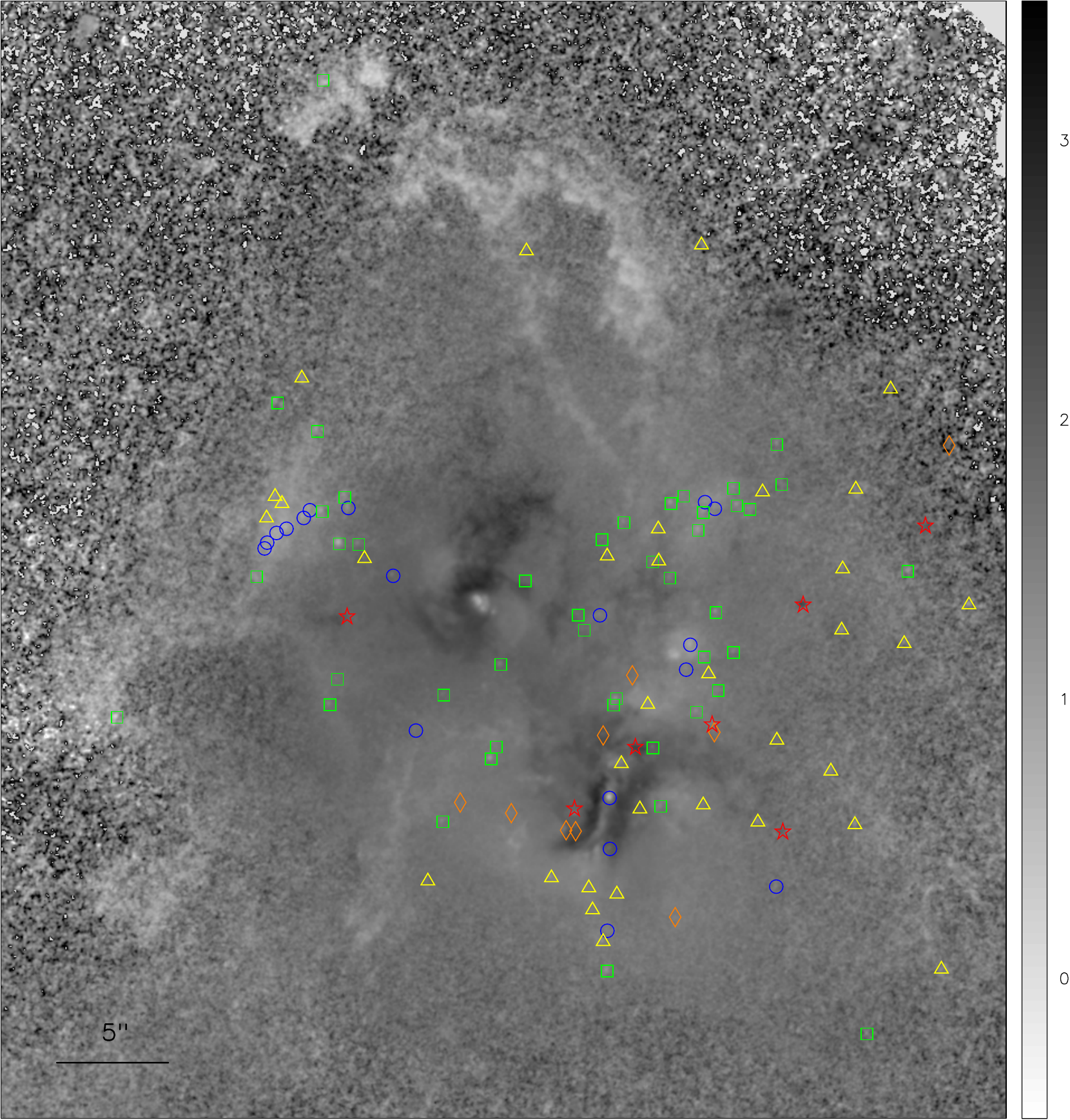} 
%Huge at 111MB! Insert when done. 
\else
\begin{figure}[!bt]
\center
\includegraphics[scale=0.7,angle=0]{fig26} 
%Huge at 111MB! Insert when done. 
\fi
\caption{\figcapBminusI \label{fig:BminusI}}
\end{figure}
\fi %close \ifnum\Mode=0 
%%%%%%%% End Figure %%%%%%%%%%%

%%%%%%%% Begin Figure %%%%%%%%%%%
%B-I vs. M_B color-magnitude diagram with BC model curves
\def\figcapClustersColorMag{
\footnotesize
{$\rm M_B~vs.~B-I$} color-magnitude diagram for SCs detected in Mrk 266.
Evolutionary tracks computed using population synthesis models \citep{2003MNRAS.344.1000B} assuming an instantaneous starburst and Solar metallicity for cluster masses of 
$10^4$ (red), $10^5$ (green) and $10^6$ (blue) \Msun\ are shown. 
Marks on the $10^6$ \Msun\ track indicate the age of the starburst. 
The data are plotted with no extinction correction (a), and with a 
mean extinction correction of $\rm A_V = 1.2$ mag estimated from various 
measurements as described in the text (b).
The vector represents 1 mag extinction in the V band.
Only SCs with uncertainties in $\rm B - I$ less than 0.2 mag are shown.
}
\ifnum\Mode=0 %Insert Figure/Table only in [preprint] or [preprint2] modes
\placefigure{fig:ClustersColorMag}
\begin{verbatim}fig27a_27b\end{verbatim}
\else
\ifnum\Mode=2 
\begin{figure*}[!tb]
\center
\includegraphics[width=0.4\textwidth,angle=-90]{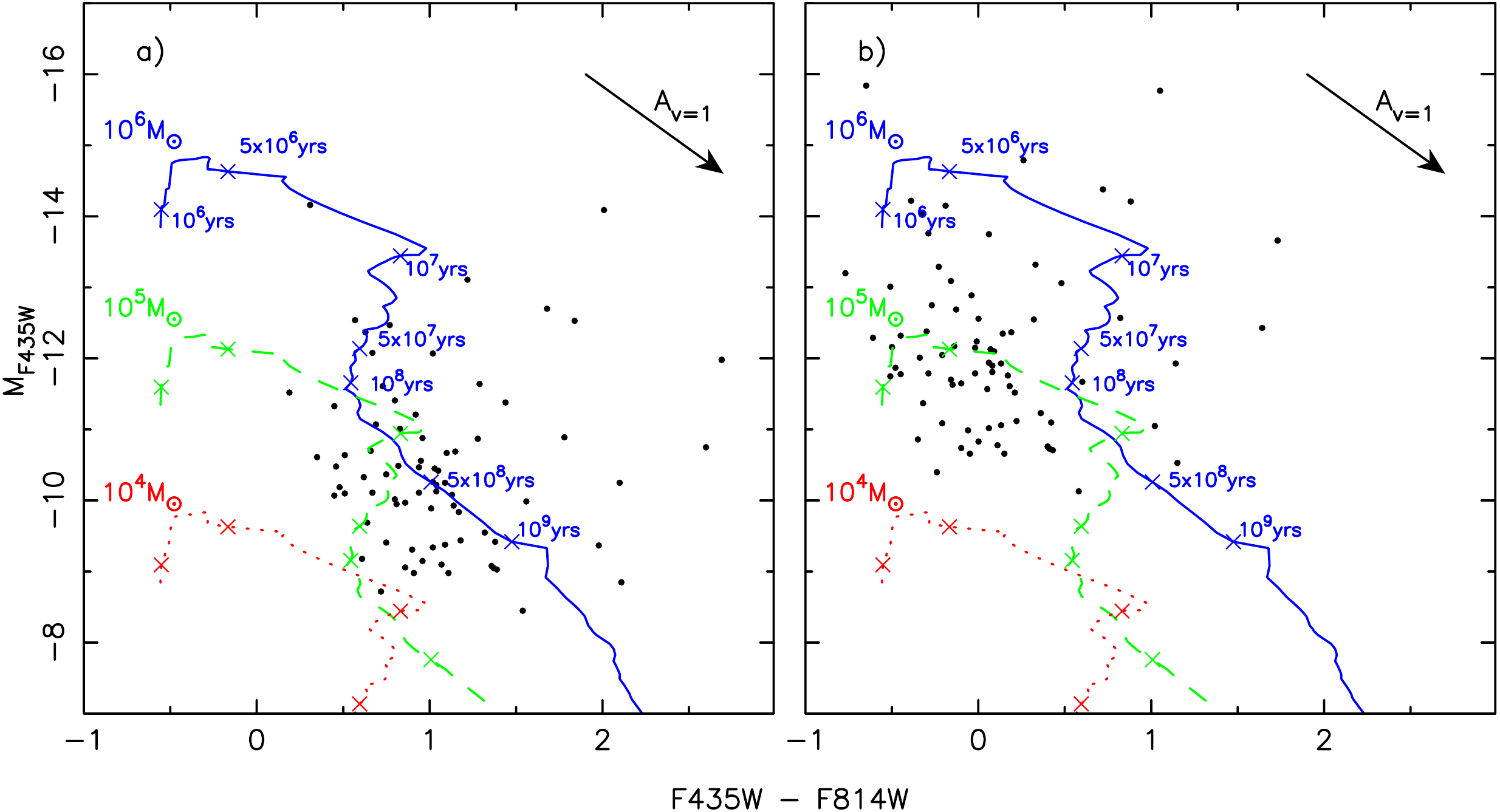}
\else
\begin{figure}[!tb]
\center
\includegraphics[width=0.5\textwidth,angle=-90]{fig27a_27b}
\fi
\caption{\figcapClustersColorMag \label{fig:ClustersColorMag}}
\ifnum\Mode=2 
\end{figure*}
\else
\end{figure}
\fi
\fi %close \ifnum\Mode=0
%%%%%%%% End Figure %%%%%%%%%%%

The very blue colors of the Northern Loop and filaments throughout 
the system are likely a signature of strong nebular [O II] $\lambda 3727 \AA$ and 
[O III] $\lambda 4364\AA$ line emission in the B band 
(see \S\ref{subsec:Superwind}). The NE nucleus, the arc 0\farcs4 to the west, 
and the bi-conic structure in which they are embedded are comparably blue, 
with (B - I) $\approx$ -0.2 mag. Similarity in color with filaments throughout 
the system that are visible in [O III] emission-line imaging \citep{2000PASJ...52..185I} 
lends support to the hypothesis presented in \S\ref{subsubsec:Outflows} that the 
circumnuclear region of Mrk 266 NE is an AGN ionization cone.

\subsubsection{Star Cluster Color-Magnitude Diagram} \label{subsubsec:SCsCM}

Figure \ref{fig:ClustersColorMag} is a $\rm B-I$ vs. $\rm M_B$ color-magnitude 
diagram with evolutionary tracks constructed using population synthesis models 
from \citet{2003MNRAS.344.1000B}.
Most of the SCs have observed (apparent) colors of 
$\rm 0.5~mag \le (B - I) \le 2.0~mag$ and $\rm -9~mag \le M_B \le -13~mag$
(Fig. \ref{fig:ClustersColorMag}a).
Although there is an age/mass degeneracy in the uncorrected measurements, at 
first glance most of the SCs appear to have ages of a few hundred Myr to 1 Gyr.
However, given the orientation of the extinction vector in Figure \ref{fig:ClustersColorMag}, 
which is nearly parallel to the age tracks in the population synthesis models, it is clear 
that the SCs are very likely to be significantly younger and/or more massive than 
they appear when uncorrected for extinction and reddening due to dust.
There are other indicators for high but variable dust extinction: 
(a) dust lanes evident in the color (B - I) map (Fig. \ref{fig:BminusI}); 
(b) the observation that $\sim$90\% of the total luminosity is generated by dust 
emitting in the thermal infrared (\S\ref{subsec:SEDanalysis}); and
(c) evidence for dust entrained in the superwind as far as 
$\sim$25\arcsec (15 kpc) from the center (\S\ref{subsec:SuperwindDust}).

Due to this high variability we can only apply a mean extinction estimate using the 
following indicators: (a) The emission-line Balmer decrement ($\rm H\alpha/H\beta$) 
measured from long-slit optical spectroscopy resulted in E(B-V) values of 
0.72 mag and 0.22 mag for Mrk 266 SW and NE, 
respectively \citep{1995ApJS...98..171V}\footnote{The identifications of the NE and 
SW nuclei of Mrk 266 were mistakenly swapped in the tables and figures of 
\citet{1995ApJS...98..129K} and \citet{1995ApJS...98..171V}.}.
(b) Long-slit spectroscopy along three position angles found $\rm E(B - V) = 0.2 - 0.5$
throughout the inner $\sim$20\arcsec \ (10 kpc) of Mrk 266  \citep{1997ApJ...474..659W}. 
These measurements indicate a characteristic extinction of 
$\rm E(B-V) = 0.4~(A_V = 1.2~mag)$ in Mrk 266, which is similar to the 
sample mean of $\rm E(B-V) = 0.5 (\pm 0.2$) mag observed for a large sample
of local, luminous starburst galaxies \citet{1994ApJ...429..582C}. 
To estimate the total extinction at particular wavelengths, we assume 
$\rm R_V = 3.1$ as found in the general ISM of our Galaxy. However, 
this can be considered a minimum, because regions containing young star 
clusters in Mrk 266 (and other LIRGs) could have a significantly larger value of
$\rm R_V\approx 6$ as measured in Galactic star-forming regions such as 
the Orion Nebula and giant molecular clouds \citep[e.g.,][]{2004AJ....128.2144M}.

%%%%%%%% Begin Figure %%%%%%%%%%%
%B-band and I-band cluster luminosity functions
%Mrk266ClusterLF
\def\figcapClusterLumFuncs{
\footnotesize
Luminosity functions of the extracted star clusters in the F435W (B) band and 
F814W (I) band before (blue) and after (black) correction for foreground contamination 
and completeness as described by \citet{2012AJ....143...16M} and Vavilkin et al. (2012, in preparation).
The solid line represents a linear fit (via $\chi^2$ minimization) to the corrected (black) 
histogram bins over the ranges $\rm -9.75 \le M_B \le -12.75$ and $\rm -10.75 \le M_I \le -13.25$.
Alpha indicates the slope of the SC luminosity function in each bandpass
over these ranges of $\rm M_B$ and $\rm M_I$, which have {\it not} been 
corrected for the estimated mean extinction (see text).
}
\ifnum\Mode=0 %Insert Figure/Table only in [preprint] or [preprint2] modes
\placefigure{fig:SCLumFuncs}
\begin{verbatim}fig28a_28b\end{verbatim}
\else
\ifnum\Mode=2 
\begin{figure*}[!tb]
\center
\includegraphics[width=0.32\textwidth,angle=-90]{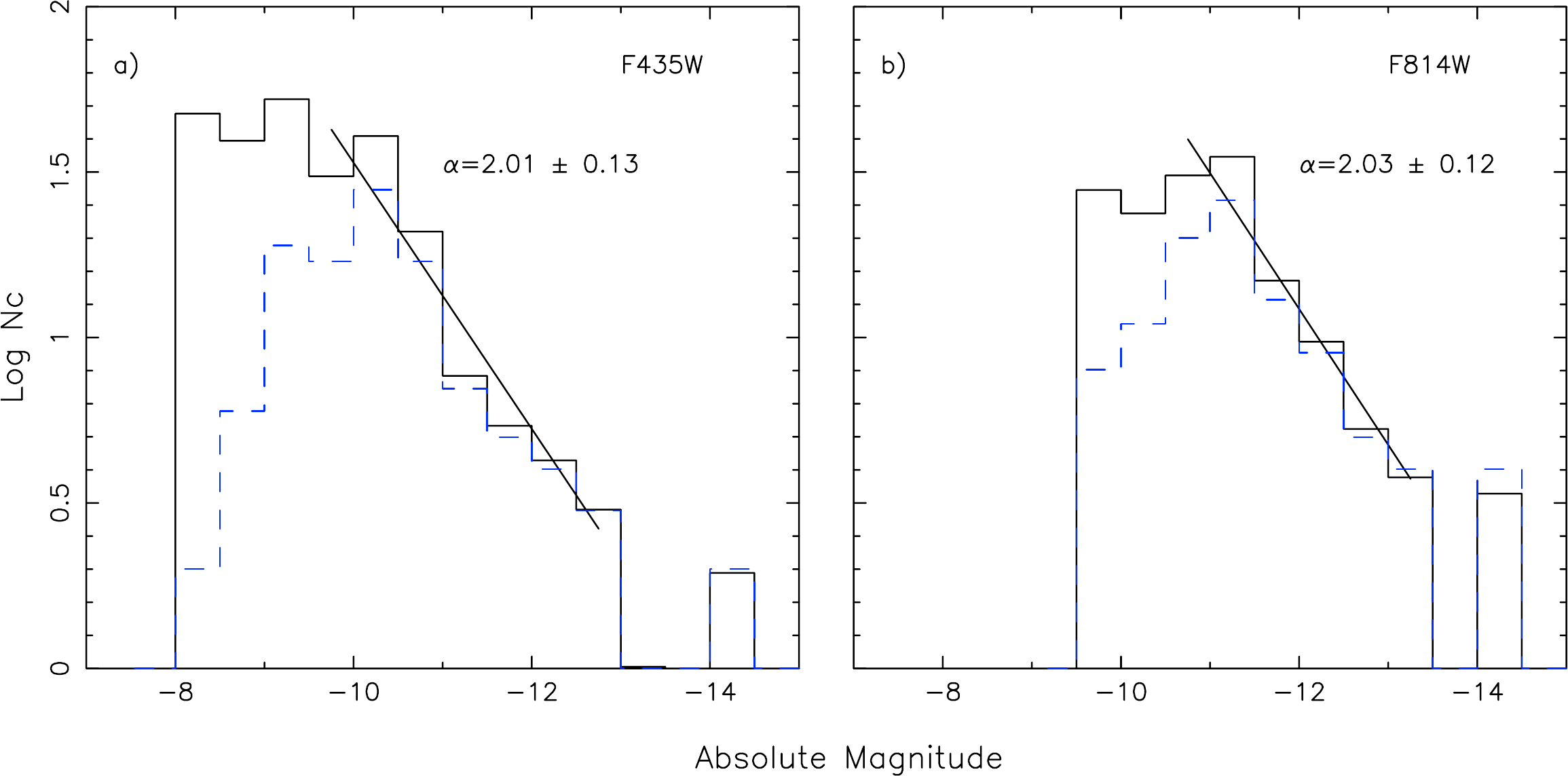}
\else
\begin{figure}[!t]
\center
\includegraphics[width=0.5\textwidth,angle=-90]{fig28a_28b}
\fi
\caption{\figcapClusterLumFuncs \label{fig:SCLumFuncs}}
\ifnum\Mode=2 
\end{figure*}
\else
\end{figure}
\fi
\fi %close \ifnum\Mode=0 
%%%%%%%% End Figure %%%%%%%%%%%

Application of a mean extinction correction of $\rm A_V = 1.2$ mag
to the observed colors 
(Fig. \ref{fig:ClustersColorMag}a)\footnote{The Galactic foreground 
extinction in the direction of Mrk 266 is only E(B-V) = 0.013 mag 
\citep{1998ApJ...500..525S}; with caveats 
documented in NED, this corresponds to $\rm A_V = 0.04$ mag.} 
shifts the points to an intrinsic mean color of (B - I) $\sim$ 0.0 mag and to a 
mean luminosity of $\rm M_B \sim$ -12 mag (Fig. \ref{fig:ClustersColorMag}b).
Comparison of the mean extinction-corrected data to the model curves  
(Fig. \ref{fig:ClustersColorMag}b) leads to the following results for 
the detected cluster population on average:
(a) most of the SCs have ages between 1 and 50 Myr;
(b) although the current data cannot remove the age/mass degeneracy 
for the reddest clusters (B - I $\ga$ 0.5 mag), about half of the objects are 
very blue (B - I $\la$ 0.2 mag), very young (1 - 5 Myr), and are
consistent with masses of $\sim 10^5$ \Msun;
(c) the 11 most luminous SCs have $\rm M_B \la -13.5$ mag and are located 
near the (blue) track for $10^6$ \Msun\ clusters with ages of only $\sim$ 1-10 Myr;
(d) the two very luminous objects with $\rm M_B \approx$ -15.8 mag 
are located 0\farcs75 E (the red cluster with B - I $\approx$ 1.1) and 
0\farcs91 W (the blue cluster with B - I $\approx$ -0.6) of the SW 
nucleus, respectively (see Fig. \ref{fig:BminusI}); and finally
(e) if a mean extinction of $\rm A_V = 1.8$ mag ($\rm R_V\approx 6$) is more 
applicable than $\rm A_V = 1.2$ mag in regions containing the blue SCs,
as argued above, than most of the SCs are on average only 
$\sim$1 Myr old and have masses spanning $\sim10^5$ to $10^6$ \Msun.
These results indicate the detected SCs are massive and likely relatively young. 
However, they represent only the ``tip of  the iceberg'' in terms 
of the extensive starburst which is mostly obscured 
by dust in optical and UV observations.

\subsubsection{Star Cluster Luminosity Functions}
\label{subsubsec:SCsOtherGalaxies}

Figure \ref{fig:SCLumFuncs} is a plot of the luminosity function (LF) 
derived for the SCs detected in the B and I band images.
The index ($\alpha$) of a power law fit to the LF, after correction of the 
cluster counts for foreground contamination and completeness 
using techniques described in \citet{2012AJ....143...16M} and 
Vavilkin et al. (2012, in preparation), is $\alpha = -2.0 \pm 0.1$ in both the B and I bands.

It is revealing to compare the power-law slope of the star cluster luminosity function (LF) 
in Mrk 266 with the star cluster LFs observed in other merger systems.
In the well-studied Antennae Galaxy, the cluster LF has 
$\alpha_V = -2.13 \pm 0.07$ \citep{2010AJ....140...75W}.
In the early-stage merging LIRG system II Zw 96, 
$\alpha_B = -1.1 \pm 0.3$ was found \citep[][]{2010AJ....140...63I}.
For clusters in two well-known ULIRGs that appear to be in intermediate and 
late stages of the merger process, the following slopes were measured by 
\citet{1998ApJ...492..116S}: $\alpha_B = -1.6 \pm 0.1$ for Mrk 463, and 
$\alpha_B = -1.8 \pm 0.2$ for Mrk 231.
For SCs in the late-stage (single nucleus) merger system IC 883,
\citet[][]{2012AJ....143...16M} derived LF indices ($\alpha_B = -2.2 \pm 0.2$ and 
$\alpha_I = -2.0 \pm 0.2$) that are remarkably similar to the values found in Mrk 266.
In another late-stage LIRG merger remnant, NGC 34 (Mrk 938), the star cluster LF 
has $\alpha_V = -1.75 \pm 0.1$ \citep{2007AJ....133.2132S}.
(Both IC 883 and NGC 34 are in the GOALS sample, with 
$\rm L_{ir}= 5.4\times10^{11}$ and $2.5\times10^{11}$ \Lsun, respectively.)
It is also noteworthy that the LF slopes in Mrk 266 and these other (U)LIRGs
are similar to the LFs of young star clusters in the irregular galaxies SMC and LMC, 
which have mean $\alpha = -2.0 \pm 0.2$ \citep{2006A&A...450..129G}.
The specific frequency of young star clusters is defined as 
$\rm T_N = N_{cl}\times10^{(0.4(M_B + 15))}$, where $\rm  N_{cl}$ is the
number of clusters above the specified luminosity limit normalized by the 
B-band luminosity of the host galaxy \citep[e.g.,][]{1999A&A...345...59L}.
Correcting for completeness in the cluster detection, Mrk 266 has  
$\rm T_N = 0.91$, which is similar to completeness-corrected $\rm T_N$ 
values observed in the cluster-rich LIRG sample of Vavilkin et al. (2012, in preparation).
We conclude that $\alpha$ and $\rm T_N$ for star clusters in Mrk 266 
are similar to other LIRGs.

Various disruption mechanisms have been described in the literature to model 
the luminosity, mass and age distributions of star clusters in interacting 
galaxies \citep[e.g.,][]{2007AJ....133.1067W}.
Given that Mrk 266 has one of the most powerful superwinds known in the local 
universe (\S\ref{subsec:Superwind}), this raises the question of whether 
star clusters are being disrupted by this outflow/feedback process to a 
larger degree than in systems such as the Antennae with much less powerful 
superwinds and lower star formation rates. 
The dearth of detected SCs in the interaction zone of Mrk 266 noted above, 
despite relatively low extinction and molecular gas spanning part of this 
region (\S\ref{subsec:MolGas}), and despite the presence of SCs in other 
regions at equally large distances from the galaxy centers, may be a sign of 
cluster disruption or suppression of star formation in this narrow band 
between the two galaxies. A possible explanation would be  
fast shocks and extensive turbulence at the interface of the colliding galaxies 
or colliding superwinds expanding from the two galaxies, as discussed in 
\S\ref{subsec:Center}. However, there is no signature of cluster disruption
on a global scale in Mrk 266, as might be revealed through an $\alpha$ or 
$\rm T_N$ value that is significantly different than the SC populations
observed in (U)LIRGs with less powerful superwinds. 

\subsection{Molecular Gas Properties}
\label{subsec:MolGas}

\subsubsection{The Integrated Emission from CO, NCH, and HCO$^{+}$}
\label{subsubsec:MolGasInt}

%%%%%%%% Begin Figure %%%%%%%%%%%
%Nobeyama molecular gas contoured on X-ray, U, B and H2 
%Mrk266_CO_On_XraysToRadio
\def\figcapMolGasA{
\footnotesize
The integrated emission of Mrk 266 in CO(1-0) (red contours), 
HCN(1-0) (cyan contours), and HCO$^{+}$(1-0) (yellow contours) with a 
spatial resolution of 4\farcs3 x 3\farcs4 (PA$=-31.5$\arcdeg) from the 
{\it Nobeyama Millimeter Array} \citep{2009AJ....137.3581I}  are shown 
superposed on grayscale representations of the following images:
(a) {\it Chandra} 0.4-7 keV X-ray image, (b) {\it XMM-OM} U band,
(c) {\it HST} ACS 0.44 \micron, (d) 1.6 \micron\ (2MASS H), (e) 1.6 \micron\ (HST NICMOS),
(f) 8.0 \micron\ (IRAC), (g) $\rm H_2~S(3)~9.7$ \micron\ (IRS), (h) 11.3 \micron\ PAH (IRS),
and (i) 20 cm (VLA). The 2MASS 1.6 \micron\ (H) image is included to show 
the accuracy of the spatial registration of the 1.6 \micron\  NICMOS image.
All panels display a field of 20\arcsec x 20\arcsec. The scale bar is 5\arcsec. 
}
\ifnum\Mode=0 %Insert Figure/Table only in [preprint] or [preprint2] modes
\placefigure{fig:MolGasA}
\begin{verbatim}fig29a_29i\end{verbatim}
\else
\ifnum\Mode=2 
\begin{figure*}[!htb]
\else
\begin{figure}[!htb]
\fi
\center
\includegraphics[width=6.0truein,angle=0]{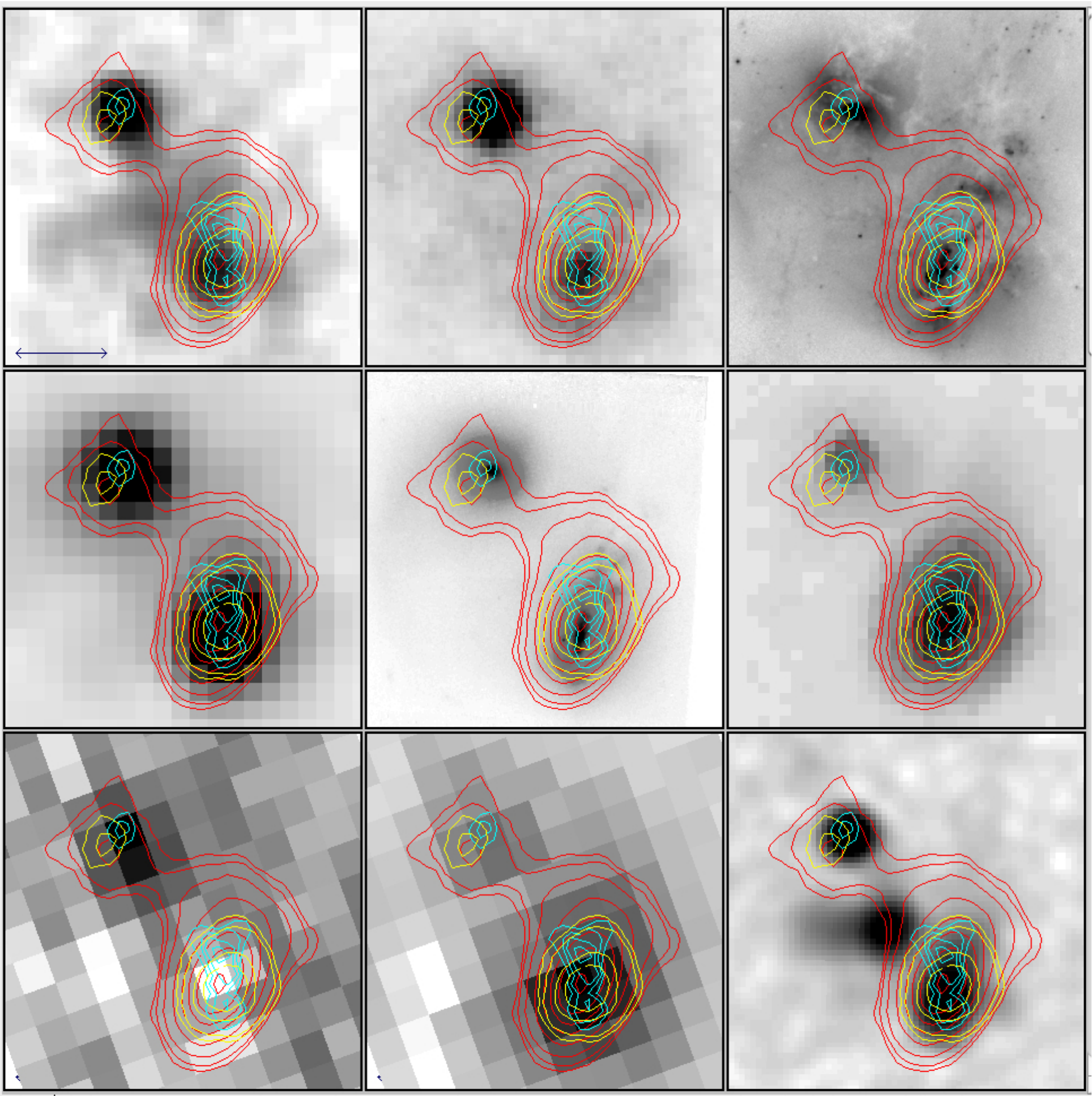}
\footnotesize
%These \parbox coordinates apply to \includegraphics[width=6.0truein]
\put(-425,422) {\parbox{10cm}{(a) 0.4-0.7 keV (Chandra)}}
\put(-282,422) {\parbox{10cm}{(b) U (XMM-OM)}}
\put(-140,422) {\parbox{10cm}{(c) 0.44 \micron\ (HST ACS)}}
\put(-425,278) {\parbox{10cm}{(d) 1.6 \micron\ (2MASS H)}}
\put(-282,278) {\parbox{10cm}{(e) 1.6 \micron\ (HST NICMOS)}}
\put(-140,278) {\parbox{10cm}{(f) 8  \micron\ (Spitzer IRAC)}}
\put(-425,133) {\parbox{10cm}{(g) S(3) H2 9.7  \micron\ (Spitzer IRS)}}
\put(-282,133) {\parbox{10cm}{(h) 11.3  \micron\ PAH (Spitzer IRS)}}
\put(-140,133) {\parbox{10cm}{(i) 20 cm (VLA)}}
\put(-410,300) {\parbox{10cm}{5\arcsec}}
\caption{\figcapMolGasA \label{fig:MolGasA}}
\ifnum\Mode=2 
\end{figure*}
\else
\end{figure}
\fi
\fi %close \ifnum\Mode=0 
%%%%%%%% End Figure %%%%%%%%%%%

Interferometric observations of Mrk 266 in HCN(1-0), HCO$^{+}$(1-0) and CO(1-0) with 
the {\it Nobeyama Millimeter Array} were presented by \citet{2009AJ....137.3581I}.
In this section these data are used to explore the distribution of molecular gas with 
respect to various features in the system, and to derive the physical conditions of
the cold molecular gas. 
Figure \ref{fig:MolGasA} presents contours of the integrated emission from these 
three molecular transitions overlaid on images ranging from X-rays through radio continuum.
There are intriguing differences between the spatial distributions of these molecular species. 
Although the SW galaxy contains about 5 times more flux in CO (1-0) than its companion, 
the NE galaxy is clearly detected in all three molecular transitions. 
The bulk of the integrated CO (1-0) emission in the system is aligned with the
major axis of the SW galaxy, with a centroid near the X-ray/infrared nucleus.
CO (1-0) is also detected in a bridge connecting the two galaxies, 
which is discussed further below (\S\ref{subsubsec:MolGasKin}). 
The HCN gas in the SW galaxy shows double structure 
with peaks located $\approx$1.2 kpc (2\arcsec) to the NE and SW of the region where 
the CO (1-0) and  HCO$^{+}$(1-0) are most heavily concentrated at the location of 
the X-ray/IR/radio nucleus. If this morphology is due to limb-brightening in a rotating disk,
the HCN disk is tilted $\sim$10\arcdeg\ with respect to the axis of the 
CO (1-0) and 1.6 \micron\ emission (Fig. \ref{fig:MolGasA}e). 
Most of the CO emission in the NE galaxy is offset by $\sim$1 kpc (1\farcs5)
to the SE of the X-ray/optical nucleus. The CO (1-0) in this region is spatially
coincident with the peak of the HCO$^{+}$ (1-0), whereas the dense gas traced by HCN
is concentrated very close to the nucleus about 1 kpc to the NW.
Figure \ref{fig:MolGasA}c shows that the centroid of the 
HCN emission is aligned within 0\farcs2 of the eastern point source in the 
{\it HST} B-band image, which we interpret to be the
AGN of Mrk 266 NE (\S\ref{subsubsec:Outflows}).

%%%%%%%% Begin Figure %%%%%%%%%%%
%CO (1-0) channel maps superposed on NICMOS H band
%Mrk266_COchannelsOnH
\def\figcapMolGasB{
\footnotesize
The spatial distribution of the CO (1-0) emission as a function of radial velocity (frequency) 
using a channel map from the {\it Nobeyama Millimeter Array} \citep{2009AJ....137.3581I}. 
The CO data are contoured in red and superposed on a grayscale representation of the {\it HST} 
1.6 \micron\ image to illustrate the kinematics of the CO with respect 
to the old stars in the galaxies. For comparison, the integrated emission from HCN (1-0) is 
plotted (cyan contours) in panel (h), and the integrated emission
from HCO$^{+}$(1-0) is shown (yellow contours) in panel (i).
The mean radial velocity (and frequency) is shown in each panel. 
All panels display a 24\arcsec x 24\arcsec\ field, and the scale bar is 5\arcsec.
}
\ifnum\Mode=0 %Insert Figure/Table only in [preprint] or [preprint2] modes
\placefigure{fig:MolGasB}
%\begin{verbatim}Mrk266_COchannelsOnH\end{verbatim}
\begin{verbatim}fig30a_30i\end{verbatim}
\else
\ifnum\Mode=2 
\begin{figure*}[!tb]
\else
\begin{figure}[!tb]
\fi
\center
\includegraphics[width=6.0truein,angle=0]{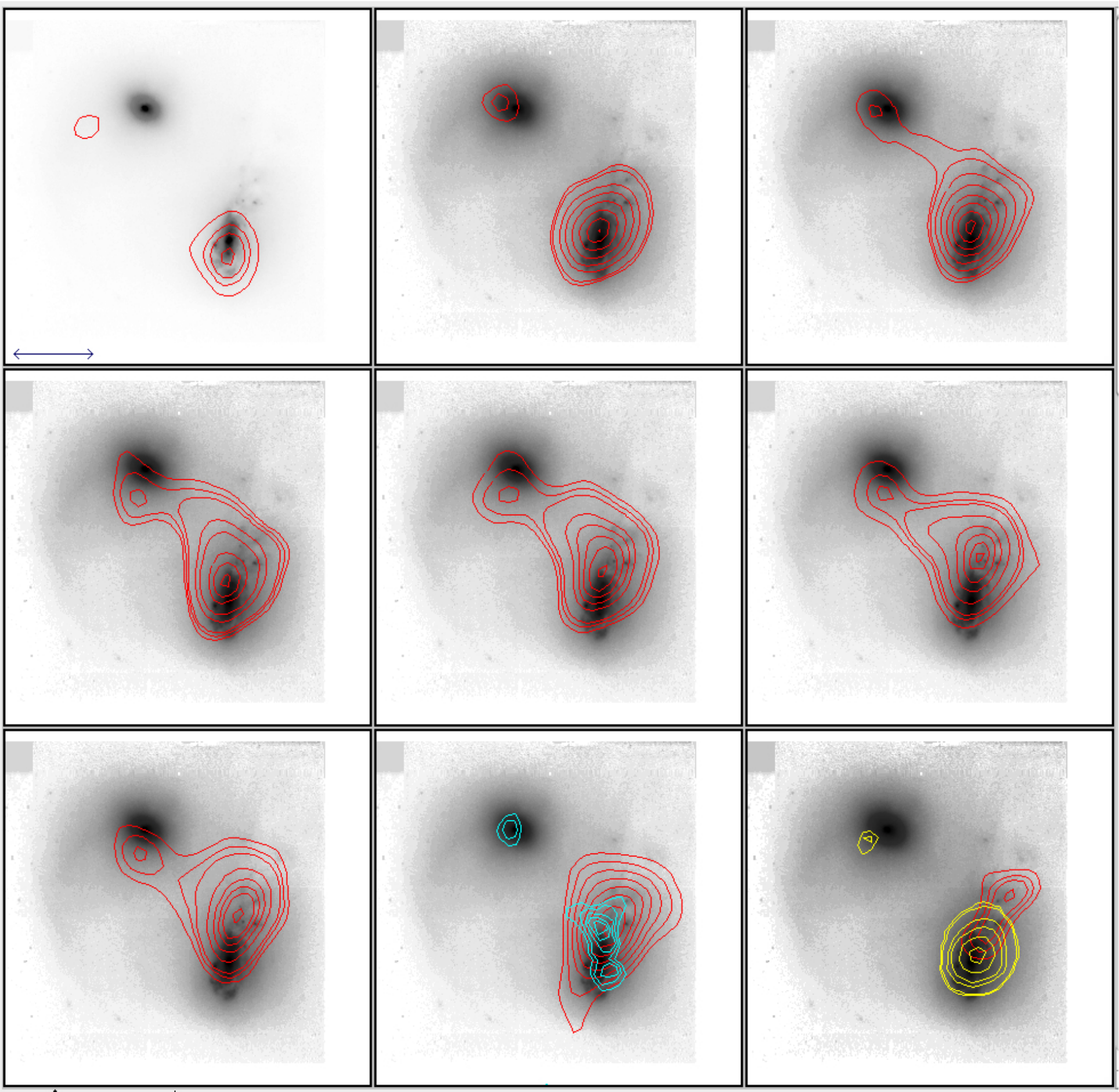} 
\footnotesize
\put(-426,409) {\parbox{10cm}{(a) 7850 km/s (112.25 GHz)}}
\put(-283,409) {\parbox{10cm}{(b) 8097 km/s (112.16 GHz)}}
\put(-140,409) {\parbox{10cm}{(c) 8152 km/s (112.14 GHz)}}
\put(-426,269) {\parbox{10cm}{(d) 8207 km/s (112.12 GHz)}}
\put(-283,269) {\parbox{10cm}{(e) 8235 km/s (112.11 GHz)}}
\put(-140,269) {\parbox{10cm}{(f) 8290 km/s (112.09 GHz)}}
\put(-426,129) {\parbox{10cm}{(g) 8317 km/s (112.08 GHz)}}
\put(-283,129) {\parbox{10cm}{(h) 8427 km/s (112.04 GHz)}}
\put(-140,129) {\parbox{10cm}{(i) 8455 km/s (112.03 GHz)}}
\put(-415,292) {\parbox{10cm}{5\arcsec}}
\caption{\figcapMolGasB \label{fig:MolGasB}}
\ifnum\Mode=2 
\end{figure*}
\else
\end{figure}
\fi
\fi %close \ifnum\Mode=0 
%%%%%%%% End Figure %%%%%%%%%%%

Applying the formulation of \citet{2005ApJS..159..197E} to compute $L^{\prime}_{CO}$, 
and assuming a CO luminosity to $\rm H_2$ gas mass conversion factor 
$\alpha = 1$ \Msun\ $\rm(K~km~s^{-1}~pc^{2})^{-1}$ suggested by observations of ULIRGs
\citep[][]{1998ApJ...507..615D}, results in estimates of the warm molecular gas mass 
for the two galaxies: $3.4\times10^{9}$ and $7.0\times10^{8}$ \Msun\ for Mrk 266 
SW and NE, respectively. The sum of these components corresponds to a total mass of
 $\rm M(H_2) = 4.1x10^{9}$ \Msun. However,  the maps also include
an extended component, and the integrated CO(1-0) line flux of the total 
system is 180 $\rm Jy~km~s^{-1}$ \citep{2009AJ....137.3581I}.
The corresponding total molecular hydrogen gas mass of Mrk 266 from the 
{\it Nobeyama} observations is thus $\rm M(H_2) = 7.0x10^{9}$ \Msun. 
Within the uncertainties, this is consistent with 206 $\rm Jy~km~s^{-1}$ measured with 
single-dish observations \citep{1986ApJ...305L..45S} and a corresponding estimate of 
$\rm M(H_2) = 2.5\times10^{9}$ \Msun\ when adjusted to $\alpha = 1$. 
If the CO to $\rm H_2$ conversion factor appropriate for Mrk 266 is similar to GMCs in our 
Milky Way \citep[$\alpha = 4 \pm 1$,][]{1991ApJ...368L..15R}, the actual gas masses 
may be $\sim$4 times larger than the estimates computed here. 
(\citet{1986ApJ...305L..45S} derived $\rm M(H_2) = 1.4\times10^{10}$ \Msun\ 
assuming $\alpha = 4$.) Assuming a weighted mean dust temperature 
of 50 K in the CO-emitting regions based on SED fitting 
(\S\ref{subsubsec:SED_Dust}) and using the observed CO (1-0) line widths,
applying the formulation of \citet{2005ApJS..159..197E} enables estimation 
of the radius of the unresolved CO-emitting regions, the dynamical
masses of the galaxies, as well as the column densities.
These results are presented in Table \ref{tbl:MolGas}.

%%%%%%%% Begin Table %%%%%%%%%%%
\def\tableMolGas{
\begin{deluxetable}{llll}
\ifnum\Mode=2
\renewcommand\arraystretch{0.5}% (MyValue=1.0 is for standard spacing)
\tabletypesize{\scriptsize}
\setlength{\tabcolsep}{0.00in} %Tighten up the columns. See AASTeX FAQ
\tablewidth{\columnwidth}
\else
\renewcommand\arraystretch{0.6}% (MyValue=1.0 is for standard spacing)
\tabletypesize{\small} 
\setlength{\tabcolsep}{0.40in} %Tighten up the columns. See AASTeX FAQ
%\tablewidth{\textwidth}
\fi
\tablecaption{Molecular Gas Properties of Mrk 266 \label{tbl:MolGas}}
\tablehead{
\colhead{Parameter} & 
\colhead{SW} & 
\colhead{NE} & 
\colhead{Total} \\
\colhead{(1)} & 
\colhead{(2)} & 
\colhead{(3)} &
\colhead{(4)}
}
\tablecolumns{4}
\startdata
%%%%%%%%%
CO (1-0) flux $\rm[Jy~km~s^{-1}]$\tablenotemark{a}              &  88           & 18        &  180 \\    
CO (1-0) FWHM $[km~s^{-1}]$\tablenotemark{a} &  500         & 250      &  \nodata \\                     
$\rm L^{\prime}_{CO}~[K~km~s^{-1}~pc^{2}]$           &  $\rm 3.4x10^{9}$ & $\rm 7.0x10^{8}$ &  $\rm 7.0x10^{9}$ \\
$\rm M(H_2)]$\ [\Msun]                       &  $\rm 3.4x10^{9}$ & $\rm 7.0x10^{8}$ &  $\rm 7.0x10^{9}$ \\
$\rm R (CO)$\ [pc]                               &  208 & 134 &  \nodata \\
$\rm M_{Dynamical}$\ [\Msun]            &  $\rm 1.2x10^{10}$ & $\rm 1.9x10^{9}$ & \nodata \\
$\rm M(H_2)/M_{Dynamical}$             &  0.28 & 0.37 & \nodata \\
$\rm Column~Density~[cm^{-2}]$        &  $\rm 1.6x10^{24}$ & $\rm 7.9x10^{23}$ & \nodata
\enddata
\tablenotetext{a}{
The CO (1-0) line fluxes and line widths are from \citet{2009AJ....137.3581I};
all other parameters are derived here for the first time.
~\\
}
\tablecomments{
\footnotesize
Column (1): Observed or derived parameter.
Columns (2)-(4): For the SW galaxy, NE galaxy, and the total system, respectively---
the value of the parameter in Column (1). 
}
\end{deluxetable}
}
\ifnum\Mode=0
\placetable{tbl:MolGas}
\else
\tableMolGas
\fi
%%%%%%%% End Table %%%%%%%%%%%

\subsubsection{Spatial Distribution and Kinematics of CO (1-0)}
\label{subsubsec:MolGasKin}

Figure \ref{fig:MolGasB} shows the spatial distribution of CO (1-0) in velocity channels 
ranging from $\rm 7850$ to $\rm 8455~km~s^{-1}$, in judiciously chosen steps varying 
from 30 to 250 $\rm km~s^{-1}$, produced from the FITS channel map provided 
by \citet{2009AJ....137.3581I}.
The CO (1-0) contours are superposed on the {\it HST} NICMOS 1.6 \micron\ image
for comparison with the highest resolution tracer available for the (old) stellar mass in the 
galaxies, and to minimize confusion from dust obscuration and young 
star-forming regions that dominate the imagery at shorter wavelengths.

The CO (1-0) emission in Mrk 266 SW shows a strong velocity gradient 
ranging from $\rm 7850~km~s^{-1}$ on the SE side to $\rm 8235~km~s^{-1}$ on the NW
side of the disk. At progressively higher velocities, the centroid of the CO (1-0) emission 
is located at larger radii, extending as far as $\approx$5\arcsec\ (3 kpc) NW from the SW nucleus.
The molecular gas in this off-nuclear region has a radial velocity 
$\rm \approx450~km~s^{-1}$ greater than the SW nucleus, and this gas 
coincides spatially with the most luminous region of off-nuclear star formation 
that is prominent in optical and UV images (see Fig. \ref{fig:NineBands}). 
Although the density of SCs is not significantly higher than in the surrounding areas, 
this region contains many of the most luminous SCs (Fig. \ref{fig:ClustersBband}).

The low $\rm L_{CO}/L_{ir}$ ratio in Mrk 266 NE is consistent with its
apparent near-IR Hubble type of S0/a (pec), suggesting that its  
pre-merger gas content was relatively low and incapable of sustaining a high SFR. 
Figure \ref{fig:MolGasB} shows that a large quantity of molecular gas detected 
over the velocity range $\rm 8150-8320~km~s^{-1}$ is bridging between the galaxies.
The highest concentration of CO on the NE side of the bridge is centered 
$\approx$1 kpc (1\farcs5) to the SE of the nucleus of Mrk 266 NE.
As shown in Figure \ref{fig:MolGasA}i, the CO (1-0) in this off-nuclear region 
coincides with the peak of the HCO$^{+}$ (1-0) emission, whereas the HCN (1-0) 
emission is concentrated at the NE nucleus.
This result is consistent with research that has called into question the use of 
HCN as a tracer of dense gas fueling star formation, suggesting that in some cases 
HCN is energized by an X-ray Dissociation Region (XDR) associated with an 
AGN \citep[e.g.,][]{2008Ap&SS.313..331G}.

The distribution of CO (1-0) in Mrk 266 suggests the system is currently in a 
short-lived phase in which $\sim$40\%  of the cold molecular gas 
is falling toward the center of mass of the system, or transferring between the galaxies
along a tidal bridge as predicted by merger simulations \citep[e.g.,][]{2006MNRAS.373.1013C}.
Mrk 266 appears to be in a similar evolutionary stage as three other LIRGs in the GOALS 
sample that have a large fraction of their total CO emission located between the nuclei:
VV 114 \citep{1994ApJ...430L.109Y}, NGC 6090 \citep{1999AJ....117.2632B}, and 
NGC 6240 \citep{2010A&A...524A..56E}. However, unlike VV 114 where 
the 1.4 GHz radio continuum emission has a similar spatial distribution as the CO, 
the enhanced radio continuum emission between the nuclei of Mrk 266 
peaks $\approx$2\arcsec (1.2 kpc) south of the CO bridge (Fig. \ref{fig:MolGasA}i).
Another difference is that whereas the bulk of the HCO$^+$ in VV 114 is 
located between the nuclei \citep{2007AJ....134.2366I}, in Mrk 266
the HCO$^+$  is still bound to the galaxies. 

\subsection{Wider Implications}
\label{subsec:Implications}

The proceeding sections have shown that Mrk 266 exhibits the simultaneous 
occurrence of a variety of phenomena including dual AGNs, 
an outflow containing a likely radiative bow shock in the narrow-line region 
of the NE nucleus, a starburst with a population of young star clusters,
a galactic scale ``blow-out'' expelling dust in a superwind, 
shocked gas between the galaxies, and molecular gas falling 
toward the center of mass or transferring between the galaxies.
In this closing section, Mrk 266 is compared with other local major mergers and (U)LIRGs, 
and we attempt to place the system into a tentative sequence with a small number of 
other confirmed dual AGNs (with kpc scale nuclear separations) that may represent stages 
in the formation of binary AGNs (accreting SMBHs with pc-scale orbital separations).

\subsubsection{Emission Between Interacting Galaxies}
\label{subsubsec:Between}

Interacting systems exhibiting substantial radio continuum, X-ray, or molecular gas 
emission between the galaxies are rare, but a few have been studied in detail. 
By comparing these systems with Mrk 266 we may be able to gain further
insight into this phenomenon.

{\bf The Taffy Galaxy (VV 254):}
This interacting pair has $\approx$50\% of its total 1.5 GHz flux
emerging between the companions, which has been interpreted as
synchrotron radiation from cosmic ray electrons trapped in magnetic field lines and 
stretched like taffy as the galaxies pass through each other \citep{1993AJ....105.1730C}.
Imaging in 450 and 850 \micron\ continuum emission and in multiple
transitions of CO indicate the bridge between the galaxies consists of interstellar dust and 
gas extruded from the galaxy disks via direct cloud-cloud collisions \citep[][]{2007AJ....134..118Z}.
Mid-infrared observations have detected substantial PAH emission in the 
bridge \citep{1999AJ....118.2132J}.

The regions between the components of Taffy and Mrk 266 both contain large fractions 
of their total radio continuum and CO emission, 
yet they contain little or no evidence for luminous star formation.
However, unlike Taffy, there is no detected infrared enhancement between the 
galaxies of Mrk 266. In addition, while the radio continuum and CO emission 
are roughly spatially aligned and uniformly distributed between the Taffy galaxies, 
in Mrk 266 the central radio emission is much more centrally concentrated and the 
bridge of CO (1-0) emission is spatially decoupled from the radio continuum 
and X-ray emission. The absence of a discrete mid-IR or far-IR source
between the galaxies of Mrk 266 suggests shocks may have 
destroyed most of the dust grains in this region.
The collision in the Taffy system was estimated to occur $\sim2\times10^7$ yr prior to 
its current configuration \citep{1993AJ....105.1730C}. Its global star formation rate is 
not significantly higher than in non-interacting spiral galaxies \citep{2007AJ....134..118Z}, 
and no evidence for an AGN has been published for either galaxy.

{\bf The Antennae (NGC 4038/4039):} This galaxy pair is well known for having
$\sim$30\% of its total bolometric luminosity, $\sim$40\% of its total CO (1-0) flux 
\citep{2003ApJ...588..243Z}, and a high concentration of young star clusters 
\citep{1999AJ....118.1551W} located in an off-nuclear region between the colliding disks.
We have shown that in Mrk 266 the molecular gas bridging the galaxies is decoupled 
from the central radio/X-ray source, and there is no evidence for star formation 
between the galaxies. Compared to Mrk 266, the Antennae exhibit:
(a) $\sim$5 times lower luminosity in the far-infrared and in CO (1-0);
(b) a cooler effective dust temperature;
(c) a large region of diffuse X-rays, but a weaker superwind;
(d) a pair of long tidal tails rather than diffuse, asymmetric tidal debris; and
(e) low-luminosity LINER and starburst (H II) nuclei.

Early simulations suggested we are viewing the Antennae galaxies soon after their 
first encounter \citep[e.g.,][]{1996ApJ...462..576D}, but a recent model that 
accounts for the starburst in the overlap region constrains the system 
to $\sim$40 Myr after the second encounter and only $\sim$50 Myr before the 
galaxy centers fully merge \citep{2010ApJ...715L..88K}. 
The close projected separation and small relative velocity of the nuclei 
imply Mrk 266 is in a similar stage of the merger process as the Antennae. 
The lack of well defined tidal tails in Mrk 266 may be due to any number of conditions,
including a relatively small disk mass in the NE galaxy (\S\ref{subsubsec:galfit});
a retrograde encounter or an orbital geometry (e.g., a smaller perigalactic distance) 
that produces more diffuse tidal debris \citep[][]{1996ApJ...462..576D}; or
a large ratio of dark matter halo mass to visible (bulge plus disk) mass 
that can cause tails produced early in the encounter to fall back onto the 
galaxies quickly, resulting in late-stage tidal debris resembling 
loops and shells \citep{1998ApJ...494..183M}. 
It is noteworthy that the extent of the diffuse optical emission in Mrk 266 
($\approx 103$ kpc) is similar to the end-to-end length of the tails in the Antennae.
A detailed simulation of kinematic data for Mrk 266 is needed 
to constrain its true dynamical age.

{\bf NGC 6240:} Mrk 266 appears to be in a similar evolutionary stage as the
well-known system NGC 6240, in which X-ray observations by {\it Chandra} 
have also revealed dual AGNs \citep{2003ApJ...582L..15K} and a large fraction 
of the molecular gas is located between the nuclei \citep{1999Ap&SS.266..157T}.
Although the total infrared luminosity is about half that of NGC 6240, Mrk 266 
has a much warmer dust temperature ($\rm f_{60\mu m}/f_{100\mu m} = 0.72$) 
compared with NGC 6240 ($\rm f_{60\mu m}/f_{100\mu m} = 0.26$).
Likewise, decomposition of the infrared SEDs indicates warm (cool) dust components 
of 225 K (73 K) in Mrk 266 compared to 193 K (61 K) in NGC 6240 (\S\ref{subsubsec:SED_Dust}). 
The 0.4-10 keV X-ray luminosity of Mrk 266,
$\rm 9.5\times10^{41}~ergs~s^{-1}$, is 57\% that of NGC 6240, and within 
the uncertainties of the X-ray flux absorption corrections their intrinsic
$\rm L_x/L_{ir}$ ratios are similar ($\rm5 - 7\times10^{-4}$).
The projected separation of the nuclei in NGC 6240 is $\approx$1 kpc (1\farcs8). 

{\bf Stephan's Quintet:}
This is a well studied compact group containing a large intergalactic 
shock region with substantial radio continuum, X-rays, infrared and molecular gas
emission. Unlike Mrk 266, the main shock region of Stephan's Quintet exhibits
broad ($\rm 870~km~s^{-1}$) rotational lines of warm $\rm H_2$ with an order of 
magnitude higher surface brightness than the soft X-ray emission  \citep{2006ApJ...639L..51A}. 
This dichotomy may be explained by a difference in the shock conditions
in the two systems. Detailed modeling suggests the shock velocity 
in Stephan's Quintet is $\rm\le300~km~s^{-1}$ \citep{2010ApJ...710..248C}.
As noted by \citet{2000ApJ...535..735D}, if the intergalactic shock front in Mrk 266 is 
produced by the collision of superwinds expanding outward from the two galaxies
at $\rm\sim300~km~s^{-1}$, adding this $\rm\Delta V \sim600~km~s^{-1}$ to the collisional 
velocity of the two galaxies can produce a net shock velocity approaching 
$\rm\sim1000~km~s^{-1}$. This would heat the gas far above the dissociation 
temperature of $\rm H_2$ \citep{2003RMxAC..15..131L}, which is consistent 
with a temperature of $\sim10^7$ K deduced from modeling of the soft X-ray data
(\S\ref{subsubsec:CenterXrays}). Another difference between these systems is that
although $\sim$40\% of the CO (1-0) in Mrk 266 is located in a bridge between 
the galaxies where little or no star formation is taking place, in Stephan's Quintet 
CO (1-0) is detected in regions where star formation is occurring as traced by 
24 \micron\ dust emission and UV continuum emission from young, massive 
stars \citep{2010ApJ...710..248C}.

In summary, the radio continuum and X-ray emission between the galaxies in Mrk 266 
is more similar to the large-scale shock detected in Stephan's Quintet 
than the off-nuclear star formation observed in the Antennae. 
The lack of bright $\rm H_2$ line emission and infrared emission from dust grains 
in the central region of Mrk 266 may be due to the presence of a very fast shock 
that effectively destroys most of the dust grains, sublimates the molecular gas, 
and inhibits star formation. The ridge of $\sim10^7$ K gas between the galaxies 
of Mrk 266, which is nearly $\sim10$ times hotter than in other regions of the diffuse 
X-ray nebula and has a correspondingly short cooling time, is consistent with this scenario.
Since the region with enhanced soft X-rays and radio continuum emission between the galaxies 
is not well aligned with the molecular gas bridge (Fig. \ref{fig:MolGasA}) and the physical 
processes generating these features are likely quite different (shocks generated via 
superwinds and gas transfer via gravitation tidal forces), we suspect that these two short-lived 
phases are not directly linked. However, their simultaneous appearance is not a mere 
coincidence, as both phenomena are causally related to the ongoing galaxy merger.

\subsubsection{Will Mrk 266 become a ULIRG?} \label{subsubsec:Evolution}

It is well established that most local LIRGs and effectively all ULIRGs involve 
interacting and merging gas-rich spiral galaxies \citep[e.g.,][]{1996ARA&A..34..749S},
and the ULIRG phenomenon occurs predominantly in late-stage mergers when 
dust is heated to high temperatures by being transported by gravitational torques 
(with the rest of the ISM) close to the central power sources 
\citep[e.g.,][]{1991AJ....101.2034M,1998ApJ...500..619M}.
Clearly ULIRGs must pass through a LIRG phase as the star formation rate and/or AGN 
fueling rate increases the dust heating to produce a luminosity of $\rm L_{ir} \ga 10^{12}$ \Lsun.
At least two lines of evidence -- one based upon merger simulations and one based on 
molecular gas properties -- suggest that Mrk 266 will transform into a ULIRG.

Numerical simulations show the timescale for major mergers (from initial approach through final
coalescence of the nuclei) averages $\sim 2$ Gyr, but spans $1-8$ Gyr, depending on the
circularity of the orbit, the virial radius and circular velocity \citep[e.g.,][]{2008ApJ...675.1095J}. 
Comparison of the observed properties of Mrk 266 with hydrodynamical simulations
\citep{1996ApJ...471..115B,1996ApJ...464..641M,2006ApJS..163....1H,2006MNRAS.373.1013C} 
suggests we are viewing this system about 50\% to 70\% of the 
way through the merger process.
The presence of an asymmetric halo of low surface-brightness optical emission spanning
over 100 kpc, a relatively small projected nuclear separation of 6 kpc, and a low relative 
velocity of the nuclei of $\rm\Delta V = 135~km~s^{-1}$ \citep{2006MNRAS.368..461K} indicate 
the two galaxies are likely undergoing their second or third encounter, with each passage 
damping the relative velocities of the nuclei, strong dynamical friction causing the orbit to 
decay, and tidal torques driving strong inflows of gas.
As pointed out by \citet{1999Ap&SS.266..195M}, the timescale for violent relaxation is $\ga1$ Gyr
in the outer regions (low $\rm V_c$), but only $\sim$ 100 My in the inner regions (high $\rm V_c$).
Thus, depending on whether the nuclei are on the verge of free-fall (which can occur as rapidly 
as $\approx \frac{1}{4}$ of a rotation period) or whether they will orbit one more time before 
fully merging, the estimated time to coalescence is about 50 to 250 Myr. 
(We are referring here to the merger of the stellar systems, not the SMBHs. 
The latter is discussed in \S\ref{subsubsec:SMBBH}.)
Based on Monte Carlo simulations of merger models to match the observed distribution of
nuclear separations in ULIRGs \citep{1996AJ....111.1025M}, \citet{1999Ap&SS.266..195M} 
concluded that most ULIRGs are in the final 20\% of the merger process. 
Thus, Mrk 266 appears to be in an interesting evolutionary stage when the nuclei are about 
to enter the final phase of coalescence characteristic of ULIRGs. 

The total cold molecular gas mass of Mrk 266 is similar to that of ULIRGs 
observed locally and at high redshifts \citep[e.g.,][]{2005ARA&A..43..677S}.
Given that $\approx$40\% of the total CO (1-0) emission is in a bridge between 
the galaxies, there is a large reservoir of cold molecular gas available to form 
more stars or to fuel the AGNs as the galaxies and nuclei fully coalesce.
With $\rm L^{\prime}_{CO} = 6.7\times10^9~L_{\odot}$ and $\rm L_{ir}/L^{\prime}_{CO} = 50.4$,
Mrk 266 falls directly on the well known correlation between 
$\rm log(L_{ir}/L^{\prime}_{CO})~and~log(L_{ir})$ \citep{2005ARA&A..43..677S}.
Within the evolutionary scenario of \citet{1988ApJ...325...74S,1988ApJ...328L..35S},
we expect Mrk 266 will evolve into a ULIRG initially with a relatively ``cool'' global far-infrared 
dust temperature similar to Arp 220, followed by a ``warm'' ULIRG phase exemplified by 
Mrk 463 \citep{1991AJ....102.1241M,2008MNRAS.386..105B}.
Given the new evidence for dust entrainment in the extensive superwind of 
Mrk 266 (\S\ref{subsec:SuperwindDust}), this system illustrates that the ``blow-out phase'' 
when large quantities of dust are expelled to reveal previously obscured optical/UV AGNs 
can begin during an intermediate stage of the merger before the nuclear coalescence 
phase when ULIRGs are most likely to occur.

ULIRGs have a much lower space density than LIRGs locally and at intermediate redshifts 
\citep{2005ApJ...632..169L}. In addition, the mean luminosity of ULIRGs in the local universe 
is lower than ULIRGs and sub-millimeter galaxies at high redshifts 
\citep[e.g.,][]{2005ARA&A..43..677S}. The resolution of these observations may well lie in 
the pre-merger gas content of the participant galaxies. 
The higher that gas content, the larger the burst of star formation and subsequent dust heating.  
In the case of Mrk 266, the NE component contains $\approx5$ times less molecular gas than 
its companion; this is consistent with its apparent morphological type (S0/a) and
correspondingly high bulge-to-disk ratio found at 1.6 \micron.
Using the well-established correlation of $\rm L_{CO}$ with $\rm L_{ir}$ 
\citep{1986ApJ...304..443Y}, a merging pair involving two late-type spirals with
molecular gas masses similar to Mrk 266 SW (SBb) would have $\rm L_{CO}$ a few 
times higher than is observed in the actual Mrk 266 (SBb + S0/a) pair, resulting in a total
$\rm L_{ir}$ at the high end of the local ULIRG distribution \citep{2005ARA&A..43..677S}. 

In the hierarchical model, a galaxy's Hubble type is a byproduct of its merger 
history \citep[e.g.,][]{2000RSPTA.358.2063S}. 
In this scenario, although merging played an important role in galaxy evolution 
over all cosmic time, high-z galaxies and proto-galaxies were 
very different than those observed at intermediate and low redhifts, and prior 
to $\rm z \sim 1$ the Hubble sequence was not yet in place.
However, from $\rm z \sim 1$ to the present, because many late-type spiral galaxies 
merged and transformed into early-type spirals, S0, or elliptical 
galaxies, the fraction of late-type spirals has decreased substantially
\citep[e.g.,][]{2010ApJ...714L..47O}.
Locally at $\rm z\sim0$, this makes pairings of gas-rich, late-type galaxies less likely 
than pairings of one late-type with one early-type galaxy (or of two early-type galaxies).
Therefore, at low redshifts a smaller fraction of major mergers will have 
sufficient gas to fuel star formation or AGN accretion at the level required to reach the 
ULIRG threshold during early to intermediate phases of the merger. As argued above, 
however, this would not prevent a ULIRG from developing later if one of the 
companions has $\rm\ga 7\times10^9$ \Msun\ of molecular gas available 
to be driven toward the center of mass of the system to fuel two massive AGNs as 
they inspiral toward final coalescence. Mrk 266 is apparently in this physical situation.

\subsubsection{Evolution from Dual to Binary AGNs}
\label{subsubsec:SMBBH}

In this final section we turn attention to the significance of Mrk 266 in context with 
other confirmed dual AGN systems in an attempt to understand the low detection rate 
of binary AGNs with sub-parsec scale orbits found in recent surveys. We also propose a
putative evolutionary sequence involving a quantitative parameter available for a small 
number of major merger systems with confirmed dual AGNs and candidate binary AGNs.

Current  theory predicts the following evolutionary sequence 
\citep[e.g.,][and references therein]{2009arXiv0906.4339C}:
(a)  pairs of massive galaxies separated by many times the diameter of their luminous 
disks and bulges will merge in $\sim$1 Gyr (a few dynamical time-scales) due to 
dynamical friction between their massive dark matter halos;
(b) the two pre-existing SMBHs in their cores 
(with $\rm M_{SMBH}/M_{bulge}\approx0.1$\%)
will form {\it dual AGNs} as tidal torques drive the ISM inward to fuel
accretion at a higher rate than existed prior to the encounter; 
during this stage the nuclei are merging from a separation of $\ga1$ kpc to $\sim$10 pc; 
(c) in many cases conditions may be conducive to forcing the black holes pairs to
{\it inspiral} over a period of $\sim100$ Myr to a separation of $\la 10$ pc, at which point
they enter Keplerian orbits and may be detectable as {\it binary AGNs}; 
(e) the SMBHs will eventually either completely coalesce or else recoil, resulting in the 
most powerful sources of gravitational waves predicted to exist.
Considering the relatively small number of dual AGNs confirmed to date, it is tempting to 
conclude that it is rare for both SMBHs to have accretion rates high enough to 
power two luminous AGNs simultaneously.
However, given that the column density of gas and dust in (U)LIRGs is high enough to make
the central $\sim100$ pc optically thick in the mid-infrared, and even in the far-infrared 
in many cases \citep[e.g.,][]{2005ApJ...630..167T}, many more dual AGNs 
will likely be detected with observations in other spectral regions.
Indeed, X-ray surveys suggest $\approx$60\% of optically classified 
LINERs are true AGNs \citep{2006A&A...460...45G}, and high resolution radio continuum 
imaging finds a high fraction (74\%) of AGNs among optically classified LINERs
in the 1 Jy ULIRG sample \citep{2003A&A...409..115N}.

Along with Mrk 266 and NGC 6240, the following local (U)LIRGs have also been confirmed 
to contain dual AGNs based on detection of two compact, luminous sources of hard X-rays: 
Mrk 171 (Arp 299 = NGC 3690 + IC 694) \citep{2004ApJ...600..634B} 
and Mrk 463 \citep{2008MNRAS.386..105B}. 
Mrk 171 is similar to NGC 6240 in terms of infrared luminosity 
($\rm L_{ir} = 8.5\times10^{11}$ \Lsun), and projected nuclear separation (0.9 kpc, 3\farcs8).
However, with $\rm L_x (2-10~keV) = 2.7\times10^{41}~ergs~s^{-1}$ (uncorrected for 
absorption), Mrk 171 has $\rm L_x/L_{ir} = 8.4\times10^{-5}$,
which is an order of magnitude smaller than Mrk 266 and NGC 6240.
With $\rm L_x (2-10~keV) = 3.8\times10^{42}~ergs~s^{-1}$, Mrk 463 has a hard
X-ray luminosity 9 times greater than Mrk 266 and 16 times greater than Mrk 171.
Mrk 463 has $\rm L_x/L_{ir} = 1.9\times10^{-3}$, which is $\sim$2-5 times larger
than NGC 6240 and Mrk 266, and $\sim$23 times larger than Mrk 171.
A remarkable dual AGN is 3C 75, in which both nuclei are emitting powerful 
double-sided radio jets \citep{1985ApJ...294L..85O}; this system consists of a 
pair of elliptical galaxies with projected separation 6.4 kpc (14\arcsec).

There is growing evidence that the current small number of confirmed dual AGNs will 
increase, primarily with observations of hard X-rays. For example, in a local sample of 
hard X-ray selected AGNs, 10\% have been found to reside in dual systems with 
separations $< 100$ kpc \citep{2012ApJ...746L..22K}.\footnote{However, it should be 
noted that many pairs with such wide separations may not be gravitationally bound, 
and a hard X-ray selected sample is naturally biased toward a much higher AGN fraction 
than an infrared-selected sample. The number of confirmed dual AGNs is 
currently less than 3\% of the GOALS sample.}
Dual AGNs and binary AGN candidates are also being found 
in surveys probing to higher redshifts. 
The $\rm z=0.36$ system COSMOS~J100043.15+020637.2 has been confirmed to contain 
two luminous AGNs with a separation of 2.5 kpc (0\farcs5) 
\citep{2009ApJ...702L..82C}. In CXOC J100043.1+020637, \citet{2010ApJ...717..209C} 
have interpreted the presence of optical morphology indicating a major merger, 
a large velocity offset, and an X-ray iron line with an inverted P-Cygni profile 
as a recoiling SMBH or a slingshot effect in a triple system.
The published SED indicates $\rm L_x/L_{ir} \approx 0.018$, which is orders of 
magnitude larger than in the local (U)LIRGs. 
Four more dual AGNs with kpc-scale separations and $0.13 < z <  0.21$
have also been discovered recently in the SDSS \citep{2010ApJ...715L..30L}.

%%%%%%%% Begin Table %%%%%%%%%%%
\def\tableDualAGNs{
\ifnum\Mode=2
\begin{deluxetable*}{llrllrccl}
\renewcommand\arraystretch{0.5}% (MyValue=1.0 is for standard spacing)
\tabletypesize{\tiny}
\setlength{\tabcolsep}{0.00in} %Tighten up the columns. See AASTeX FAQ
%\tablewidth{\columnwidth}
\else
\begin{deluxetable}{llrllrcll}
\renewcommand\arraystretch{1.0}% (MyValue=1.0 is for standard spacing)
\tabletypesize{\scriptsize}
\setlength{\tabcolsep}{0.02in} %Tighten up the columns. See AASTeX FAQ
%\tablewidth{\textwidth}
\fi
\tablecaption{Dual to Binary AGN Evolution Based on $\rm L_{x}(Hard)/L_{ir}$ \label{tbl:DualAGNs}}
\tablehead{
\colhead{Object} & 
\colhead{z} & 
\colhead{$\rm D_L$} &
\colhead{Sep} & 
\colhead{Sep} & 
\colhead{$\rm log(\frac{L_{ir}}{L_{\odot}})$} & 
\colhead{$\rm log(\frac{L_{x}(Hard)}{L_{ir}})$} &
\colhead{Notes} &
\colhead{Refs} \\
\colhead{} & 
\colhead{} & 
\colhead{(Mpc)} &
\colhead{(\arcsec)} & 
\colhead{(kpc)} & 
\colhead{(km/s)} & 
\colhead{} & 
\colhead{} &
\colhead{} \\
\colhead{(1)} & 
\colhead{(2)} & 
\colhead{(3)} & 
\colhead{(4)} &
\colhead{(5)} & 
\colhead{(6)} &
\colhead{(7)} &
\colhead{(8)} &
\colhead{(9)} 
}
\tablecolumns{9}
\startdata

Mrk 171 (NGC 3690 + IC 694)   & 0.01041 & \phn51.1             &    \phn3.8               & 0.9    &    11.88 & -4.1 & \nodata &1 \\
Mrk 266 (NGC 5256)  & 0.02786 &     129\phd\phn  &  10  & 6.0    &  11.53 & -3.5  &\nodata & 2 \\  
NGC 6240              & 0.02448 &     116\phd\phn  &    \phn1.8               & 0.9    &    11.93 & -3.1 &\nodata & 3  \\ 
Mrk 463                   & 0.05036 &     233\phd\phn  &    \phn3.8               & 3.8    &  11.77 & -2.8 &Young radio jet? & 4 \\    
3C 75                      & 0.02315 &  \phn97.9   &   14   & 6.4    & \nodata & \nodata & Twin radio jets & 5 \\    
COSMOS J100043.15+020637.2  & 0.36060 & 1944\phd\phn   &    \phn0.5       & 2.5    & 11.59 & -1.7 & Recoiling SMBH? & 6 \\
4C +37.11              & 0.05500 &    242\phd\phn   & \phn0.0073  & 0.008    &  \nodata & \nodata & Binary AGN & 7\\  
SDSS~J153636.22+044127.0  & 0.38930 & 2133\phd\phn   &    \phn1.0?     & $10^{-4}$~or~5.1 &  \nodata & \nodata & Binary or dual AGN? & 8,9
\enddata
\tablecomments{
\footnotesize
Basic parameters for confirmed dual AGNs and putative binary AGNs.
Column (1): Object name.
Column (2): Heliocentric redshift from NED (see citations therein).
Column (3): Luminosity distance computed by NED.
Column (4): Projected nuclear separation in arcseconds.
Column (5): Projected separation in kpc.
Column (6): Log of the infrared luminosity in Solar units.
Column (7): Log of the ratio of observed hard X-ray luminosity (2-10 keV from XMM, 
otherwise 2-7 keV from Chandra, uncorrected for absorption) to infrared luminosity.
Column (8): Notes.
Column (9): References for the redshift in column (2), the hard X-ray flux used to 
compute  the ratio in column (7), or the Notes in column (8); the codes are as follows:
1 = \citet{2004ApJ...600..634B}; 2 = this article; 3 = \citet{2003ApJ...582L..15K};
4 = \citet{2008MNRAS.386..105B}; 5 = \citet{1985ApJ...294L..85O}; 
6 = \citet{2010ApJ...717..209C}; 7 = \citet{2006ApJ...646...49R}; 
8 = \citet{2009Natur.458...53B}; 9 = \citet{2009ApJ...699L..22W}.
Measurements taken from the literature have been adjusted to
$\rm H_0 = 70~km~s^{-1}~Mpc^{-1},  \Omega_M = 0.27, \Omega_V = 0.73$.
For COSMOS J100043.15+020637.2, $\L_{ir}$ was estimated from the
{\it Spitzer} IRAC 8 \micron\ and MIPS 24 \micron\ measurements in 
\citet{2010ApJ...717..209C} utilizing equation \ref{eqn:Lir} of the current study.}
\ifnum\Mode=2  
\end{deluxetable*}
\else
\end{deluxetable}
\fi
}
\ifnum\Mode=0
\placetable{tbl:DualAGNs}
\else
\tableDualAGNs
\fi
%%%%%%%% End Table %%%%%%%%%%%

What appears to be the first bona fide {\it binary} AGN was discovered 
in 4C +37.11, which contains two compact radio sources 
with a projected separation of only 7.5 mas (7 pc).
Based on the radio source compactness, variability, motion in the jets,
and spectral shapes, \citet{2006ApJ...646...49R} concluded that both
radio cores contain AGNs orbiting with a rotational period of $1.5\times10^5$ yr
within a single elliptical galaxy host. It is notable that the sum of the 
SMBH masses in 4C +37.11, $\sim 1.5\times10^8$ \Msun, is similar to Mrk 266;
this implies that LIRGs with bulge masses similar to the companions in Mrk 266 
may end up as remnants similar to 4C +37.11.
The soft X-ray luminosity of 4C +37.11 is $\approx170$ times
larger than Mrk 266. (Hard X-ray observations of 4C +37.11 are not available.)

Very few additional binary AGNs have been confirmed, although there are a few candidates. 
Observations of double-peaked emission lines in the 
radio-quiet QSO SDSS J153636.22+044127.0 ($\rm z=0.388$)
have been interpreted as a candidate binary SMBH
with masses $\rm 8\times10^{8}$ and $\rm 2\times10^{7}$ inferred to be 
separated by only 0.02 mas (0.1 pc) \citep{2009Natur.458...53B}.
\citet{2009ApJ...699L..22W} reported detection of 
two 8.5 GHz radio continuum sources separated by 0\farcs97 (5.1 kpc) 
within the aperture of the original spectroscopic observations, suggesting
this is another kpc-scale dual AGN system. However, further observations
show the second radio source is an elliptical galaxy that is not the origin of 
the red- or blue-shifted Balmer lines \citep{2009ApJ...703..930L}.
Another possibility is that SDSS~J153636.22+044127.0 is 
similar to other ``double-peak'' AGNs such as 3C 390.3 and Arp 102B, which were 
also promoted as candidate binary AGNs but later found to be more consistent with 
a rapidly rotating accretion disk within a single broad-line region \citep{2010Natur.463E...1G}.
Additional observations are needed to distinguish between these alternate hypotheses.

Table \ref{tbl:DualAGNs} lists basic parameters for the small number of
confirmed dual AGNs and putative binary AGNs discussed above.
Due to the high degree of uncertainty in age dating mergers with factors 
such as nuclear separation, the morphology of tidal debris, and varying 
techniques applied in the simulations of specific systems, 
we propose a merger sequence based on the quantitative parameter 
$\rm L_x/L_{ir}$, where $\rm L_x$ is the 2-10 keV hard X-ray from 
{\it XMM} where available, otherwise the 2-7 keV band from {\it Chandra}.
According to the evolutionary paradigm whereby 
ULIRGs transform into UV/optical QSOs during the merger of gas-rich galaxies
\citep{1988ApJ...325...74S,1988ApJ...328L..35S}, the degree to which the 
luminous optical/UV/X-ray emission from the accretion disk is obscured should 
diminish as superwinds driven by star formation and/or AGNs diffuse or expel the 
dust from the system. 
To the extent that increasing $\rm log(L_x/L_{ir})$, given in parenthesis below, is a
reliable gauge of the degree to which the embedded AGN-powered UV/X-ray 
sources have emerged from their obscuring cocoons, a presumptive sequence is:
$\rm
Mrk~171~(-4.1) 
\to Mrk~266~(-3.5) 
\to NGC~6240~(-3.1) 
\to Mrk~463~(-2.8)
\to COSMOS~J100043.15+020637.2~(-1.7)
$\footnote{SDSS~J153636.22+044127.0 is omitted from the proposed sequence
because its nature as a true binary AGN is controversial, and the required
IR and X-ray measurements are not yet available.}. 

Despite the growing number of kpc-scale {\it dual} AGNs, the number of sub-parsec 
{\it binary} AGNs in the class of 4C +37.11 remains very small.  Currently
SDSS~J153636.22+044127.0 is the only candidate out of $\approx17,500$ AGNs 
with $\rm z < 0.70$ in the SDSS  \citep{2009ApJ...703..930L}. Similarly,
a recent study of VLBA data for over 3100 radio-luminous AGNs 
found only one binary, the previously 
known 4C +37.11 \citep{2011MNRAS.410.2113B}. Since major mergers
provide a natural process to form binary SMBHs, the apparent dearth of binary AGNs 
needs explanation.  \citet{2010Natur.463E...1G} has noted two possibilities:
either the lifetime of binary SMBHs is much shorter than predicted (i.e., they inspiral
very rapidly rather than remaining in a stalled orbit), or fueling 
of the accretion disk(s) is somehow suspended. 
The former explanation seems unlikely because the very presence of the
accretion disks forms a significant buffer against rapid inspiral and effectively 
``softens'' the SMBH merger process. A possible physical mechanism for the latter
suggestion may be that AGN jets have the potential to impart enough energy 
into the inner parsecs of a galaxy to impede gas inflow to the 
accretion disk \citep{2005A&A...435..521N}. We propose another physically
viable alternative: If dual AGNs are confirmed in many more (U)LIRGs in 
intermediate stages of the merger process with kpc scale separations 
(i.e., if most of the LINERs in the GOALS and IRAS 1 Jy ULIRG samples 
are confirmed to be ionized by a power-law source as suggested by the radio
observations of \cite{2003A&A...409..115N}), this raises the possibility that there 
will be an insufficient fuel supply remaining to power luminous binary AGNs 
late in the merger process.

The lifetime of nuclear starbursts and AGNs is 
inversely proportional to the efficiency of gas depletion and ranges
between $\sim10^7$ for ULIRGs and $\sim10^8$ yr  for LIRGs 
\citep[e.g.,][]{2005ARA&A..43..677S}. The lifetime of powerful radio-loud 
AGNs has also been estimated to be only $\sim10^7$ yr \citep[][]{2008ApJ...676..147B}.
This is similar to the dynamical time scale of the final coalescence of
the nuclei with high circular velocity at the core of a merger, but only 1-10\% 
of the $\sim 10^9$ Gyr time scale for dynamical relaxation of the 
stars in outer orbits of the remnant galaxy \citep[e.g.,][]{1999Ap&SS.266..195M}.
Therefore, it is possible that starbursts and dual AGNs may consume or expel the 
bulk of available gas relatively early in many major mergers, leaving insufficient fuel
to power QSOs that can be detected spectroscopically through the stage 
when the SMBHs inspiral to form a true binary and then coalesce or recoil.
This hypothesis counters simulations that find AGN accretion rates 
peak during the SMBH coalescence \citep[e.g.,][]{2009ApJ...707L.184J}.

\section{Summary and Conclusions} \label{sec:Summary}

We have presented the results of a multi-wavelength campaign on the 
luminous infrared galaxy (LIRG) system Markarian 266 that has yielded the 
following conclusions regarding the morphology, nuclei, dust, molecular gas,
outflows, star clusters, and radiation between the nuclei, as well as implications for 
a better understanding of the origin of ULIRGs and binary AGNs.

1.  {\it HST} images show that the ``double-nucleus galaxy'' observed in 
low-resolution images is a strongly interacting pair; we classify the SW and 
NE galaxies as SBb (pec) and S0/a (pec), respectively.  
Deeper imaging reveals asymmetric tidal debris spanning 
$\approx$100 kpc. At H band, the companions are more luminous than 
$\rm L^{*}$ galaxies, and their estimated stellar masses are 
$\rm 6.1 \times10^{10}$ \Msun\ (SW) and $\rm 4.4\times 10^{10}$ \Msun\ (NE). 
The stellar bulge luminosities imply remarkably similar nuclear 
black hole masses of $2\times10^{8}$ \Msun.

2. The NE (LINER) nucleus is emitting 6.4 times more hard X-ray (2-7 keV)
flux than the SW (Seyfert 2) nucleus, yet the luminosity of the NE galaxy 
is only $\approx$20\% of the SW galaxy 
at 24 \micron\ and in CO (1-0) molecular gas emission.  
The {\it Chandra} spectra indicate that the SW nucleus is 
likely the primary source of a bright Fe K$\alpha$ line detected in
the {\it XMM} spectrum of the total system, consistent
with a reflection-dominated X-ray spectrum of a heavily obscured AGN. 
The F(X-ray)/F([O~III]) flux ratios also indicate the SW nucleus is much 
more heavily obscured than the NE nucleus.

3. Soft (0.4-2 keV) X-ray emission extends 15 kpc to the 
north of the system, between the nuclei, and 3 kpc west of the SW nucleus. 
The smoothed X-ray structure corresponds remarkably well with 
filaments in H$\alpha$ and [O III] images. A ridge of X-ray emission
between the galaxies has an X-ray spectrum consistent with $\rm T \sim 10^7$ K 
shock-heated gas, strengthening the idea that the corresponding radio
emission is shock-induced synchrotron radiation.
The lack of detected 9.7 \micron\ $\rm H_2$ S(3) line emission supports the 
suggestion that a very fast shock at the interface of superwinds expanding 
from the two galaxies might generate the observed 
non-thermal radio continuum emission and heat the gas above the 
dissociation temperature of $\rm H_2$. The derived cooling time of the X-ray 
emitting gas between the nuclei is only $\approx$4 Myr.

4. The HST optical images reveal a circumnuclear arc containing three knots 
240 pc west of the NE nucleus. The arc is embedded within B-band emission 
with a filamentary, bi-conic morphology extending over a radius of 1.2 kpc.
This region appears to be dominated by nebular emission from  
the narrow-line region of a double-sided AGN ionization cone.
Radio continuum emission at 1.6 GHz peaks at the optical/near-IR nucleus 
and has components in alignment with the nucleus and the
circumnuclear B-band knots (PA$=$56\arcdeg).
A plausible explanation is that the optical knots are clouds entrained in a 
shock front produced by an AGN-powered collimated plasma outflow (jet). 
We liken this region to the radiative bow shock 230 pc south of the LINER in M51.

5. Mid-IR spectral diagnostics suggest the bolometric luminosity is powered by 
roughly an equal mixture of radiation from AGNs and star formation, with 
substantial scatter among the various methods.
Newly constructed SEDs give infrared luminosities of 
$\rm L_{ir} = $ (2.3, 0.7, 3.4)$\times10^{11}$ \Lsun\  for the SW, NE, 
and total system respectively. Decomposition of the SED of the total system 
indicates that cold (26 K), cool (72 K), warm (235 K) and hot ($\sim$1500 K) 
thermal dust components contribute approximately 57\%, 20\%, 12\%, and 
11\% of the bolometric luminosity, respectively. The total cold dust mass 
estimate is $\rm 1.5(\pm0.4)\times10^{8}$ \Msun.

6. {\it HST} imaging resolves the Northern Loop extending 6-20 kpc from the galaxies 
into a fragmented morphology suggestive of Rayleigh-Taylor instabilities. A few 
structures 300-600 pc in length are radially aligned with the NE nucleus, suggestive 
of bow shocks predicted by hydrodynamic simulations of galactic superwinds. 
Detection of 24 \micron\ emission in the Northern Loop implies that at least 
$\rm \sim2x10^7$\Msun\ of dust is being swept out of the system by the superwind 
during a ``blow-out'' phase that is well underway prior to the final galaxy merger.

7. Approximately 120 star clusters detected with {\it HST} are concentrated in the SW 
galaxy and at the base of the Northern Loop; most have estimated ages less 
than 50 Myr and masses of $\sim 10^5$ \Msun.
The ratios of cluster surface densities, $\rm L_{CO}$ and $\rm L_{ir}$ within 
3 kpc of each nucleus are similar (i.e., $\rm L(SW) / L(NW) \approx 4-5$).
The star cluster luminosity function is similar to what has been found in other LIRGs, 
and the unobscured clusters contribute little to powering the total infrared luminosity.

8. We conclude that Mrk 266 NE is an AGN based on the presence of:
(1) an obscured, hard X-ray point source;
(2) a radiative bow shock aligned with ionization cones
and apparent radio plasma outflow; (3) PAH emission with small equivalent 
widths and a deficiency of 6.2 and 7.7 \micron\  flux relative to 11.3 \micron\  flux;
and (4) a $\rm H_2(1-0)~S(1)~to~Br\gamma$ flux ratio of 3.5.
The spatially extended $\rm H_2$ emission demonstrates that LINERs 
can exhibit strong signatures of an AGN and shock excitation.

9. The bulk of the CO (1-0) emission in the system is aligned with the
major axis of the SW galaxy. HCN (1-0) is aligned with $\rm H_2~S(3)$ 9.7 \micron\ 
emission in the nucleus of Mrk 266 NE.  However, the CO (1-0) and HCO$+$ (1-0) 
emission peaks are located $\sim$1 kpc SE of the NE nucleus,
suggesting that either most of the cold molecular gas has already been stripped 
from the NE galaxy, or gas is being transferred from the SW galaxy to the NE galaxy.
Approximately 40\% of the total CO (1-0) emission is bridging the galaxies, likely
in-falling toward the center of mass or transferring between the galaxies.
In this regard, Mrk 266 appears to be in an evolutionary stage similar to 
VV 114, NGC 6090, and NGC 6240.

10. Two lines of evidence suggest Mrk 266 may evolve into a ULIRG:
First, in the context of merger simulations, the galaxies lie within an 
asymmetric, low surface-brightness halo of tidal debris spanning $\approx$100 kpc, 
with a projected separation of 6 kpc and a relative velocity of only 
$\rm 135~km~s^{-1}$, indicating they are likely undergoing their second or 
third encounter with only 50 to 250 Myr remaining until they merge via tidal dissipation.
Therefore, Mrk 266 appears to be in a short-lived stage when the nuclei 
are about to enter the final phase of coalescence characteristic of ULIRGs.
Second, the total cold molecular gas mass of $\rm \approx7\times10^{9}$ \Msun\ 
is similar to local and high-redshift ULIRGs. 
Since $\approx$40\% of the total CO (1-0) emission is located between the 
galaxies, this reservoir is available to form more stars and to fuel the AGNs 
as the stellar systems and nuclei inexorably coalesce.

11. We propose that Mrk 266 belongs to an evolutionary sequence in which 
{\it dual AGNs} with kpc separations represent precursors to putative {\it binary AGNs} 
with pc scale orbital radii. In this scenario, where Mrk 266 is in an intermediate phase
between Mrk 171 and NGC 6240, the global $\rm L_x/L_{ir}$ ratio increases by over 
four orders of magnitude as obscuring material is expelled by outflows to gradually 
expose previously obscured AGNs.

12. Since major mergers provide a natural process to form SMBH/AGN pairs, the scarcity
of confirmed and candidate binary QSOs in large spectroscopic and VLBI surveys 
is unexpected. While two possibilities are raised in the literature, that either the SMBHs 
inspiral very rapidly, or fueling of the accretion disks is quenched during 
the binary phase, we propose a third hypothesis. Since the gas-depletion lifetime 
of ULIRGs and LIRGs is 10-100 times shorter than the time-scale for creation of 
a binary SMBH ($\ga$1 Gyr), it is possible that, in most instances, the 
gas will be consumed by star formation and accretion during a dual AGN phase 
long before the SMBHs inspiral to sub-pc separation.

%\ifnum\Mode=2 
%\pagebreak
%\fi

\acknowledgments

This work is based on observations with the following facilities:
the {\it Spitzer Space Telescope}, which is operated by the Jet Propulsion 
Laboratory, California Institute of Technology under a contract with NASA;
the NASA/ESA {\it Hubble Space Telescope}, which is operated by the Association 
of Universities for Research in Astronomy, Inc. under NASA contract NAS5-26555;
the {\it Chandra} X-ray Observatory Center, which is operated by the 
Smithsonian Astrophysical Observatory for and on behalf of NASA under contract NAS8-03060;
and the {\it Galaxy Evolution Explorer (GALEX)}, which is operated for NASA by the 
California Institute of Technology under NASA contract NAS5-98034.
The {\it Aladin} (CDS, Strasbourg)  and {\it SAOImage DS9} tools were used for image analysis.
Support for this work was provided by the following NASA grants:
an award issued by JPL/Caltech ({\it Spitzer} PID 3672, PI J. Mazzarella); 
HST-GO10592.01-A (ACS imaging, PI A. Evans) and program 11235
(NICMOS imaging, PI J. Surace) issued by the STScI. 
This research made use of the NASA/IPAC Extragalactic Database (NED) and the
Infrared Science Archive (IRSA), which are operated by the Jet Propulsion Laboratory, 
California Institute of Technology, under contract with NASA.
We thank Masa Imanishi for providing FITS files for the Nobeyama Millimeter Array 
observations and Tsuyoshi Ishigaki for supplying optical emission-line images.
We thank Chris Mihos for helpful discussions regarding merger time-scales.
Finally, we thank the anonymous referee for very helpful 
comments that led to improvements to the manuscript.

%%%%%%%%%%%%%%%%%%%%%%%%%%%%%%%
\ifnum\Mode=2 
~\\
\fi
{\it Facilities:} \facility{CXO (ACIS-S)}, \facility{XMM-Newton (EPIC pn, OM)}, 
\facility{GALEX (FUV, NUV)}, \facility{HST(ACS, NICMOS)}, \facility{UH:2.2m ()},
\facility{Spitzer(IRAC, MIPS, IRS)}
%\ifnum\Mode=2 
%~\\
%\fi

{\it Objects:} \objectname[Mrk 266]{Mrk 266}, 
\objectname[Mrk 266 SW]{Mrk 266 SW},
\objectname[Mrk 266 NE]{Mrk 266 NE}

%%%%%%%%%%%%%%%%%%%%%%%%%%%%%%%
%Dataset: \dataset{ads/sa.spitzer\#xxxxxxxxxx}

%%%%%%%%%%%%%%%%%%%%%%%%%%%%%%%
%\begin{appendix}
%\ifnum\Mode=2 \clearpage \fi

%\ifnum\Mode=2 
%~\\
%\fi
\section{Appendix}
\label{sec:Appendix}

\subsection{Photometric Data for Mrk 266}

Figure \ref{fig:Regions} displays regions parameterized in
Table \ref{tbl:Regions} that were used to perform aperture 
photometry tabulated in Tables \ref{tbl:SED_Tot} and \ref{tbl:SED_Regions},
as plotted in Figure \ref{fig:SEDs} (\S\ref{subsubsec:SEDdata}).

%%%%%%%% Begin Figure %%%%%%%%%%%
%Regions overlaid on 24um, B, X-rays, 20 cm
\def\figcapRegions{
\footnotesize
The regions used to perform aperture photometry of the nuclei, central region,
and Northern Loop superposed on the 24 \micron\ {\it Spitzer} MIPS image (upper left),
the HST ACS B band image (upper right),  the {\it Chandra} full-band X-ray image (lower left),
and the 20 cm radio continuum image (lower right). All four images are shown at the same scale 
and orientation.
%the {\it Spitzer} MIPS 24 \micron\ image after 
%subtraction of a model consisting of two point sources (lower left)
}
\ifnum\Mode=0 %Insert Figure/Table only in [preprint] or [preprint2] modes
\placefigure{fig:Regions}
\begin{verbatim}fig31a_31d\end{verbatim}
\else
%For preprint
%\ifnum\Mode=2 \onecolumn \fi %4-panel fig is too small in 2-column mode
\begin{figure}[!hbt]
\center
\ifnum\Mode=2 
\includegraphics[width=1.0\columnwidth,angle=0]{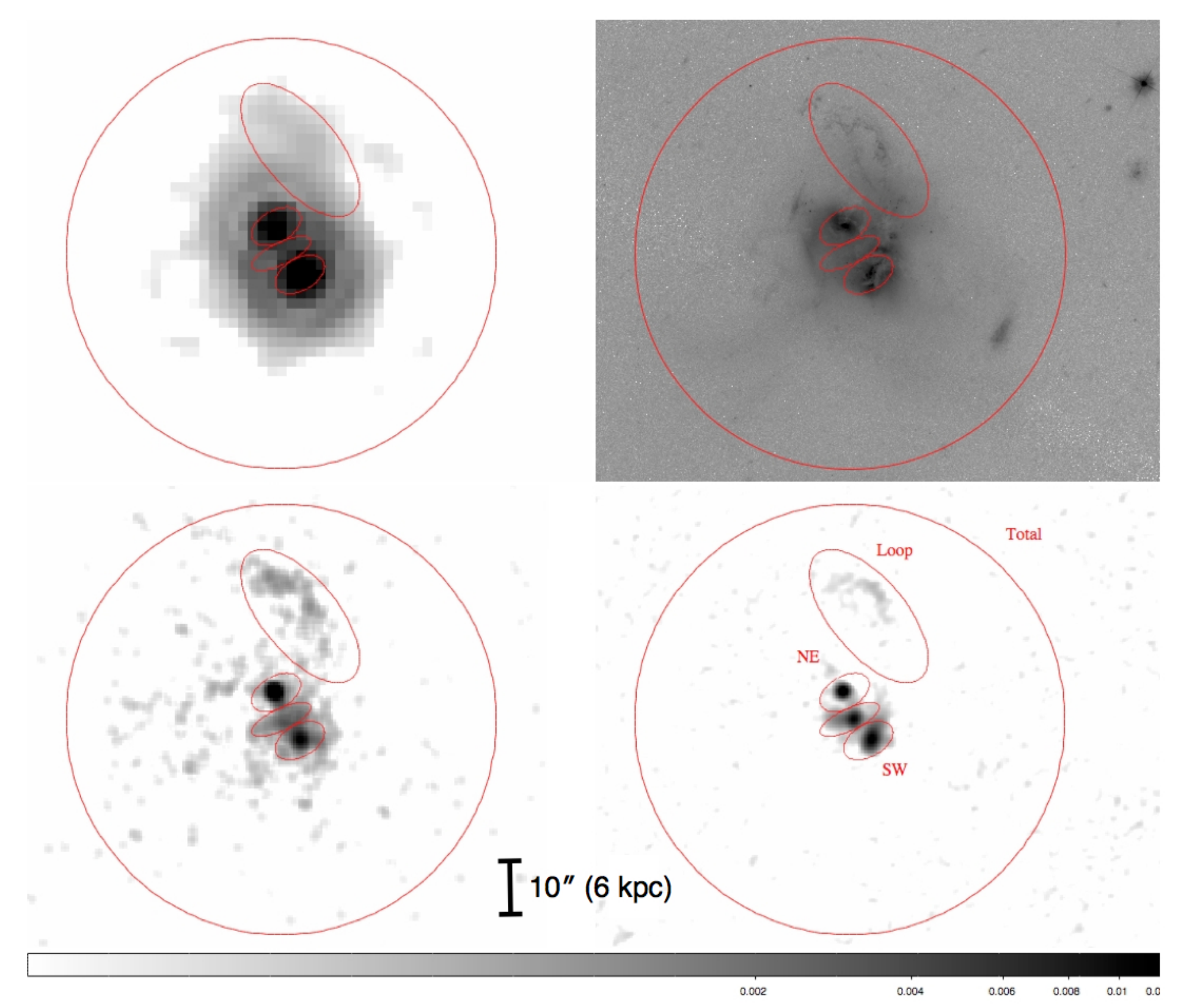} 
\else
\includegraphics[width=6.0truein,angle=0]{fig31a_31d} 
\fi
\caption{\figcapRegions \label{fig:Regions}}
\end{figure}
\fi %close \ifnum\Mode=0 
%%%%%%%% End Figure %%%%%%%%%%%

%%%%%%%% Begin Table %%%%%%%%%%%
\def\tableRegions{
\begin{deluxetable}{lrrrr}
\ifnum\Mode=2
\renewcommand\arraystretch{0.5}% (MyValue=1.0 is for standard spacing)
\tabletypesize{\scriptsize} 
\setlength{\tabcolsep}{0.01in} %Tighten up the columns. See AASTeX FAQ
\tablewidth{\columnwidth}
\fi
\renewcommand\arraystretch{1.0}% (MyValue=1.0 is for standard spacing)
\tabletypesize{\small} 
\setlength{\tabcolsep}{0.3in} %Tighten up the columns. See AASTeX FAQ
%\tablewidth{\textwidth}
\tablecaption{Elliptical Apertures for Regions of Interest \label{tbl:Regions}}
\tablehead{
\colhead{Region} & \colhead{RA, DEC (J2000)} & \colhead{a\arcsec} & \colhead{b\arcsec} & \colhead{PA\arcdeg}\\
\colhead{(1)} & \colhead{(2)} & \colhead{(3)} & \colhead{(4)} & \colhead{(5)} 
}
\tablecolumns{5}
\startdata
SW       & (13:38:17.28,+48:16:32.1) & \phn5.0 & \phn3.0 & 120\\
NE        & (13:38:17.79,+48:16:41.2) & \phn5.0 & \phn3.0 & 120 \\
Between Galaxies & (13:38:17.51,+48:16:36.8) & \phn6.0 & \phn2.0 & 115\\
Northern Loop    & (13:38:17.54,+48:16:56.0) & 17.0      & \phn7.5 & \phn40\\
Total System      & (13:38:17.66,+48:16:35.8) & 40.0       & 40.0      & \nodata
\enddata
%\caption{Caption X\label{tbl:Xrays}}
%\tablenotetext{a}{}.
\tablecomments{\footnotesize
Parameters for elliptical apertures used for photometric measurements,
as illustrated in Fig. \ref{fig:Regions}.}
\end{deluxetable}
}
\ifnum\Mode=0
\placetable{tbl:Regions}
\else
\tableRegions
\fi
%%%%%%%% End Table %%%%%%%%%%%

%%%%%%%% Begin Table %%%%%%%%%%%
\def\tableSEDtotal{
\begin{deluxetable}{lcccr}
\ifnum\Mode=2
\renewcommand\arraystretch{0.5}% (MyValue=1.0 is for standard spacing)
\tabletypesize{\tiny}
\setlength{\tabcolsep}{0.0in} %Tighten up the columns. See AASTeX FAQ
\tablewidth{\columnwidth}
\else
\renewcommand\arraystretch{1.0}% (MyValue=1.0 is for standard spacing)
\tabletypesize{\footnotesize} 
\setlength{\tabcolsep}{0.2in} %Tighten up the columns. See AASTeX FAQ
\tablewidth{\textwidth}
\fi
\tablecaption{Photometric Data for Mrk 266 - Total System \label{tbl:SED_Tot}}
\tablehead{
\colhead{Bandpass} & 
\colhead{$\nu$ (Hz)} & 
\colhead{$f_\nu$ (Jy)} & 
\colhead{$\sigma_{f_\nu}$ (Jy)} &
\colhead{Ref.} \\
\colhead{(1)} & 
\colhead{(2)} & 
\colhead{(3)} &
\colhead{(4)} &
\colhead{(5)} 
}
\tablecolumns{5}
\startdata
2-10 keV (ASCA)     	&{\tt 	1.45E+18	}&{\tt 	2.48E-08	}&{\tt 	5.0E-09	}&{\tt 	2	}\\
2.0-7.0 keV (XMM)	&{\tt 	1.09E+18	}&{\tt 	2.11E-08	}&{\tt 	6.3E-09	}&{\tt 	1	}\\
0.7-7 keV (ASCA)    	&{\tt 	9.31E+17	}&{\tt 	3.65E-08	}&{\tt 	7.3E-09	}&{\tt 	2	}\\
0.5-2.0 keV (XMM)	&{\tt 	3.02E+17	}&{\tt 	5.96E-08	}&{\tt 	1.2E-08	}&{\tt 	1	}\\
GALEX FUV	&{\tt 	1.96E+15	}&{\tt 	7.94E-04	}&{\tt 	3.1E-05	}&{\tt 	1	}\\
XMM-OM UVW2	      &{\tt 	1.41E+15	}&{\tt 	2.06E-03	}&{\tt 	8.0E-04	}&{\tt 	1	}\\
GALEX NUV	&{\tt 	1.32E+15	}&{\tt 	1.63E-03	}&{\tt 	3.4E-05	}&{\tt 	1	}\\
XMM-OM UVW1	      &{\tt 	1.03E+15	}&{\tt 	2.12E-03	}&{\tt 	2.6E-04	}&{\tt 	1	}\\
XMM-OM U	&{\tt 	8.71E+14	}&{\tt 	3.87E-03	}&{\tt 	1.4E-03	}&{\tt 	1	}\\
HST ACS F435W (B)	&{\tt 	6.89E+14	}&{\tt 	8.92E-03	}&{\tt 	4.1E-05	}&{\tt 	1	}\\
V (Johnson)           	&{\tt 	5.42E+14	}&{\tt 	1.82E-02	}&{\tt 	8.6E-04	}&{\tt 	3	}\\
LW2 (ISOCAM)        	&{\tt 	4.44E+13	}&{\tt 	1.14E-01	}&{\tt 	1.4E-02	}&{\tt 	4	}\\
R (Johnson)           	&{\tt 	4.28E+14	}&{\tt 	2.86E-02	}&{\tt 	8.0E-04	}&{\tt 	3	}\\
HST ACS F814W (I)	&{\tt 	3.68E+14	}&{\tt 	2.97E-02	}&{\tt 	5.5E-05	}&{\tt 	1	}\\
2MASS J (1.2 $\mu$m)	&{\tt 	2.43E+14	}&{\tt 	5.44E-02	}&{\tt 	6.5E-04	}&{\tt 	1	}\\
2MASS H (1.6 $\mu$m)	&{\tt 	1.80E+14	}&{\tt 	6.42E-02	}&{\tt 	8.8E-04	}&{\tt 	1	}\\
2MASS Ks (2.2 $\mu$m)	&{\tt 	1.39E+14	}&{\tt 	6.02E-02	}&{\tt 	8.3E-04	}&{\tt 	1	}\\
Spitzer 3.6 $\mu$m	&{\tt 	8.44E+13	}&{\tt 	3.82E-02	}&{\tt 	5.7E-03	}&{\tt 	1	}\\
Spitzer 4.5 $\mu$m	&{\tt 	6.67E+13	}&{\tt 	3.05E-02	}&{\tt 	4.6E-03	}&{\tt 	1	}\\
Spitzer 5.8 $\mu$m	&{\tt 	5.23E+13	}&{\tt 	6.80E-02	}&{\tt 	1.0E-02	}&{\tt 	1	}\\
Spitzer 8.0 $\mu$m	&{\tt 	3.81E+13	}&{\tt 	1.79E-01	}&{\tt 	2.7E-02	}&{\tt 	1	}\\
IRAS 12 $\mu$m     	&{\tt 	2.50E+13	}&{\tt 	3.20E-01	}&{\tt 	1.8E-02	}&{\tt 	5	}\\
Spitzer 24 $\mu$m	&{\tt 	1.27E+13	}&{\tt 	9.55E-01	}&{\tt 	1.9E-01	}&{\tt 	1	}\\
IRAS 25 $\mu$m     	&{\tt 	1.20E+13	}&{\tt 	1.07E+00	}&{\tt 	2.8E-02	}&{\tt 	5	}\\
IRAS 60 $\mu$m     	&{\tt 	5.00E+12	}&{\tt 	7.25E+00	}&{\tt 	3.3E-02	}&{\tt 	5	}\\
Spitzer 70 $\mu$m	&{\tt 	4.20E+12	}&{\tt 	8.95E+00	}&{\tt 	1.8E+00	}&{\tt 	1	}\\
IRAS 100 $\mu$m    	&{\tt 	3.00E+12	}&{\tt 	1.01E+01	}&{\tt 	1.4E-01	}&{\tt 	5	}\\
Spitzer 160 $\mu$m	&{\tt 	1.92E+12	}&{\tt 	6.41E+00	}&{\tt 	1.3E+00	}&{\tt 	1	}\\
170 $\mu$m (ISO)   	&{\tt 	1.76E+12	}&{\tt 	8.73E+00	}&{\tt 	2.6E+00	}&{\tt 	6	}\\
SCUBA 850 $\mu$m  &{\tt 	3.53E+11	}&{\tt 	8.20E-02	}&{\tt 	1.7E-02	}&{\tt 	7	}\\
20.0 GHz (OVRO)     	&{\tt 	2.00E+10	}&{\tt 	1.49E-02	}&{\tt 	1.3E-03	}&{\tt 	8	}\\
4.85 GHz            	      &{\tt 	4.85E+09	}&{\tt 	4.70E-02	}&{\tt 	7.1E-03	}&{\tt 	9	}\\
1.46 GHz (VLA)      	&{\tt 	1.46E+09	}&{\tt 	1.07E-01	}&{\tt 	6.2E-03	}&{\tt 	8	}\\
Texas 365 MHz       	&{\tt 	3.65E+08	}&{\tt 	3.53E-01	}&{\tt 	4.3E-02	}&{\tt 	10	}\\
151 MHz (6C)        	&{\tt 	1.52E+08	}&{\tt 	5.90E-01	}&{\tt 	4.0E-02	}&{\tt 	11	}
\enddata
%\caption{Caption X\label{tbl:Xrays}}
\tablecomments{
\footnotesize
Reference codes in Column (5):
1=New measurements in this article;
2=\citet{2005ApJS..161..185U};
3=\citet{1977ApJS...35..171H};
4=\citet{2007AJ....134.2006R};
5=\citet{2003AJ....126.1607S};
6=\citet{2004A&A...422...39S};
7=\citet{2000MNRAS.315..115D};
8=\citet{1987ApJ...313..651E};
9=\citet{1991ApJS...75....1B};
10=\citet{1996AJ....111.1945D};
11=\citet{1990MNRAS.246..256H}.}
\end{deluxetable}
}

\ifnum\Mode=0
\placetable{tbl:SED_Tot}
\else
\tableSEDtotal
\fi
%%%%%%%% End Table %%%%%%%%%%%

%%%%%%%% Begin Table %%%%%%%%%%%
\def\tableSEDregions{
\begin{deluxetable}{lcccr}
\ifnum\Mode=2
\renewcommand\arraystretch{0.5}% (MyValue=1.0 is for standard spacing)
\tabletypesize{\tiny}
\setlength{\tabcolsep}{0.0in} %Tighten up the columns. See AASTeX FAQ
\tablewidth{\columnwidth}
\else
\renewcommand\arraystretch{0.8}% (MyValue=1.0 is for standard spacing)
\tabletypesize{\footnotesize} 
\setlength{\tabcolsep}{0.3in} %Tighten up the columns. See AASTeX FAQ
%\tablewidth{\textwidth}
\fi
\tablecaption{Photometric Data for Mrk 266 - Components \label{tbl:SED_Regions}}
\tablehead{
\colhead{Bandpass} & 
\colhead{$\nu$ (Hz)} & 
\colhead{$f_\nu$ (Jy)} & 
\colhead{$\sigma_{f_\nu}$ (Jy)} &
\colhead{Ref.} \\
\colhead{(1)} & 
\colhead{(2)} & 
\colhead{(3)} &
\colhead{(4)} &
\colhead{(5)} 
}
\tablecolumns{5}
\startdata
\cutinhead{Mrk 266 SW}
2.0-7.0 keV	&{\tt 	1.09E+18	}&{\tt 	2.02E-09	}&{\tt 	6.1E-10	}&{\tt 	1	}\\
0.5-2.0 keV	&{\tt 	3.02E+17	}&{\tt 	7.28E-09	}&{\tt 	1.5E-09	}&{\tt 	1	}\\
XMM-OM UVW2	&{\tt 	1.41E+15	}&{\tt 	1.68E-04	}&{\tt 	1.1E-05	}&{\tt 	1	}\\
XMM-OM UVW1	&{\tt 	1.03E+15	}&{\tt 	2.72E-04	}&{\tt 	5.4E-05	}&{\tt 	1	}\\
XMM-OM U	&{\tt 	8.71E+14	}&{\tt 	5.12E-04	}&{\tt 	2.2E-05	}&{\tt 	1	}\\
HST ACS F435W (B)	&{\tt 	6.89E+14	}&{\tt 	1.31E-03	}&{\tt 	1.2E-05	}&{\tt 	1	}\\
HST ACS F814W (I)	&{\tt 	3.68E+14	}&{\tt 	5.03E-03	}&{\tt 	2.3E-05	}&{\tt 	1	}\\
2MASS J (1.2 $\mu$m)	&{\tt 	2.43E+14	}&{\tt 	1.04E-02	}&{\tt 	2.5E-04	}&{\tt 	1	}\\
NICMOS H (1.6 $\mu$m)	&{\tt 	1.87E+14	}&{\tt 	1.45E-02	}&{\tt 	1.7E-04	}&{\tt 	1	}\\
2MASS H (1.6 $\mu$m)	&{\tt 	1.80E+14	}&{\tt 	1.46E-02	}&{\tt 	3.3E-04	}&{\tt 	1	}\\
2MASS Ks (2.2 $\mu$m)	&{\tt 	1.39E+14	}&{\tt 	1.53E-02	}&{\tt 	3.1E-04	}&{\tt 	1	}\\
Spitzer 3.6 $\mu$m	&{\tt 	8.44E+13	}&{\tt 	1.50E-02	}&{\tt 	2.3E-03	}&{\tt 	1	}\\
Spitzer 4.5 $\mu$m	&{\tt 	6.67E+13	}&{\tt 	1.47E-02	}&{\tt 	2.2E-03	}&{\tt 	1	}\\
Spitzer 5.8 $\mu$m	&{\tt 	5.23E+13	}&{\tt 	4.59E-02	}&{\tt 	6.9E-03	}&{\tt 	1	}\\
Spitzer 8.0 $\mu$m	&{\tt 	3.81E+13	}&{\tt 	1.28E-01	}&{\tt 	1.9E-02	}&{\tt 	1	}\\
Spitzer 24 $\mu$m	&{\tt 	1.27E+13	}&{\tt 	5.91E-01	}&{\tt 	1.2E-01	}&{\tt 	1	}\\
Spitzer 70 $\mu$m\tablenotemark{a}	&{\tt 	4.20E+12	}&{\tt 	6.01E+00	}&{\tt 	1.2E+00	}&{\tt 	1	}\\
Spitzer 160 $\mu$m\tablenotemark{a}	&{\tt 	1.92E+12	}&{\tt 	4.30E+00	}&{\tt 	8.6E-01	}&{\tt 	1	}\\
2 cm	&{\tt 	1.50E+10	}&{\tt 	4.90E-03	}&{\tt 	7.0E-04	}&{\tt 	12	}\\
6 cm	&{\tt 	4.89E+09	}&{\tt 	1.01E-02	}&{\tt 	5.0E-04	}&{\tt 	12	}\\
20 cm	&{\tt 	1.47E+09	}&{\tt 	3.24E-02	}&{\tt 	1.4E-03	}&{\tt 	12	}\\
\cutinhead{Mrk 266 NE}	
2.0-7.0 keV	&{\tt 	1.09E+18	}&{\tt 	1.29E-08	}&{\tt 	3.9E-09	}&{\tt 	1	}\\
0.5-2.0 keV	&{\tt 	3.02E+17	}&{\tt 	9.93E-09	}&{\tt 	2.0E-09	}&{\tt 	1	}\\
XMM-OM UVW2	&{\tt 	1.41E+15	}&{\tt 	2.45E-04	}&{\tt 	2.0E-05	}&{\tt 	1	}\\
XMM-OM UVW1	&{\tt 	1.03E+15	}&{\tt 	3.63E-04	}&{\tt 	9.1E-05	}&{\tt 	1	}\\
XMM-OM U	&{\tt 	8.71E+14	}&{\tt 	6.49E-04	}&{\tt 	3.0E-05	}&{\tt 	1	}\\
HST ACS F435W (B)	&{\tt 	6.89E+14	}&{\tt 	1.59E-03	}&{\tt 	1.5E-05	}&{\tt 	1	}\\
HST ACS F814W (I)	&{\tt 	3.68E+14	}&{\tt 	6.02E-03	}&{\tt 	2.8E-05	}&{\tt 	1	}\\
2MASS J (1.2 $\mu$m)	&{\tt 	2.43E+14	}&{\tt 	9.62E-03	}&{\tt 	2.4E-04	}&{\tt 	1	}\\
NICMOS H (1.6 $\mu$m)	&{\tt 	1.87E+14	}&{\tt 	1.22E-02	}&{\tt 	1.6E-04	}&{\tt 	1	}\\
2MASS H (1.6 $\mu$m)	&{\tt 	1.80E+14	}&{\tt 	1.22E-02	}&{\tt 	3.0E-04	}&{\tt 	1	}\\
2MASS Ks (2.2 $\mu$m)	&{\tt 	1.39E+14	}&{\tt 	1.09E-02	}&{\tt 	2.6E-04	}&{\tt 	1	}\\
Spitzer 3.6 $\mu$m	&{\tt 	8.44E+13	}&{\tt 	7.35E-03	}&{\tt 	1.1E-03	}&{\tt 	1	}\\
Spitzer 4.5 $\mu$m	&{\tt 	6.67E+13	}&{\tt 	5.90E-03	}&{\tt 	8.9E-04	}&{\tt 	1	}\\
Spitzer 5.8 $\mu$m	&{\tt 	5.23E+13	}&{\tt 	9.57E-03	}&{\tt 	1.4E-03	}&{\tt 	1	}\\
Spitzer 8.0 $\mu$m	&{\tt 	3.81E+13	}&{\tt 	2.23E-02	}&{\tt 	3.3E-03	}&{\tt 	1	}\\
Spitzer 24 $\mu$m	&{\tt 	1.27E+13	}&{\tt 	3.28E-01	}&{\tt 	6.6E-02	}&{\tt 	1	}\\
Spitzer 70 $\mu$m\tablenotemark{a}	&{\tt 	4.20E+12	}&{\tt 	1.74E+00	}&{\tt 	7.0E-01	}&{\tt 	1	}\\
Spitzer 160 $\mu$m\tablenotemark{a}	&{\tt 	1.92E+12	}&{\tt 	1.25E+00	}&{\tt 	5.0E-01	}&{\tt 	1	}\\
2 cm	&{\tt 	1.50E+10	}&{\tt 	4.10E-03	}&{\tt 	6.0E-04	}&{\tt 	12	}\\
6 cm	&{\tt 	4.89E+09	}&{\tt 	1.22E-02	}&{\tt 	4.0E-04	}&{\tt 	12	}\\
20 cm	&{\tt 	1.47E+09	}&{\tt 	2.35E-02	}&{\tt 	1.0E-03	}&{\tt 	12	}\\
\cutinhead{Between the Nuclei}									
2.0-7.0 keV	&{\tt 	1.09E+18	}&{\tt 	9.19E-10	}&{\tt 	2.8E-10	}&{\tt 	1	}\\
0.5-2.0 keV	&{\tt 	3.02E+17	}&{\tt 	8.60E-09	}&{\tt 	1.7E-09	}&{\tt 	1	}\\
2 cm	&{\tt 	1.50E+10	}&{\tt 	2.70E-03	}&{\tt 	5.0E-04	}&{\tt 	12	}\\
6 cm	&{\tt 	4.89E+09	}&{\tt 	6.00E-03	}&{\tt 	2.7E-03	}&{\tt 	12	}\\
20 cm	&{\tt 	1.47E+09	}&{\tt 	2.15E-02	}&{\tt 	9.0E-04	}&{\tt 	12	}\\
\cutinhead{Northern Loop}									
0.5-2.0 keV	&{\tt 	3.02E+17	}&{\tt 	4.63E-09	}&{\tt 	9.3E-10	}&{\tt 	1	}\\
XMM-OM UVW2	&{\tt 	1.41E+15	}&{\tt 	1.94E-04	}&{\tt 	1.7E-06	}&{\tt 	1	}\\
XMM-OM UVW1	&{\tt 	1.03E+15	}&{\tt 	2.10E-04	}&{\tt 	4.8E-06	}&{\tt 	1	}\\
XMM-OM U	&{\tt 	8.71E+14	}&{\tt 	4.28E-04	}&{\tt 	3.0E-05	}&{\tt 	1	}\\
HST ACS F435W (B)	&{\tt 	6.89E+14	}&{\tt 	8.84E-04	}&{\tt 	1.6E-05	}&{\tt 	1	}\\
HST ACS F814W (I)	&{\tt 	3.68E+14	}&{\tt 	2.50E-03	}&{\tt 	2.1E-05	}&{\tt 	1	}\\
Spitzer 24 $\mu$m	&{\tt 	1.27E+13	}&{\tt 	3.62E-02	}&{\tt 	7.2E-03	}&{\tt 	1	}\\
20 cm	&{\tt 	1.47E+09	}&{\tt 	2.60E-03	}&{\tt 	5.0E-05	}&{\tt 	12	}\\
\enddata
\tablenotetext{a}{\footnotesize 
The 70 and 160 \micron\ flux densities for the SW and NE galaxies
were estimated using the 8 and 24 \micron\ measurements as described in the text.}
\tablecomments{
\footnotesize
The reference codes in Column (5) are as follows:
1=New measurements  in this article; 12=\citet{1988ApJ...333..168M}.
}
\end{deluxetable}
}
\ifnum\Mode=0
\placetable{tbl:SED_Regions}
\else
\tableSEDregions
\fi

%%%%%%%% End Table %%%%%%%%%%%

\subsection{Clarification of Regions Covered in Published {\it Spitzer} IRS Data for 
Mrk 266} \label{subsec:PublishedIRSdata}

As noted in \S\ref{sec:IRSdata}, mid-IR spectroscopy of Mrk 266 obtained in 
staring mode with the {\it Spitzer} IRS is complicated due to aperture affects.
To facilitate interpretation of the IRS spectroscopy 
published to date, the available observations are summarized here, and
Table \ref{tbl:IRSpublished} provides clarification of which emission features 
have been measured for each component of the system.

In the low spectral resolution (R 65-130) observations (AOR \#3755264; GTO program 
P00014; PI J. Houck), the Short-Low (SL, 5.2-14.5 \micron) aperture of the 
IRS contained Mrk 266 NE (off center), and missed both Mrk 266 SW and the central 
radio/X-ray source; the Long-Low (LL, 14--38 \micron) slit contained Mrk 266 SW and the
radio/X-ray source between the nuclei, but it missed Mrk 266 NE.
Neither aperture had the orientation required to sample the Northern Loop.
The low-resolution IRS spectrum displayed in \citet{2006ApJ...653.1129B} 
is therefore a mixture of emission from Mrk 266 NE in the 5-14 \micron\ region with
emission from Mrk 266 SW (and the radio/X-ray source between the nuclei)
in the 14-38 \micron\ region.

AOR \#3755264 also contained high-resolution observations (spectral resolution $\sim$600).
\citet[][Fig. 8]{2007ApJ...664...71D} plotted the SH and LH apertures of 
these observations as an overlay on a 20 cm image, although they did not present the 
resulting spectra; their diagram clearly illustrates the aperture affects that complicate the 
interpretation of IRS spectra for this system.
The SH slit contained only the radio source between the nuclei (missing both nuclei), 
while the LH slit contained 266 SW and the central radio source but it missed Mrk 266 NE.
Observations with the SH (9.9 - 19.6 \micron) module were also
obtained on 2005 January 10 in {\it Spitzer} program P03237 
(PI. E. Sturm, AOR \#10510592); these data cover the 
SW nucleus, but not the NE nucleus.
The available high-resolution spectra were published 
in an atlas of {\it Spitzer} spectra for a sample
of starburst galaxies \citep{2009ApJS..184..230B}.

%%%%%%%% Begin Table %%%%%%%%%%%
\def\tableIRSpublished{
\ifnum\Mode=2
\begin{deluxetable}{lrcc}
\renewcommand\arraystretch{0.5}% (MyValue=1.0 is for standard spacing)
\tabletypesize{\footnotesize}
\setlength{\tabcolsep}{0.00in} %Tighten up the columns. See AASTeX FAQ
\tablewidth{\columnwidth}
\else
\begin{deluxetable}{lrcc}
\renewcommand\arraystretch{0.9}% (MyValue=1.0 is for standard spacing)
\tabletypesize{\footnotesize}
\setlength{\tabcolsep}{0.3in} %Tighten up the columns. See AASTeX FAQ
%\tablewidth{\textwidth}
\fi
\tablecaption{Regions Covered in Published IRS Measurements \label{tbl:IRSpublished}}
\tablehead{
\colhead{Feature} & 
\colhead{Flux ($\rm W~m^{-2}$)} & 
\colhead{EQW ($\rm \mu m$)} & 
\colhead{Ref.} \\
\colhead{(1)} & 
\colhead{(2)} & 
\colhead{(3)} &
\colhead{(4)} 
}
\tablecolumns{4}
\startdata
%%%%%%%%%
\cutinhead{Mrk 266 SW}
%The Short-High (SH, 9.9-19.6~\micron) aperture of the Sturm AOR \#10510592 covered the
%SW nucleus; the pointing in GTO AOR \#3755264 did not (but covered the central source).
$[S~IV]~10.51$\micron~(SH) &{\tt $9.00(\pm 0.38)\times10^{-21}$}  & \nodata &{\tt B09} \\
$PAH~10.6$\micron~(SH) & {\tt $0.25(\pm 0.05)\times10^{-20}$}   & \nodata & {\tt B09} \\
$PAH~10.7$\micron~(SH) & {\tt $0.12(\pm 0.03)\times10^{-20}$}  & \nodata & {\tt B09} \\
$PAH~11.0$\micron~(SH) &{\tt $1.09(\pm 0.05)\times10^{-20}$}  & \nodata & {\tt B09} \\
$PAH~11.3$\micron~(SH) &{\tt $14.90(\pm 0.74)\times10^{-20}$}  & \nodata & {\tt B09} \\
$PAH~12.0$\micron~(SH) &{\tt $0.73(\pm 0.10)\times10^{-20}$}  & \nodata & {\tt B09} \\
$H_2~(S2)~12.28$\micron~ (SH) &{\tt $3.72(\pm 0.32)\times10^{-21}$}  & \nodata  & {\tt B09} \\
$HI~(7-6)~12.37$\micron~(SH) &{\tt $0.69(\pm 0.05)\times10^{-21}$} & \nodata  & {\tt B09} \\
$PAH~12.74$\micron~(SH) &{\tt $7.13(\pm 0.36)\times10^{-20}$}  & \nodata & {\tt B09} \\
$[Ne II]~12.81$\micron~(SH)&{\tt $57.04(\pm 1.64)\times10^{-21}$}  & \nodata  & {\tt B09} \\
$PAH~13.5$\micron~(SH) &{\tt $0.13(\pm 0.06)\times10^{-21}$}   & \nodata & {\tt B09} \\
$PAH~14.2$\micron~(SH) &{\tt $0.57(\pm 0.06)\times10^{-20}$}  & \nodata & {\tt B09} \\
$[Ne~V]~14.32$\micron~(SH)&{\tt $7.96(\pm 0.17)\times10^{-21}$}   & \nodata  & {\tt B09} \\
$[Cl~II]~14.38$\micron~(SH)&{\tt $0.46(\pm 0.07)\times10^{-21}$}    & \nodata  & {\tt B09} \\
$[Ne~III]~15.55$\micron~(SH)&{\tt $27.95(\pm 0.76)\times10^{-21}$}   & \nodata  & {\tt B09} \\
$PAH~16.45$\micron~(SH) &{\tt $0.84(\pm 0.08)\times10^{-20}$}  & \nodata & {\tt B09} \\
$PAH~17.0$\micron~(SH) &{\tt $1.07(\pm 0.14)\times10^{-20}$}  & \nodata & {\tt B09} \\
$H_2~(S1)~17.03$\micron ~(SH)&{\tt $8.40(\pm 0.97)\times10^{-21}$}  & \nodata  & {\tt B09} \\
$[P~III]~17.89$\micron~(SH)&{\tt $0.67(\pm 0.15)\times10^{-21}$}  & \nodata  & {\tt B09} \\
$[Fe~II]~17.95$\micron~(SH)&{\tt $0.48(\pm 0.14)\times10^{-21}$}  & \nodata  & {\tt B09} \\
$[S~III]~18.71$\micron~(SH)&{\tt $24.33(\pm 0.69)\times10^{-21}$}  & \nodata  & {\tt B09} \\
%%%%%%%%%
\cutinhead{Mrk 266 SW + Source Between Nucleus}
$[Ne~III]~15.55$\micron  ~(LL)\tablenotemark{a}   & \nodata  & \nodata & {\tt B06} \\
$PAH~17.0$\micron~(LL)   & {\tt $0.51\times10^{-19}$} & {\tt 0.403}  &{\tt B06} \\
$[S~III]~18.71$\micron  ~(LL)\tablenotemark{a}   & \nodata  & \nodata & {\tt B06} \\
$[Fe~III]~22.93$\micron~(LH)&{\tt $2.15(\pm 0.29)\times10^{-21}$}  & \nodata  & {\tt B09} \\
$[Ne~V]~24.31$\micron~(LH)\tablenotemark{b}&{\tt $11.12(\pm 0.59)\times10^{-21}$} & \nodata  & {\tt B09} \\
$[O~IV]~25.89$\micron~(LH)&{\tt $52.94(\pm 5.66)\times10^{-21}$} & \nodata  & {\tt B09} \\
$[O~IV]~25.89$\micron  ~(LL)\tablenotemark{a}   & \nodata  & \nodata & {\tt B06} \\
$[Fe~II]~25.98$\micron~(LH)&{\tt $4.35(\pm 0.70)\times10^{-21}$} & \nodata  & {\tt B09} \\
$S_2~(S0)~28.22$\micron~(LH)&{\tt $<1.60\times10^{-21}$} & \nodata  & {\tt B09} \\
$[S~III]~33.48$\micron~(LH)&{\tt $50.92(\pm 2.91)\times10^{-21}$}  & \nodata  & {\tt B09} \\
$[S~III]~33.48$\micron  ~(LL)\tablenotemark{a}   & \nodata  & \nodata & {\tt B06} \\
$[Si~II]~34.81$\micron~(LH)&{\tt $87.02(\pm 4.44)\times10^{-21}$}  & \nodata  & {\tt B09} \\
\cutinhead{Mrk 266 NE}
$PAH~6.2$\micron ~(SL)  & {\tt $0.59\times10^{-19}$}  & {\tt 0.619} & {\tt B06} \\
$PAH~7.7$\micron   ~(SL)  & {\tt $0.92\times10^{-19}$}  & {\tt 0.467} & {\tt B06} \\
$PAH~8.6$\micron   ~(SL)  & \nodata  & {\tt 0.126} & {\tt B06} \\
$H_2~S(3)~9.665$\micron  ~(SL)\tablenotemark{a}   & \nodata  & \nodata & {\tt B06} \\
$[S~IV]~10.51$\micron~(SH)\tablenotemark{a}      & \nodata  & \nodata &{\tt B06} \\
$H_2~S(2)~12.279$\micron  ~(SL)\tablenotemark{a}   & \nodata  & \nodata & {\tt B06} \\
$[Ne II]~12.81$\micron  ~(SL)\tablenotemark{a}   & \nodata  & \nodata & {\tt B06} \\
$PAH~11.3$\micron  ~(SL)  & {\tt $0.47\times10^{-19}$}  & {\tt 0.422} & {\tt B06} \\
$PAH~14.2$\micron  ~(SL) & {\tt $0.08\times10^{-19}$}  & {\tt 0.065} & {\tt B06}
\enddata
\tablenotetext{a}{\small These features are visible in the low-res spectrum published by
\citet{2006ApJ...653.1129B}, but fluxes were not published.}
\tablenotetext{b}{\small This is consistent with a [Ne V] 24.32 \micron\ flux of
$1.19(\pm0.06)\times10^{-20}~W~cm^{-2}$ measured from the same 
LH aperture by \citet[][Table 2]{2007ApJ...664...71D}.}
\tablecomments{
\footnotesize
Column (1): Spectral line identification followed in parentheses by the IRS module used:
SL $=$ 5.2-14.5~\micron, 3\farcs7x57\arcsec, $\rm R\sim60-127$; 
LL $=$ 14.0-38.0~\micron, 10\farcs5x168\farcs3, $\rm R\sim57-126$;
SH $=$ 9.9-19.6~\micron, 4\farcs7x11\farcs3, $\rm R\sim600$; 
LH $=$ 18.7-37.3~\micron, 11\farcs1x22\farcs3, $\rm R\sim600$.
Column (2): Flux in $\rm W~cm^{-2}$.
Column (3): Equivalent width in \micron. 
Column (4): Reference: B06 $=$ \citet{2006ApJ...653.1129B}; 
B09 $=$ \citet{2009ApJS..184..230B}. 
}
\ifnum\Mode=2  
\end{deluxetable}
\else
\end{deluxetable}
\fi
}
\ifnum\Mode=0
\placetable{tbl:IRSpublished}
\else
\tableIRSpublished
\fi
%%%%%%%% End Table %%%%%%%%%%%

\ifnum\Mode=2 
~\\
~\\
\fi

\subsection{Evidence That Fe K$\alpha$ Originates in Mrk 266 SW}
\label{subsec:FeKdetails}

Another clue to the origin of the Fe K$\alpha$ emission in Mrk 266 comes from analysis of the X-ray 
data in combination with optical spectroscopy. The ratio of X-ray (2-10 keV) continuum flux to
(dereddened) $\rm [O~III] \lambda 5007$ emission-line flux has been shown to be anti-correlated with 
$\rm N_H$ and with the equivalent width of the Fe K$\alpha$ line, thus providing an indicator for 
Compton-thick AGNs \citep{1999ApJS..121..473B}.
Using the $F_x(hard)$ values measured from {\it Chandra} (Table \ref{tbl:XrayData}) and the 
de-reddened [O III] $\lambda$5007 line fluxes from \citet{1995ApJS...98..171V} 
($\rm 7.8\times10^{-14}~and~8.5\times10^{-15}~erg~s^{-1}$ for the SW and NE nuclei), 
the $\rm F_x/F_{[O~III]}$ ratios are 0.28 and 16 for the SW and NE nuclei, respectively.
A ratio of $\rm F_x/F_{[O~III]} \la 0.1$ indicates a highly absorbed source
($\rm N_H \ga 10^{24}~ cm^{-2}$), whereas $\rm F_x/F_{[O~III]} >> 1$ is a signature of a relatively
unobscured (Type 1) AGN \citep[][]{1999ApJS..121..473B}. This result adds compelling evidence 
that the Fe K$\alpha$ line detected in the XMM spectrum of Mrk 266 originates from the 
SW (Seyfert 2) nucleus. Although Mrk 266 SW is a strong candidate for containing a Compton-thick 
X-ray source, observations at higher energies and greater sensitivity are required for confirmation.

\subsection{Star Clusters Photometry (\S\ref{subsec:SCs})}
\label{subsec:SCphotometry}

%%%%%%%% Begin Table %%%%%%%%%%%
\def\tableSCdata{
%\section{SC Measurements}
%\label{app:SCs}
\ifnum\Mode=2
\LongTables %\LongTables only works with emulateapj.cls
\begin{deluxetable}{cccccccccc}
\tabletypesize{\tiny}
\renewcommand\arraystretch{0.5}% (MyValue=1.0 is for standard spacing)
\setlength{\tabcolsep}{3pt} %Tighten up the columns. See AASTeX FAQ
%\tablewidth{\columnwidth}
\else
\begin{deluxetable}{cccccccccc}
\renewcommand\arraystretch{0.6}% (MyValue=1.0 is for standard spacing)
\tabletypesize{\small}
\setlength{\tabcolsep}{0.15in} %Tighten up the columns. See AASTeX FAQ
%\tablewidth{\textwidth}
\fi
\tablecaption{Star Cluster Photometry \label{tbl:SCdata}}
\tablehead{
\colhead{SC$\#$} &
\colhead{$\delta$ R.A.} & \colhead{$\delta$ Dec} &
\colhead{$\rm B$} & \colhead{$\sigma$} & \colhead{$M_{\rm B}$} &
\colhead{$\rm I$} & \colhead{$\sigma$} &
\colhead{$\rm B-I$} & \colhead{$\sigma$} \\
\colhead{} &
%\colhead{$hh~mm~ss.ss$} & \colhead{$dd~mm~ss.s$} &
\colhead{ss.ss} & \colhead{mm~ss.s} &
\colhead{mag} & \colhead{mag} & \colhead{mag} &
\colhead{mag} & \colhead{mag} &
\colhead{mag} & \colhead{mag} \\
\colhead{(1)} & \colhead{(2)} & \colhead{(3)} & \colhead{(4)} & \colhead{(5)} &
\colhead{(6)} & \colhead{(7)} & \colhead{(8)} & \colhead{(9)} & \colhead{(10)}
}
\tablecolumns{10}
\startdata
001 & 17.19 & 16  32.4 & 21.45 & 0.03 & -14.16 & 21.11 & 0.05 & 0.31 & 0.06   \\
002 & 17.35 & 16  31.9 & 21.52 & 0.04 & -14.09 & 19.48 & 0.04 & 2.01 & 0.06   \\
003 & 17.14 &  16  33.9 & 22.50 & 0.04 & -13.11 & 21.25 & 0.08 & 1.22 & 0.09   \\
004 & 17.35 &  16  30.9 & 22.91 & 0.13 & -12.70 & 21.20 & 0.08 & 1.68 & 0.15   \\
005 & 18.40 &  16  43.8 & 23.07 & 0.01 & -12.54 & 22.47 & 0.03 & 0.57 & 0.03   \\
006 & 17.39 &  16 30.9 & 23.08 & 0.07 & -12.53 & 21.21 & 0.05 & 1.84 & 0.09   \\
007 & 17.16 &  16 36.8 & 23.14 & 0.05 & -12.47 & 22.34 & 0.09 & 0.77 & 0.10   \\
008 & 17.17 &  16 36.5 & 23.24 & 0.06 & -12.37 & 22.58 & 0.08 & 0.63 & 0.10   \\
009 &  17.72 &  16 34.1 & 23.53 & 0.02 & -12.08 & 22.83 & 0.03 & 0.67 & 0.04   \\
010 &  17.02 &  16 36.6 & 23.54 & 0.09 & -12.07 & 22.49 & 0.13 & 1.02 & 0.16   \\
011 &  17.08 &  16 34.7 & 23.63 & 0.04 & -11.98 & 20.91 & 0.02 & 2.69 & 0.04   \\
012 &  16.96 &  16 32.0 & 23.84 & 0.09 & -11.77 & 23.17 & 0.20 & 0.64 & 0.22   \\
013 &  16.15 &  16 39.9 & 23.97 & 0.01 & -11.64 & 22.65 & 0.01 & 1.29 & 0.01   \\
014 &  16.79 &  16 44.4 & 24.00 & 0.03 & -11.61 & 23.24 & 0.04 & 0.73 & 0.05   \\
015 &  16.83 &  16 39.2 & 24.09 & 0.10 & -11.52 & 23.87 & 0.16 & 0.19 & 0.19   \\
016 &  17.68 &  16 38.4 & 24.20 & 0.04 & -11.41 & 23.37 & 0.16 & 0.80 & 0.16   \\
017 &  17.45 &  16 28.8 & 24.23 & 0.13 & -11.38 & 22.76 & 0.14 & 1.44 & 0.19   \\
018 &  18.64 &  16 44.5 & 24.28 & 0.06 & -11.33 & 23.80 & 0.07 & 0.45 & 0.09   \\
019 &  17.06 &  16 31.9 & 24.30 & 0.15 & -11.31 & 23.23 & 0.22 & 1.04 & 0.27   \\
020 &  16.80 &  16 36.2 & 24.40 & 0.07 & -11.21 & 23.45 & 0.13 & 0.92 & 0.15   \\
021 &  16.91 &  16 45.6 & 24.54 & 0.03 & -11.07 & 23.82 & 0.04 & 0.69 & 0.05   \\
022 &  16.71 &  16 40.7 & 24.60 & 0.03 & -11.01 & 23.74 & 0.04 & 0.83 & 0.05   \\
023 &  18.16 &  16 42.3 & 24.71 & 0.07 & -10.90 & 25.32 & 0.81 & -0.64 & 0.81  \\
024 &  17.22 &  16 35.2 & 24.72 & 0.08 & -10.89 & 22.91 & 0.12 & 1.78 & 0.14   \\
025 &  16.62 &  16 45.5 & 24.73 & 0.04 & -10.88 & 23.74 & 0.04 & 0.96 & 0.06   \\
026 &  17.27 &  16 27.4 & 24.74 & 0.11 & -10.87 & 23.43 & 0.16 & 1.28 & 0.19   \\
027 &  17.29 &  16 28.4 & 24.76 & 0.26 & -10.85 & 23.47 & 0.32 & 1.26 & 0.41   \\
028 &  17.57 &  16 42.1 & 24.83 & 0.19 & -10.78 & 23.83 & 0.19 & 0.97 & 0.27   \\
029 &  17.19 &  16 30.1 & 24.85 & 0.12 & -10.76 & 25.81 & 1.05 & -0.99 & 1.06  \\
030 &  16.32 &  16 41.0 & 24.86 & 0.02 & -10.75 & 22.23 & 0.01 & 2.60 & 0.02   \\
031 &  16.77 &  16 45.2 & 24.91 & 0.04 & -10.70 & 24.22 & 0.06 & 0.66 & 0.07   \\
032 &  16.97 &  16 44.4 & 24.92 & 0.07 & -10.69 & 23.74 & 0.05 & 1.15 & 0.09   \\
033 &  17.09 &  16 37.9 & 24.92 & 0.15 & -10.69 & 23.32 & 0.16 & 1.57 & 0.22   \\
034 &  16.53 &  16 31.3 & 24.94 & 0.03 & -10.67 & 23.81 & 0.04 & 1.10 & 0.05   \\
035 &  17.23 &  16 40.6 & 24.95 & 0.08 & -10.66 & 24.73 & 0.22 & 0.19 & 0.23   \\
036 &  18.48 &  16 45.2 & 24.97 & 0.05 & -10.64 & 24.43 & 0.07 & 0.51 & 0.09   \\
037 &  18.45 &  16 36.6 & 25.00 & 0.08 & -10.61 & 24.33 & 0.21 & 0.64 & 0.22   \\
038 &  18.69 &  16 44.3 & 25.00 & 0.07 & -10.61 & 24.62 & 0.13 & 0.35 & 0.15   \\
039 &  17.94 &  16 31.3 & 25.05 & 0.06 & -10.56 & 24.07 & 0.07 & 0.95 & 0.09   \\
040 &  18.38 &  16 45.9 & 25.12 & 0.04 & -10.49 & 24.27 & 0.08 & 0.82 & 0.09   \\
041 &  16.76 &  16 45.6 & 25.13 & 0.07 & -10.48 & 24.64 & 0.11 & 0.46 & 0.13   \\
042 &  18.47 &  17 04.5  & 25.14 & 0.03 & -10.47 & 24.17 & 0.03 & 0.94 & 0.04   \\
043 &  18.73 &  16 44.9 & 25.16 & 0.11 & -10.45 & 24.10 & 0.11 & 1.03 & 0.16   \\
044 &  16.97 &  16 43.0 & 25.19 & 0.07 & -10.42 & 24.11 & 0.08 & 1.05 & 0.11   \\
045 &  16.70 &  16 37.2 & 25.20 & 0.13 & -10.41 & 24.57 & 0.25 & 0.60 & 0.28   \\
046 &  16.92 &  16 42.2 & 25.24 & 0.06 & -10.37 & 24.46 & 0.07 & 0.75 & 0.09   \\
047 &  17.21 &  16 24.6 & 25.28 & 0.04 & -10.33 & 24.63 & 0.06 & 0.62 & 0.07   \\
048 &  16.77 &  16 38.7 & 25.32 & 0.14 & -10.29 & 24.72 & 0.31 & 0.57 & 0.34   \\
049 &  17.16 &  16 28.1 & 25.34 & 0.28 & -10.27 & 24.01 & 0.34 & 1.30 & 0.44   \\
050 &  17.22 &  16 26.0 & 25.35 & 0.07 & -10.26 & 24.30 & 0.08 & 1.02 & 0.11   \\
051 &  16.42 &  16 30.9 & 25.36 & 0.05 & -10.25 & 23.23 & 0.03 & 2.10 & 0.06   \\
052 &  18.32 &  16 43.7 & 25.36 & 0.09 & -10.25 & 24.48 & 0.22 & 0.85 & 0.24   \\
053 &  17.20 &  16 43.2 & 25.36 & 0.06 & -10.25 & 24.24 & 0.10 & 1.09 & 0.12   \\
054 &  17.00 &  16 34.6 & 25.38 & 0.22 & -10.23 & 24.57 & 0.58 & 0.78 & 0.62   \\
055 &  18.57 &  16 51.2 & 25.39 & 0.04 & -10.22 & 24.32 & 0.03 & 1.04 & 0.05   \\
056 &  18.56 &  16 44.9 & 25.42 & 0.07 & -10.19 & 24.91 & 0.10 & 0.48 & 0.12   \\
057 &  16.77 &  16 32.1 & 25.48 & 0.27 & -10.13 & 24.40 & 0.33 & 1.05 & 0.43   \\
058 &  18.66 &  16 45.6 & 25.48 & 0.13 & -10.13 & 24.41 & 0.11 & 1.04 & 0.17   \\
059 &  17.13 &  16 44.7 & 25.50 & 0.06 & -10.11 & 24.80 & 0.10 & 0.67 & 0.12   \\
060 &  15.85 &  16 42.5 & 25.50 & 0.04 & -10.11 & 24.53 & 0.04 & 0.94 & 0.06   \\
061 &  18.77 &  16 42.3 & 25.51 & 0.08 & -10.10 & 24.97 & 0.12 & 0.51 & 0.14   \\
062 &  16.20 &  16 33.6 & 25.53 & 0.04 & -10.08 & 24.37 & 0.03 & 1.13 & 0.05   \\
\ifnum\Mode=2 
\else
\tablebreak
\fi
063 &  18.41 &  16 37.7 & 25.53 & 0.09 & -10.08 & 24.57 & 0.20 & 0.93 & 0.22   \\
064 &  16.72 &  16 35.3 & 25.53 & 0.18 & -10.08 & 23.99 & 0.19 & 1.51 & 0.26   \\
065 &  18.36 &  16 45.4 & 25.54 & 0.07 & -10.07 & 25.06 & 0.16 & 0.45 & 0.17   \\
066 &  17.00 &  16 43.0 & 25.59 & 0.10 & -10.02 & 24.76 & 0.16 & 0.80 & 0.19   \\
067 &  17.63 &  16 31.7 & 25.60 & 0.16 & -10.01 & 24.06 & 0.19 & 1.51 & 0.25   \\
068 &  19.08 &  16 13.2 & 25.62 & 0.04 & -9.99  & 24.03 & 0.02 & 1.56 & 0.04   \\
069 &  16.63 &  16 46.2 & 25.64 & 0.11 & -9.97  & 24.75 & 0.13 & 0.86 & 0.17   \\
070 &  17.33 &  16 40.6 & 25.65 & 0.18 & -9.96  & 24.68 & 0.17 & 0.94 & 0.25   \\
071 &  18.73 &  16 43.8 & 25.65 & 0.22 & -9.96  & 25.84 & 0.51 & -0.22 & 0.56  \\
072 &  16.64 &  16 38.9 & 25.66 & 0.09 & -9.95  & 24.82 & 0.14 & 0.81 & 0.17   \\
073 &  16.09 &  16 31.2 & 25.68 & 0.04 & -9.93  & 24.51 & 0.04 & 1.14 & 0.06   \\
074 &  17.94 &  16 37.0 & 25.70 & 0.10 & -9.91  & 24.86 & 0.30 & 0.81 & 0.32   \\
075 &  18.06 &  16 35.4 & 25.70 & 0.12 & -9.91  & 25.18 & 0.29 & 0.49 & 0.31   \\
076 &  15.58 &  16 41.0 & 25.72 & 0.05 & -9.89  & 24.68 & 0.04 & 1.01 & 0.06   \\
077 &  20.51 &  16 18.3 & 25.77 & 0.05 & -9.84  & 24.57 & 0.04 & 1.17 & 0.06   \\
078 &  17.70 &  16 34.7 & 25.77 & 0.15 & -9.84  & 25.10 & 0.33 & 0.64 & 0.36   \\
079 &  18.29 &  16 43.1 & 25.80 & 0.12 & -9.81  & 24.49 & 0.22 & 1.28 & 0.25   \\
080 &  18.74 &  16 43.6 & 25.82 & 0.20 & -9.79  & 26.29 & 0.62 & -0.50 & 0.65  \\
081 &  18.50 &  16 48.8 & 25.92 & 0.08 & -9.69  & 25.25 & 0.08 & 0.64 & 0.11   \\
082 &  16.72 &  16 45.3 & 25.95 & 0.22 & -9.66  & 25.45 & 0.27 & 0.47 & 0.35   \\
083 &  16.75 &  16 37.9 & 26.02 & 0.25 & -9.59  & 24.63 & 0.19 & 1.36 & 0.31   \\
084 &  16.50 &  16 46.1 & 26.06 & 0.07 & -9.55  & 24.71 & 0.06 & 1.32 & 0.09   \\
085 &  17.31 &  16 39.9 & 26.08 & 0.20 & -9.53  & 25.29 & 0.19 & 0.76 & 0.28   \\
086 &  16.56 &  16 45.3 & 26.08 & 0.12 & -9.53  & 25.45 & 0.20 & 0.60 & 0.23   \\
087 &  16.86 &  16 45.9 & 26.12 & 0.13 & -9.49  & 25.54 & 0.23 & 0.55 & 0.26   \\
088 &  18.54 &  16 45.3 & 26.13 & 0.14 & -9.48  & 25.81 & 0.28 & 0.29 & 0.31   \\
089 &  16.09 &  16 46.2 & 26.17 & 0.09 & -9.44  & 24.96 & 0.07 & 1.18 & 0.11   \\
090 &  18.01 &  16 28.7 & 26.19 & 0.11 & -9.42  & 24.78 & 0.09 & 1.38 & 0.14   \\
091 &  16.85 &  16 38.1 & 26.19 & 0.40 & -9.42  & 25.87 & 0.63 & 0.29 & 0.75   \\
092 &  16.44 &  16 48.2 & 26.20 & 0.08 & -9.41  & 25.42 & 0.11 & 0.75 & 0.14   \\
093 &  19.64 &  16 13.9 & 26.23 & 0.07 & -9.38  & 25.11 & 0.05 & 1.09 & 0.09   \\
094 &  17.86 &  16 32.2 & 26.24 & 0.13 & -9.37  & 24.23 & 0.08 & 1.98 & 0.15   \\
095 &  16.14 &  16 42.6 & 26.27 & 0.09 & -9.34  & 25.22 & 0.09 & 1.02 & 0.13   \\
096 &  17.20 &  16 26.4 & 26.29 & 0.16 & -9.32  & 26.50 & 0.79 & -0.24 & 0.81  \\
097 &  19.40 &  16 36.0 & 26.30 & 0.09 & -9.31  & 25.37 & 0.09 & 0.90 & 0.13   \\
098 &  16.90 &  16 27.0 & 26.30 & 0.18 & -9.31  & 24.45 & 0.12 & 1.82 & 0.22   \\
099 &  16.73 &  16 35.7 & 26.31 & 0.41 & -9.30  & 24.08 & 0.21 & 2.20 & 0.46   \\
100 &  18.68 &  16 50.1 & 26.43 & 0.09 & -9.18  & 25.79 & 0.12 & 0.61 & 0.15   \\
101 &  17.50 &  16 16.2 & 26.46 & 0.09 & -9.15  & 25.47 & 0.08 & 0.96 & 0.12   \\
102 &  16.45 &  16 28.4 & 26.50 & 0.11 & -9.11  & 26.66 & 0.39 & -0.19 & 0.41  \\
103 &  15.87 &  16 39.3 & 26.51 & 0.09 & -9.10  & 25.41 & 0.09 & 1.07 & 0.13   \\
104 &  14.99 &  16 28.8 & 26.53 & 0.09 & -9.08  & 25.14 & 0.07 & 1.36 & 0.11   \\
105 &  18.08 &  16 17.9 & 26.55 & 0.10 & -9.06  & 25.66 & 0.10 & 0.86 & 0.14   \\
106 &  20.16 &  16 24.9 & 26.56 & 0.10 & -9.05  & 25.16 & 0.06 & 1.37 & 0.12   \\
107 &  18.69 &  16 45.9 & 26.56 & 0.36 & -9.05  & 25.34 & 0.27 & 1.19 & 0.45   \\
108 &  16.44 &  16 35.0 & 26.58 & 0.13 & -9.03  & 25.16 & 0.13 & 1.39 & 0.18   \\
109 &  17.49 &  16 11.3 & 26.63 & 0.09 & -8.98  & 25.69 & 0.10 & 0.91 & 0.13   \\
110 &  15.71 &  16 24.7 & 26.63 & 0.10 & -8.98  & 25.49 & 0.09 & 1.11 & 0.13   \\
111 &  15.77 &  16 44.5 & 26.76 & 0.12 & -8.85  & 24.62 & 0.05 & 2.11 & 0.13   \\
112 &  17.22 &  16 44.0 & 26.77 & 0.21 & -8.84  & 25.77 & 0.41 & 0.97 & 0.46   \\
113 &  16.42 &  16 46.4 & 26.85 & 0.16 & -8.76  & 25.87 & 0.19 & 0.95 & 0.25   \\
114 &  16.04 &  16 21.8 & 26.89 & 0.12 & -8.72  & 26.14 & 0.16 & 0.72 & 0.20   \\
115 &  16.78 &  16 57.1 & 27.07 & 0.18 & -8.54  & 25.73 & 0.12 & 1.31 & 0.22   \\
116 &  18.37 &  16 40.5 & 27.12 & 0.88 & -8.49  & 24.25 & 0.28 & 2.84 & 0.92   \\
117 &  15.93 &  16 50.7 & 27.13 & 0.17 & -8.48  & 25.98 & 0.13 & 1.12 & 0.21   \\
118 &  15.67 &  16 48.2 & 27.16 & 0.13 & -8.45  & 25.59 & 0.09 & 1.54 & 0.16   \\
119 &  17.56 &  16 56.9 & 27.23 & 0.22 & -8.38  & 25.73 & 0.13 & 1.47 & 0.26  
\enddata
\tablecomments{
\footnotesize
Column (1): Star cluster (candidate) number.
Cols. (2) and (3): J2000 $\delta$ R.A. and $\delta$ Dec of the object.
To save space, offsets in seconds of time in R.A. and offsets in arcminutes and arcseconds 
in Dec are listed. Full coordinates are obtained by adding these values to 
R.A.=13h38m and Dec=48d (e.g., 13h38m17.19s; 48d16m32.4s).
The $1\sigma$ uncertainty in the coordinates is $\pm 0\farcs3$. 
The objects can be identified as
``Mrk 266 SC$\#\#\#$'' or as ``GOALS Jhhmmssss+ddmmsss.''
For example: Mrk 266 SC001 = GOALS J13381719+4816324.
Col. (4): apparent mag in the HST ACS F435W (B) filter.
Col. (5): $1\sigma$ uncertainty in the B mag.
Col. (6): absolute B mag.
Col. (7): apparent mag in the HST ACS F814W (I) filter.
Col. (8): $1\sigma$ uncertainty in the I mag.
Col. (9): B - I color mag.
Col. (10): $1\sigma$ uncertainty in the B - I color mag.
Absolute B mag and B-I colors have been corrected for foreground Galactic extinction 
using values from \citet{1998ApJ...500..525S} as provided by NED:
$\rm A_B = 0.056$ mag; $\rm A_I = 0.025$ mag.
These data have {\it not} been corrected for extinction within 
Mrk 266 (see \S\ref{subsubsec:SCsCM}).
}
\ifnum\Mode=2  
\end{deluxetable}
\else
\end{deluxetable}
\fi
}
\ifnum\Mode=0
\placetable{tbl:SCdata}
\else
\tableSCdata
\fi
%%%%%%%% End Table %%%%%%%%%%%

%%%%%%%%%%%%%%%%%%%%%%%%%%%%%%%
%BIBLIOGRAPHY/REFERENCES

%%%%%%%%%%%%%%%%%%%%%%%%%%%%%%%
%Preprint modes (embedded tables & figures with no separate figure captions)
%use \end{document} here; [manuscript] mode gets \end{document} at EOF.
\ifnum\Mode>0 
\end{document} 
\fi %[preprint] or [preprint2] mode only
%%%%%%%%%%%%%%%%%%%%%%%%%%%%%%%
%Everything below only prints in manuscript (submission) mode
\centerline{FIGURES}
\label{sec:figcaps}

\setcounter{figure}{0}
\begin{figure}[t]
\center
\includegraphics[width=6.0truein,angle=0]{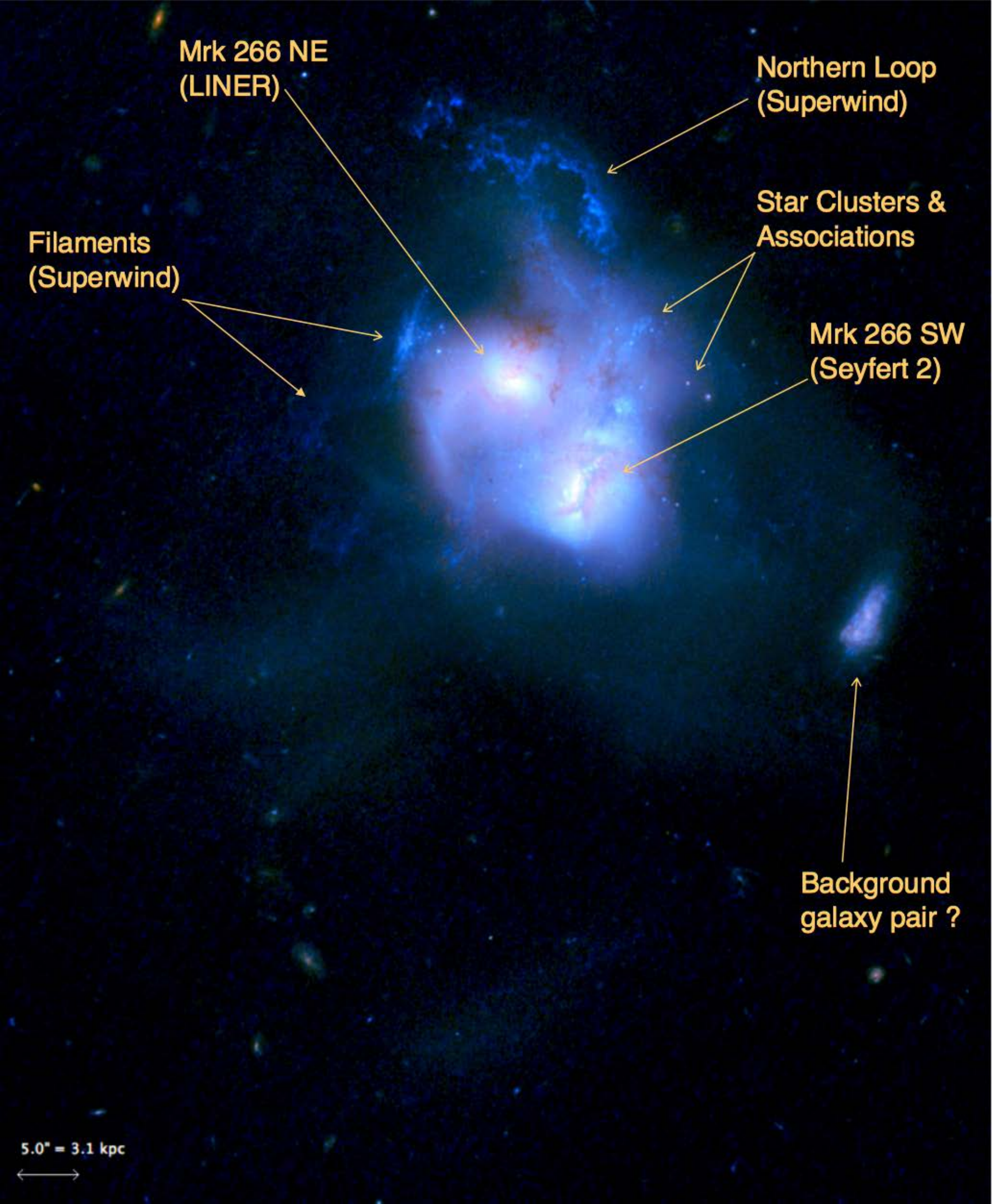}
\caption[f1]{\figcapHSTcolor \label{fig:HST_ACS4x}}
\end{figure}
\clearpage

\begin{figure}[t]
\center
\includegraphics[width=6.0truein,angle=0]{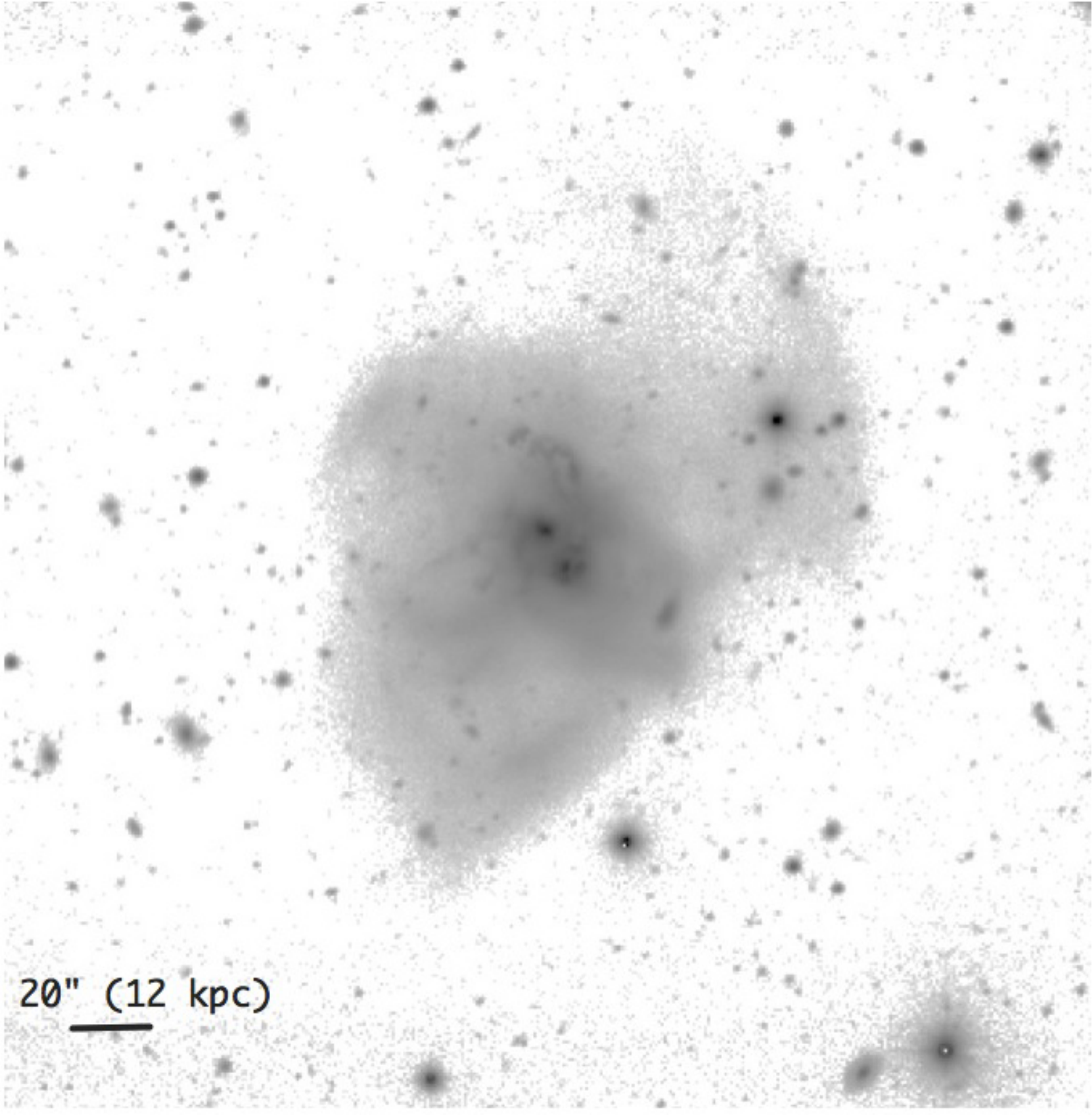}
\caption[f2]{\figcapUH \label{fig:UH88image}}
\end{figure}
\clearpage

\begin{figure}[t]
\center
\includegraphics[width=6.0truein,angle=0]{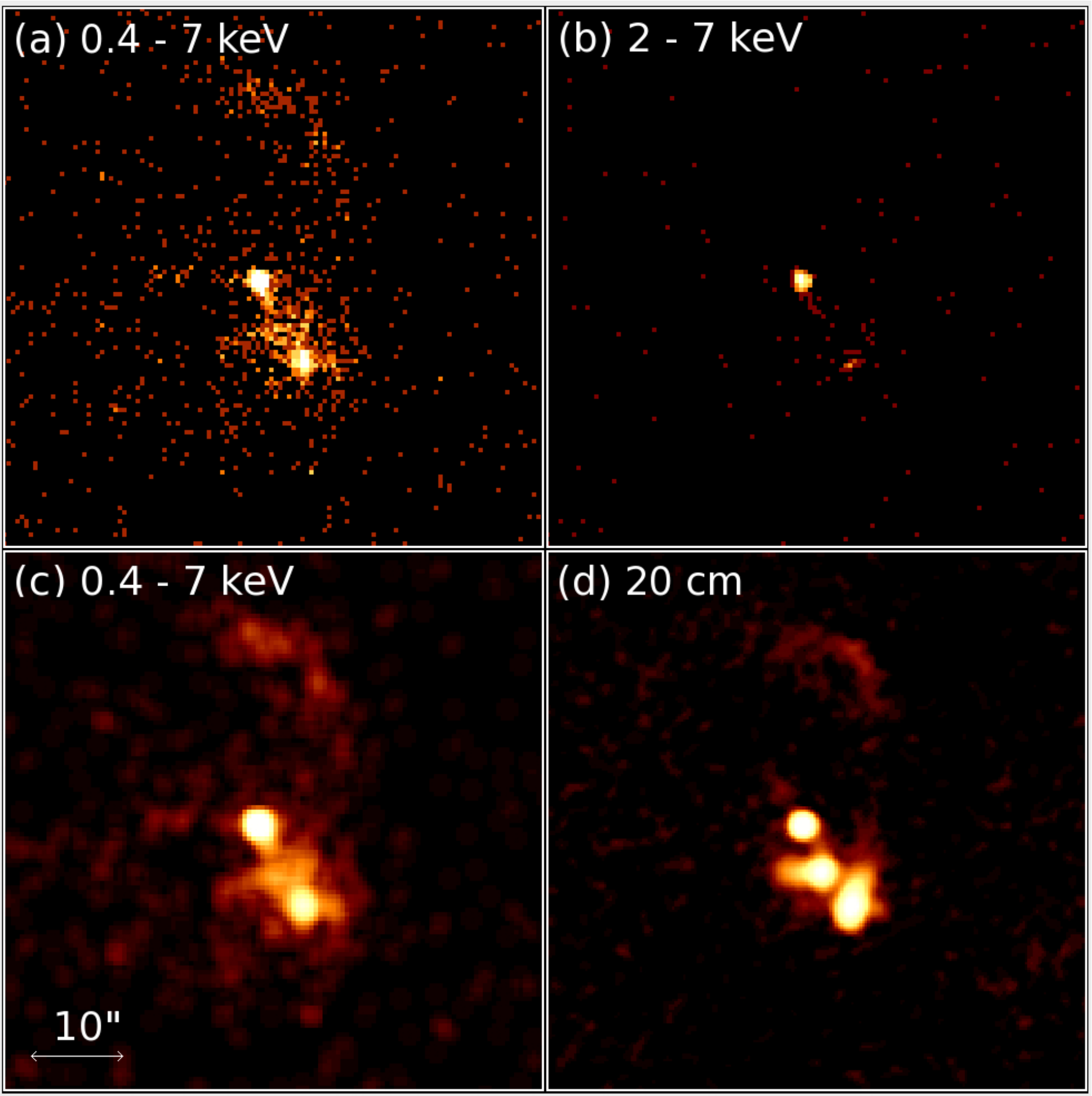}
\caption[f3]{\figcapChandraImages \label{fig:ChandraImages}}
\end{figure}
\clearpage

\begin{figure}[t]
\center
\includegraphics[width=6.0truein,angle=0]{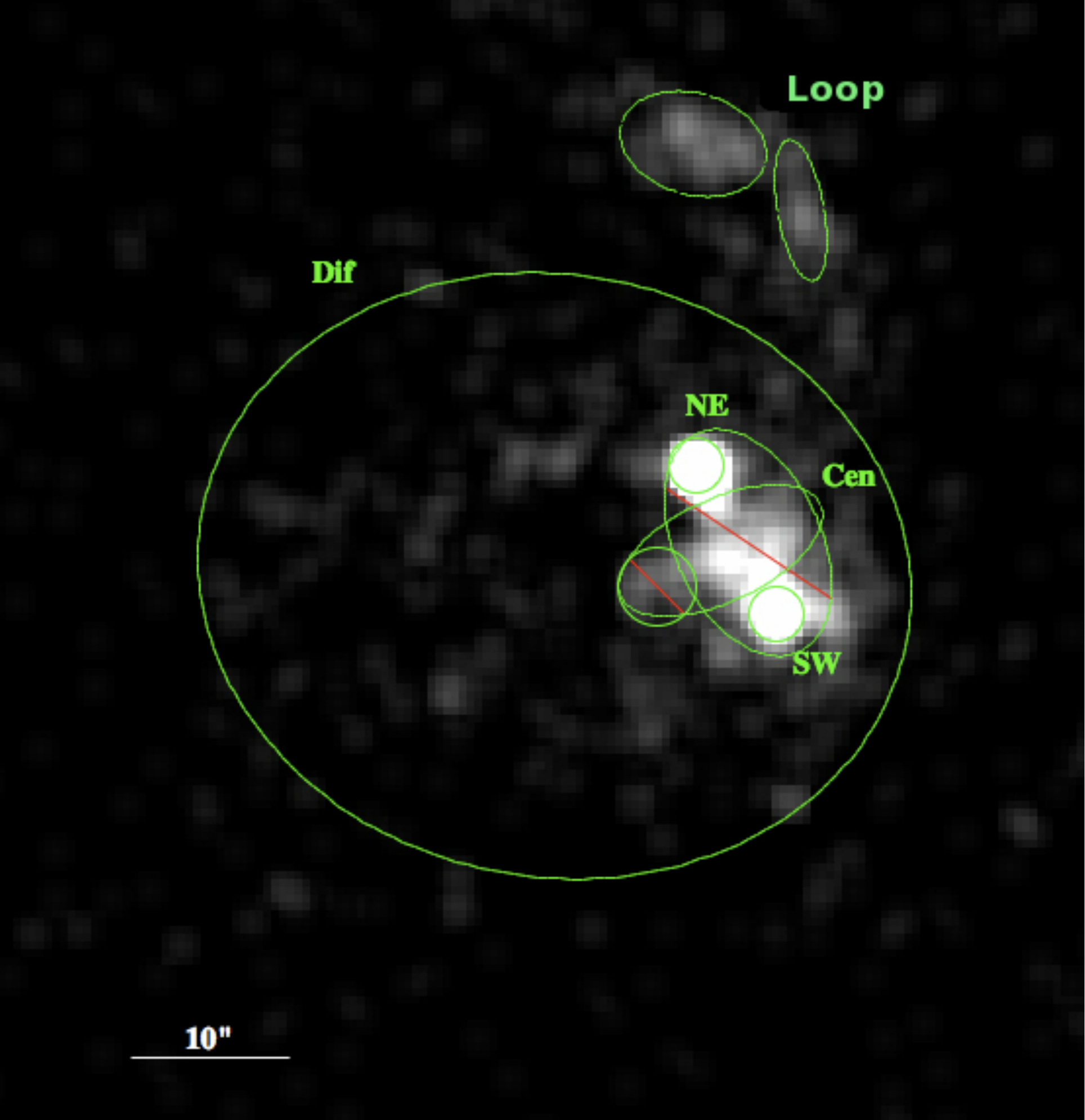}
\caption[f4]{\figcapXrayRegions \label{fig:XrayRegions}}
\end{figure}
\clearpage

\begin{figure}[t]
\center
\includegraphics[width=6.5truein]{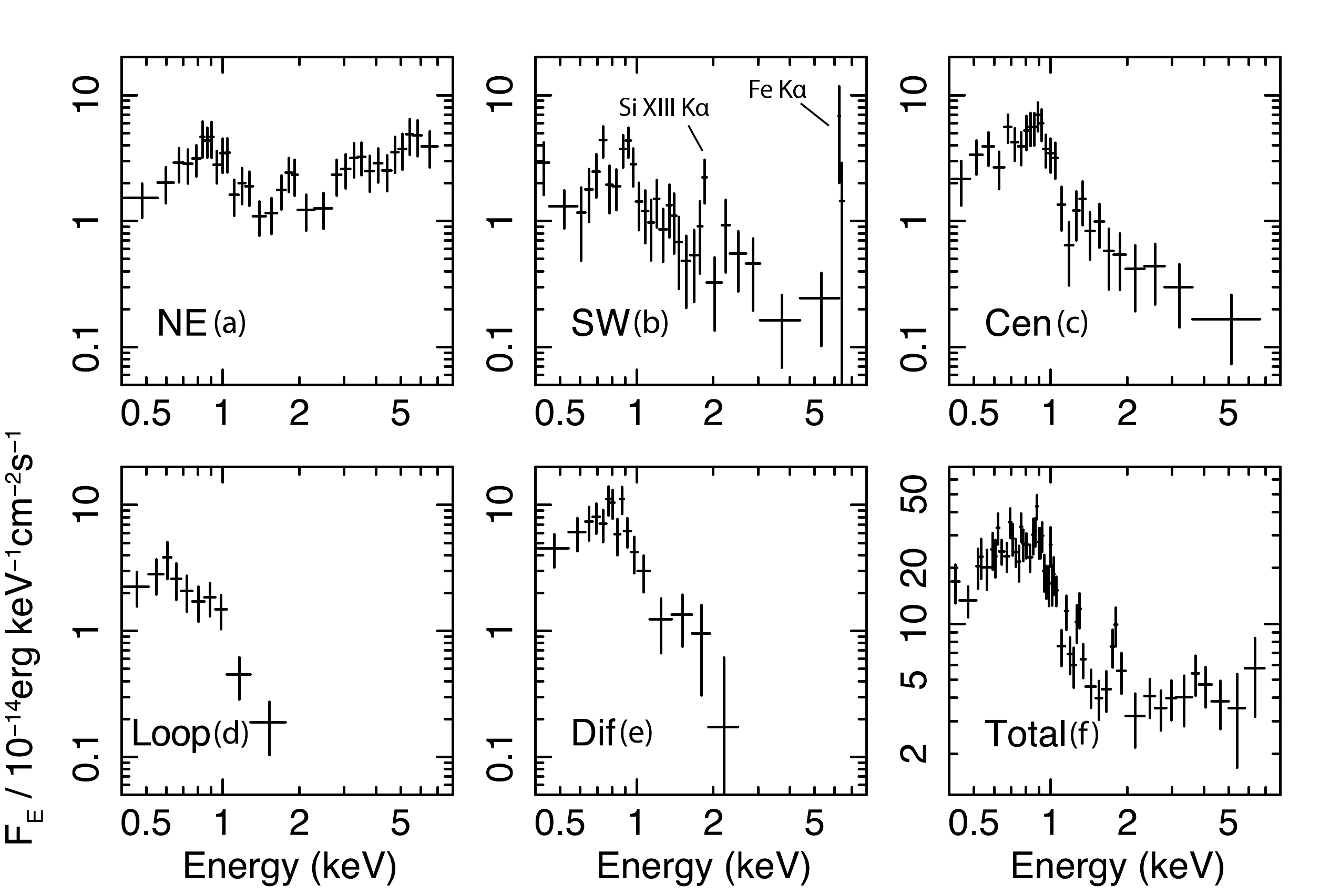}
\caption[f5]{\figChandraSpectra \label{fig:ChandraSpectra}}
\end{figure}
\clearpage

\begin{figure}[t]
\center
\includegraphics[scale=1.0]{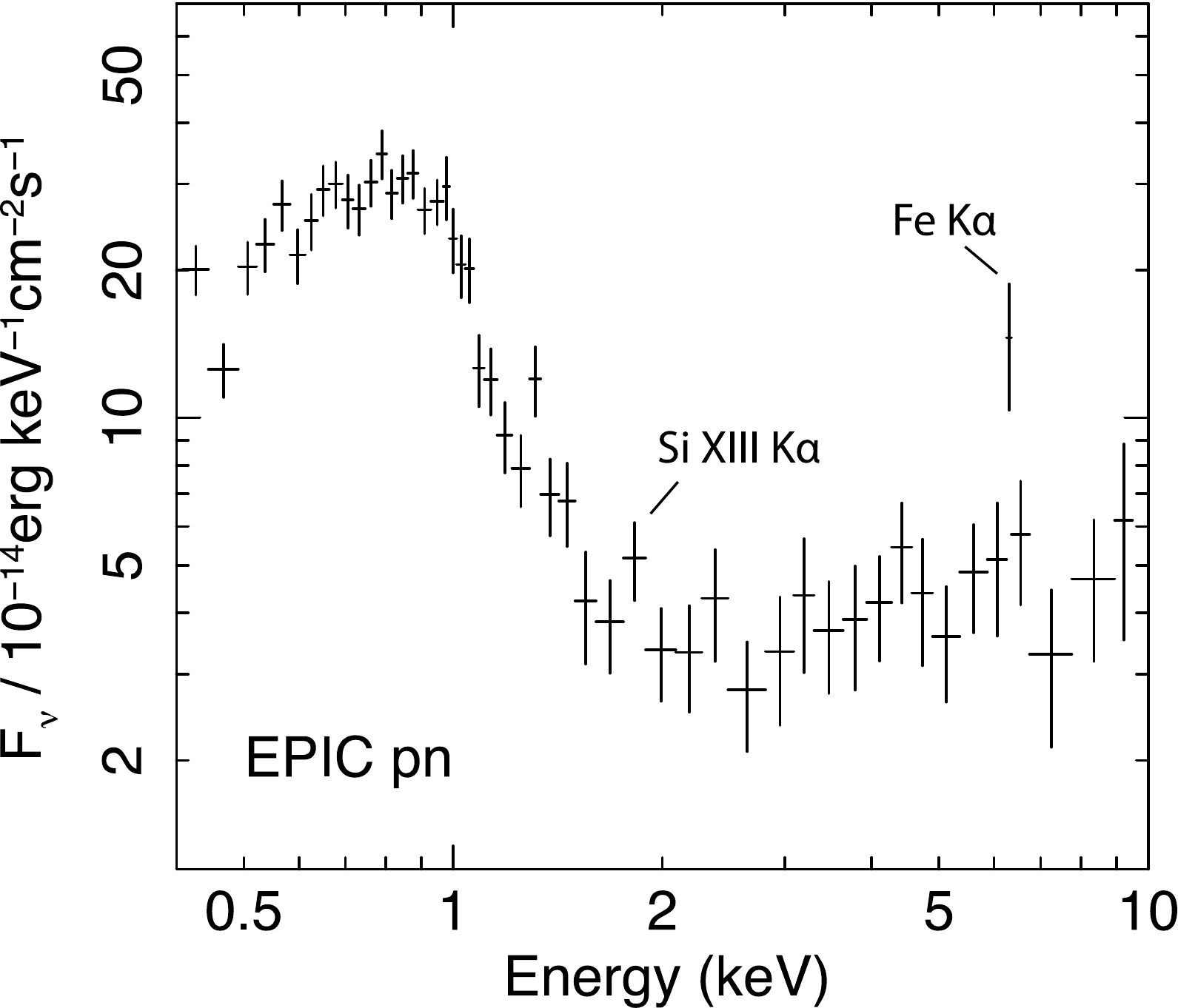} 
\caption[f06]{\figcapXMMSpectrum \label{fig:XMMSpectrum}}
\end{figure}

\begin{figure}[t]
\center
\includegraphics[scale=0.75,angle=0]{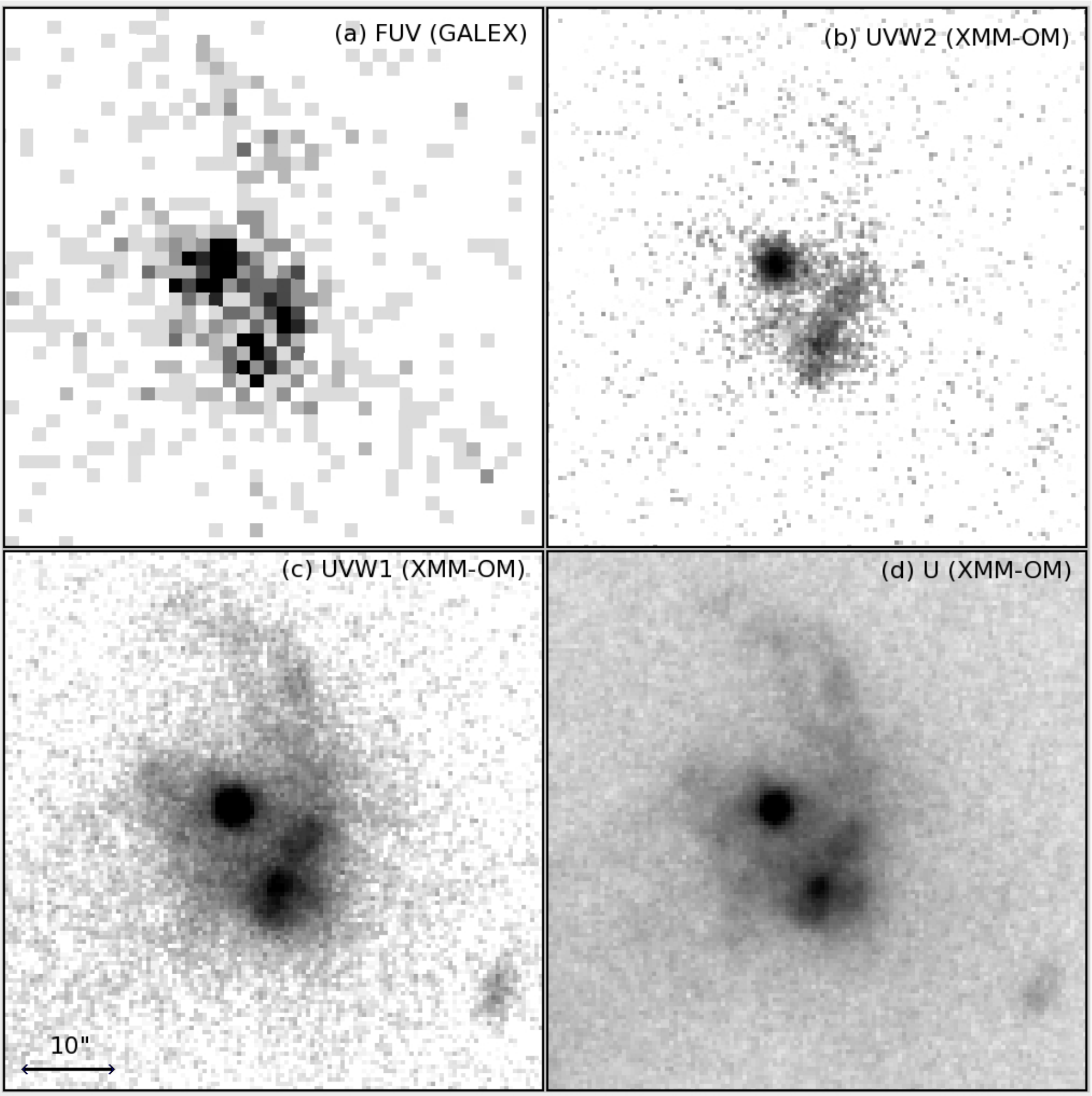}
\caption[f07]{\figcapUVGALEXXMM \label{fig:UV_GALEX_XMM}}
\end{figure}
\clearpage

\begin{figure}[t]
\center
\includegraphics[width=6.5truein,angle=0]{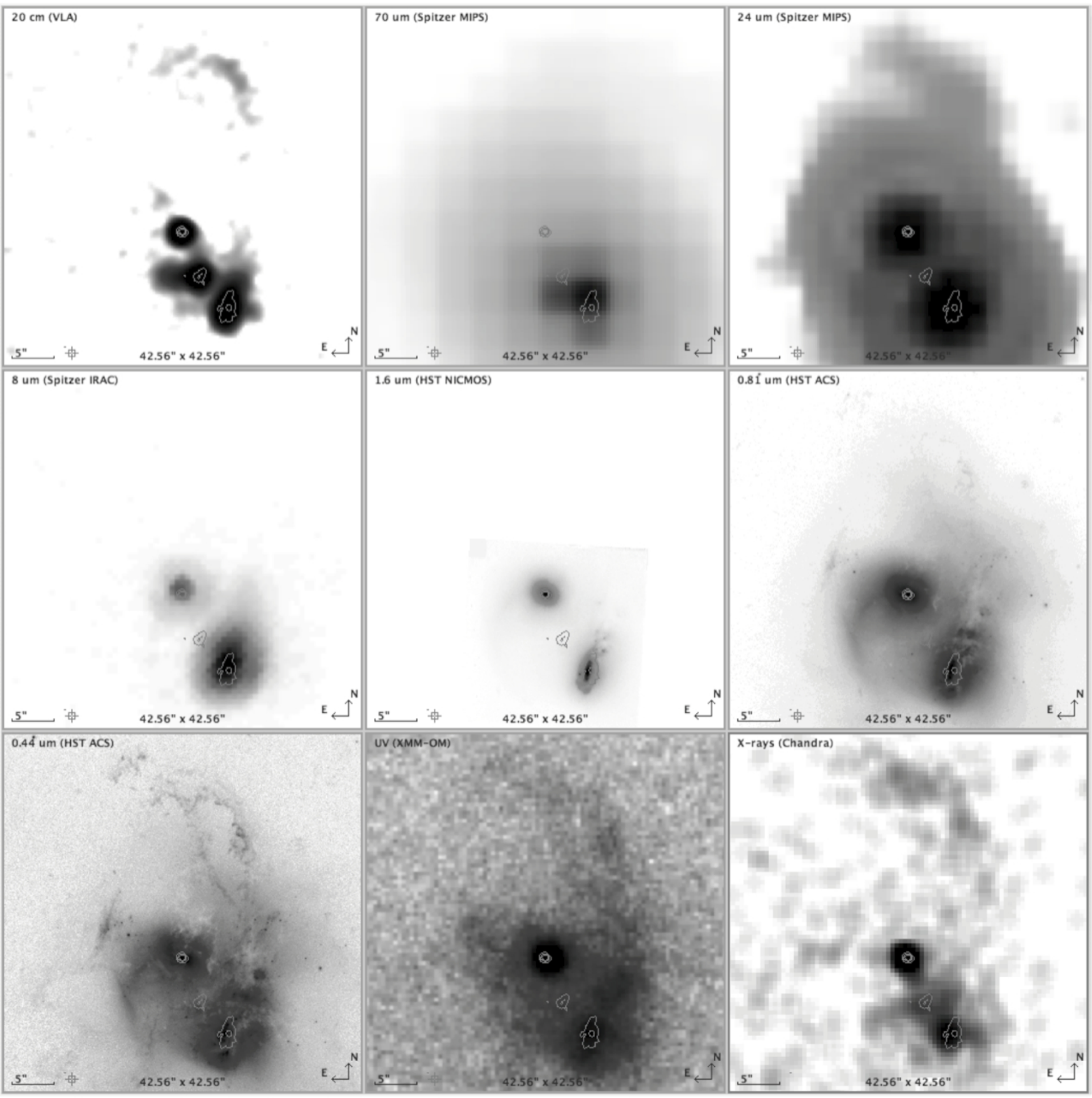}
\put(-329,453) {\parbox{10cm}{(a)}} \put(-173,453) {\parbox{10cm}{(b)}} \put(-16,453) {\parbox{10cm}{(c)}}
\put(-329,298) {\parbox{10cm}{(d)}} \put(-173,298) {\parbox{10cm}{(e)}} \put(-16,298) {\parbox{10cm}{(f)}}
\put(-329,143) {\parbox{10cm}{(g)}} \put(-173,143) {\parbox{10cm}{(h)}} \put(-16,143) {\parbox{10cm}{(i)}}
\caption[f08]{\figcapNineBands \label{fig:NineBands}}
\end{figure}
\clearpage

\begin{figure}[t]
\center
\includegraphics[width=1.0\textwidth,angle=0]{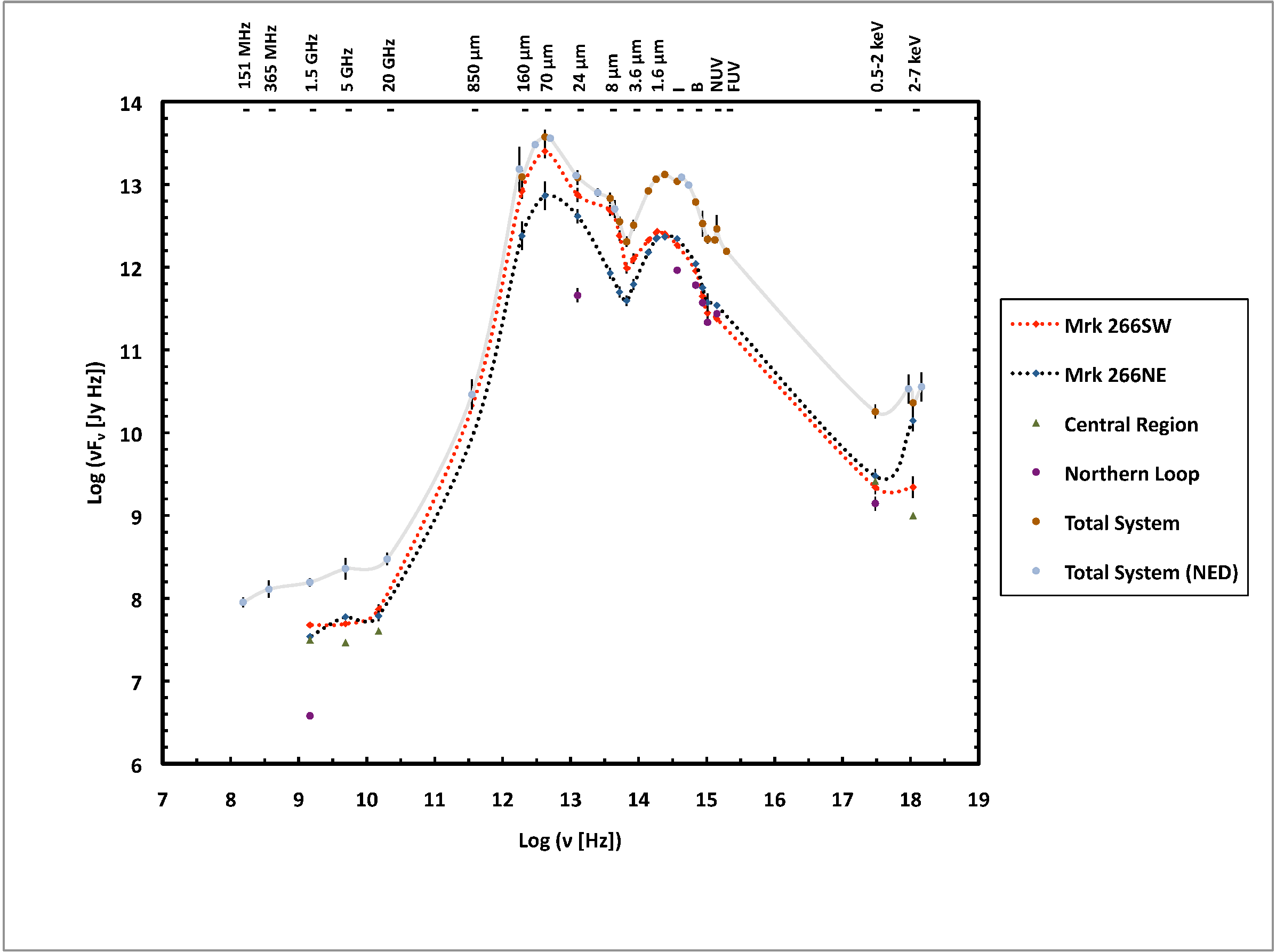}
\caption[f09]{\figcapSEDs \label{fig:SEDs}}
\end{figure}
\clearpage

\begin{figure}[t]
\center
\includegraphics[width=6.0truein,angle=0]{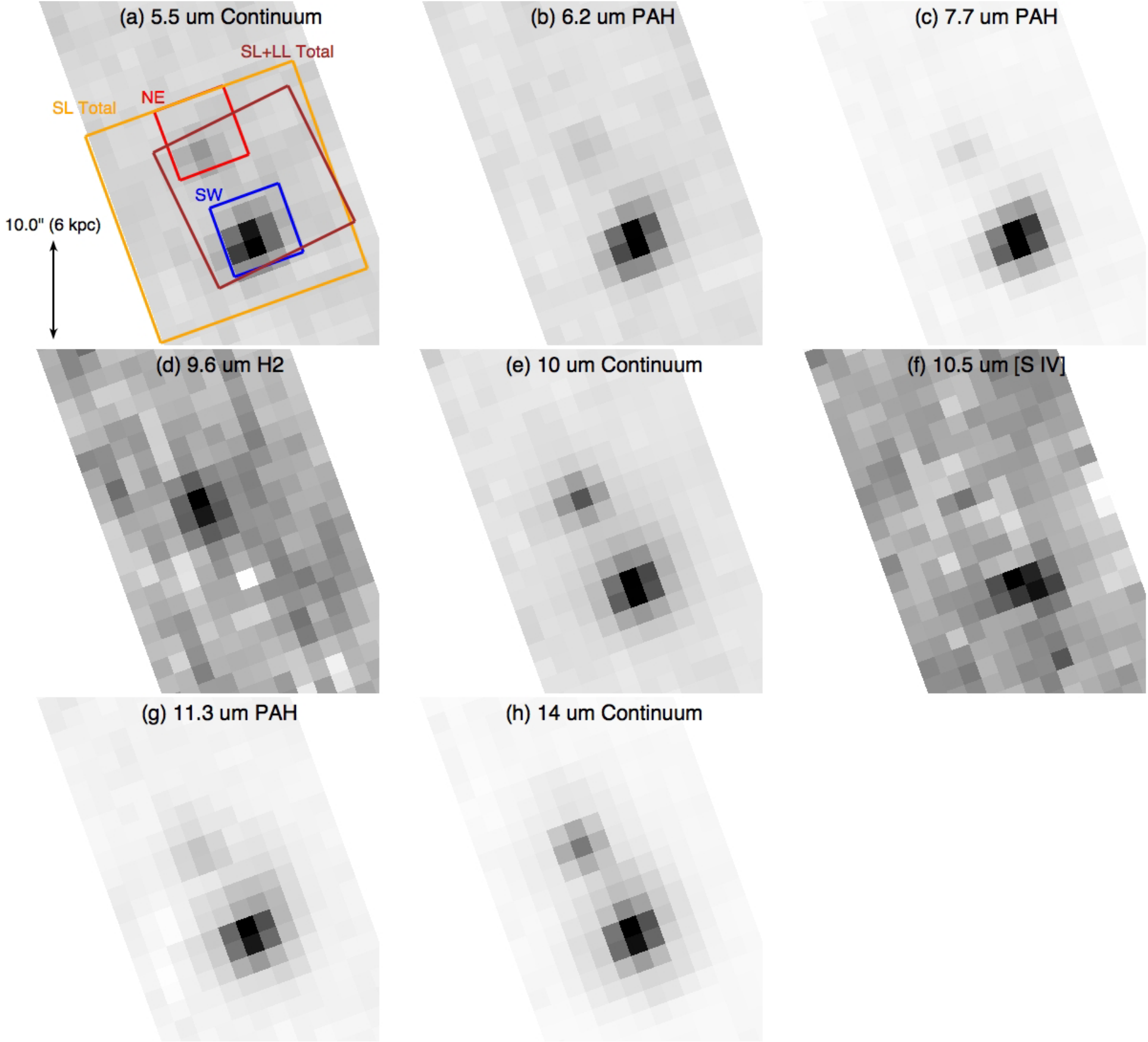}
\caption[f10]{\figcapIRScube \label{fig:IRScube}}
\end{figure}
\clearpage

\begin{figure}[t]
\center
\includegraphics[width=6.0truein,angle=0]{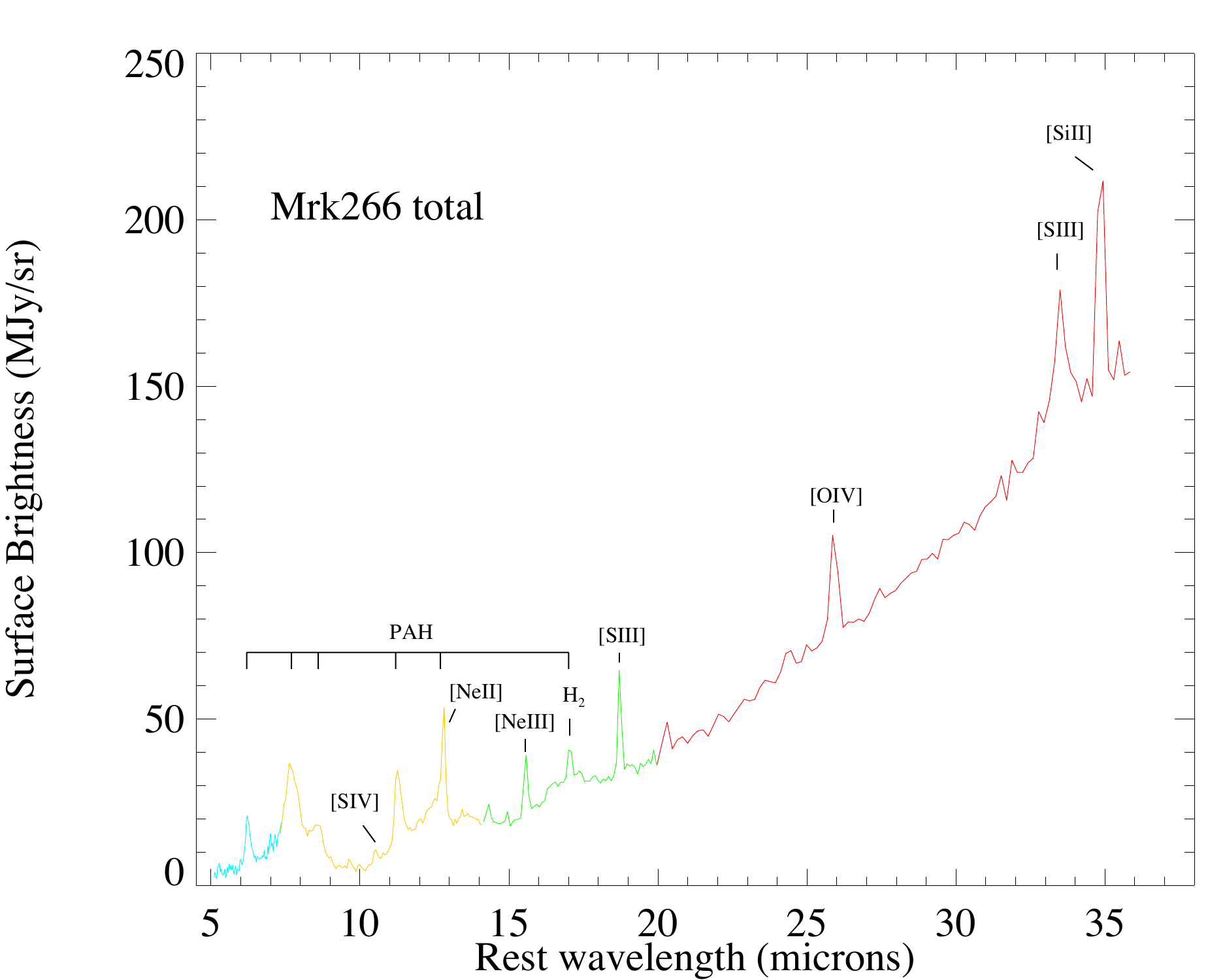}
\caption[f11]{\figcapIRScubeSLLLTotal \label{fig:IRScube_SLLL_Total}}
\end{figure}
\clearpage

\begin{figure}[t]
\center
\includegraphics[scale=0.6,angle=0]{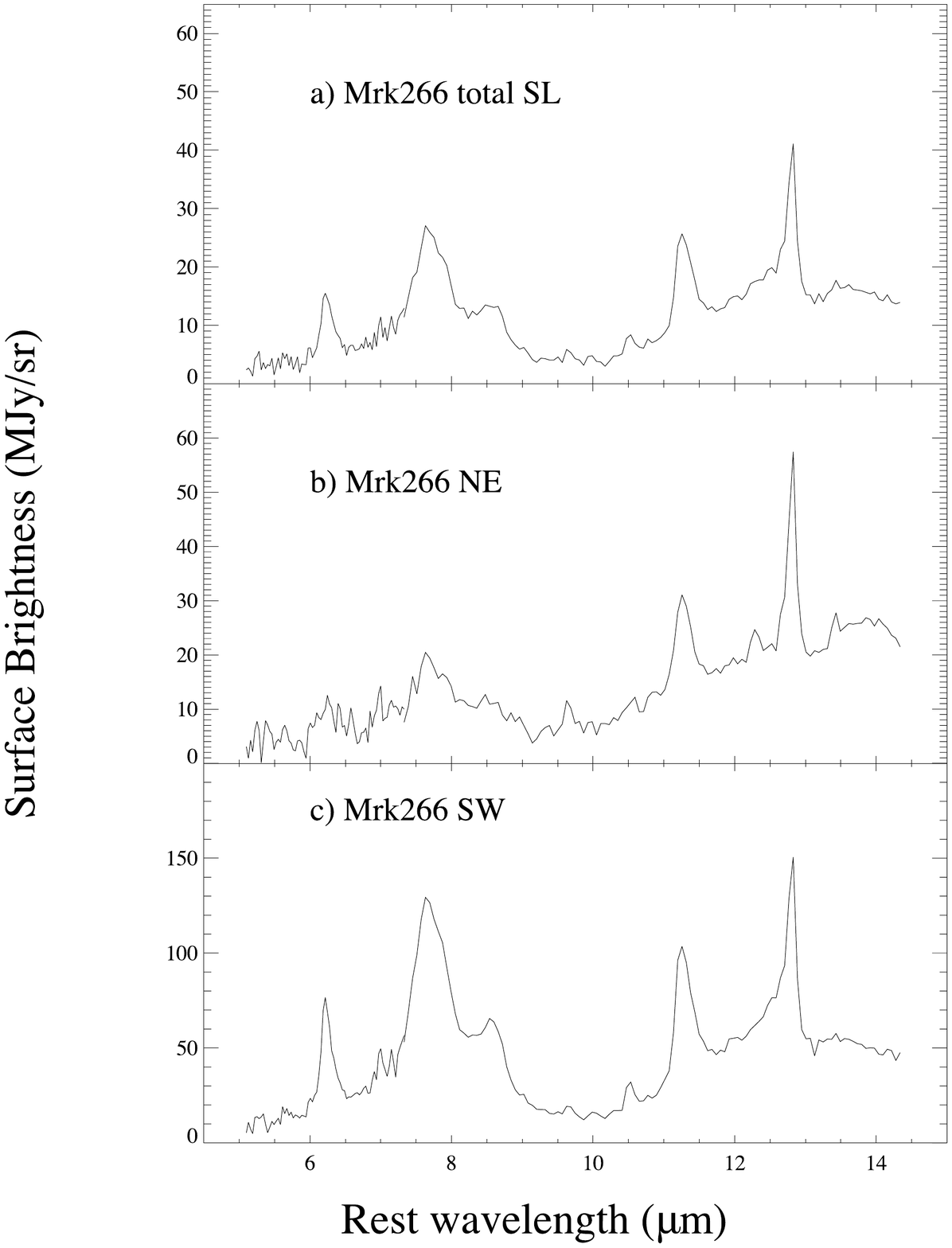}%In 1-column mode, it must be scaled down
\caption[f12]{\figcapIRScubeSL \label{fig:IRScube_SL}}
\end{figure}
\clearpage

\begin{figure}[t]
\center
\includegraphics[width=6.0truein,angle=0]{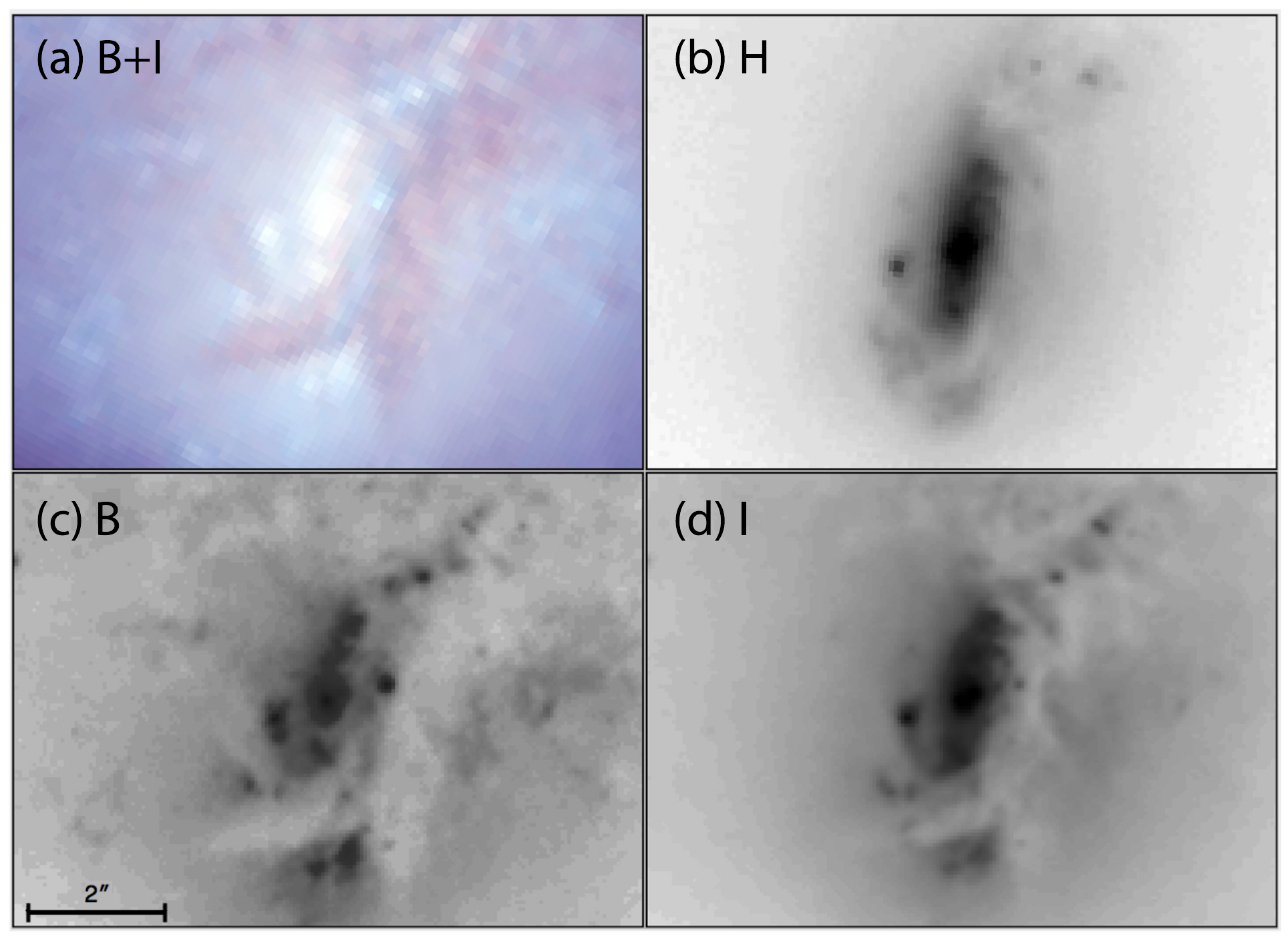} 
\caption[f13]{\figcapSW \label{fig:Mrk266SW}}
\end{figure}

\begin{figure}[t]
\center
\includegraphics[width=6.0truein,angle=0]{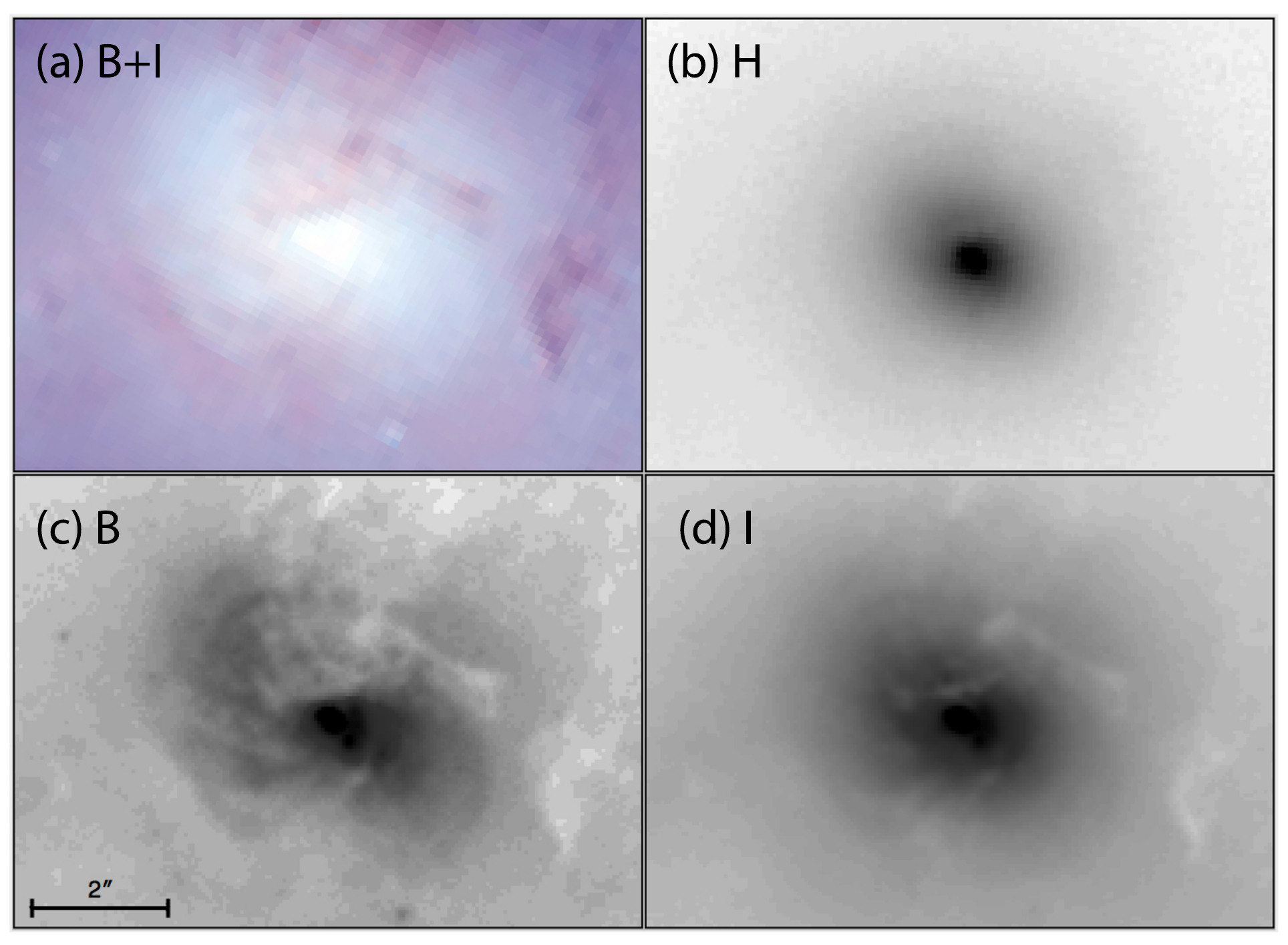} 
\caption[f14]{\figcapNE \label{fig:Mrk266NE}}
\end{figure}
\clearpage

\begin{figure}[t]
\center
\includegraphics[width=6.0truein,angle=0]{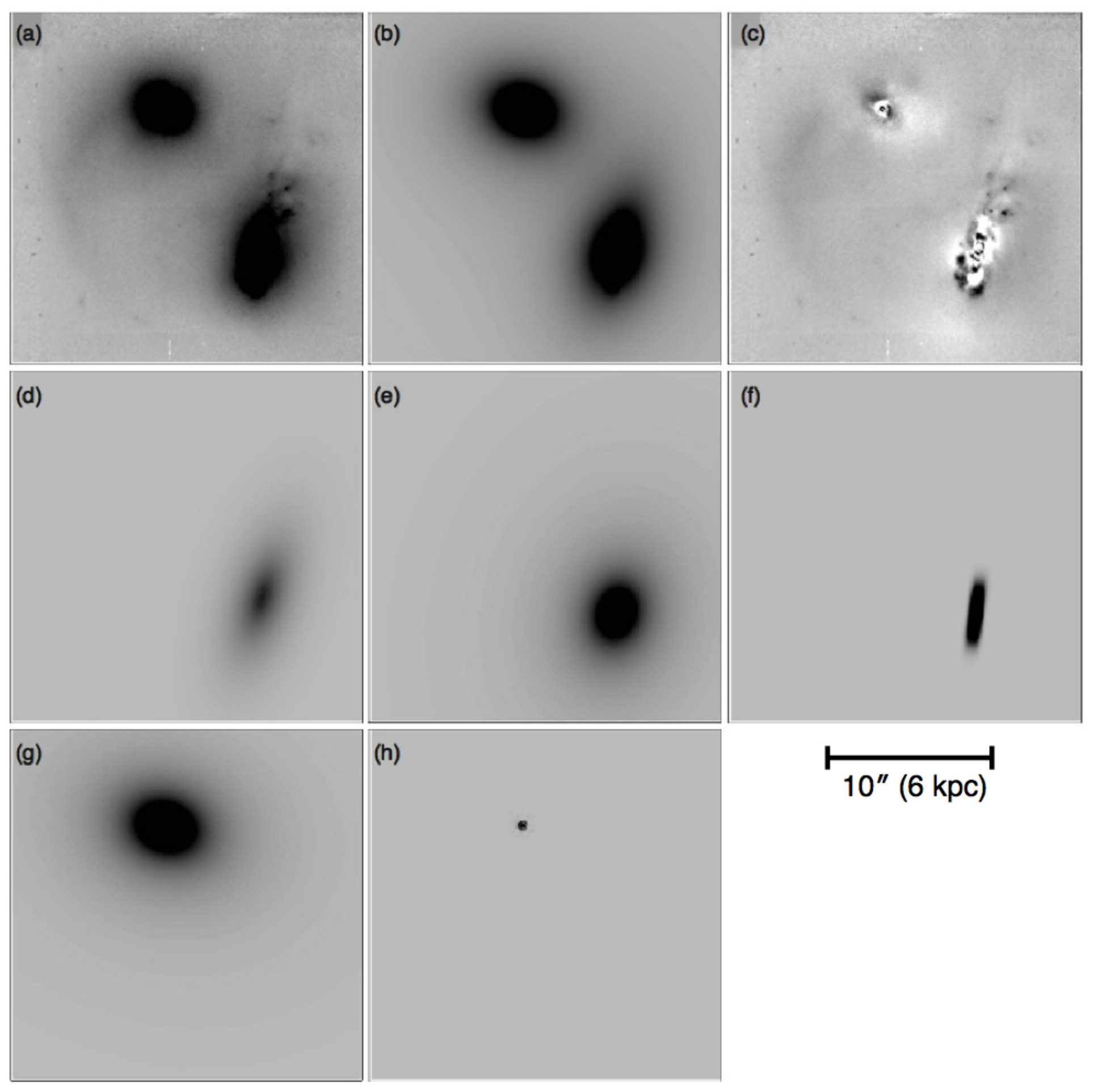} 
\caption[f15]{\figcapGalfitH \label{fig:galfitH}}
\end{figure}
\clearpage

\begin{figure}[t]
\center
\includegraphics[width=4.5truein,angle=0]{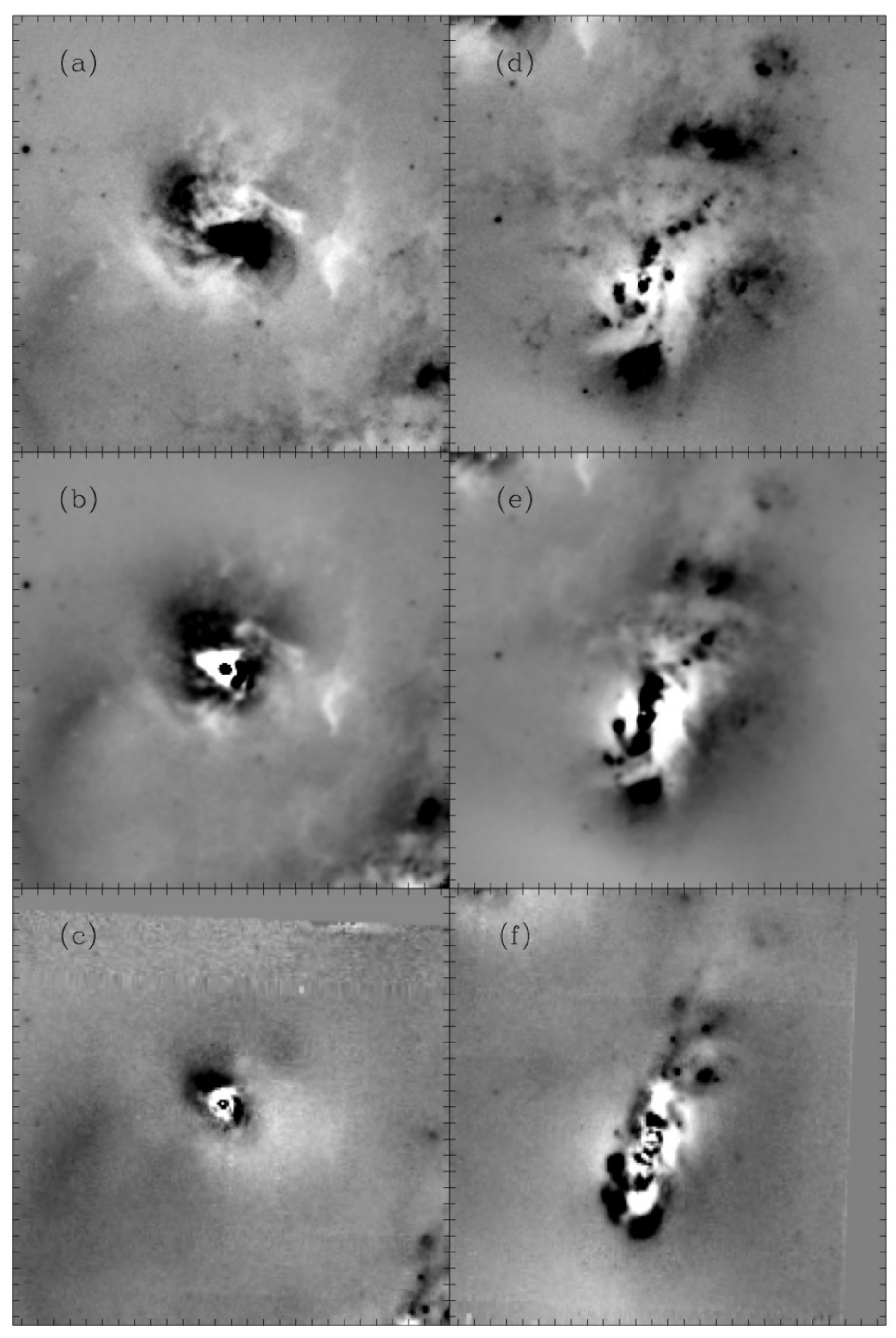} 
\caption[f16]{\figcapGalfitBIHres \label{fig:galfitBIHres}}
\end{figure}
\clearpage

\begin{figure}[t]
\center
\includegraphics[scale=0.45,angle=0]{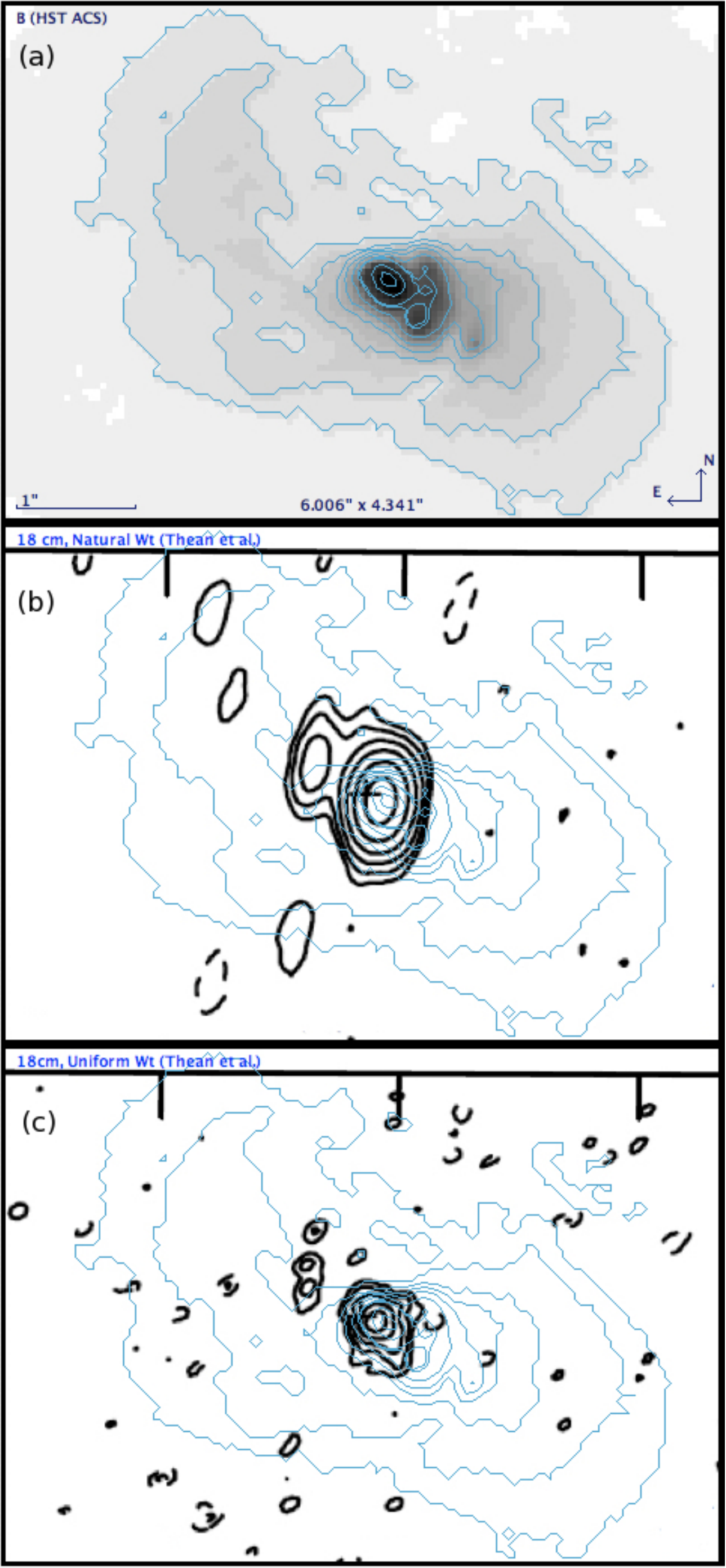} 
\caption[f17]{\figcapNEoutflow \label{fig:NEinBand18cm}}
\end{figure}

\begin{figure}[t]
\includegraphics[width=0.5\columnwidth,angle=0]{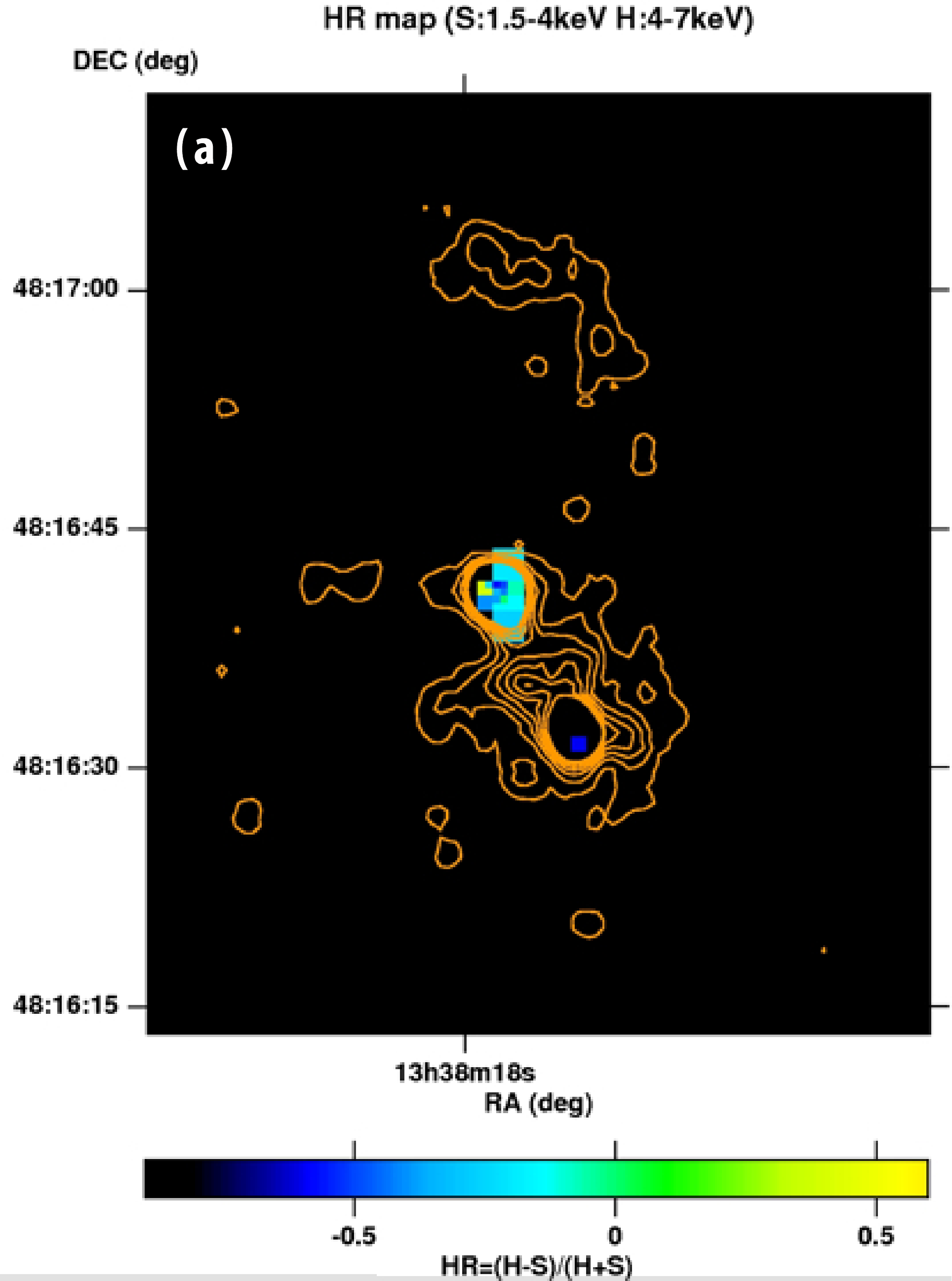}
\includegraphics[width=0.5\columnwidth,angle=0]{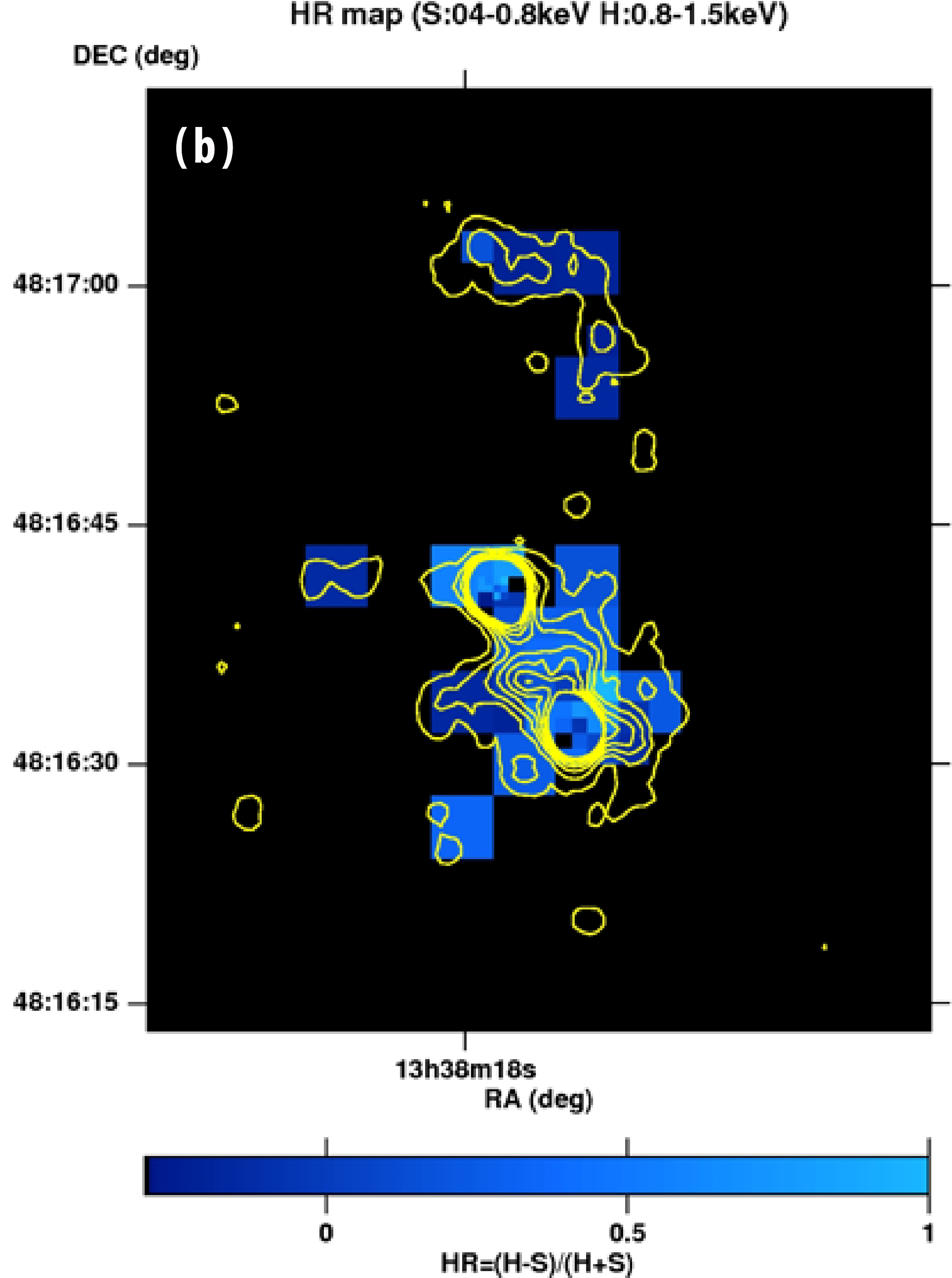}
\caption[f18]{\figcapXrayHardness \label{fig:ChandraHardness}}
\end{figure}
\clearpage

\begin{figure}[t]
\center
\includegraphics[width=6.0truein,angle=0]{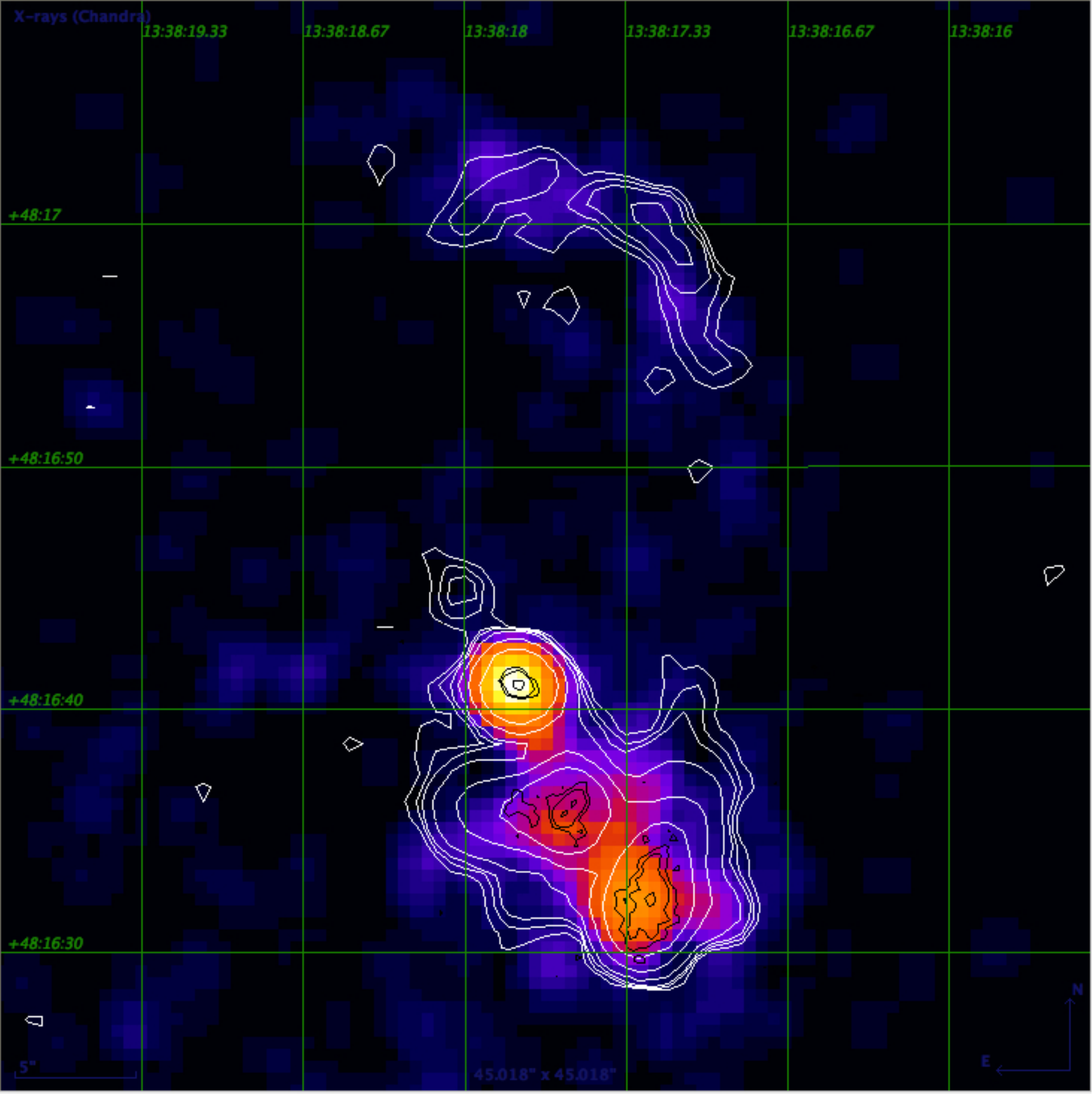} 
\caption[f19]{\figcapChandraVLA \label{fig:ChandraVsRadio}}
\end{figure}
\clearpage

\begin{figure}[t]
\center
\includegraphics[scale=0.5,angle=0]{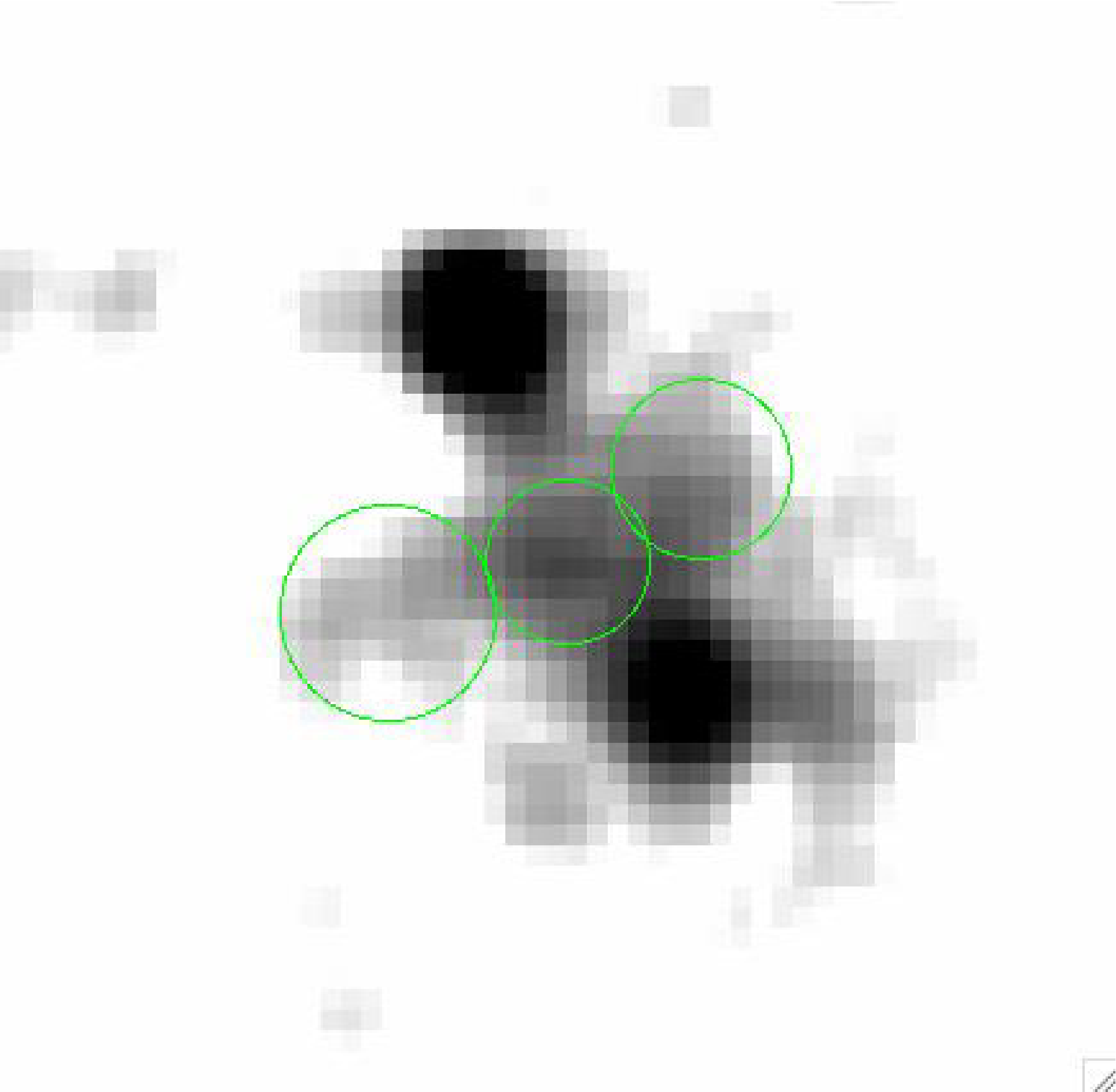}
\caption[f20]{\figcapCenterXrays \label{fig:CenterXrays}}
\end{figure}
\clearpage

\begin{figure}[t]
\center
\includegraphics[width=6.0truein,angle=0]{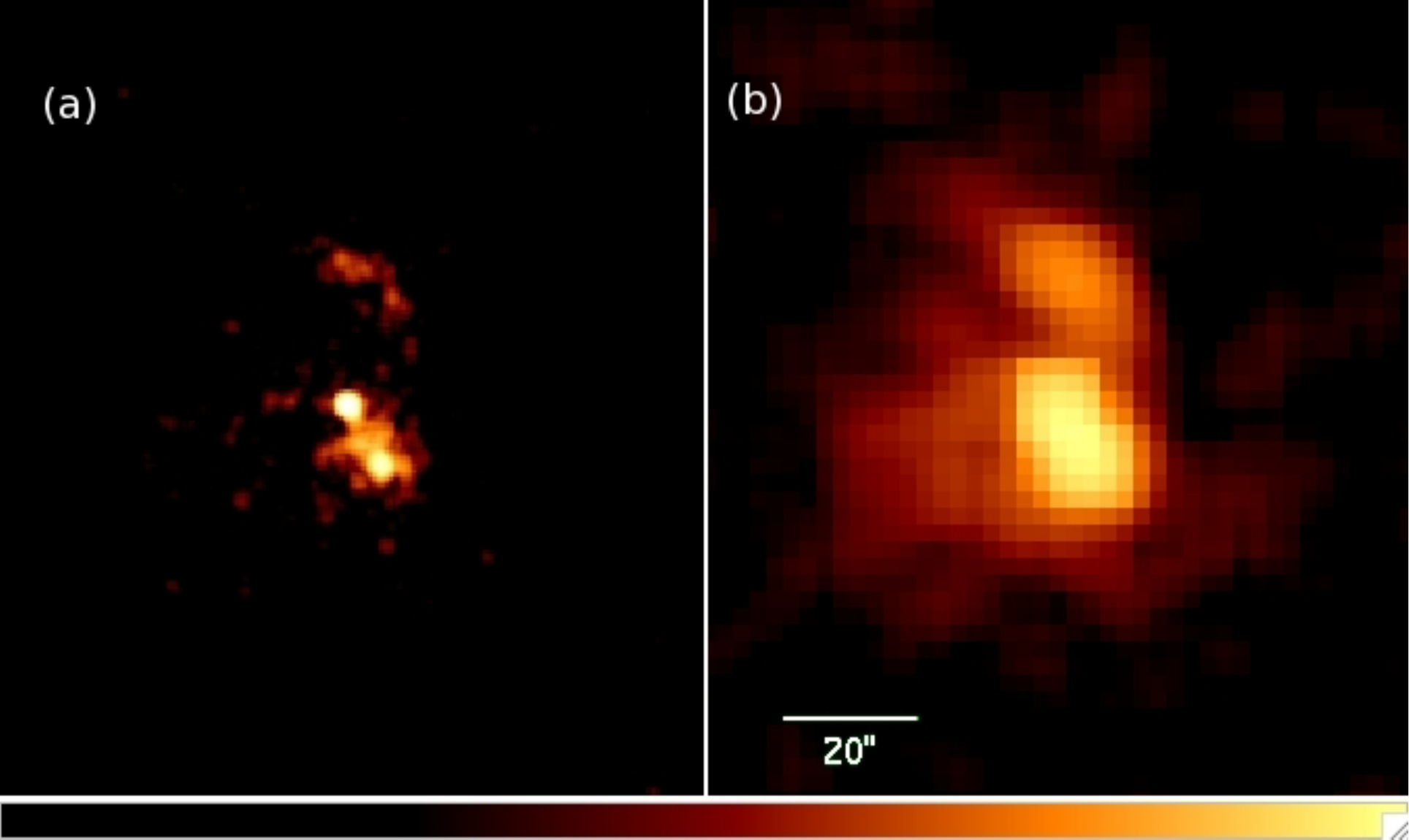}
\caption[f21]{\figcapDiffuseXrays \label{fig:DiffuseXrays}}
\end{figure}
\clearpage

\begin{figure}[t]
\includegraphics[width=6.0truein,angle=0]{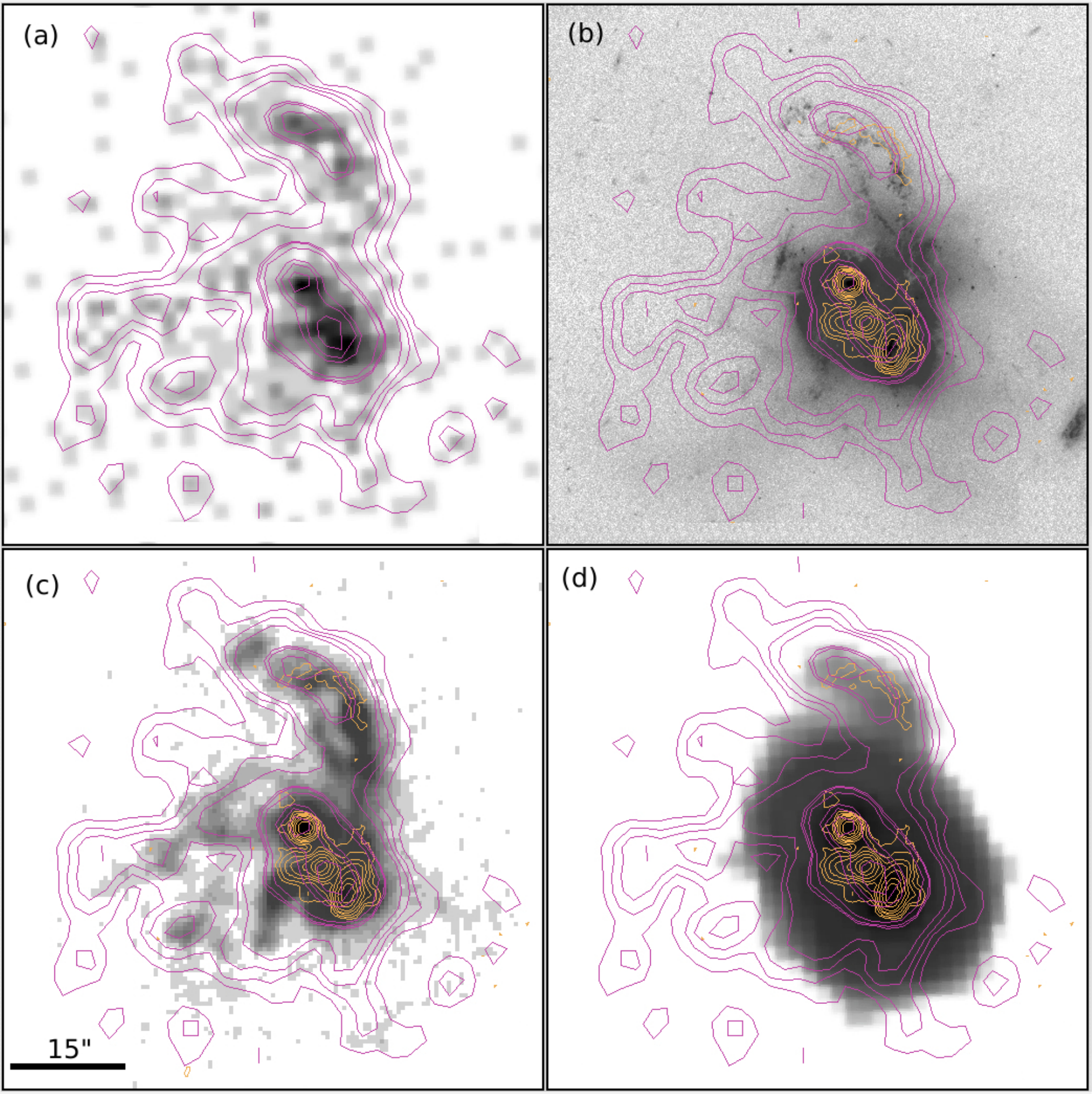}
\caption[f22]{\figcapSoftXrays \label{fig:SmoothXrays}}
\end{figure}
\clearpage

\begin{figure}[t]
\center
\includegraphics[width=5.8truein,angle=0]{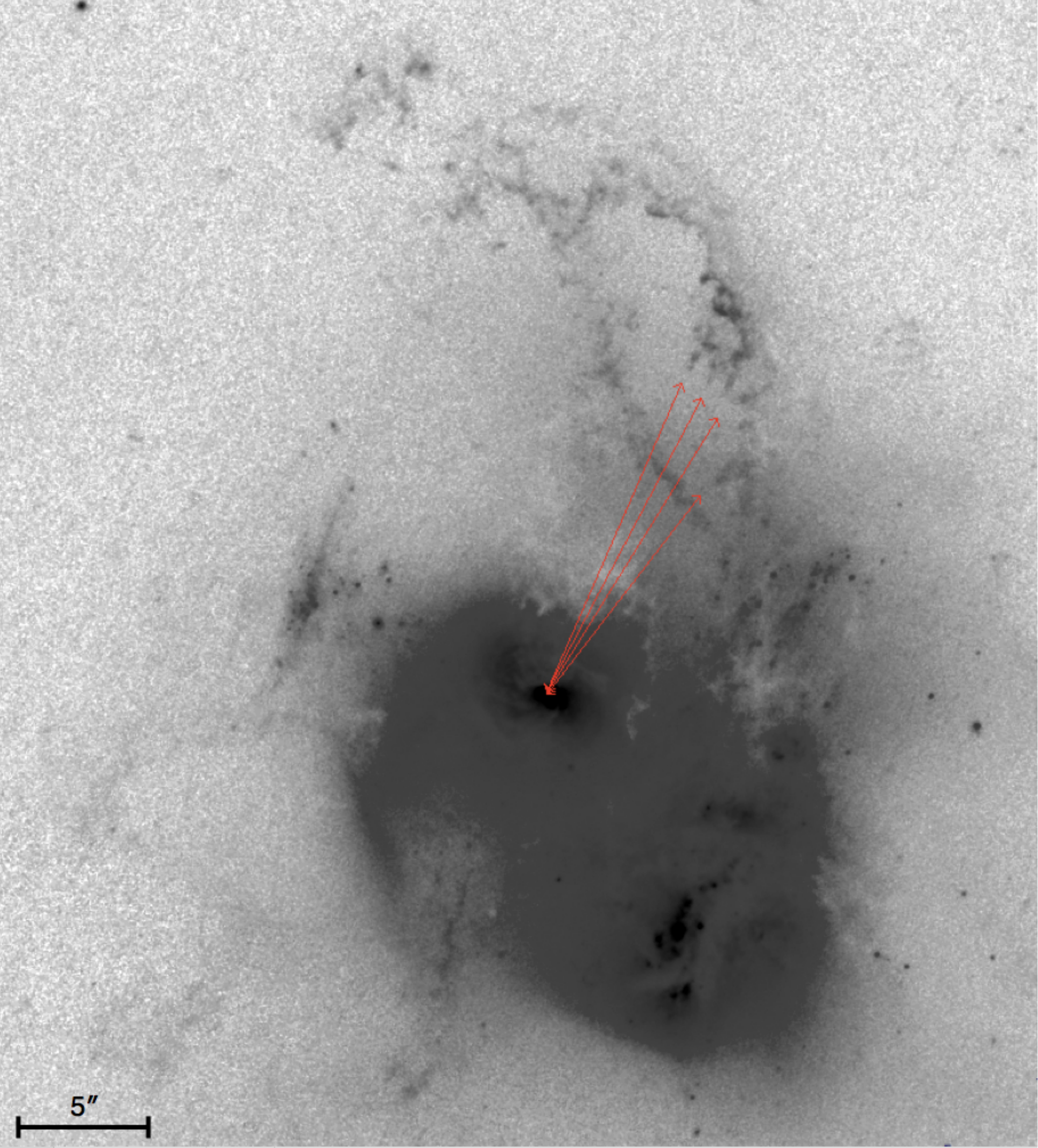} 
\caption[f23]{\figcapLoopZoom \label{fig:NorthernLoopZoom}}
\end{figure}
%\clearpage

\begin{figure}[t]
\center

\includegraphics[width=4.8truein,angle=0]{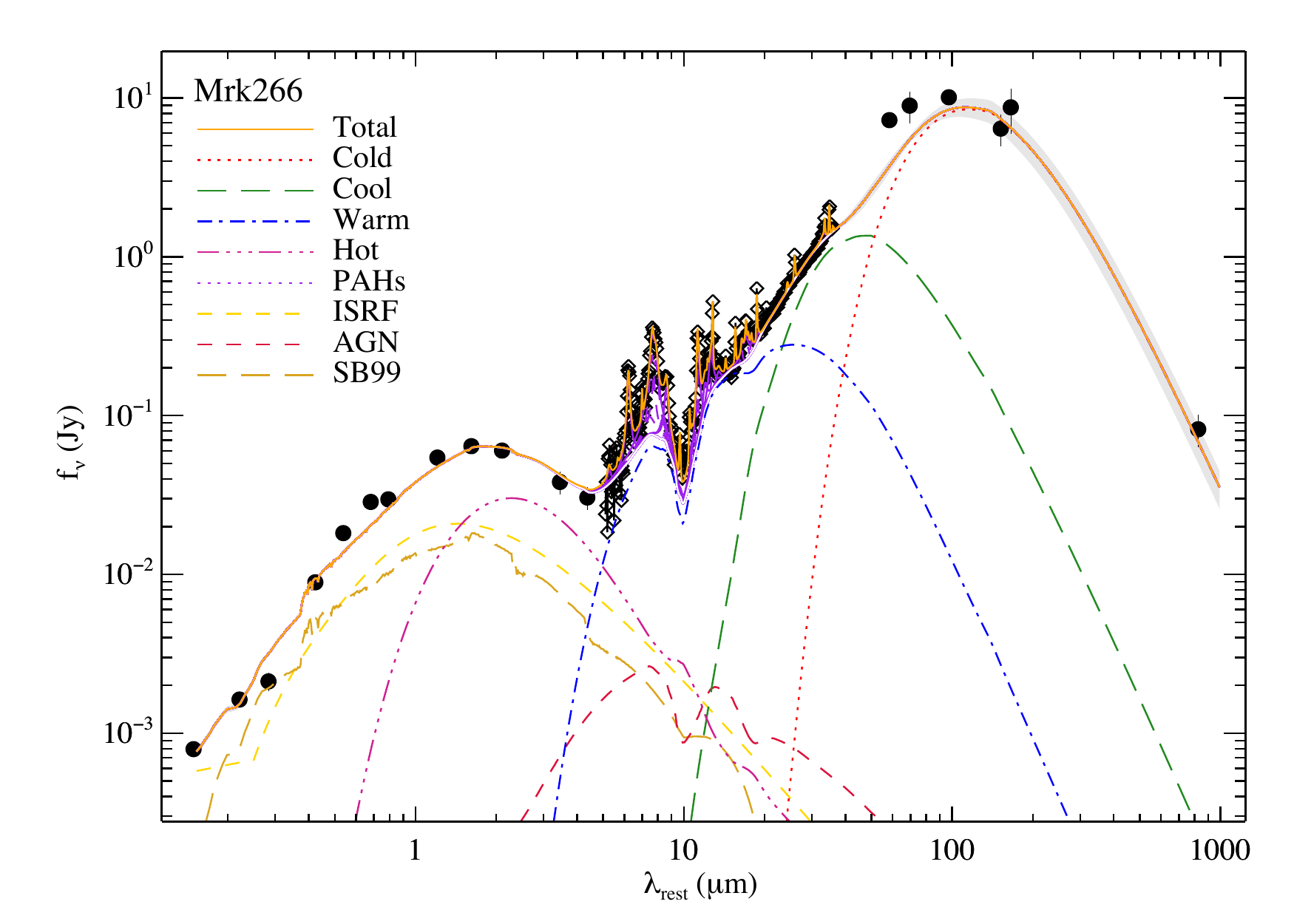}
\caption[f24]{\figcapSEDcafe \label{fig:SEDcafe}}
\end{figure}
\clearpage

\begin{figure}[t]
\center
\includegraphics[width=3.0truein,angle=0]{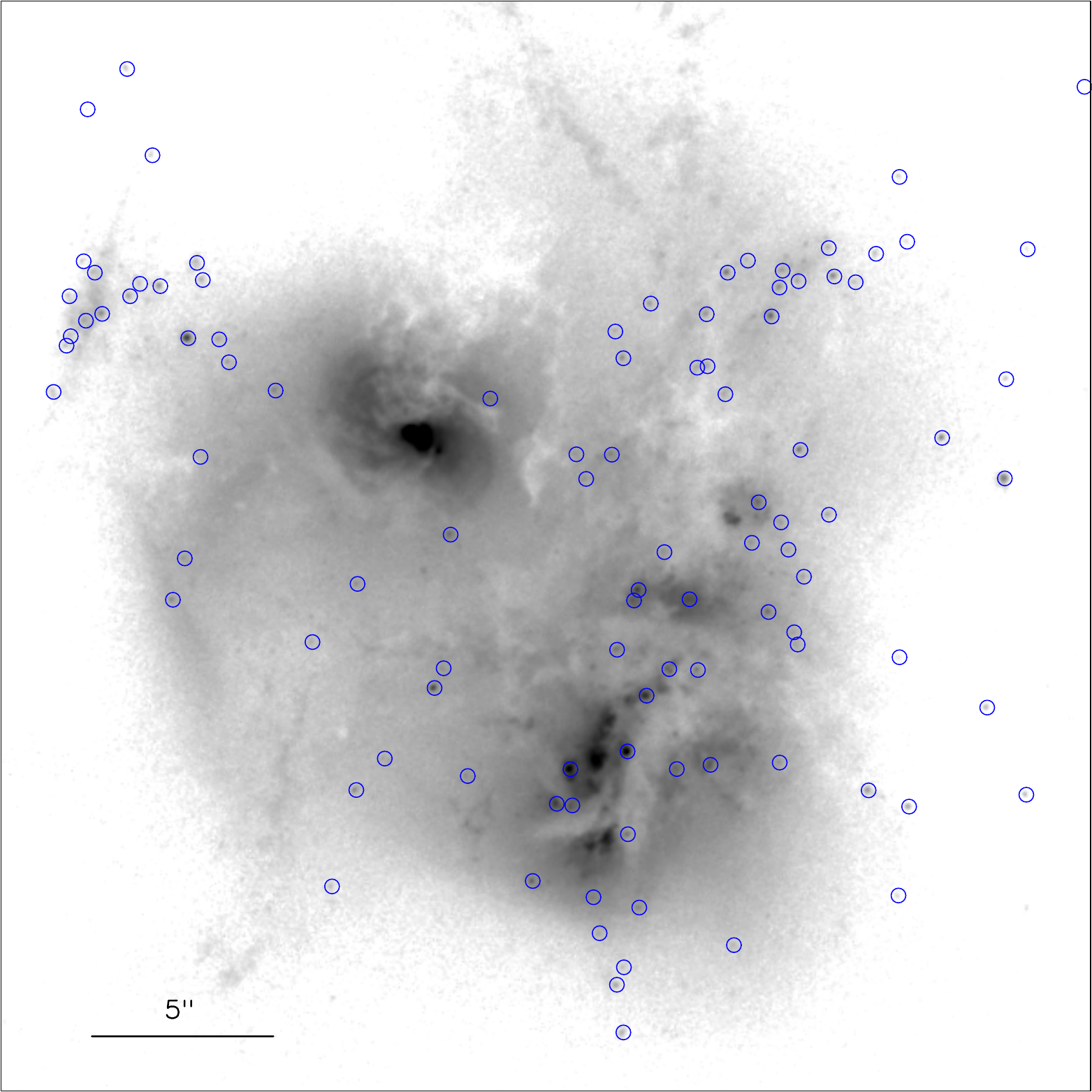}  \put(-20,10) {\parbox{10cm}{(a)}}
\hskip 0.05in
\includegraphics[width=3.0truein,angle=0]{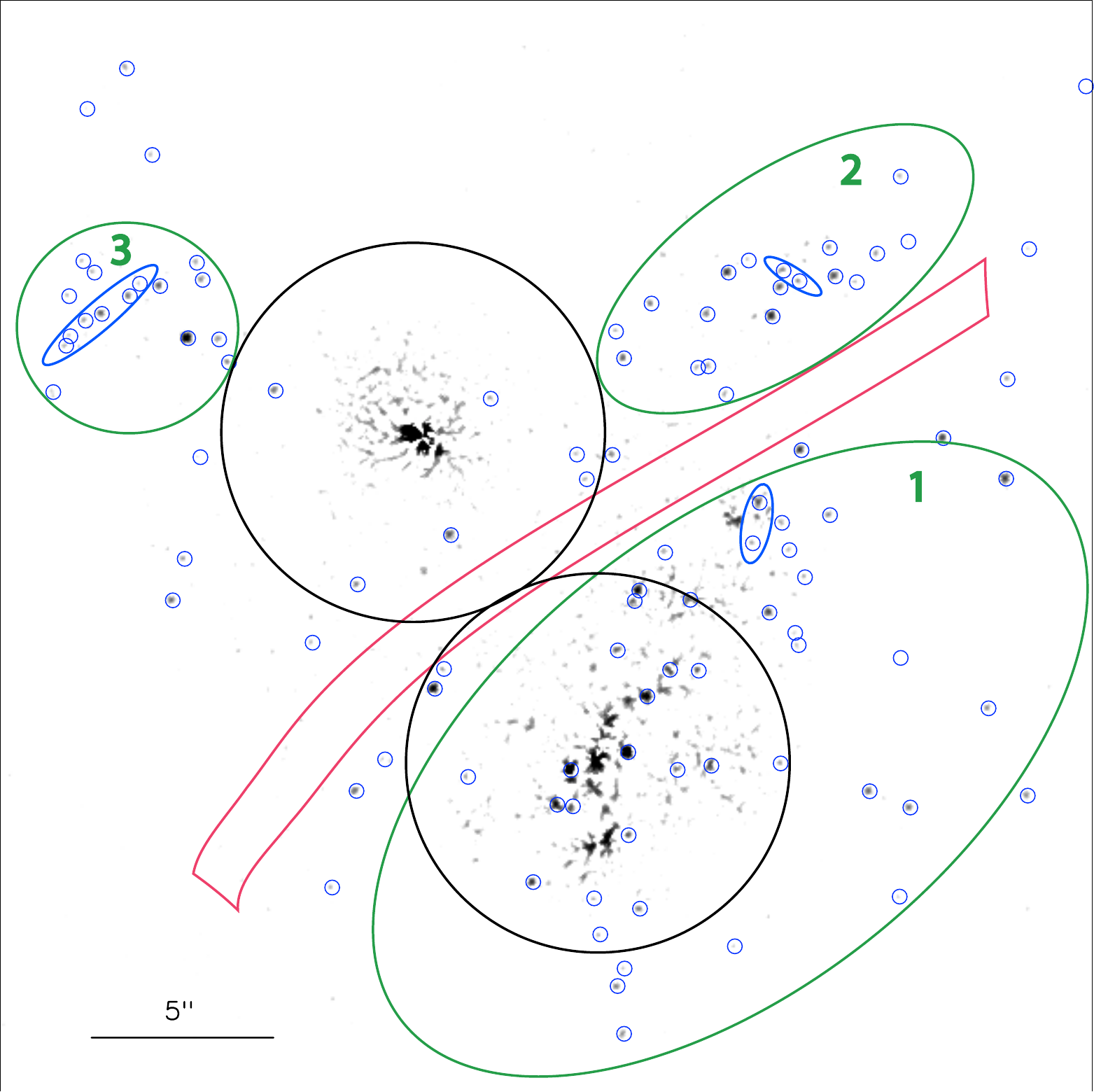} \put(-20,10) {\parbox{10cm}{(b)}} \\
\caption[f25]{\figcapClustersBband \label{fig:ClustersBband}}
\end{figure}
\clearpage

\begin{figure}[t]
\center
\includegraphics[width=6.0truein,angle=0]{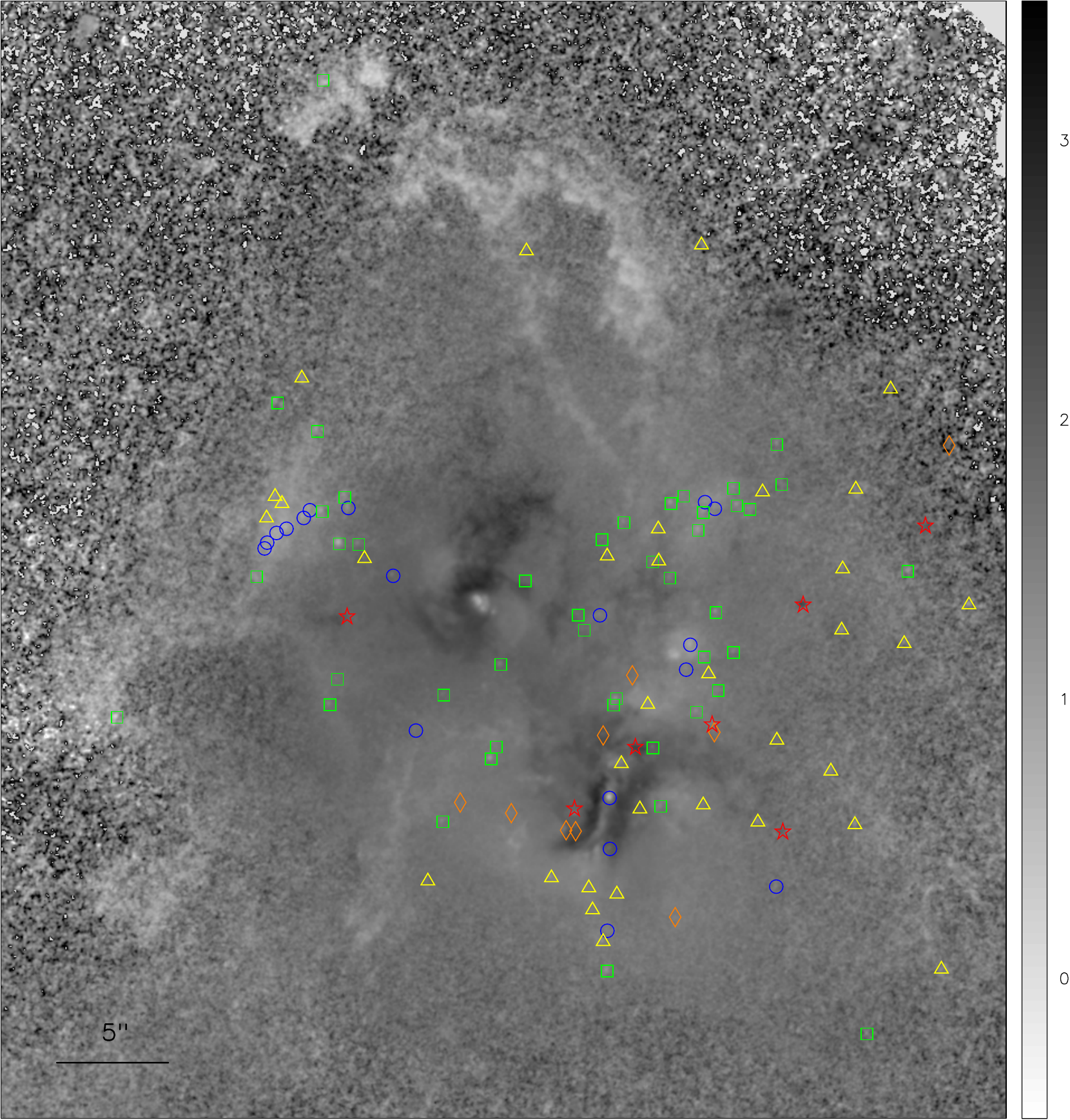} 
\caption[f26]{\figcapBminusI \label{fig:BminusI}}
\end{figure}
\clearpage

\begin{figure}[t]
\center
\includegraphics[scale=0.7,angle=-90]{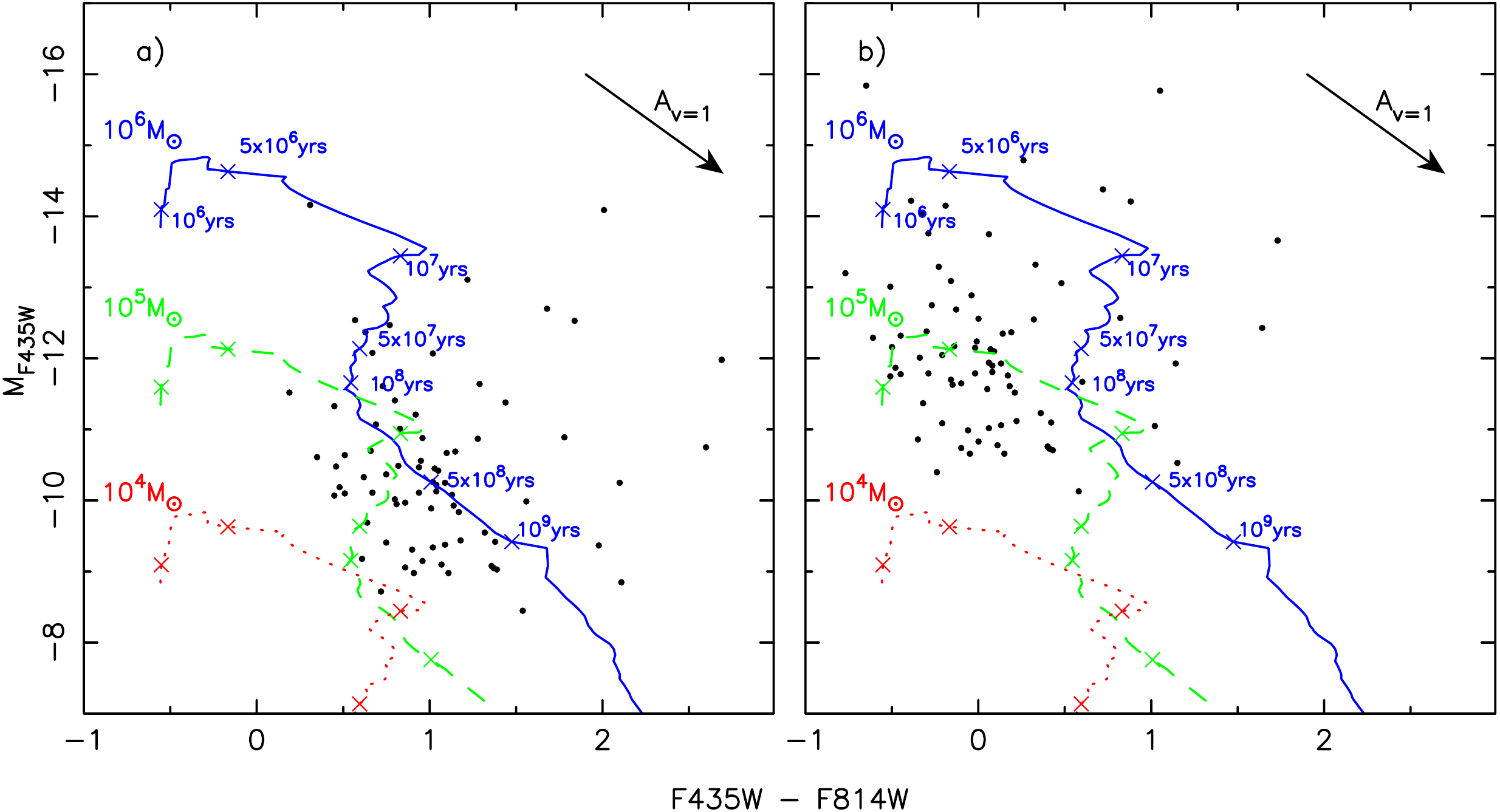}
\caption[f27]{\figcapClustersColorMag \label{fig:ClustersColorMag}}
\end{figure}
\clearpage

\begin{figure}[t]
\center
\includegraphics[scale=0.7,angle=-90]{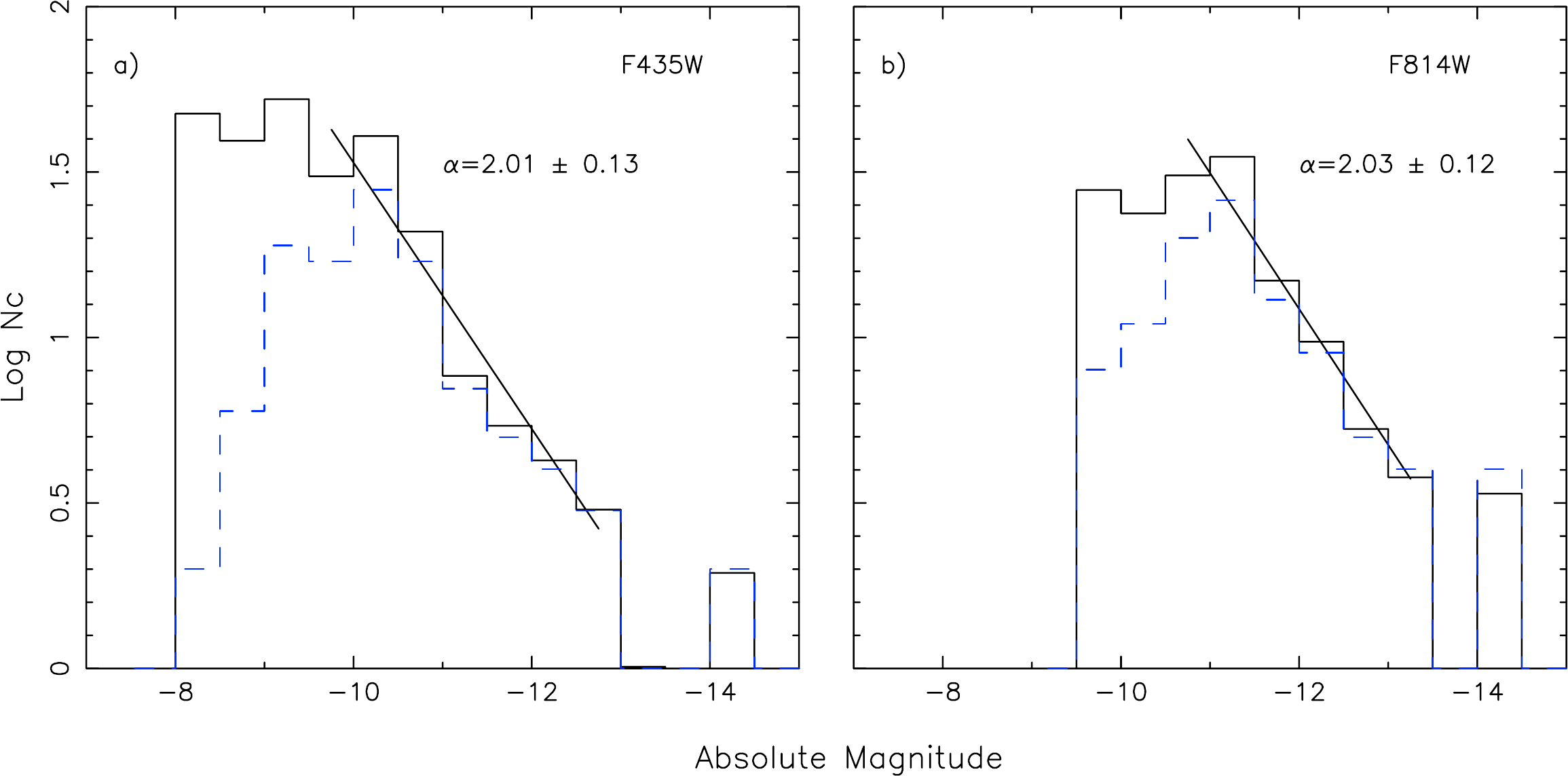}
\caption[f28]{\figcapClusterLumFuncs \label{fig:SCLumFuncs}}
\end{figure}
\clearpage

\begin{figure}[t]
\center
\includegraphics[width=6.0truein,angle=0]{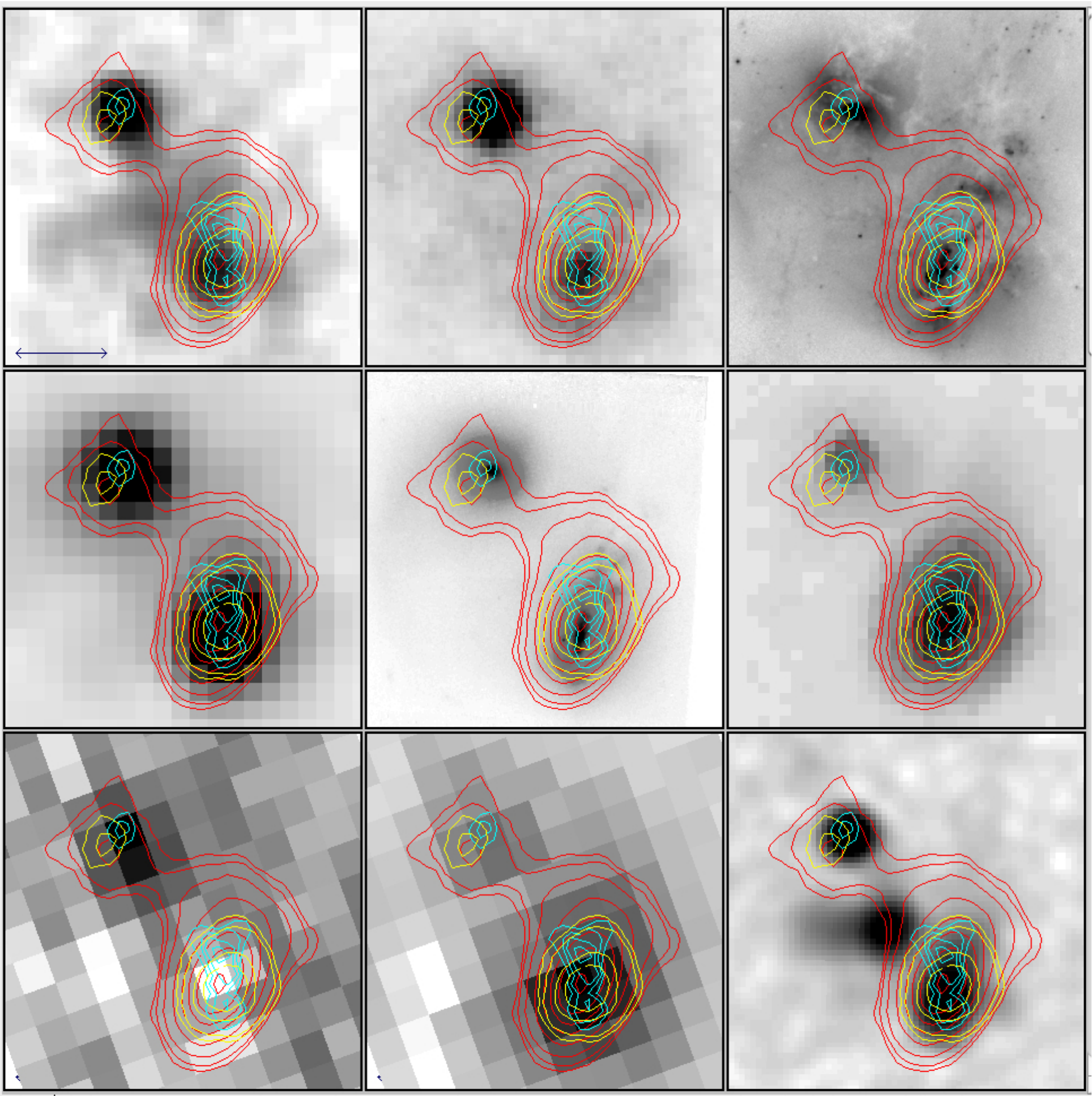}
\scriptsize
\put(-425,422) {\parbox{10cm}{(a) 0.4-0.7 keV (Chandra)}}
\put(-282,422) {\parbox{10cm}{(b) U (XMM-OM)}}
\put(-140,422) {\parbox{10cm}{(c) 0.44 \micron\ (HST ACS)}}
\put(-425,278) {\parbox{10cm}{(d) 1.6 \micron\ (2MASS H)}}
\put(-282,278) {\parbox{10cm}{(e) 1.6 \micron\ (HST NICMOS)}}
\put(-140,278) {\parbox{10cm}{(f) 8  \micron\ (Spitzer IRAC)}}
\put(-425,133) {\parbox{10cm}{(g) S(3) H2 9.7  \micron\ (Spitzer IRS)}}
\put(-282,133) {\parbox{10cm}{(h) 11.3  \micron\ PAH (Spitzer IRS)}}
\put(-140,133) {\parbox{10cm}{(i) 20 cm (VLA)}}
\put(-410,300) {\parbox{10cm}{5\arcsec}}
\caption[f29]{\figcapMolGasA \label{fig:MolGasA}}
\end{figure}
\clearpage

\begin{figure}[t]
\center
\includegraphics[width=6.0truein,angle=0]{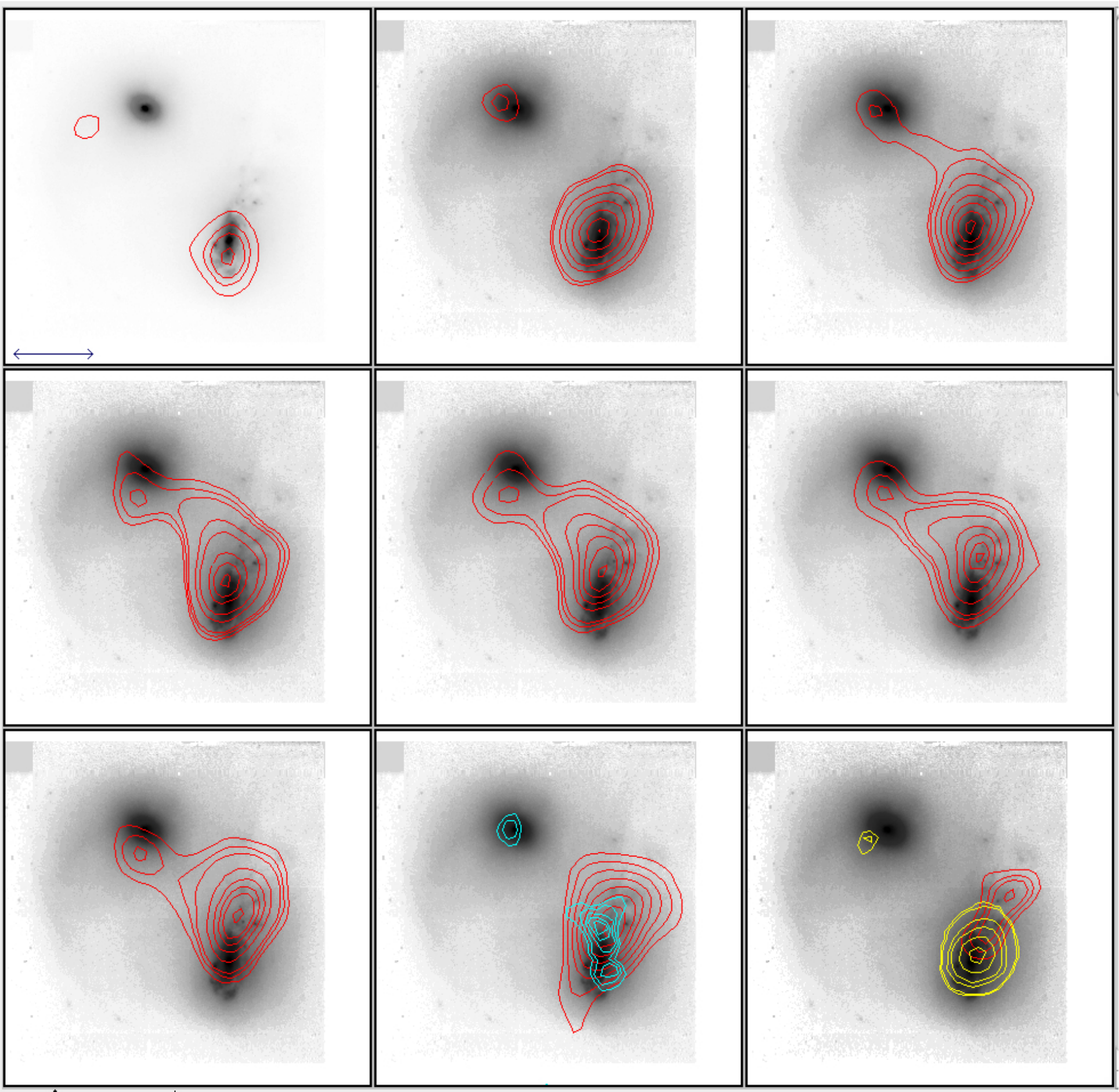} 
\footnotesize
\put(-426,409) {\parbox{10cm}{(a) 7850 km/s (112.25 GHz)}}
\put(-283,409) {\parbox{10cm}{(b) 8097 km/s (112.16 GHz)}}
\put(-140,409) {\parbox{10cm}{(c) 8152 km/s (112.14 GHz)}}
\put(-426,269) {\parbox{10cm}{(d) 8207 km/s (112.12 GHz)}}
\put(-283,269) {\parbox{10cm}{(e) 8235 km/s (112.11 GHz)}}
\put(-140,269) {\parbox{10cm}{(f) 8290 km/s (112.09 GHz)}}
\put(-426,129) {\parbox{10cm}{(g) 8317 km/s (112.08 GHz)}}
\put(-283,129) {\parbox{10cm}{(h) 8427 km/s (112.04 GHz)}}
\put(-140,129) {\parbox{10cm}{(i) 8455 km/s (112.03 GHz)}}
\put(-415,292) {\parbox{10cm}{5\arcsec}}
\caption[f30]{\figcapMolGasB \label{fig:MolGasB}}
\end{figure}
\clearpage

%APPENDIX Fig.
\begin{figure}[t]
\center
\includegraphics[width=6.0truein,angle=0]{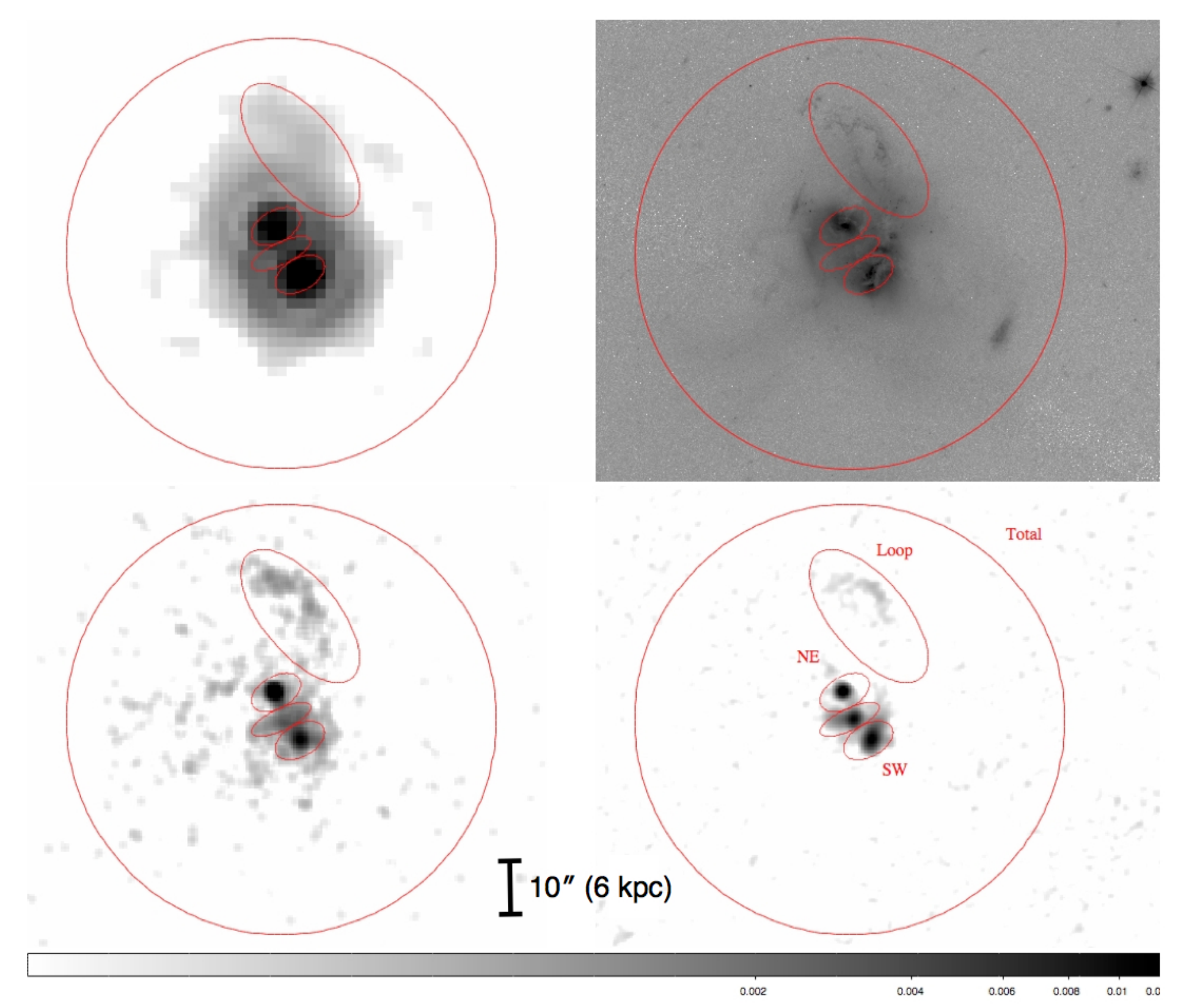} 
\caption[f31]{\figcapRegions \label{fig:Regions}}
\end{figure}
\clearpage

%%%%%%%%%%%%%%%%%%%%%%%%%%%%%%%
%Figures - In [manuscript] mode figures cannot be
%          embedded, but get included in tarball as 
%          separate PS files.
%%%%%%%%%%%%%%%%%%%%%%%%%%%%%%%
%Tables - In [manuscript] mode the LaTeX code must 
%be inserted here, not with \include or \includegraphics.
%NOTE: However, if the LaTeX is embedded in the manuscript, 
%and does not format with correct line spacing and font size,
%format it separately.

\centerline{TABLES}
\label{sec:tables}

\center
%%% Table 1 %%%
\tableXrayRegions
\clearpage

%%% Table 2 %%%
\tableXrayData
\clearpage

%%% Table 3 %%%
\tableIRScube
\clearpage

%%% Table 4 %%%
\tableGALFIT
\clearpage

%%% Table 5 %%%
\tableAGNfractions
\clearpage

%%% Table 6 %%%
\tableBowKnots
\clearpage

%%% Table 7 %%%
\tableCenterXrays
\clearpage

%%% Table 8 %%%
\tableSEDfitting
\clearpage

%%% Table 9 %%%
\tableMolGas
\clearpage

%%% Table 10 %%%
\tableDualAGNs
\clearpage

%%% Table 11 %%%
\tableRegions
\clearpage

%%% Table 12 %%%
\tableSEDtotal
\clearpage

%%% Table 13 %%%
\tableSEDregions
\clearpage

%%% Table 14 %%%
\tableIRSpublished
\clearpage

%%% Table 15 %%%
\tableSCdata
\clearpage

\end{document}